\documentclass{optica-article}

\journal{opticajournal}
\articletype{Review Article}
\usepackage{wrapfig}
\usepackage{lineno}
\linenumbers
\usepackage{braket} 
\usepackage{array}  % added by Jehyung for tables
\usepackage{graphicx} % added by Jehyung for tables
\def\ketbra#1#2{{\vert#1\rangle\!\langle#2\vert}} 
\usepackage{threeparttable}% added by Tobi, enables footnotes in tables via \tnote{}

\nolinenumbers

\begin{document}

\title{Quantum dots for photonic quantum information technology}

\author{Tobias Heindel,\authormark{1} Je-Hyung Kim,\authormark{2} Niels Gregersen,\authormark{3} Armando Rastelli,\authormark{4} and Stephan Reitzenstein\authormark{1,*}}

\address{\authormark{1} Institut für Festkörperphysik, Technische Universität Berlin, Hardenbergstraße 36, 10623 Berlin, Germany\\
\authormark{2}Department of Physics, Ulsan National Institute of Science and Technology (UNIST), Ulsan, 44919 Republic of Korea\\
\authormark{3}DTU Electro, Department of Electrical and Photonics Engineering, Technical University of Denmark, Ørsteds Plads, Building 343, DK-2800 Kongens Lyngby, Denmark\\
\authormark{4}Institute of Semiconductor and Solid State Physics,
Johannes Kepler University Linz, Altenbergerstr. 69, 4040 Linz, Austria}

\email{\authormark{*}stephan.reitzenstein@physik.tu-berlin.de} %% email address is required; see note below about the corresponding author designation



\begin{abstract}
The generation, manipulation, storage, and detection of single photons play a central role in emerging photonic quantum information technology. Individual photons serve as flying qubits and transmit the relevant quantum information at high speed and with low losses, for example between individual nodes of quantum networks. Due to the laws of quantum mechanics, the associated quantum communication is fundamentally tap-proof, which explains the enormous interest in this modern information technology. On the other hand, stationary qubits or photonic states in quantum computers can potentially lead to enormous increases in performance through parallel data processing, to outperform classical computers in specific tasks when quantum advantage is achieved. In this review, we discuss in depth the great potential of semiconductor quantum dots in photonic quantum information technology. In this context, quantum dots form a key resource for the implementation of quantum communication networks and photonic quantum computers because they can generate single photons on-demand. Moreover, these solid-state quantum emitters are compatible with the mature semiconductor technology, so that they can be integrated comparatively easily into nanophotonic structures such as resonators and waveguide systems, which form the basis for quantum light sources and integrated photonic quantum circuits. After a thematic introduction, we present modern numerical methods and theoretical approaches to device design and the physical description of quantum dot devices. We then present modern methods and technical solutions for the epitaxial growth and for the deterministic nanoprocessing of quantum devices based on semiconductor quantum dots. Furthermore, we present the most promising concepts for quantum light sources and photonic quantum circuits that include single quantum dots as active elements and discuss applications of these novel devices in photonic quantum information technology. We close with an overview of open issues and an outlook on future developments.
\end{abstract}

\tableofcontents

%%%%%%%%%%%%%%%%%%%%%%%%%%  body  %%%%%%%%%%%%%%%%%%%%%%%%%%
\section{Introduction }\label{sec:intro}
%ToDo Stephan\\

%\emph{This section gives an introduction to the topic. It starts with a general background and motivation, describes the need of quantum light sources, explains the general properties of quantum dots (QDs), and introduces the most important parameters of QD devices for quantum information technology}\\

%Section content:
%\begin{itemize}
%    \item general motivation
%    \item envisioned applications in photonic quantum technologies
%    \item QD device concepts and requirements / challenges
%    \item overview of the key QD parameters
%    \item overview of the content of the article 
%\end{itemize}

Ever since Richard Feynman's famous proposal 40 years ago to use quantum physics to build computers with ultimate performance, scientists worldwide have been fascinated by this prospect~\cite{Feynman1982,Preskill2021}. For a long time, the development of corresponding concepts was mainly in the area of theory and basic research~\cite{Deutsch1985,Shor1999}, but in recent years there have been breathtaking advances in application-oriented quantum technology. Quantum computers are no longer just the dream of many scientists, but are now being further developed by global players on an almost industrial scale~\cite{Arute2019,Gibney2020,Petar2021}. The latest generations have even achieved the quantum advantage~\cite{Harrow2017} for special problems such as (Gaussian) boson sampling with 100 photonic inputs~\cite{Zhong2020} and using 53 qubits to sample the output of a pseudo-random quantum circuit~\cite{Arute2019}. Applications in the field of quantum communication are currently evolving with a similar dynamic. Interestingly, the development of quantum cryptography was mainly triggered by the prospect of implementing Shor's quantum algorithm for efficient prime factorization of large numbers in a quantum computer~\cite{Deutsch1985}, which makes classic encryption methods vulnerable. Simple point-to-point quantum communication systems are already well-established and commercially available~\cite{Lo2014}, and more complex 
quantum networks are emerging worldwide~\cite{Sasaki2011, Stucki2011, Wang2014,Dynes2019}. Even satellite-based quantum links have already been established that allow quantum cryptography over distances of more than 1200 km~\cite{Yin2017}.

A key resource of all photonic quantum technology systems are single photons. As flying qubits, they serve as carriers of quantum information. For example, in the famous BB84 quantum key distribution (QKD) protocol~\cite{Bennett2014}, the polarization degree of freedom is used to encode the information about the secret key exchanged between the sender (Alice) and the receiver (Bob). The same applies to more complex QKD concepts in the field of long-distance quantum communication. For example, measurement-device-independent QKD (MDI-QKD) relies on indistinguishable photons for quantum data exchange~\cite{Lo2012}, and the quantum repeater concept uses entanglement distribution across nodes of a network 
to extend the communication distance and the data rate compared to simple point-to-point QKD protocols~\cite{Briegel1998,Loock2020}. Similarly, single-photon states that are purposefully manipulated, stored, and detected are the basis of photonic quantum computers~\cite{Madsen2022}. Photonic cluster states could be of particular importance in the future, especially 2D cluster states that are largely immune to decoherence and can pave the way to powerful fault-tolerant photonics quantum computers~\cite{Raussendorf2001}.

Against this background, it is clear that sources of single photons and entangled photon pairs are central building blocks of applications in photonic quantum technology. Ideally, they deliver the photons on-demand at the desired wavelength. In current applications, however, probabilistic photon sources are mostly used that do not emit photons in a deterministic manner. An example is represented by heavily attenuated lasers for BB84-like QKD using decay-state protocols~\cite{Lo2005}, where the mean number of photons per pulse is below one. Due to the underlying classic photon statistics, however, each pulse contains a number of photons that is distributed according to the Poisson distribution and, in addition to individual photons, often also contains the vacuum state (no photon) or several photons. Photon sources based on non-linear emission processes such as parametric down-conversion behave similarly~\cite{Burnham1970} and are also frequently applied in quantum communication and quantum computing settings~\cite{Gisin2002,OBrien2007}. These are very attractive sources of entangled photon pairs, but the number of pairs per pulse is also affected by statistical fluctuations. Quasi-deterministic operation can only be achieved through comparatively complex heralding~\cite{Barz2010}. In both cases, it is of great practical advantage that the sources (laser or parametric down-conversion) can be operated at room temperature.

In contrast, there is the class of non-classical light sources that can emit individual photons and entangled photon pairs deterministically, i.e. at the ``push of a button''. Such quantum emitters typically have an extension in all space dimensions in the range of or smaller than the de Broglie wavelength of the enclosed charge carriers, leading to discrete electronic energy levels so that one and only one photon is emitted under suitable experimental conditions in the usually radiative recombination process~\cite{Gerard99,Michler2000}. Biexcitons consisting of two bound electrons in the conduction band and two bound holes in the valence band are of particular interest in the context of this article because they can generate entangled photon pairs~\cite{Benson2000,Akopian2006,Young2006} (see Section~\ref{sec:entangledpair} for details).

Compared to other quantum emitters such as nitrogen vacancy centers in diamond~\cite{Kurtsiefer2000}, defect centers in SiC~\cite{Castelletto2020} and in 2D transition metal dichalcogenides~\cite{Tonndorf2015}, quantum dots (QDs) have the enormous advantage that their material basis makes them compatible with common processes and technical solutions in modern III/V optoelectronics. For example, sophisticated epitaxial processes are used to growth high-quality semiconductor heterostructures with QDs as the active medium~\cite{Pohl2013}. Furthermore, the emission wavelength of the QDs can cover in a wide spectral range from about 300 nm to beyond 1.55 $\mu$m by a suitable choice of the materials and by strain engineering~\cite{Arakawa2020}. In particular, this wavelength range includes the telecom O-band at 1.3~$\mu$m and the C-band at 1.55~$\mu$m, which are of crucial importance for fiber-based quantum communication~\cite{Vajner2022}.

Due to its discrete electronic energy levels, a QD already represents an almost ideal 2-level system or, in the case of the biexciton (XX) cascade, a 3-level or 4-level system, so that it can act as a source of single photons or entangled photon pairs. The foundation for discovering and studying these exciting properties was laid in the 1990s and early 2000s when single-QD spectroscopy was developed. Early work on single QD properties includes the first studies on isolated GaAs QDs~\cite{Brunner1994} and Stranski-Krastanov (S-K) QDs~\cite{Marzin1994} and the observation of some Coulomb effects of particles~\cite{Landin1998}. In addition, important electronic features such as the splitting of the excitonic fine structure~\cite{Bayer2002} and hidden symmetries in the QD energy levels~\cite{Bayer2000} had been identified and studied in detail. Going beyond such fundamental investigations, a number of challenges need to be addressed in order to be able to use QDs as quantum light sources (QLSs)\footnote{We refer to quantum light sources in the general context of non-classical light sources that emit single photons, entangled photon pairs, or photonic cluster states. In contrast, we define the subset of sources that emit single photons as single-photon sources.} in photonic quantum information technology, as discussed for instance also in a recent review article by X. Zhou et al.~\cite{Zhou2022}. On the one hand, photonic structures are required that direct the emitted photons in the intended direction with high extraction efficiency~\cite{Senellart2017}, so that they can be coupled directly into a glass fiber for applications in quantum communication, for example. Similarly, in the field of integrated quantum photonics, the photons have to be guided into integrated waveguide systems with high efficiency~\cite{Rodt2021, Moody2022}. In order to take these requirements into account, various concepts for the efficient light extraction and transmission of photons from QDs have been developed. These include, for example, micropillar cavities~\cite{Gerard1998, Moreau2001, Somaschi2016, Ding2016, Schlehahn2016}, circular Bragg gratings (CBGs)~\cite{Sapienza2015, Liu2019,Wang2019}, photonic wires~\cite{Claudon2010} and microlenses~\cite{Gschrey2015} for photon extraction normal to the sample surface and ridge waveguides~\cite{Jns2015} and photonic crystal waveguides~\cite{Arcari2014} for lateral photon guiding in integrated quantum photonic circuits (IQPCs). These quantum devices must be modeled numerically with high 
accuracy in order to achieve optimal performance, for example in terms of photon extraction efficiency. For this purpose, optimization calculations are carried out in multidimensional parameter spaces using modern algorithms and numerical methods such as finite difference time domain method~\cite{Zadeh2016} and the finite element method~\cite{Garcia-Santiago2021}, where Bayesian optimization shows superior performance~\cite{Schneider2019}. 

\begin{figure}[tb]
  \includegraphics[width=\linewidth]{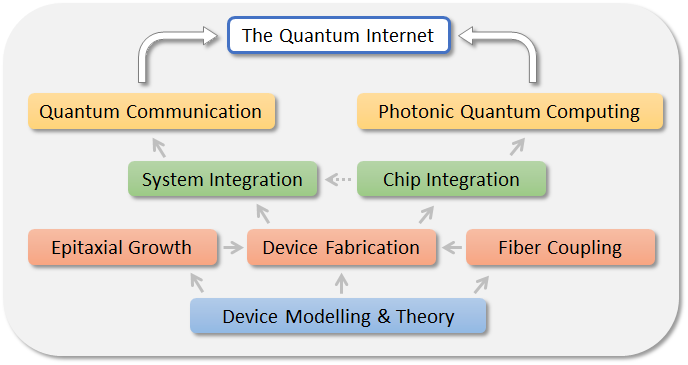}
  \caption{Schematic overview of the development and application of QD-based quantum devices in quantum information technology. Theory and numerical modelling is used to predict and optimize the device performance. QDs are then epitaxially grown, integrated into nanophotonic devices to enhance their optical properties and fiber-coupled for user-friendly operation. At the next level, QD quantum devices are integrated into larger systems to implement, for instance, quantum communication networks. In the other hand, on-chip integration of QDs is a cornerstone in photonic quantum computing. Ultimately, the overarching goal is to combine quantum networks and photonic quantum processing units in a global quantum internet.}
  \label{fig:Fig_overview}
\end{figure}

The technological implementation of these concepts is usually very demanding and can only be achieved with highly optimized nanoprocessing concepts. Especially in the field of lithography, new approaches are needed to integrate individual QDs with nm accuracy and spectral matching in resonator structures and waveguide systems. For this purpose, high-precision deterministic lithography processes have been developed in recent years~\cite{Rodt2020}, which are now used very successfully for the realization of QD based quantum devices. Other important and current aspects regarding the application in quantum technology are the spectral control of the QD emission via external variables such as strain tuning~\cite{Ding2010} and the direct fiber coupling of the sources for user-friendly integration in quantum networks~\cite{Bremer2022b}.

Against this background, this review article gives a comprehensive overview of the development and various application perspectives of semiconductor QDs in the field of photonic quantum information technology, as illustrated in Fig.\ \ref{fig:Fig_overview}. The article is aimed at students and scientists who want to get a well-founded insight into the basics of QDs, highly optimized device concepts, modern nanoprocessing technologies and the optical and quantum-optical properties of corresponding QLSs and IQPCs. Furthermore, open questions are discussed, and future development directions are presented, which should pave the way for the application of QD quantum devices in photonic quantum information technology.

The article is structured as follows. In Section~\ref{sec:application_scenarios} we first introduce application scenarios of QDs in quantum information technology and the associated requirements and key parameters.  
We then introduce the theoretical concepts needed to understand light emission and the numerical modeling methods used to predict the performance of QLSs based on QDs in Section~\ref{sec:theory}. 
Numeric optimization is often the basis for sample growth and device nanofabrication which are discussed in Section~\ref{sec:epi} and Section~\ref{sec:fab} along with modern fiber-coupling solutions. Section~\ref{sec:optprop} presents the optical and quantum optical properties of state-of-the-art QD-based QLSs in the most relevant spectral ranges and introduces advanced QD device concepts for acting as spin-photon interfaces and photonic cluster state generators. In the following Section~\ref{sec:integration} we introduce and discuss recent advanced in the realization of QD-based IQPCs, before presenting system integration and first applications of single QD devices in photonic quantum technologies in Section~\ref{sec:applications}. The article closes with an outlook onto open questions and future research directions towards a global quantum internet, and a conclusion in  Section~\ref{sec:outlook} and in Section~\ref{sec:conclusion}, respectively. 

\section{Application scenarios and requirements}\label{sec:application_scenarios}

This section introduces envisaged application scenarios of QDs in photonic quantum information technologies. In the broader context of the quantum internet, this includes the field of quantum communication on the one hand and the area of photonic quantum computers on the other hand as key building blocks of a global quantum network. Furthermore, interfaces between stationary and flying qubits are presented, which are required to connect different nodes in large-scale quantum networks. We also discuss the associated requirements as the basis for the following sections, in which we present the design, fabrication, optical properties and first applications of QD quantum devices in photonic quantum information technologies.

\subsection{Quantum communication}\label{sec:application_scenarios_QKD}
Quantum communication is currently considered to be the quantum information technology with the highest short-term application potential and large impact on the secure exchange of sensitive data. As one of the most studied cryptographic primitives, quantum key distribution enables the generation of a secret and random bit-string shared between two authenticated parties. Once distributed, this key can be used to encrypt data, with its security being protected by the laws of quantum mechanics rather than computational complexity as in classical schemes. Using the so-called one-time-pad scheme for data encryption, even information-theoretical security is possible~\cite{Shor2000,Gisin2002}.

\begin{figure}[ht]
  \includegraphics[width=\linewidth]{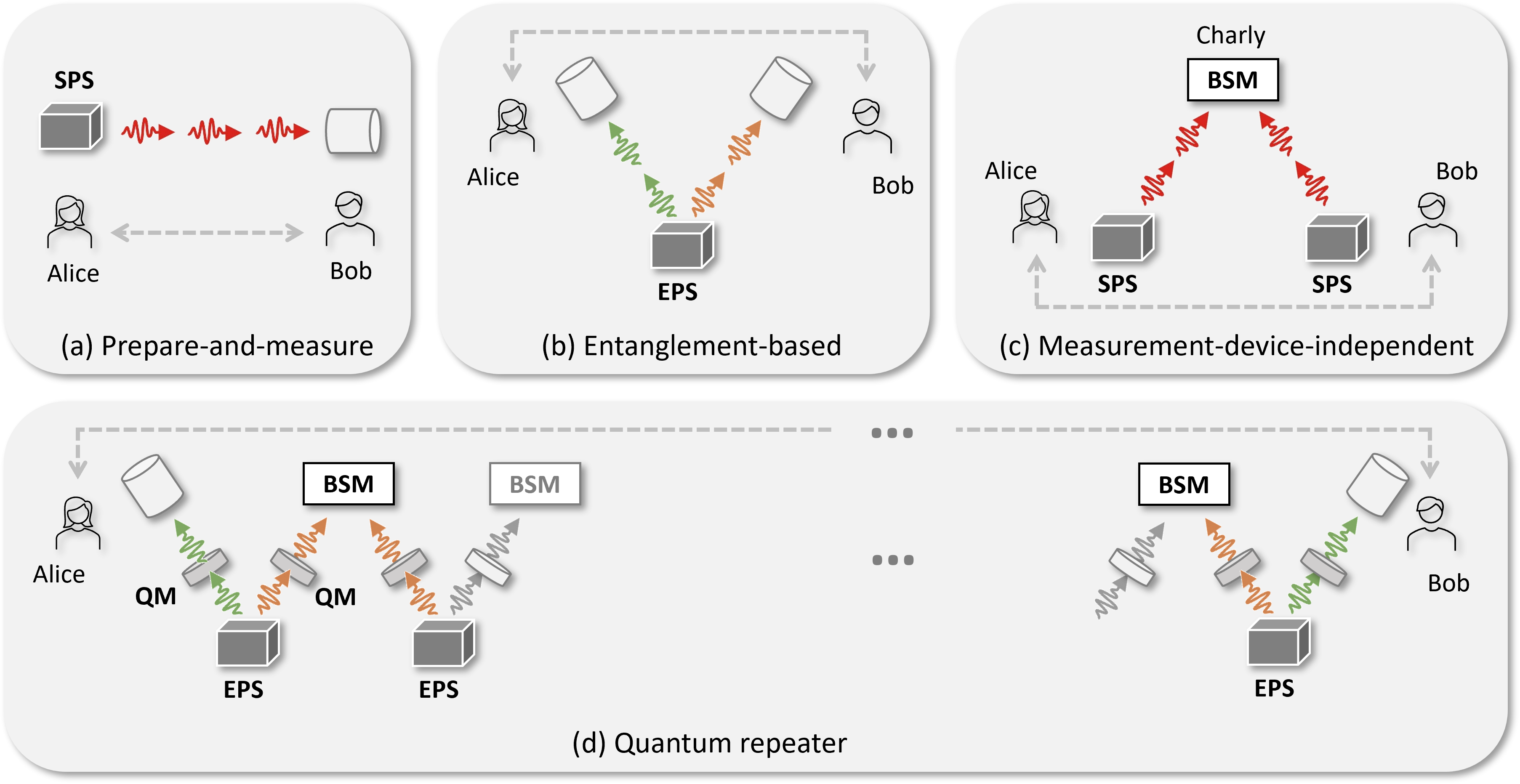}
  \caption{Quantum communication concepts. Depending on the type of quantum resource, different types of QKD scenarios are possible: (a) Prepare-and-measure based QKD protocols using single-photon states, (b) entanglement based QKD protocols using polarization entangled photon pairs, (c) device-independent QKD protocols requiring indistinguishable photons from remote sources, and (d) the quantum repeater concept for long-distance QKD.}
  \label{fig:Fig_QKD_schemes}
\end{figure}

The first QKD protocol, known as BB84, was proposed by Charles H. Bennett and Gilles Brassard in 1984 \cite{Bennett1984} and uses the quantum mechanical properties of single photons to establish security\footnote{The concept of quantum cryptography was born already earlier, within the idea of conjugate coding by S. Wiesner in the late 1960s, work which has not been published until 1983 \cite{Wiesner} (see Bennett et al. \cite{Bennett1992a} for a historical review).}. Here, the polarization of single photons is used to encode the bits in different, randomly chosen bases in a so-called prepare-and-measure type configurations (see Fig.~\ref{fig:Fig_QKD_schemes}\,(a)). The sending party (Alice) first prepares qubit states randomly in four different states, sends them to the receiving party (Bob) via a quantum channel where the states are detected, again in randomly set basis settings. Eavesdropping attempts of an adversary would lead to an increase in errors in the bit sequence (e.g. 25\% for an simple intercept-resend strategy) and can thus be detected by comparing a subset of the results. The BB84 protocol can be subdivided into five basic steps, common to most QKD protocols: Qubit exchange, sifting, parameter estimation, and the classical post-processes error correction and privacy amplification. These steps result in a final secure or secret key rate, being the most important figure of merit for the benchmarking of QKD systems. Errors occurring in the key after the sifting step, i.e. the sifted key, are quantified by the quantum bit error ratio (QBER)\footnote{A more widely used term in the literature is the quantum bit error rate in units of $s^{-1}$. As the QBER entering the key rate equations must be a probability, it is sometimes beneficial to use the quantum bit error ratio for consistency reasons.} - the probability that a bit value of Alice and Bob differs, even though they used the same measurement basis. It should be noted, that the steps mentioned above require an authenticated classical channel between both parties, which implies that a small amount of secret key is required already before the first quantum-key exchange \cite{Fung2010}. For this reason, QKD is also referred to as a “secret growing scheme".

Prepare-and-measure type QKD as introduced with the BB84 protocol is however not the only possible choice. As proposed in the E91 protocol by Artur Ekert in 1991 \cite{Ekert1991}, also entangled photon pair sources can be used for implementations of QKD (see Fig. \ref{fig:Fig_QKD_schemes}(b)). Here, Alice and Bob independently perform a measurement on one photon of an entangled two-photon state using randomly selected bases. Keeping only results in which both used the same basis, both parties obtain a perfectly correlated bit-string. By quantifying the remaining degree of entanglement after the photon transmission, e.g.\ by verifying the violation of the Bell-type Clauser, Horne, Shimony, and Holt (CHSH) inequality \cite{Clauser1969}, eavesdropping attempts can be uncovered. Alternatively, one can also use the distributed entangled photons directly for measurements in the BB84 bases, compare some of the results and deduce the security from the identified error rates just as in the BB84 protocol - a protocol known as BBM92 \cite{Bennett1992}. Both, prepare-and-measure and entanglement-based QKD can be enhanced in their performance, compared to implementations using attenuated lasers, if deterministic QD-based QLSs are employed. Recent progress in this direction using QD-based QLSs is reviewed in Sections~\ref{sec:application_SPS_QKD} and \ref{sec:application_EPS_QKD}.

While the quantum cryptographic protocols discussed above can be proved secure in an information theoretical sense, device imperfections in physical realizations can compromise the protocol's security by introducing loop-holes or side-channel attacks (see Ref.\ \cite{Xu2020} for an in-depth review on quantum hacking strategies). 
For this reason, device-independent (DI) QKD protocols have been invented, which are constructed such that imperfections of the technical realization do not compromise the protocols' security, representing a major advantage for practical applications \cite{Diamanti2016}. Full-fledged implementations of DI-QKD are extremely challenging to realize~\cite{Acin2007} and require loophole-free Bell-state measurements (BSMs) across remote locations with high entanglement fidelity~\cite{Zhang2022,Nadlinger2022,Wen-Zhao2022}. On the other hand, already partially device-independent protocols eliminating attacks on specific devices, are very useful. An example are MDI-QKD protocols~\cite{Lo2012,Braunstein2012}, for which the protocol security can be guaranteed independent of the measurement device, i.e.\ the detection setup. Here, Alice and Bob each send single indistinguishable photons to a central receiver station (Charly), where both photons are projected into an entangled two-photon state via a BSM (cf.\ Fig.\ \ref{fig:Fig_QKD_schemes}(c)). To date, MDI-QKD has mostly been implemented using weak coherent pulses~\cite{Rubenok2013,Liu2013,DaSilva2013,Cao2020,Semenenko2020,Wei2020}, for which the underlying Poissonian photon statistic fundamentally limited the achievable two-photon interference (TPI) visibility to 50\%. Exceeding this classical limit increases the efficiency of the BSMs \cite{Lee2021}. Thus, for implementations using deterministic QLSs based on QDs, substantial advances can be expected. In addition, as MDI-QKD protocols are intrinsically suited for star-like network topologies, MDI-QKD is particularly useful for the realization of scalable multi-user QKD networks in metropolitan areas \cite{Tang2016}. An experimental demonstration of this type of quantum network with sub-Poissonian QLSs would be a major step forward. Recent progress in this direction will be reviewed in Section~\ref{sec:applications_advanced_QKD}.

To cover arbitrary distances in quantum-secured communication, QKD links as discussed above can in principle be chained using intermediate trusted nodes~\cite{Chen2021}, which however reduce the overall security in the end-to-end connection. An elegant solution for transferring quantum information over arbitrary distances without compromises in the security are quantum repeaters. Here, the quantum channel is divided into shorter segments using entangled photon pair sources and entanglement swapping as key resources (cf. Fig.\ \ref{fig:Fig_QKD_schemes}(d)).
The first quantum repeater scheme, known as BDCZ protocol, was proposed by Briegel, Dür, Cirac, and Zoller in 1998, to overcome the exponential scaling of errors in the quantum channel due to depolarization and transmission losses \cite{Briegel1998,Dur1999}. The BDCZ protocol enables the distribution of a maximally entangled photon pair, e.g.\ the well-known Einstein-Podolski-Rosen (EPR) state~\cite{Einstein1935}, over arbitrary distances. The entangled photon pair can then be used directly, to realize QKD protocols (e.g.\ the E91 protocol), or, to teleport a quantum state from one end to the other. To distribute the entanglement, Briegel et al.\ proposed to use multiple EPR sources along the quantum channel each sending entangled photons in opposite directions, thus dividing the complete quantum channel into shorter segments. At intermediate nodes, photons from two neighboring EPR sources are stored in a quantum memory and then used for swapping the entanglement to the two outer photons via a joint BSM of the photons at the intermediate station. By repeating the swapping in a nested fashion, arbitrary distances can in principle be covered. To implement this scheme at reasonable levels of photon loss tolerance, however, quantum memories and coherent spin-photon interfaces are required at the intermediate nodes, which makes the practical implementation very complex. To implement the BDCZ quantum repeater protocol, QD-based entangled photon pair sources can be used in combination with suitable spin-photon interfaces (see Section~\ref{sec:spin_photon} and Ref.~\cite{Loock2020} for a comparison of different quantum emitter platforms). Alternatively, all-photonic measurement-based quantum repeater schemes also promise quantum communication at arbitrary scales without requiring quantum memories~\cite{Zwerger2012,Azuma2015,Borregaard2020}. These types of repeater protocols require photonic cluster states as key resources, which have already been generated using QD-QLSs~\cite{Schwartz2016,Istrati2020} as discussed in Section~\ref{sec:cluster}.

\subsection{Photonic quantum computing} \label{sec:application_scenarios_PQC}
\begin{figure}[ht]
  \includegraphics[width=\linewidth]{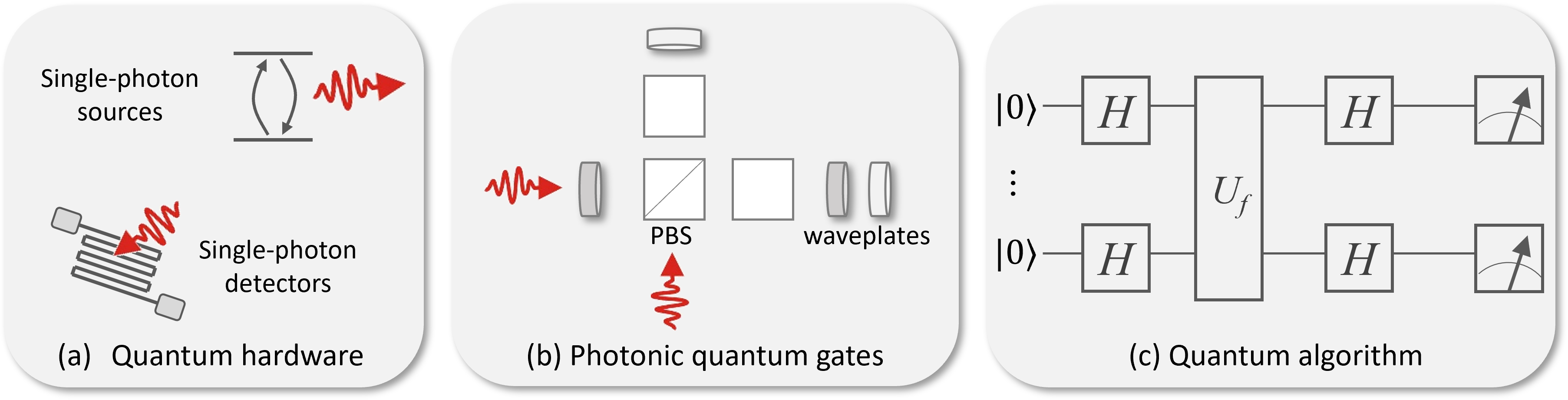}
  \caption{Realization of photonic quantum computers: (a) Single-photon sources and detectors providing the essential quantum hardware, (b) photonic quantum gates based on linear optics, and (c) quantum algorithms to solve target problems.}
  \label{fig:Fig_PQC}
\end{figure}
The realization of photonic quantum processors and eventually photonic quantum computers is another appealing application scenario of QD-based quantum devices, and Fig.\ \ref{fig:Fig_PQC} illustrates important building blocks for such quantum information systems. In this context, photons have distinct advantages as a qubit source over other qubit platforms. First, they have a variety of degrees of freedom encoding quantum states such as polarization, path, time-bin, and frequency. It is also possible to utilize their high-dimensional or continuous variables, such as orbital angular momenta and spatial modes. Second, photons do not suffer from decoherence and merely interact with the environment. Third, there exist well-developed technologies for generating, manipulating, and measuring photons in free space, fiber optics, and integrated chips. Therefore, photons provide excellent quantum information carriers. 
From these advantages, the field of photonic quantum computing is growing rapidly. Although fault-tolerant quantum computing still requires significantly more quantum resources with higher accuracy, photonic quantum computing showed its potential by solving specific problems beyond classical computers~\cite{Zhong2020,Madsen2022}, which could be useful to simulate complex molecular interactions~\cite{Huh2015} and find eigenvalues~\cite{Peruzzo2014}. Therefore, the applications of photonic quantum computing range from simulating new materials and drugs to solving optimization and factorial problems. 
While using photons leads to fewer concerns about decoherence issues, a major challenge in photonic quantum computing and simulation is implementing quantum gates, since direct interactions between photons are quite difficult to establish. Furthermore, photons are subject to loss and other errors during the operation, and therefore it is necessary to develop methods for improving efficiency and correcting these errors in order to secure the reliability of photonic quantum computers.
In 2001, a scheme was proposed for linear optics quantum computing that does not require direct interaction but introduces nonlinearities in the quantum interference and measurement~\cite{Knill2001}. Since then, significant advances have been made in the key building blocks of photonic quantum computing, including high-quality QLSs, efficient photon detection, and fast photonic gates. In particular, deterministically operating semiconductor QDs are starting to outperform existing heralded QLSs based on spontaneous parametric down-conversion process in terms of brightness, single-photon purity, and indistinguishability as well as high fidelity for entangled photon pairs~\cite{Chen2018}.
As single photons from QDs can be collected efficiently, see Section~\ref{sec:optprop}, and commercialized high-efficiency superconducting nanowire single-photon detectors become available, the single-photon detection rate from a source to a detector now reaches over 10 MHz\cite{Barbiero2022}. Achieving high efficiency allows the use of quantum gates with a high success rate and an improved signal-to-noise ratio. Therefore, the QD QLSs are highly suitable for measurement-based quantum computing with minimized loss and errors.  
However, in addition to efficiency, other important properties of photonic quantum computing include stability, scalability, and compatibility with other components in the quantum system. As the size of the quantum system increases, multiple single photons need to interact and be entangled. Therefore, it is necessary to eliminate frequency jitter over time and between different emitters. The lifetime-limited linewidth is required to ensure a long coherence time. Quantum memory is also essential for several tasks in photonic quantum computing, including quantum error correction and quantum algorithms, such that a quantum memory can store the quantum information that is being protected from errors and store intermediate results during the process~\cite{Bussières2014, Glaudell2016}. This stored information needs to be retrieved later effectively, requiring efficient spin-photon interfaces. The ground state spin of QDs has shown a spin coherence time of up to a few microseconds \cite{Stockill2016}, and it could be prolonged with low-strain GaAs QDs\cite{Gillard2021, Zhai2020a} (see Section~\ref{sec:epi}). Besides, incorporating QD spin qubits can provide nonlinearity based on spin-photon entanglement (see Section~\ref{sec:spin_photon}) and brings new functionalities in photonic quantum computation, such as sequential entangler~\cite{Istrati2020}, single-photon transistor~\cite{Sun2018} and deterministic quantum gates~\cite{Fan2021}. Of particular interest are 2D photonic cluster states, which allow for efficient one-way photonic quantum computing~\cite{Raussendorf2001} (see Section~\ref{sec:cluster}). To implement practical photonic quantum systems, the integration of QLSs and memories with classical photonic integrated circuits is a crucial aspect. This integration combines the strengths of quantum and classical photonic technologies to create systems with improved functionality and performance. The quantum resources bring their inherent quantum properties, such as coherence and entanglement, while the classical photonic chips provide stable, compact, and  programmable platforms, required for practical applications (see Section~\ref{sec:integration}). By meeting all these requirements, it is possible to develop photonic quantum computing systems that are robust, scalable and capable of performing complex quantum algorithms. 

\subsection{Building blocks of the quantum internet} 

\begin{figure}[ht]
  \includegraphics[width=\linewidth]{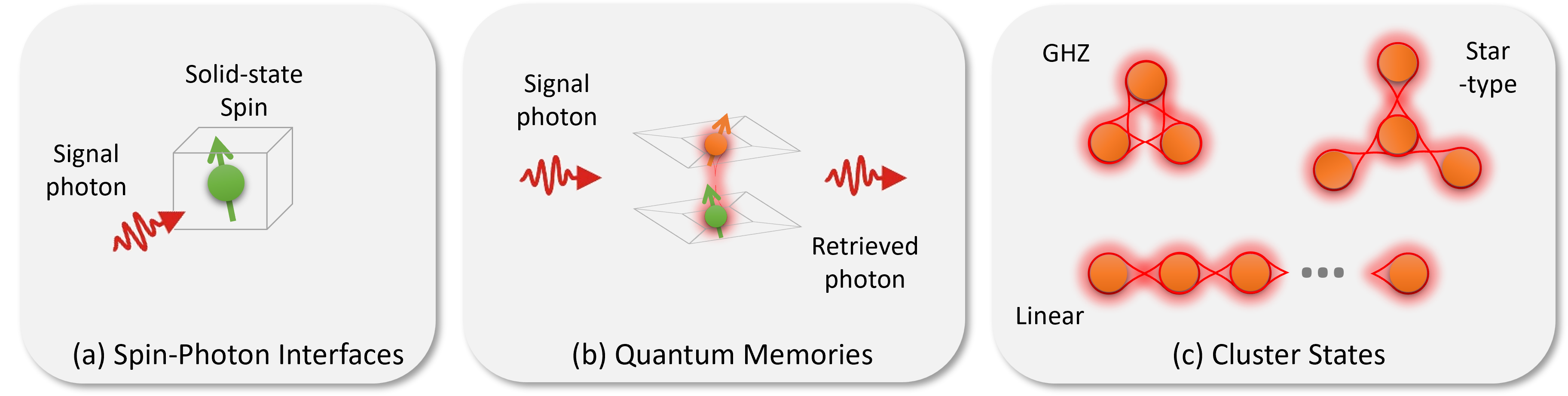}
  \caption{Building blocks for a future quantum internet: (a) Spin-photon interfaces converting quantum information from flying to stationary qubits, (b) quantum memories storing and releasing quantum information on-demand, and (c) cluster states as key resources for fault-tolerant quantum computing.}
  \label{fig:Fig_BuildingBlocks}
\end{figure}
The overarching goal of quantum information technology is the development of a global quantum internet~\cite{Kimble2008}. Such a network consists of quantum nodes, which can represent quantum computers, interconnected by quantum channels, in which information distributed via single photons acting as flying qubits. In this way, distributed quantum computing can be performed in the future. In close connection with quantum communication and photonic quantum computing introduced in the two previous sections, the implementation of large-scale quantum networks, and finally the quantum internet, requires coherent interfaces between stationary and flying qubits to connect different quantum nodes. In the same context, sources of photonic cluster states should also be mentioned, which represent a powerful resource for measurement-based quantum computing and loss-tolerant quantum communication, and which form further important application perspectives for QDs. The concepts of these building blocks, illustrated in Fig.~\ref{fig:Fig_BuildingBlocks}, is discussed in the following.

Quantum memories and the related spin-photon interfaces have the task of storing a quantum state for as long as possible in order to read it out at a later point in time. They are central elements of quantum repeater networks and quantum computers, which explains the enormous research activities in this field. On the one hand, such coherent interfaces must interact with their environment for writing and reading, but on the other hand they must also be decoupled from it in order to avoid decoherence of the stored quantum state. So far, the best coherence times have been achieved in atomic quantum memories, which, however, are hardly compatible with scalable component technologies. QDs could be an interesting alternative in this context. An additional electron or hole can be localized in a QD by targeted doping. A quantum state can be encoded in its spin degree of freedom, which can be initialized and retrieved via optical excitation. In addition, electrically addressable quantum dot molecules promise increased functionality and better storage properties.

Stationary qubits reside in local devices, such as the memory or processor of a quantum computer. Flying qubits are typically photons that carry the quantum information through the air, a vacuum of space, or through fiber optic networks. Thus, interfaces between stationary and flying qubits are key building blocks of quantum networks. There are different proposals to implement them, including spin-photon interfaces which can be realized by atoms or QDs (see Section~\ref{sec:spin_photon}. 

Another important resource in photonic quantum computing and quantum networks are photonic cluster states. Such states are highly entangled states of multiple qubits which allow for one-way, and measurement-based quantum computing~\cite{Raussendorf2001, Raussendorf2003} and loss-tolerant quantum communication~\cite{Azuma2015}. However, such applications require cluster states of two dimensions or higher. For instance, for the realization of topologically fault-tolerant cluster state quantum computation, at least three dimensions a necessary~\cite{Raussendorf2003}. Interestingly, there exist proposals for the deterministic generation of 1D~\cite{Lindner2009}, 2D~\cite{Economou2010} and multidimensional~\cite{Shi2021} photonic cluster states using semiconductor QDs. 

\subsection{Requirements and key parameters of quantum dot  quantum devices}

In order to meet the envisaged applications in photonic quantum information technology, the QD quantum devices must meet a number of stringent requirements, the most important of which are briefly introduced below and will be discussed in detail in this review article:

\begin{itemize}
    \item \textbf{Emission wavelength}: While for proofs of principle on QD QLSs the emission wavelength is/was     of secondary importance and was usually in the range of 900 – 950 nm, quantum applications require specific target wavelengths. Especially for use in fiber-based quantum networks, emission wavelengths in the telecom O-band at 1.3~$\mu$m and in the C-band at 1.55~$\mu$m are aimed for, which are characterized by minimal dispersion and minimal attenuation, respectively~\cite{Schubert2006}. For free-space quantum communication, shorter wavelengths are generally preferred~\cite{Lanning2021} and the first quantum satellites operate around 800~nm. In addition, for coupling to available atomic-based or rare-earth-ion-based quantum memories, specific wavelengths are needed, see e.g.\ Ref.~\cite{Neuwirth2021}. An example is represented by the D1 transition in Rb vapors~\cite{Wolters2017}. 
    \item \textbf{Single-photon purity}: A central parameter of all SPSs is multi-photon emission suppression which is quantified via the autocorrelation function at zero time delay $g^{(2)}(0)$, which should be as close to zero as possible. In the context of QLSs, this property is often referred to as "single-photon purity".
    \item \textbf{Emission linewidth and indistinguishability}: Quantum emitters are characterized by discrete emission lines, which ideally should only be homogeneously broadened 
    due to the finite lifetime for spontaneous emission $\tau_{\rm r}$. In this case, the linewidth $\Gamma$ results from a Fourier transformation of the spontaneous decay and is given by $\Gamma = \hbar /\tau_{\rm r}$. In practice, pure dephasing~\cite{Krummheuer2002}, inelastic interaction with phonons~\cite{Iles-Smith2017}, and the Coulomb interaction of the confined carriers with charged states in the vicinity of the QD lead to an additional broadening~\cite{Kuhlmann2015}. 
    All these mechanisms have an adverse effect on the indistinguishability of the photons, as discussed further in Section~\ref{sec:theory_decoh}.
    While phonon-related broadening can be limited by operating at sufficiently low temperature, charge noise is often the dominant inhomogeneous-broadening mechanism and is therefore particularly problematic for quantum functionalities such as the entanglement distribution in quantum repeater networks, which are based on ''Hong-Ou-Mandel'' (HOM)-like TPI~\cite{Vajner2022}.
    \item  \textbf{Entanglement fidelity}: QDs can generate entangled photons on-demand at high photon flux. This can happen via the biexciton-exciton (XX-X) cascade, which leads to polarization-entangled photon pairs~\cite{Benson2000,Akopian2006,Huber2018}. Furthermore, time-bin entanglement and hyper-entanglement are possible to achieve using QDs~\cite{Jayakumar2014,Prilmuller2017}. In all cases, the entanglement fidelity
    with respect to a maximally entangled state
    is an important parameter that should be as close to one as possible for quantum applications. 
    
    \item \textbf{Spin coherence}:
    The spin coherence is an important parameter of QDs with regard to the realization of spin-photon interfaces and in the generation of photonic cluster states. Via the spin degree of freedom, quantum information can be stored on a timescale of the spin coherence time. The aim is to achieve the highest possible spin coherence time, for which, for example, the spin of a confined electron must be decoupled from the solid-state environment as efficiently as possible to generate long-lived stationary qubits. At the same time, efficient photonic coupling to the environment is usually required to realize spin-photon interfaces that interact with flying qubits.
    
    \item \textbf{Preparation and quantum efficiency}:
    In the ideal case, the state preparation efficiency $\eta_{\rm prep}$, i.e.\ the probability of initializing a QD in the desired state upon excitation, e.g.\ a charged exciton or a biexciton state (depending on the application), should be unity. In reality, $\eta_{\rm prep}<1$ because of possible random fluctuations in the charges captured by (or generated in) the QD~\cite{Davano2014} and/or other effects such as electron-phonon interaction limiting the population-inversion efficiency~\cite{Carmele2019}. The former limit is usually referred to as ``blinking'' and can be strongly reduced in charge-tunable devices~\cite{Zhai2020a}, while the latter can be reduced by sophisticated excitation schemes~\cite{Gustin2019}. Also, the emission (or quantum) efficiency $\eta_{\rm em}$, i.e.\ the probability that the recombination results in a photon (or photon pair) in the desired optical mode is limited due to possible non-radiative decay channels as well as radiative side-channels, such as phonon side-bands and radiative Auger~\cite{Lbl2020} channels. Compared to other quantum emitters, QDs have in general very high quantum efficiency. 
    
    \item  \textbf{Photon extraction and coupling efficiency}: For the on-demand character of the QD QLSs, it is necessary to generate a usable photon or entangled photon pair with each trigger impulse. In order to come close to this goal, deterministic excitation concepts are used on the one hand and device geometries are developed on the other hand, which couple the photons generated by the QD into certain modes or emit them in the desired direction, with the emission into loss channels being suppressed as much as possible. Furthermore, effects of cavity quantum electrodynamics (cQED) come into play in resonator-based device concepts, which can accelerate the spontaneous emission of the QDs in the regime of weak coupling in order to improve  the photon extraction efficiency as well as the indistinguishability~\cite{Gerard1998,Santori2002a, Kaer2013, Ding2016}.

    \item  \textbf{Device fabrication}: In addition to the physical aspects mentioned, the device fabrication itself is a very important aspect in the development of QD-based QLSs for applications in photonic quantum technology. For example, QDs often have to be integrated into nanophotonic devices with nm accuracy and spectral matching, which in view of the self-organized growth of QDs with indeterminate (lateral) position and spectral location necessarily requires deterministic production methods. This is particularly essential for upscaling to complex quantum networks and highly integrated IQPCs based on QLSs with identical properties. Compatibility with quantum memories is also required for certain applications, which can be achieved using hybrid concepts. Finally, a high functional integration density is aimed at for IQPCs, which, in addition to the sources, also includes on-chip detectors~\cite{Schwartz2018}.
\end{itemize}

\section{Theory and modeling of quantum dot devices} \label{sec:theory}

While numerous design strategies exist for controlling the light emission from QD-based QLSs, they all require accurate modeling and careful optical engineering to achieve high performance. In this section, we review the theory of light emission from QDs. Furthermore, we discuss the numerical simulation techniques used to model the collection efficiency and the photon indistinguishability. We exemplify with an analysis of the performance of the micropillar SPS.

\subsection{Theory of quantum dot states} \label{sec:theory_qd_states}

The direct bandgap of In(Ga)As QDs (and GaAs QDs) enables efficient spontaneous emission using the radiative transition from the conduction band to the valence band at the $\Gamma$ point of the Brillouin zone. The zinc-blende crystal structure results in three valence bands, the heavy hole, the light hole and the split-off band \cite{Yu2010}. However, spin orbit coupling and the aspect ratio of pyramidal-shaped QDs shift the energies of the latter two, such that light emission predominantly takes place through transitions to the heavy hole band. For standard sized QDs, the extension of the electron and hole wavefunctions is dominated by the strong confinement of the 3D potential landscape, while Coulomb interaction instead plays a perturbative role for the energy levels \cite{Lodahl2015}. While QDs generally feature an advanced energy level structures, the lowest energy s-shell is typically used for light emission.

\begin{figure}
	\includegraphics{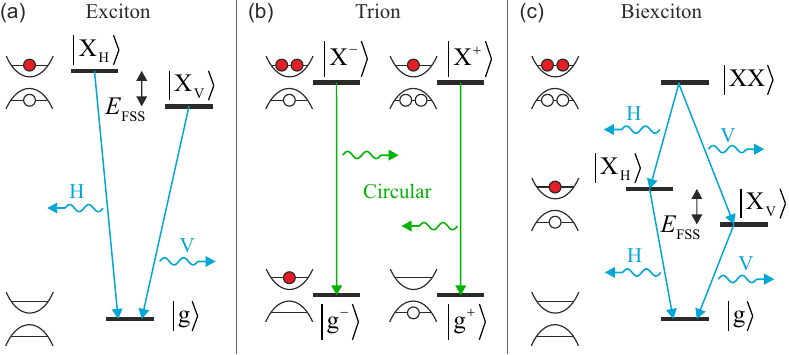}
	\caption{The electron (hole) configurations represented by red full (empty) circles for the $s$ shell in the conduction (valence) band and associated photon emission transitions. (a) The exciton state generates a linearly polarized photon. (b) The trion state produces a circularly polarized photon. (c) The biexciton configuration generates polarization entangled linearly polarized photon pair through the biexciton-exciton emission cascade.}
	\label{fig_QD_Conf}
\end{figure}

The most basic optical excitation is the exciton state configuration consisting of a single electron in the conduction band and a single hole in the valence band, shown in Fig.~\ref{fig_QD_Conf}(a). In terms of the vertical orientation $S$ along the quantization growth axis of the electron ($S_{\rm e} = \ket{\uparrow}$ or $\ket{\downarrow}$) and hole spin state ($S_{\rm h} = \ket{\Uparrow}$ or $\ket{\Downarrow}$), the relevant optically bright states of the exciton are $\ket{\uparrow \Downarrow}$ and $\ket{\downarrow \Uparrow}$, whereas emission from the dark states $\ket{\uparrow \Uparrow}$ and $\ket{\downarrow \Downarrow}$ is forbidden due to lack of angular momentum conservation. In the presence of electron-hole exchange interaction, the bright energy eigenstates are $\ket{{\rm X}_{\rm H}} = \frac{1}{\sqrt{2}} (\ket{\uparrow \Downarrow} + \ket{\downarrow \Uparrow})$ and $\ket{{\rm X}_{\rm V}} = \frac{1}{\sqrt{2}} (\ket{\uparrow \Downarrow} - \ket{\downarrow \Uparrow})$, which produce photons linearly polarized in the horizontal (H) and vertical (V) direction during the radiative transition to the ground state $\ket{\rm g}$. The pyramidal shape of the QD lifts the energy degeneracy between the two states, which are separated by the fine structure splitting $E_{\rm FSS}$ of typically 10 - 100 $\mu$eV for InGaAs QD~\cite{Seguin2005}, while very low $E_{\rm FSS}$ in the range of the homogenous linewidth are usually observed for highly symmetric GaAs QDs~\cite{Huo2013}.

By adding a single charge to the exciton state, the charged exciton or trion states depicted in Fig.\ \ref{fig_QD_Conf}(b) are obtained. The negatively (positively) charged trion state $\ket{{\rm X}^-}$ ($\ket{{\rm X}^+}$) consists of two electrons (holes) in the conduction (valence) band and a single hole (electron) in the valence (conduction) band, and the corresponding spin configurations are $\ket{{\rm X}^-} = \frac{1}{\sqrt{2}} (\ket{\uparrow \downarrow } - \ket{\downarrow \uparrow}) S_{\rm h}$ and $\ket{{\rm X}^+} = \frac{1}{\sqrt{2}} (\ket{\Uparrow \Downarrow } - \ket{\Downarrow \Uparrow}) S_{\rm e}$. The trion is a fermion, and the superpositions arise due to the requirement of an anti-symmetric state for identical particles. For the trion state, recombination of an electron-hole pair leads to emission of a circularly polarized photon leaving the system in the charged ground state $\ket{\rm g^\pm}$. The additional charge of the trion state represents an important asset for spin physics \cite{Warburton2013} enabling e.g.\ generation of the photonic cluster states \cite{Lindner2009, Economou2010, Schwartz2016, Shi2021}, see Section~\ref{sec:cluster}. 

Finally, by adding another electron-hole pair to the excitonic state, the biexciton configuration $\ket{\rm XX} = \ket{\uparrow \downarrow \Uparrow \Downarrow}$ including two electron and holes is obtained. The biexciton can decay to the ground state via either of the two channels illustrated in Fig.\ \ref{fig_QD_Conf}(c) producing a pair of linearly polarized photons. In the absence of fine structure splitting $E_{\rm FSS}$ = 0, the two decay channels are indistinguishable resulting in the entangled photon pair state $\frac{1}{\sqrt{2}} (\ket{\rm HH} + \ket{\rm VV})$, and control of the fine structure splitting \cite{Huber2018} is thus an essential tool for entangled photon pair generation. Coulomb interaction typically lowers the energy of the initial photon emitted from the biexciton by a few meV, and for the entangled photon pair source a broadband photonic design approach is thus needed to ensure collection of both the biexciton and exciton photons.% \textcolor{red}{NIELS: There is some overlap with the introduction in this paragraph. What do we do about that?} SR: fixed

\subsection{Modeling of the light-matter interaction} \label{sec:theory_light_matter}

The Lorentz force governs the interaction between the electromagnetic field and the charge states of the QD, typically modeled as a two-level system with an excited (ground) state wavefunction $\Psi _{\rm e}$ ($\Psi _{\rm g}$). Since typical QDs are small compared to the wavelength, the electric field $\textbf{E}(\textbf{r},t)$ can be considered constant $\textbf{E}(t)$ over the QD. In this dipole approximation, the interaction Hamiltonian $\hat{H}_{\rm int} = -\hat{\textbf{d}} \cdot \textbf{E}(t)$ \cite{Novotny2012} is the product of the electric field and the dipole moment operator $\hat{\textbf{d}} = -e \hat{\textbf{r}}$, where $e$ is the electronic charge and $\hat{\textbf{r}}$ is the position operator. The ability of the QD to emit light is then quantified by the dipole moment $\textbf{d} = \bra{\Psi_{\rm e}} \hat{\textbf{d}} \ket{\Psi _{\rm g}}$. The simplest model for describing the electronic wavefunction is the single-band effective mass approximation \cite{Coldren2012}, where the wavefunction $\Psi = F(\textbf{r}) u(\textbf{r})$ using the Bloch theorem is given as the product of a slowly varying envelope function $F(\textbf{r})$ and the periodic electronic Bloch function $u(\textbf{r})$. In this approximation, the envelope function is a solution to the time independent Schr\"odinger equation
\begin{align}
	- \frac{\hbar ^2}{2m_0} \nabla \cdot \left(\frac{1}{m_{\rm eff} (\textbf{r})} \nabla F(\textbf{r})\right) + V(\textbf{r}) F(\textbf{r}) = E F(\textbf{r}),
\end{align}
where $m_0$ ($m_{\rm eff}$) is the electron mass (effective mass), $V(\textbf{r})$ is the energy potential, and $E$ is the eigenstate energy. The dipole moment for an interband transition can then be written as $\textbf{d} = \braket{F_{\rm e} | F_{\rm g}} \bra{u_{\rm e}} \hat{\textbf{d}} \ket{u_{\rm g}} $, where the Bloch matrix element $\bra{u_{\rm e}} \hat{\textbf{d}} \ket{u_{\rm g}}$ depends only on the properties of the bulk material \cite{Coldren2012}.

QD-based QLSs generally exploit the spontaneous emission process of the weak coupling regime of cQED for deterministic light emission. Here, the spontaneous emission rate $\Gamma$ for an emitter at the position $\textbf{r}_0$ with transition energy $\hbar \omega_0$ is derived using Fermi's golden rule \cite{Loudon2000} as the product of the dipole moment and the local photonic density of states $\rho _{\rm L}$ \cite{Novotny2012}, 
\begin{align}
\Gamma  &= \frac{{\pi {\omega _0}}}{{\hbar {\varepsilon _0}}}{\left| {\bf{d}} \right|^2}{\rho _{\rm L}} ( {{{\bf{n}}_{\rm{d}}},{{\bf{r}}_0},{\omega _0}} ), \label{eq_FMG}
\end{align}
where $\textbf{n}_{\rm d} = \textbf{d} / \left| \textbf{d} \right|$ is the dipole moment orientation. The local density of states $\rho_{\rm L}$ fully describes the photonic environment at the position of the emitter and includes e.g. cavity effects accelerating the spontaneous emission rate through Purcell enhancement or photonic bandgap or dielectric screening effects suppressing the rate. Even though the spontaneous emission rate is fully described using Eq.\ \eqref{eq_FMG}, the light emission process is typically simulated numerically using a classical optical calculation by exploiting an equivalence principle \cite{Novotny2012}. To see this, we consider the classical optical dyadic Green's function $\overleftrightarrow{\textbf{G}} ({\bf{r}}, {\bf{r}}_0)$ in the frequency domain defined as the solution to the wave equation
\begin{align}
\nabla  \times \nabla  \times \overleftrightarrow{\textbf{G}} ({\bf{r}}, {\bf{r}}_0) - {\varepsilon}({\bf{r}}) \frac{\omega_0^2}{c^2} \overleftrightarrow{\textbf{G}} ({\bf{r}}, {\bf{r}}_0) = \overline{\overline I} \delta \left( {{\bf{r}} - {\bf{r}}_0} \right).
\end{align}
Physically, the dyadic Green's function represents the classical electric field at the position \textbf{r} generated by a dipole \textbf{d} at the position $\textbf{r}_0$ such that 
\begin{align}
{\bf{E}}({\bf{r}}) = {\omega _0^2}{\mu _0}  \overleftrightarrow{\textbf{G}} ({\bf{r}},{\bf{r}}_0) {\bf{d}}. \label{eq_E_GF}
\end{align} 
In terms of the Green's function, the power $P$ emitted by a classical dipole at $\textbf{r}_0$ oscillating at the frequency $\omega_0$ then becomes
\begin{align}
P = \frac{\omega_0}{2} \text{Im} \left[\textbf{d}^* \cdot \textbf{E}(\textbf{r}_0)\right] =  \frac{{\omega _0^3}{\mu _0} \left| \textbf{d} \right|^2} {2} \text{Im} \left(\textbf{n}_{\rm d}^* \cdot \overleftrightarrow{\textbf{G}} ({\bf{r}}_0,{\bf{r}}_0)  \cdot \textbf{n}_{\rm d} \right). \label{eq_classical_power}
\end{align}
On the other hand, the local photonic density of states $\rho_{\rm L}$ describing the spontaneous emission rate can be written \cite{Novotny2012} in terms of the dyadic Green's function as
\begin{align}
\rho _{\rm L} ({{{\bf{n}}_{\rm{d}}}, {{\bf{r}}_0},{\omega _0}} ) = \frac{2 \omega_0}{\pi c^2} \text{Im} \left(\textbf{n}_{\rm d}^* \cdot \overleftrightarrow{\textbf{G}} ({\bf{r}}_0,{\bf{r}}_0)  \cdot \textbf{n}_{\rm d} \right). \label{eq_ldos}
\end{align}
Normalizing the spontaneous emission rate (classical power) to its value $\Gamma_{\rm Bulk}$ ($P_{\rm Bulk}$) in a bulk medium, we then obtain from Eqs.\ (\ref{eq_classical_power},\ref{eq_ldos}) the equivalence principle 
\begin{align}
\frac{\Gamma}{\Gamma_{\rm Bulk}} = \frac{P}{P_{\rm Bulk}}, 
\end{align}
stating that the normalized spontaneous emission rate of a two-level system equals the normalized power emitted by a classical dipole oscillating at the same frequency. This equivalence principle allows for the spontaneous emission rate to be computed using a standard solver of Maxwell's equations by considering a classical dipole at the position of the emitter, and has been used extensively \cite{Gur2021a, Hayrynen2016, Hayrynen2017, Lecamp2007b, MangaRao2007, Gregersen2010a,Wang2020b, Wang2021, Osterkryger2019,Gaal2022} to compute Purcell enhancement. 

\subsection{Optical simulations of quantum dots in nanophotonic devices} \label{sec:theory_simulations}

\begin{figure} \centering
	\includegraphics{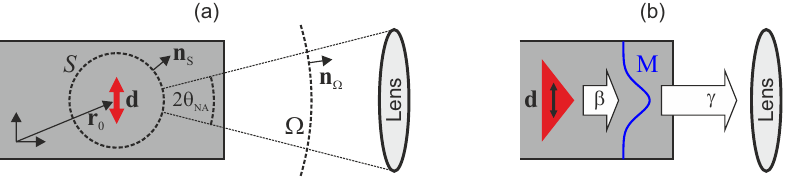}
	\caption{Light emission from a QD inside a generic nanophotonic structure. (a) A classical point dipole (red arrow) inside a volume \textit{V} emits light through the surface \textit{S}. The light emitted within a cone of polar angle $\theta_{\rm NA}$ defined by the NA of the lens is detected by the collection optics. (b) The spontaneous emission $\beta$ factor describes the light emitted from the QD (red triangle) into an optical mode M of interest (blue Gaussian). The transmission coefficient $\gamma$ describes the fraction of light transmitted from the optical mode M to the lens.}
	\label{fig_extraction}
\end{figure}

Let us now consider the general scenario illustrated in Fig.\ \ref{fig_extraction}(a) consisting of a QD modeled using a classical dipole placed at the position $\textbf{r}_0$ inside a generic photonic structure. The extraction efficiency $\eta_{\rm ext}$ is given by 
\begin{align}
	\eta_{\rm ext} = \frac{P_\text{Lens}}{P}, \label{eq_eff_def}
\end{align}
where $P_\text{Lens}$ is the total power detected by the collection lens having a finite numerical aperture (NA). Calculation of the efficiency $\eta_{\rm ext}$ requires an optical simulation predicting the classical electromagnetic field in the vicinity of the emitter. The electrical field may be determined using Eq.\ \eqref{eq_E_GF} from the optical dyadic Green's function, which describes the response of the photonic environment. Analytical expressions for the optical Green's function only exist in very few cases, and generally a numerical solution of Maxwell's equations is needed. 

\begin{figure} \centering
	\includegraphics{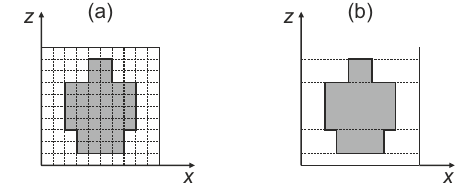}
	\caption{Spatial discretization methods: (a) Spatial grid employed by the finite difference method. (b) Uniform layer decomposition along the $z$ axis used in the modal method.}
	\label{fig_num_method}
\end{figure}

Popular numerical simulation techniques used to model light emission in QD-based QLSs include the finite difference time domain technique \cite{Taflove2004,Lavrinenko2015} and the finite element method approach \cite{Jin2014,Lavrinenko2015} in the frequency domain. Here, the computational domain is expanded on a spatial grid as illustrated in Fig.\ \ref{fig_num_method}(a), and Maxwell's equations are either solved directly in the finite difference time domain technique or reformulated as a vectorial wave equation in the finite element method. Advantages of these spatial grid expansion methods include their availability in commercial user-friendly software packages and their ability to handle arbitrary geometries without simplifying symmetries. A disadvantage is the necessity to implement absorbing boundary conditions around the limited computational domain to correctly model an open geometry and light emission into the far field. 

As an alternative to using a spatial grid, the optical field may be expanded on the optical modes of the geometry: The modal method \cite{Lavrinenko2015,Gur2021a,Hayrynen2016,Hayrynen2017} is a frequency domain technique, where the geometry is divided into layers uniform along a propagation $z$ axis as depicted in Fig.\ \ref{fig_num_method}(b). The electric field $\textbf{E}(\textbf{r}_\perp,z)$ inside a layer is then expanded on the eigenmodes $\textbf{e}_j(\textbf{r}_\perp)$ of the layer, determined assuming uniformity along the propagation axis. The field in a specific layer is written as
\begin{align}
	\textbf{E}(\textbf{r}_\perp,z) = \sum_{j} a_j \textbf{e}_j(\textbf{r}_\perp) e^{i \beta_j z} + \sum_{j} b_j \textbf{e}_j(\textbf{r}_\perp) e^{-i \beta_j z},
\end{align}
where $a_j$ ($b_j$) is the amplitude coefficient of the forward (backward) propagating eigenmode $j$ and the summation includes discrete guided modes as well as the continuum of radiation modes. The fields at each side of a layer interface are then connected using a scattering matrix formalism \cite{Li1996a,Lavrinenko2015}. Advantages of the modal method include the option for directly implementing a true open boundary condition \cite{Hayrynen2016,Hayrynen2017,Gur2021a} as well as direct access to the optical modes of the photonic structure, facilitating the understanding of the governing physics. However, the modal method suffers from convergence issues \cite{DeLasson2018} when considering large 3D geometries without rotational symmetry.

As discussed in section~\ref{sec:theory_light_matter}, the simulation is then performed by computing the optical field generated by the classical dipole given by Eq.\ \eqref{eq_E_GF}. A numerical difficulty in the evaluation of the power $P$ emitted by the dipole at the position $\textbf{r}_0$ is the divergence of Re$(\textbf{E}(\textbf{r}_0))$. For this reason, instead of directly using Eq.\ \eqref{eq_classical_power}, the power is often computed by considering a small sphere with surface $S$ centered on the dipole as illustrated in Fig.\ \ref{fig_extraction}(a). The power is then computed by integrating the Poynting vector over the surface $S$ with normal unit vector \textbf{n}$_S$ as  
\begin{align}
	P  = {\frac{1}{2}\int {{\mathop{\rm Re}\nolimits} \left( {{\bf{E}} \times {{\bf{H}}^*}} \right) \cdot {{\bf{n}}_{S} }d\text{S} } }. \label{eq_ptotal}
\end{align}
To determine the photon collection efficiency, the power $P_{\text{Lens}}$ collected by the first lens is typically evaluated by computing the far field using a near field to far field transformation \cite{Balanis2012}. To model the collection of a finite NA lens, the Poynting vector is then integrated over the solid unit angle $\Omega$ of the cone shown in Fig.\ \ref{fig_extraction}(a) with unit normal vector $\textbf{n}_\Omega$ as
\begin{align}
P_{\text{Lens}}  = {\frac{1}{2}\int_{\theta  < {\theta _{{\rm{NA}}}}} {{\mathop{\rm Re}\nolimits} \left( {{\bf{E}} \times {{\bf{H}}^*}} \right) \cdot {{\bf{n}}_\Omega }d \Omega } }, \label{eq_plens}
\end{align}
where the integration is limited to the polar angle $\theta_{\rm NA}$ defined by the numerical aperture of the lens.

Even though the collection efficiency $\eta_{\rm ext}$ is correctly modeled using Eqs.\ (\ref{eq_eff_def},\ref{eq_ptotal},\ref{eq_plens}), this approach does not always provide direct insight into the physics governing the light extraction. 
In many QLs designs, the light is transmitted to the lens predominantly via a single optical mode M of interest as illustrated in Fig.\ \ref{fig_extraction}(b), and the photonic structure is then engineered to direct light from this mode towards the collection optics. In this case, a single mode description of the efficiency $\eta_{\rm ext,S}$ given by $\eta_{\rm ext,S} = \beta \gamma$ may provide an excellent description of the light emission. 
Here, the spontaneous emission factor $\beta = \Gamma_\text{M} / \Gamma = P_\text{M} / P$ factor describes the fraction of the spontaneous emission (or total power) emitted into the optical mode M, % over the total power $P = P_\text{M} + P_\text{R}$, where $P_\text{R}$ is the power into all other modes. 
whereas the transmission $\gamma = P_\text{Lens,M} / P_\text{M}$ is the power detected by the lens $P_\text{Lens,M}$ from the mode M alone. In terms of the Purcell factor $F_\text{P}$, the spontaneous emission $\beta$ factor can be written as~\cite{Barnes2002} 
\begin{align}
	\beta = \frac{\Gamma_\text{M}}{\Gamma_\text{M}+\Gamma_\text{B}} = \frac{F_\text{P} \Gamma_\text{Bulk}}{F_\text{P} \Gamma_\text{Bulk}+\Gamma_\text{B}}, \label{eq_beta_def}
\end{align}
where the total spontaneous emission rate has been written as a sum of the rate $\Gamma_\text{M}$ into the mode M and the background spontaneous emission $\Gamma_\text{B}$ into all other modes. We observe that the $\beta$ factor can be increased either by introducing cavity effects and Purcell enhancement of the spontaneous emission into the cavity mode or by controlling the background emission $P_\text{B}$ using e.g.\ dielectric screening \cite{Bleuse2011} or photonic bandgap effects \cite{Lecamp2007b,MangaRao2007,Arcari2014}. Similarly, the transmission $\gamma$ can be analyzed and optimized, e.g.\ using tapering strategies \cite{Gregersen2008,Gregersen2010a}. As an example, the performance of the micropillar SPS is analyzed in terms of $\beta$ and $\gamma$ in Section \ref{sec:theory_mp}.

\begin{figure} \centering
	\includegraphics{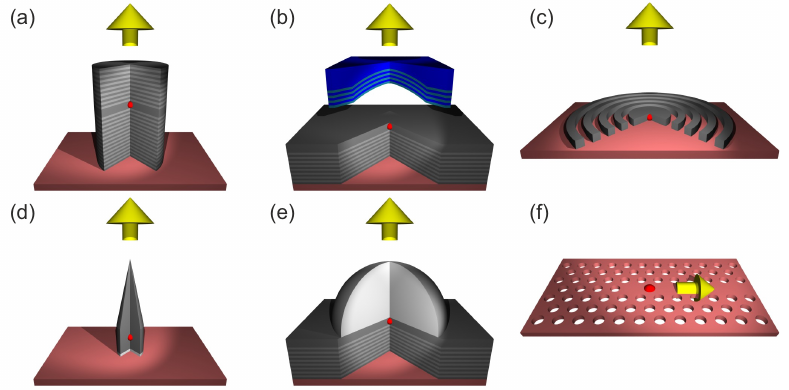}
	\caption{Artistic illustrations of popular QLS designs, with 90$^\circ$ cuts to illustrate the inner region and the position of the QD (red sphere). (a) The micropillar cavity \cite{Somaschi2016, Ding2016, Schlehahn2016}, (b) the open cavity geometry \cite{Tomm2021}, (c) the "bullseye" circular Bragg grating \cite{Sapienza2015,Yao2018, Liu2019,Wang2019}, (d) the photonic nanowire \cite{Claudon2010,Reimer2012}, (e) the microlens \cite{Gschrey2015,Thoma2016} and (f) the photonic crystal waveguide \cite{Arcari2014}.}
	\label{fig_qls_designs}
\end{figure}

An overview of the most successful QLS design approaches is presented in Fig.\ \ref{fig_qls_designs}. The micropillar and the open cavity are narrow-band designs relying on Purcell enhancement to increase the $\beta$ factor. Whereas the micropillar design \cite{Somaschi2016, Ding2016, Schlehahn2016} in Fig.\ \ref{fig_qls_designs}(a) allows for monolithic integration and represents a mature technology, the cavity resonance frequency is not tunable. This limitation is overcome in the fully tunable open cavity geometry \cite{Tomm2021} shown in Fig.\ \ref{fig_qls_designs}(b), even though this design is more sensitive to mechanical vibrations. Broadband approaches include the "bullseye" CBG, the photonic nanowire and the microlens designs. Even though the CBG design \cite{Sapienza2015, Liu2019,Wang2019,Yao2018} depicted in Fig.\  \ref{fig_qls_designs}(c) does feature significant Purcell enhancement, the light extraction mechanism does not rely on the resonant effect and remains high over a wavelength range much broader than the resonance \cite{Yao2018}. Similarly, the photonic nanowire \cite{Claudon2010,Reimer2012} illustrated in Fig.\ \ref{fig_qls_designs}(d) exploits suppression of the background emission rate $\Gamma_\text{B}$ to increase the $\beta$ factor, whereas the broad-band microlens \cite{Gschrey2015,Thoma2016} in Fig.\ \ref{fig_qls_designs}(e) benefits from a classical hemispherical lens effect to direct the light towards the collection optics. Finally, the photonic crystal waveguide geometry \cite{Arcari2014} shown in Fig.\ \ref{fig_qls_designs}(f) exploits the slow light effect near the waveguide band edge to efficiently couple light into the planar waveguide, and light is subsequently extracted typically using grating out-couplers.

\subsection{Modeling of decoherence effects} \label{sec:theory_decoh}

The photon indistinguishability quantified by the visibility of the two-photon interference $V_{\rm TPI}$ takes a value between 0 for distinguishable photons and 1 for perfectly indistinguishable photons. In addition to efficient emission of single photons, most quantum information protocols require also high indistinguishability of the emitted photons. However, for QDs embedded in a solid-state environment, the indistinguishability is compromised by several physical mechanisms leading to decoherence in the emission process.  Numerically, it can be determined from the second-order correlation of the electromagnetic field operator as \cite{Kaer2013a,Kiraz2004} 
\begin{align}
V_{\rm TPI} = \frac{{\int_0^\infty  {\int_0^\infty  {{{\left| {\left\langle {{{\hat a}^\dag }(t + \tau )\hat a(t)} \right\rangle } \right|}^2}dt} d\tau } }}{{\int_0^\infty  {\int_0^\infty  {\left\langle {{{\hat a}^\dag }(t)\hat a(t)} \right\rangle \left\langle {{{\hat a}^\dag }(t + \tau )\hat a(t + \tau )} \right\rangle dt} d\tau } }},
\end{align}
where $\hat{a}^\dag$ ($\hat{a}$) is the creation (annihilation) operator for the output electric field.  Modeling of the indistinguishability then requires an evaluation of the two-time correlation function ${\left\langle {{{\hat a}^\dag }(t + \tau )\hat a(t)} \right\rangle }$. 

\begin{figure} \centering
	\includegraphics{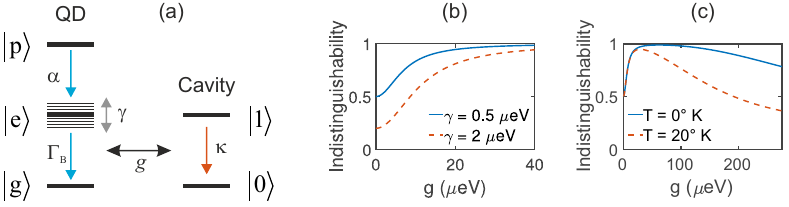}
	\caption{(a) The QD three-level system coupled to an optical cavity with light-matter interaction strength $g$. $\alpha$ and $\Gamma_{\rm B}$ are the decay rates of the respective transitions, whereas $\kappa$ is the cavity escape rate and $\gamma$ is the dephasing rate. (b) The indistinguishability $V_{\rm TPI}$ as function of light-matter interaction strength $g$ for constant pure dephasing rates $\gamma$ of 0.5 $\mu$eV (full) and 2 $\mu$eV (dashed) and $\alpha = \infty$. Indistinguishability vs.\ $g$ where $\gamma$ = 0.5 $\mu$eV + $\gamma_{\rm Ph}$ is the sum of a constant dephasing rate and the phonon contribution given by Eq.\ \eqref{eq_gamma_phonon} for T = 0$^\circ$ K (full) and T = 20$^\circ$ K. Parameters: $\alpha = \infty$, $\eta$ = 0.032 ps$^2$, $\omega_{\rm c}$ = 0.95 meV, $\kappa$ = 100 $\mu$eV, $\Gamma_{\rm B}$ = 1 $\mu$eV and $B$~=~1.}
	\label{fig_id}
\end{figure}

We now consider a QD placed spectrally on resonance inside an optical cavity. For the general case of a QD pumped using non-resonant excitation, we can model the QD as a three-level system with ground state $\ket{\rm g}$, excited state $\ket{\rm e}$ and a pump state $\ket{\rm p}$ as shown in Fig.\ \ref{fig_id}(a). The pump state relaxes to the excited state with a rate $\alpha$ and can subsequently relax to the ground state through spontaneous emission into non-cavity modes with a rate $\Gamma_{\rm B}$. Additionally, we consider a coupling to an optical cavity described by a light-matter coupling constant $g$ and a cavity leakage rate $\kappa$. Finally, the QD is subject to a decoherence mechanism with a rate $\gamma$ as discussed below, leading to uncertainty of the excited state energy level.
The coherent interaction between the QD and the cavity is described in a rotating frame by a Jaynes-Cummings \cite{Gerry2008} Hamiltonian as $\hat{H} = \hbar g (\hat{a}^\dag \hat{\sigma}_{\rm ge} + \hat{a} \hat{\sigma}_{\rm eg})$, where  $\hat{\sigma}_{\rm ij} = \ketbra{\rm i}{\rm j}$ is the dipole operator. The interaction with the environment is then modeled using a master equation formalism \cite{Carmichael1999} for the reduced density operator $\hat{\rho}$ describing the emitter alone. The master equation is given by
\begin{align}
	\frac{d}{dt} \hat{\rho} =  \alpha \mathcal{L}_{\hat{\sigma}_{\rm ep}} (\hat{\rho}) + \Gamma_{\rm Bulk} \mathcal{L}_{\hat{\sigma}_{\rm ge}} (\hat{\rho}) + 2\gamma \mathcal{L}_{\hat{\sigma}_{\rm eg} {\hat{\sigma}_{\rm ge}}} (\hat{\rho}) + \kappa \mathcal{L}_{\hat{a}} (\hat{\rho})   -\frac{i}{\hbar} \left[  \hat{H}, \hat{\rho}\right] , \label{eq_ME}
\end{align}
where loss is included using the Lindblad operator defined as $\mathcal{L}_{\hat{x}}(\hat{\rho}) = \hat{x} \hat{\rho} \hat{x}^\dag - \left(\hat{x}^\dag \hat{x} \hat{\rho} +  \hat{\rho} \hat{x}^\dag \hat{x}\right)/2 $. On the right-hand side of Eq.\ \eqref{eq_ME}, the first two terms describe respectively the p$\rightarrow$e relaxation and the e$\rightarrow$g transition through spontaneous emission, whereas the third and fourth terms represent the pure dephasing mechanism and the cavity leakage of photons, respectively.

\subsubsection{Markovian decoherence: Time jitter and pure dephasing}

In the weak coupling regime of cQED, the QD-cavity coupling strength $g$ is small compared to the total decoherence rate $\gamma _{\rm T} = \gamma + \left(\Gamma + \kappa \right)/2$. This allows for the cavity mode to be adiabatically \cite{Kaer2013a} eliminated from Eq.\ \eqref{eq_ME} leading to an equation for the emitter alone of the form
\begin{align}
	\frac{d}{dt} \hat{\rho} =  \alpha \mathcal{L}_{\hat{\sigma}_{\rm ep}} (\hat{\rho}) + \Gamma_{\rm Bulk} \mathcal{L}_{\hat{\sigma}_{\rm ge}} (\hat{\rho}) + 2\gamma \mathcal{L}_{\hat{\sigma}_{\rm eg} {\hat{\sigma}_{\rm ge}}} (\hat{\rho}) + F_{\rm p} \Gamma_{\rm Bulk} \mathcal{L}_{\hat{\sigma}_{\rm ge}} (\hat{\rho}), \label{eq_ME2}
\end{align}
where the last term on the right-hand side now describes spontaneous emission into the cavity enhanced by the Purcell factor $F_{\rm p} = 4g^2 / \left( \kappa \Gamma_{\rm Bulk}  \right)$. 
Since the field operator for the cavity mode was eliminated, the light emission is typically described in terms of the emitter dipole operator $\hat{\sigma}_{\rm ge}$, where it is assumed that all light emitted from the QD reaches the beamsplitter in a HOM-TPI configuration, and the indistinguishability is then evaluated using the correlation function $\langle \hat{\sigma}_{\rm ge}(t + \tau) \hat{\sigma}_{\rm ge}(t) \rangle$.  
The solution of Eq.\ \eqref{eq_ME} provides the one-time expectation value $\langle \hat{\sigma}_{\rm ge}(t) \rangle$ of the dipole operator, and the two-time expectation value $\langle \hat{\sigma}_{\rm ge}(t + \tau) \hat{\sigma}_{\rm ge}(t) \rangle$ is subsequently obtained using the quantum regression theorem \cite{Carmichael1999} valid for a Markovian environment without memory effects. 

Writing the total spontaneous emission rate as $\Gamma _{\rm T} = \left(F_{\rm p}+1 \right)\Gamma _{\rm Bulk}$, the indistinguishability then takes the form
\begin{align}
	V_{\rm TPI} = \frac{\Gamma _{\rm T}}{\Gamma _{\rm T} + 2\gamma} \frac{\alpha}{\alpha + \Gamma _{\rm T}}, \label{eq_ID1}
\end{align}
describing both the influence of pure dephasing and the p$\rightarrow$e relaxation process in the Markovian regime. Physically, a dominating mechanism for pure dephasing is excited state energy variations due to a fluctuating charge environment \cite{Berthelot2006}. 
The indistinguishability predicted by Eq.\ \eqref{eq_ID1} is presented in Fig.\ \ref{fig_id}(b) for $\alpha = \infty$. We observe that increasing the spontaneous emission rate using Purcell enhancement serves to improve not only the efficiency but also the indistinguishability in the presence of pure dephasing. 
However, for non-resonant excitation with a finite pump relaxation rate $\alpha < \infty$, a trade-off occurs due to the second fraction in Eq.\ \eqref{eq_ID1}. Even though the relaxation from the pump to the excited state does not in itself introduce decoherence in the emission process, a finite relaxation time $\alpha$ results in uncertainty in the photon emission time and reduced temporal overlap for photons impinging on the beamsplitter. Increasing the spontaneous emission rate through Purcell shortens the pulse duration in time and thus amplifies this detrimental effect, which is referred to as time jitter. 
The reduction of the indistinguishability due to time jitter can be avoided either using resonant excitation \cite{He2013,Somaschi2016,Ding2016} or by accelerating the p$\rightarrow$e relaxation time $\alpha$, e.g.\ using a stimulation pulse \cite{Wei2022,Sbresny2022}. 
Thus, Purcell enhancement is generally beneficial to the performance within the Markovian regime. However, this picture relies on decoupling of the Purcell enhancement and the dephasing rate, which is not always a good approximation, as discussed below.

\subsubsection{Non-Markovian decoherence: Phonons}

For a QD in a solid-state material, interaction between the QD and quantized lattice vibrations, phonons, in the bulk environment  \cite{Ramsay2010, Kaer2012, Kaer2013a,Iles-Smith2017a, Denning2020a} represents an additional fundamental decoherehence mechanism. The dominating coupling is with longitudinal acoustic phonons \cite{Besombes2001,Favero2005,Ramsay2010} and leads to an additional contribution to the Hamiltonian given by 
\begin{align}
	\hat{H} = \ketbra{\rm e}{\rm e} \sum_{\textbf{k}}\hbar g_{\textbf{k}} \left(\hat{b}_{\textbf{k}} ^\dag + \hat{b}_{\textbf{k}} \right) +  \sum_{\textbf{k}} \hbar \nu_{\textbf{k}} \hat{b}_{\textbf{k}} ^\dag \hat{b}_{\textbf{k}}, \label{eq_H_ph}
\end{align}
where $\hat{b}_{\textbf{k}}^\dag$ ($\hat{b}_{\textbf{k}}$) is the creation (annihilation) operator for the phonon mode with frequency $\nu_{\textbf{k}}$ and wave vector \textbf{k} the output electric field, and $g_{\textbf{k}}$ is the QD-phonon coupling strength. Whereas the QD-phonon interaction is completely described by Eqs.\ (\ref{eq_ME},\ref{eq_H_ph}), the calculation of the two-time expectation function is challenging: The interaction with the phonons again leads to excited state energy fluctuations this time due to a deformation of the energy potential producing non-Markovian memory effects in the environment in the short-time limit of $\sim$ 5 ps, such that the quantum regression theorem can no longer be used to determine the two-time correlation functions. Accurate modeling of the phonon interaction has been performed using an exact diagonalization technique \cite{Kaer2012,Kaer2013a}, which however, quickly becomes numerically demanding with increasing size of the Hilbert space. More recently, a numerically exact method based on a time-evolving matrix product operator \cite{Strathearn2018} was proposed, however this method is also computationally demanding.

Significant simplification was achieved with the introduction of the polaron transformation \cite{Iles-Smith2017a}, allowing for a formulation of a Born-Markov master equation in the polaron frame and subsequently for approximate analytic expressions for the efficiency and indistinguishability. In the weak coupling regime and in the limit of weak QD-phonon interaction, the dephasing rate $\gamma_{\text{Ph}}$ of the zero-phonon line becomes
\begin{align}
	\gamma_{\text{Ph}} = 2\pi \left( \frac{g B}{\kappa} \right)^2 J_{\rm Ph}(2 gB) \text{coth} \left(\frac{\hbar g B}{k_{\rm B} T}\right) \label{eq_gamma_phonon},
\end{align}
where $B$ is the Franck-Condon factor describing the overlap of the lattice configurations of the excited and ground states and $J_{\rm Ph}(\omega)$ is the phonon spectral density. In the case of a spherical QD placed in a bulk material, the phonon spectral density is given by $J_{\rm Ph}(\omega) = \eta \omega^3 \text{exp}(-\omega^2/ \omega_{\rm c} ^2)$ \cite{Nazir2016}, where $\eta$ is the interaction strength and $\omega_{\rm c}$ is the cutoff frequency inversely proportional to the QD length. 

Since the phonon interaction strength depends on the QD dressed state energy separation \cite{Iles-Smith2017a}, which depends in turn on the QD-cavity coupling strength $g$, the phonon-induced decoherence cannot be accurately modeled using a fixed pure dephasing rate independent of the remaining parameters. 
The indistinguishability computed from Eqs.\ (\ref{eq_ID1},\ref{eq_gamma_phonon}) is presented in Fig.\ \ref{fig_id}(c) as a function of the light-matter interaction strength $g$. We observe that the increase in the total emission rate through Purcell enhancement is initially beneficial for indistinguishability. However, as $g$ increases, the phonon-induced decoherence rate $\gamma_{\rm Ph}$ increases, and we observe a reduction in the indistinguishability with increasing $g$ even at 0$^\circ$ K. 
As $g$ increases even further, the strong coupling regime of cQED is reached. Here, the emission spectrum is split into two hybrid polariton states and phonon-induced transitions between the polariton states occur \cite{Denning2020a, Iles-Smith2017a} which is detrimental to the indistinguishability. 

We conclude that whereas the implementation of Purcell enhancement appears beneficial to both increase the spontaneous emission $\beta$ factor in Eq.\ \eqref{eq_beta_def} and to overcome pure dephasing in Eq.\ \eqref{eq_ID1}, phonon-induced decoherence results in an inherent trade-off between the achievable efficiency and indistinguishability in the cQED  design approach, as exemplified below for the micropillar geometry. 

\subsection{Performance of the micropillar single-photon source} \label{sec:theory_mp}

\begin{figure} \centering
	\includegraphics{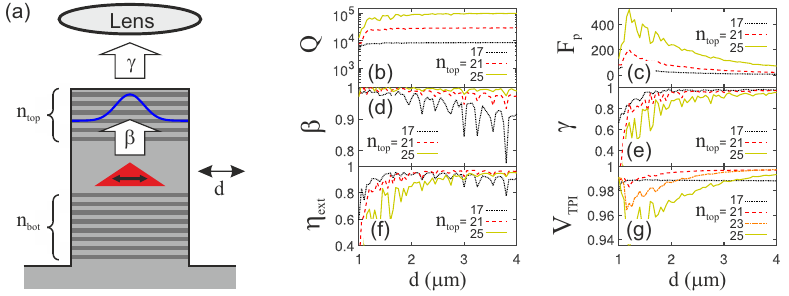}
	\caption{(a) Sketch of the micropillar geometry consisting of a QD sandwiched between vertical DBRs. (b) The Q-factor, (c) the Purcell-factor $F_{\rm P}$, (d) the spontaneous emission $\beta$-factor, (e) the transmission $\gamma$, (f) the collection efficiency $\eta_{\rm ext}$ and (g) the photon indistinguishability $V_{\rm TPI}$ as function of pillar diameter $d$ for varying number of top DBR layer pairs $n_{\rm top}$ with $n_{\rm bot}$ = 40. See Ref.~\cite{Wang2020b} for additional geometrical parameters.}
	\label{fig_micropillar}
\end{figure}

We now consider the specific example of the micropillar cavity-based SPS. The geometry features a QD placed in a vertical $\lambda$ cavity generally featuring an asymmetric distributed Bragg reflector (DBR) configuration illustrated in Fig.\ \ref{fig_micropillar}(a) enabling light emission through the top. The micropillar has been subject to intense numerical investigation, e.g.\ to predict and explain fundamental physics such as the diameter-dependent variations in the Q-factor \cite{Lalanne2004,Gregersen2010} and to increase its Q/V-ratio for strong-coupling experiments \cite{Vuckovic2002,Lermer2012} using Bloch-wave engineering \cite{Lalanne2003}. 

The performance of the micropillar for the SPS application was analyzed in Refs.\ \cite{Wang2020b,Wang2021}, and the main figures of merit are presented in Fig.\ \ref{fig_micropillar}(b-g). The micropillar geometry relies on cavity QED effects to ensure high efficiency, and the Q-factor is presented in Fig.\ \ref{fig_micropillar}(b) as a function of pillar diameter for increasing number $n_{\rm top}$ of top DBR layer pairs. In addition to an increase in the overall Q-factor, oscillatory variations are observed in the low-diameter high-Q limit resulting from interaction with higher-order Bloch modes \cite{Reitzenstein2009,Gregersen2010}. These oscillations are directly observed in the Purcell-factor shown in Fig.\ \ref{fig_micropillar}(c), where an additional reduction in $F_{\rm P}$ occurs due to a decrease in the mode volume with $d$. As predicted from Eq.\ \eqref{eq_beta_def}, the large Purcell-factor $F_{\rm P}$ results in a large $\beta$-factor shown in Fig.\ \ref{fig_micropillar}(d). Again, significant oscillations are observed, however their origin is not the variations in the Q-factor. Instead, they are resulting from a periodic variation \cite{Wang2021} of the background spontaneous emission rate $\Gamma_{\rm B}$. In Ref.\ \cite{Gines2022}, a careful choice of pillar diameter was made to ensure a -$\beta$factor at a peak position, resulting in a final 69\% collection efficiency.

However, the performance of the micropillar suffers from fundamental trade-offs. Since a large pillar diameter results in a narrow far field emission pattern, the transmission $\gamma$ in Fig.\ \ref{fig_micropillar}(e) generally increases with $d$. Here a trade-off between high $\beta$ and $\gamma$ is observed, resulting in the collection efficiency $\eta_{\rm ext}$ shown in Fig.\ \ref{fig_micropillar}(f) taking its maximum value in the $d \in [1.5,2] \mu$m regime for $n_{\rm top}$ = 17. An initial procedure to improve the collection efficiency could be to increase the number of layer pairs in the DBRs, which would increase $\beta$ also for large diameters. However, here the onset of the strong coupling regime limits the indistinguishability in the presence of phonon-induced decoherence: We observe in Fig.\ \ref{fig_micropillar}(g) that increasing $n_{\rm top}$ beyond 21 layer pairs leads to a significant reduction in the indistinguishability. This leads to an inherent trade-off between the achievable efficiency and indistinguishability for the micropillar SPS, the product of the two taking a maximum value of $\eta_{\rm ext} V_{\rm TPI} \sim$ 0.95 \cite{Wang2020b}. Realizing QLSs with performance beyond this value requires new design concepts such as the "hourglass" geometry \cite{Osterkryger2019,Gaal2022}, which exploits suppression of the background spontaneous emission rate to increase $\beta$ further towards unity while avoiding the strong coupling regime.

\section{Methods to fabricate semiconductor quantum dots}\label{sec:epi}

After having discussed the theoretical background, we now turn to the fabrication of high-quality semiconductor QDs  with a focus on epitaxial growth. In general, there are different approaches to create QDs capable of confining the motion of free charge carriers (electrons in conduction band states and holes in valence band states) in three dimensions. Here, we restrict our attention to optically-active QDs, in which both carrier types are confined in the same region of a direct-bandgap semiconductor.

Arguably, the simplest method to obtain QDs relies on chemical synthesis of colloidal nanocrystals~\cite{Park2021}. Such QDs have been employed for pioneering demonstration of non-classical light emission~\cite{Michler2000nat} and are widely used in classical optoelectronic devices such as displays and also as fluorescent markers. Their interest in quantum technologies is however rather limited, mostly because the confined carriers are typically located close to the free surface of the nanostructures, leading to significant {\em interaction with surface states} and consequent deterioration of their quantum optical properties. In addition, the preparation from solution may lead to higher impurity concentrations compared to QDs obtained via epitaxial growth methods.

In the following, we provide a brief introduction to the basics of epitaxial growth of semiconductors in Section~\ref{sec:epitaxy} and then move to illustrate the current methods employed to make high-quality QDs with different properties for quantum information science and technology.  

\begin{figure}[h!]
\centering\includegraphics[width=12cm]{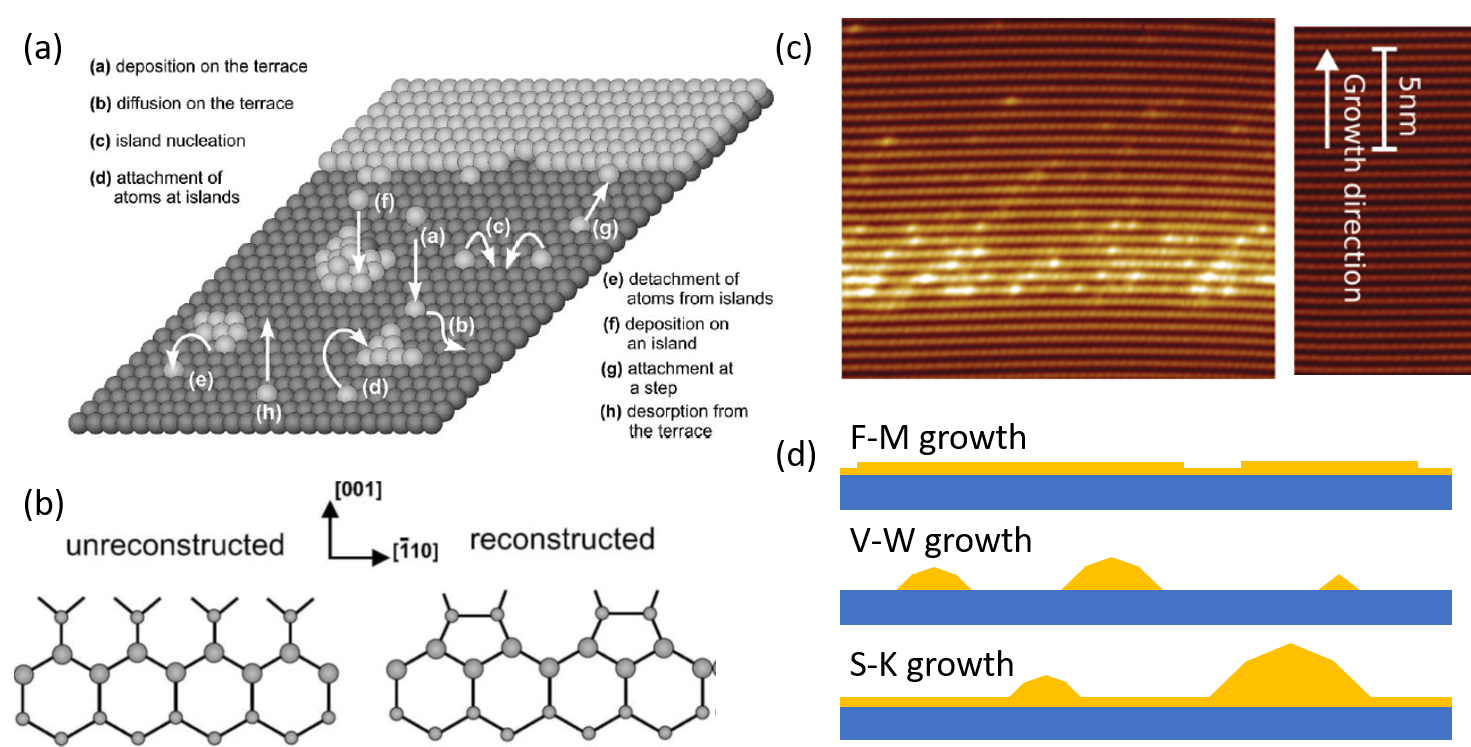}
\caption{Overview of phenomena related to epitaxial growth of semiconductors. (a) Basic processes for atoms and molecules deposited on surfaces. (b) Illustration of surface reconstruction. (c) Cross-sectional scanning tunneling microscopy (STM) image of a layer of InGaAs in a GaAs matrix showing disorder due to mixing and surface segregation. (d) Classification of growth modes. (a, b) Reprinted from Surface Science Reports, Vol 43, B. Voigtl\"ander, Fundamental processes in Si/Si and Ge/Si epitaxy studied by scanning tunneling microscopy during growth, Pages 127-254, Copyright (2001), with permission from Elsevier~\cite{VOIGTLANDER2001}, (c) Reprinted from \href{https://doi.org/10.1063/1.3577960}{\textit{Keizer et al. 2011}} \cite{Keizer2011}, with the permission of AIP Publishing.}
\label{FIG:EPI}
\end{figure}

\subsection{General concepts of epitaxial growth of semiconductors} \label{sec:epitaxy}
% Definition 
Epitaxial growth consists in the growth of crystalline layers by deposition of atoms or molecules on a clean surface of a crystalline semiconductor substrate and at sufficiently high substrate temperature to allow them to be incorporated and arranged in a periodic fashion, compatible with the crystalline structure of the substrate~\cite{Pohl2013}. For epitaxial growth to occur, the deposited material and substrate must be capable of adopting compatible crystal structures. 

% Substrates
The substrates of interest here are typically (001)- and (111)-oriented GaAs and InP (with zincblende crystal structure) and, in some case, Si. (001)-oriented substrates are the standard in electronic and optoelectronic industry, while many different orientations have been used in the past to study QD formation~\cite{Liang2006}. 
The layer growing on the substrate is called {\em epilayer}, and we distinguish between {\em homoepitaxy} and {\em heteroepitaxy}, depending on whether the epilayer consists of the same material as the substrate (beside possible doping) or a different material.  In most of the circumstances, the different materials in a {\em heterostructure} have the same crystal structure and differ in their {\em lattice constants}, giving rise to {\em strained growth}. In this case, the {\em lattice-mismatch} $\epsilon$ is defined as the relative difference in the in-plane lattice constants of epilayer $a_e$ and substrate $a_s$:
\begin{equation} \label{eq:latticemism}
    \epsilon=(a_s-a_e)/a_e
\end{equation}. 
In the absence of strain relaxation, the epilayer adapts its in-plane lattice constant to the underlying substrate, which exerts stress on the epilayer, leading to an in-plane strain $\epsilon<0$ ($\epsilon>0$) for compressive (tensile) strain.

% common growth methods
To limit impurities, deposition takes place from the vapor phase rather than from the liquid phase. The most common physical- and chemical-vapor deposition techniques used for fabricating QDs are the molecular beam epitaxy (MBE) and the metal-organic-vapor-phase-epitaxy (MOVPE), respectively~\cite{Pohl2013}. In the former, atomic or molecular beams are obtained by thermal or electron-beam heating of highest-purity source materials in ultrahigh vacuum conditions and atoms or molecules impinge on the substrate after travelling ballistically in the deposition chamber. In MOVPE, molecules containing the elements to be deposited are transported by a carrier gas into a reactor containing the heated substrate, where they decompose into the desired species and volatile molecules. The first step after introducing a substrate into an epitaxy system consists in oxide removal and growth of a {\em buffer layer} (usually homoepitaxial) with a thickness of some 100~nm to place the active structures away from the original substrate surface, which usually contains defects and impurities.

%Physical processes at surfaces
Upon {\em deposition} of the desired atoms on the surface, the adsorbed atoms (or {\em ``adatoms"}) {\em diffuse} on {\em terraces} and can {\em attach} to surface features like steps as well as other adatoms and adatom clusters, named {\em islands}. The reverse processes of adsorption and attachment are desorption and detachment, respectively. In thermal equilibrium, these processes would counterbalance each other. Crystal growth is therefore an inherently non-equilibrium process, as the net adsorption rate must be higher than the desorption rate for the crystal to grow on top of the substrate. These processes are illustrated in Fig.~\ref{FIG:EPI}(a).

% Everything happens at the surface
Although the overall system consisting of substrate, epilayer, and source vapor is not in a global thermodynamic equilibrium, surface processes can be often treated in a quasi-equilibrium framework, allowing us to see them as driven by (local) free-energy minimization. It is important to note that under common growth conditions, all relevant processes occur at the surface layers and -- in some cases -- in the first one or two subsurface monolayers. The reason is that common semiconductors are characterized by strong covalent bonds, and the energy necessary to allow an atom to move inside the ``bulk'' lattice is much larger than the energy required to break surface bonds. This means that, once atoms are buried below a few monolayers of material, they can be considered as immobile. This fact reduces considerably the complexity of the theoretical description of growth processes, which can be done either using continuum or atomistic models.

% Description of surface diffusion and driving forces
Surface diffusion is in general anisotropic. Preferential directions for diffusion can be caused by (i) {\em gradients in surface chemical potential}, which, in turn, can originate from local surface curvature, local strain, local composition fluctuations, atomic steps, as well as by (ii) {\em surface reconstructions}, i.e. the rearrangement of surface atoms in periodic structures with unit cells larger than the bulk unit cell to reduce the surface energy due to dangling bonds (see Fig.~\ref{FIG:EPI}(b)). As an example of (ii), the (4$\times$2) reconstruction of As-terminated GaAs(001) surface is characterized by dimers, making diffusion along the $[1\overline10]$ direction faster than along the perpendicular $[110]$ direction (see Fig.~\ref{FIG:EPI}(c)).
If we neglect the effect of surface reconstruction we can write the chemical potential $\mu$ of an adatom as~\cite{Songmuang2003,Ledentsov1996}:
\begin{equation}
\label{eq:chempot}
\mu(\vec{r})=\mu_0+\Omega E_s(\vec{r})+\gamma\Omega\kappa(\vec{r})-\frac{\zeta\Omega\theta(\vec{r})}{a},
\end{equation}
where $\mu_0$ is the chemical potential of adatoms on an unstressed surface, $E_s$ the elastic energy density due to local strain, $\Omega$ the atomic volume. The third term is the surface energy contribution, where $\gamma$ is the surface energy and $\kappa$ the surface curvature. Finally, the last term gives account to {\em surface segregation}, with $\zeta$ the energy benefit (per unit area) of having a surface composed of the deposited species compared to the underlying species, $a$ the lattice constant and $\theta$ varying between 0 and 1 depending on whether the adatom moves on a layer with atoms of the same species or another. Segregation leads to the swapping of deposited and surface atoms when the latter allow the surface to have lower energy. An example is represented by the overgrowth of an InAs surface with GaAs. Because of the lower surface energy of InAs compared to GaAs, In atoms tend to float on top of the Ga atoms, naturally leading to a smearing of interfaces between layers of different materials. The cross-sectional STM image of Fig.~\ref{FIG:EPI}(d), with In atoms appearing brighter than Ga atoms, clearly illustrate this phenomenon. Surface segregation also leads to inhomogeneous vertical composition profiles when species with different surface energies, such as InAs and GaAs are co-deposited to form alloys~\cite{Walther2001}.

% Examples
From Eq.~\ref{eq:chempot}, we see that the tendency of the growing layer to minimize the chemical potential allows us to describe {\em capillarity} effects, i.e. the spontaneous planarization of rough surfaces or dimples, which are characterized by negative curvature. 
In general, there is a delicate interplay between different contributions to the chemical potential, which may lead to surface roughening instead of smoothing. As an important example, elastic energy (or strain energy) can favor the occurrence of three-dimensional (3D) {\em islands}. According to nucleation theory, when the number of adatoms contained in an island exceeds a certain critical size (critical nucleus), the island becomes stable and growth leads to a drop in chemical potential. Islands can also form without any nucleation barrier~\cite{Tersoff2002}.

% Growth modes
In spite of the rich physics of surface phenomena, the result of depositing an epilayer on top of a crystalline substrate can be schematically summarized according to three growth modes, see Fig.~\ref{FIG:EPI}(d):  the Frank-van-der Merwe (F-M) or layer-by-layer growth, the Volmer-Weber (V-W) or island growth, and the Stranski-Krastanow or layer-plus-island growth mode. Whether one or the other growth mode occurs depends on the relative properties of the epilayer and substrate. In particular, a useful classification concentrates on the relation between the surface energy of the substrate material $\gamma_s$, of the epilayer $\gamma_e$, and the interface energy $\gamma_{\rm is}$. The F-M growth mode takes place whenever $\gamma_s\geq\gamma_e+\gamma_{\rm is}$, i.e. in the case in which substrate wetting is energetically favored, the V-W in the opposite case, and the S-K whenever $\gamma_{\rm is}$ increases with the epilayer thickness, so that – above a critical thickness – the inequality changes sign. All these growth modes are relevant for the growth of QDs, as we will see below. 
Before concluding this brief overview on epitaxial growth of semiconductors, we mention that {\em crystal defects} should be avoided. In general, such defects locally disrupt the crystal periodicity and introduce localized electronic states. These can act as non-radiative recombination centers, reducing the quantum efficiency of the QDs because of enhanced defect-induced non-radiative emission, or as traps for charges, leading to charge noise. Defects can be distinguished into {\em point defects}, such as intersitials, vacancies, and unintentional impurity atoms either at lattice sites or as interstitials and {\em extended defects}, such as dislocations, stacking faults, antiphase boundaries, and surfaces with associated dangling bonds, impurities and oxides. 

\begin{figure}[h!]
\centering\includegraphics[width=12cm]{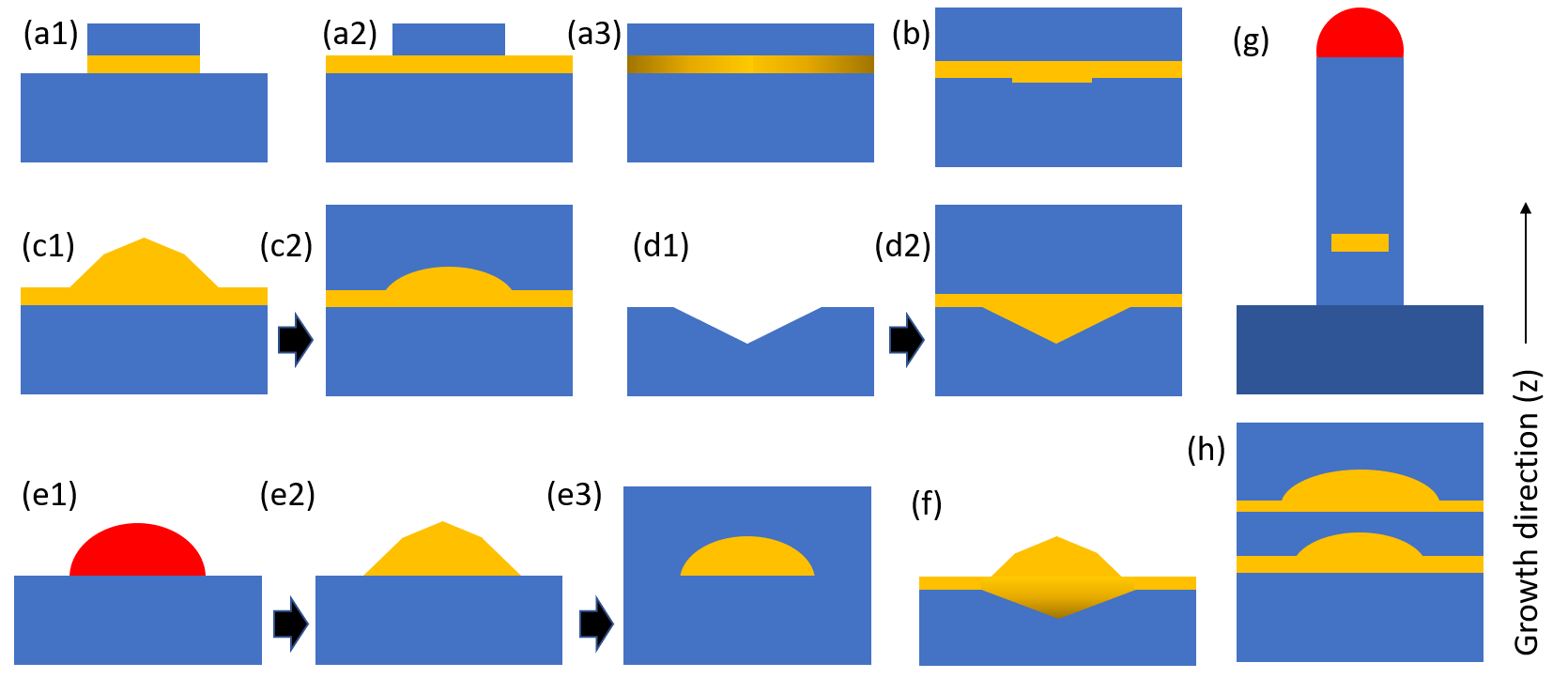}
\caption{Overview of the main types of epitaxial QDs defined in semiconductor heterostructures containing a region of a semiconductor with lower energy bandgap (QD material, in orange) embedded in a matrix with larger bandgap (barrier material, blue): (a1) and (a2) shallow-etched quantum wells, leading to in-plane confinement in addition to vertical confinement; (a3) QDs by lateral bandgap-modulation in quantum wells; (b) Natural QDs in quantum wells; (c1-c2) Self-assembled S-K QDs before and after capping; (d1-d2) QDs in nanoholes; (e1-e3) QDs by droplet epitaxy (a metal droplet (in red) is recrystallized and capped; (f) Site-controlled S-K QDs; (g) QDs in nanowires; (h) Vertically staked QDs. In all sketches the substrate is at the bottom and the growth proceeds towards the top.}
\label{FIG:QDGROWTH}
\end{figure}

\subsection{Epitaxial quantum dots}
Epitaxial QDs are mostly obtained out of heterostructures containing a region of a semiconductor with lower energy bandgap (QD material) embedded in a matrix with larger bandgap (barrier material). In our treatment, we focus on heterostructures with {\em type-I band alignment}, for which the conduction band edge and valence band edge of the QD material lie inside the energy bandgap of the barrier material. We classify the main types of epitaxial QDs with reference to Fig.~\ref{FIG:QDGROWTH}: QDs in quantum wells, obtained either by post-growth definition of in-plane confinement regions (a1-a3) or by spontaneous exciton localization (b); S-K QDs (c1-c2); QDs in nanoholes, obtained by filling self-assembled or lithographically-defined surface dimples via capillarity-driven diffusion (d1-d2); QDs obtained by the droplet epitaxy method (e1-e3); Site-controlled S-K QDs obtained by S-K growth on lithographically-patterned nanoholes (f); QDs in nanowires, obtained by vapor-liquid-solid growth (g); (h) Vertically stacked QDs, or {\em QD molecules}. 

We further distinguish among {\em site-controlled} and {\em self-assembled} QDs. The position of the former on the substrate is determined a priori, facilitating deterministic device fabrication, while the latter are randomly placed on the substrate, requiring registration methods for device fabrication (see Section~\ref{sec:deterministic_fabrication}). Since  site-controlled QDs require substrate manipulation {\em before growth} (and associated contamination/defects), it remains challenging to obtain the same high optical quality as for self-assembled QDs. This is the reason why self-assembled growth keeps being pursued by most of the research groups in spite of additional steps {\em after growth} and foreseeable limitations for scalability, see Section~\ref{sec:SCQDs}. For single-QD devices, typical inter-QD distances are of the order of the wavelength of the emitted light, allowing QDs to be individually addressed by far-field optics.  

Another important criterion for the choice of the QD ``hardware" is the spectral range of their optical transitions. This is determined by the energy bandgaps of the used materials, the extent of the confinement region, and the strain present in the structures. In general, the emission wavelength increases for decreasing energy bandgap of QD and barrier material, for increasing QD size, and for decreasing compressive strain. Each material combination allows a certain spectral range to be accessed. As for any heterostructure, the lattice-mismatch between epilayer and substrate must be limited to avoid the occurrence of dislocations. This limitation is relaxed in the case of QD in nanowires. In order of increasing wavelength, the following material combinations are commonly used: GaN/AlGaN, InGaN/GaN, InP/In(Al,Ga)P, GaAs/AlGaAs, In(Ga)As/GaAs, InAs/InP or InAs on In(Ga,Al)As. Since epitaxial QDs usually have a flat morphology (height/width ratio of the order of $\sim$0.05--0.3), the size along the growth direction is the one that mostly determines the transition energy. 

\subsubsection{Quantum dots in quantum wells}
Historically, the first methods to create 3D confinement in semiconductors were based on introducing quantum wells, in which the carrier motion is free only in the quantum well plane (for a review, see Ref.~\cite{Michler2009}). Lateral confinement could be introduced by deep or shallow etching~\cite{Bayer1998} (see sketches in Fig.~\ref{FIG:QDGROWTH}(a1, a2)), local strain modulation~\cite{Lipsanen1995}, local interdiffusion promoted by laser irradiation~\cite{Brunner1992} or, in special cases, via hydrogen irradiation~\cite{Birindelli2014} (see sketches in Fig.~\ref{FIG:QDGROWTH}(a3)). In all cases to laterally confine the carrier motion in the quantum well plane, the quantum wells needed to be located at a few tens of nanometers from the sample surface, leading to significant interaction of the confined excitons with surface states and consequent deterioration of the optical properties of the resulting QDs. Because of the limited optical quality of the resulting QDs, these methods have been useful for pioneering studies, but are now practically abandoned.
The above-mentioned methods have the appealing feature of allowing the position of the QDs to be controlled. Lateral confinement in quantum wells is also achieved without intentional lateral modulation, e.g.\ due to local random fluctuations in the thickness of the quantum well~\cite{Gammon1996}. At sufficiently low temperatures, excitons get confined in areas of the quantum well with locally larger thickness. The resulting QDs are often referred to as {\em natural QDs} and appear also as a consequence of local alloy fluctuations, which may result in regions of effective lower energy bandgap (see, e.g. Fig.~\ref{FIG:EPI}(c)). Natural QDs are characterized by poorly defined structural properties, but -- in case of lateral extensions of the order of several tens of nm -- they feature very high oscillator strengths~\cite{Andreani1999}, which are pivotal to experiments where strong light-matter interaction is needed~\cite{Peter2005}.

\begin{figure}[h!]
\centering\includegraphics[width=12cm]{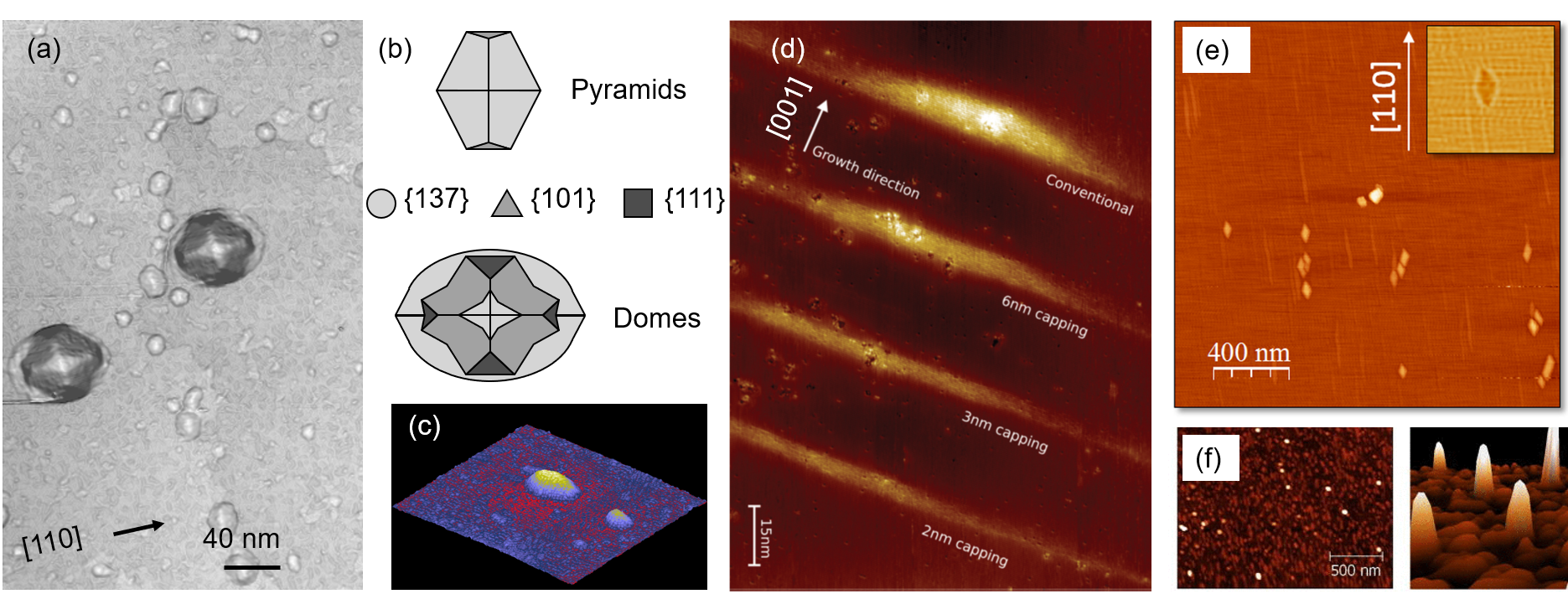}
\caption{Example of Stranski-Krastanow QDs: InGaAs QDs on GaAs(001) and InP(001) substrates. (a) STM image of InGaAs QDs obtained by depositing 1.8~monolayers of InAs on GaAs at a substrate temperature of 500$^\circ$C via MBE. (b) Two families of faceted nanocrystals are observed, ``pyramids'' and ``domes''. (c) A 3D view of a pyramid with a lateral size of $\sim$10~nm and a height of $\sim$3~nm. (d) Cross-sectional STM image of a stack of InGaAs QDs overgrown with GaAs without any growth interruption (top) or with the ``In-flush method'' after partial capping with the indicated amount of GaAs prior to complete overgrowth. (e) InAs QDs on InP(001) and (f) InAs QDs on InAlGaAs lattice-matched to InP(001) grown by MBE. (a-c) Reprinted from Journal of Crystal Growth, Vol 278, G. Costantini et al., Pyramids and domes in the InAs/GaAs(0 0 1) and Ge/Si(0 0 1) systems, Pages 38-45, Copyright (2005), with permission from Elsevier~\cite{Costantini2005}, (d) Reprinted from \href{https://doi.org/10.1063/1.3577960}{\textit{Keizer et al. 2011}} \cite{Keizer2011}, with the permission of AIP Publishing, (e) Reprinted figure with permission from \href{https://doi.org/10.1103/PhysRevApplied.8.014013}{\textit{Skiba-Szymanska et al. 2017}} ~\cite{Skiba-Szymanska2017} Copyright (2017) by the American Physical Society, (f) Reprinted from \href{https://doi.org/10.1063/1.4861940}{\textit{Yacob et al. 2014}} \cite{Yacob2014}, with the permission of AIP Publishing.}
\label{FIG:InGaAsQDs}
\end{figure}
\subsubsection{Quantum dot fabrication via the Stranski-Krastanow method}\label{sec:fab_SK}
The most common method to fabricate high quality QDs is via self-assembly of crystalline 3D islands in the Stranski-Krastanow growth mode. Although the initial S-K concept~\cite{stranski1939} did not involve strain, in the context of QD growth, the common driving force leading to a change from a layer-to-layer to a layer-plus-island growth is usually strain due to the lattice mismatch between deposited material and substrate (see Eq.~\ref{eq:latticemism}). With increasing amount of deposited material $t_e$, the elastic energy in the {\em wetting layer}  increases $\propto \epsilon t_e^2$ up to a critical thickness, above which 3D island formation is favored due to their capability of partially relaxing elastic energy through the free surfaces and substrate. (In our description of the growth modes illustrated in Fig.~\ref{FIG:EPI}(d) we can imagine that the interface energy $\gamma_{\rm is}$ increases with increasing elastic energy). From an atomistic perspective, 3D island growth is favored by the fact that deposited atoms have lower chemical potential (see second term in Eq.~\ref{eq:chempot}) on top of an island compared to the surrounding planar surfaces. In turn, this is because the local lattice constant at the island top is closer to $a_e$ compared to the surrounding regions, where it is close to the substrate lattice constant $a_s$ and strain is this larger. Early work on the growth and development of S-K QDs include the first report of intense luminescence from In-rich coherent clusters in GaAs~\cite{Goldstein1985,Leonard1993} and the first studies of such nanostructures by atomic force microscopy (AFM)~\cite{Moison1994}.

To illustrate the main properties of S-K QDs, we focus on the prototypical example of QDs obtained by depositing InAs on GaAs(001) substrates, with $\epsilon\simeq 7\%$. In this case, the wetting layer thickness is about 1.2-1.7~monolayers depending on the growth temperature~\cite{Xu2005}, corresponding to $\sim 0.5$~nm. Figure~\ref{FIG:InGaAsQDs}(a) shows an STM image of 3D islands obtained by deposition of nominally 1.8~monolayers of InAs on GaAs by MBE at a substrate temperature of $500^\circ$C followed by immediate cooling to room temperature. Under such conditions, a {\em bimodal} distribution of faceted islands is observed. The small (large) islands, with a height up to about 4~nm (15~nm) are referred to ``pyramids'' (``domes'') and are bound by relatively shallow $\{137\}$ (steep $\{110\}$ and $\{101\}$) facets (see sketches in Fig.~\ref{FIG:InGaAsQDs}). Past investigations have shown that a morphological transition occurs between pyramids and domes at a critical volume~\cite{Kratzer2006}, similar to other material systems~\cite{Costantini2004} and that part of the material in the wetting layer is consumed by the islands after their formation. From Fig.~\ref{FIG:InGaAsQDs}(a) we also see that pyramids tend to nucleate close to steps or terraces. In an elegant experiment, Bart et al.~\cite{Bart2022} have recently demonstrated that a larger roughness of the GaAs surface leads to higher density of InAs QDs due to local thickening of the wetting layer. A bimodal size distribution leads to a large {\em inhomogeneous broadening} of the emission wavelength of the resulting QD ensembles and is often undesired. For this reason, a growth interruption is usually introduced after island formation to promote the growth of domes at the expense of pyramids due to {\em ripening}. To obtain ensembles with sufficiently large interdot distance for single-QD devices, growth is usually performed at relatively high substrate temperatures (above $\sim 490^\circ$C) and low InAs deposition rate (less than $\sim 0.05$~monolayers/s) -- to reduce the island nucleation rate -- and by carefully tuning the amount of deposited InAs and/or using gradients in local thickness, roughness, or temperature~\cite{Rastelli2004,Bart2022}. Since the maximum temperature is limited by In desorption and the minimum rate by the time required to obtain QDs, it remains challenging to obtain consistently low surface densities across full wafers. It is also important to note that S-K islands resulting from InAs deposition on GaAs are unavoidably alloyed~\cite{Kegel2000} because of Ga-In {\em intermixing} occurring during growth, so that the resulting QDs are often referred as In(Ga)As or InGaAs QDs. Intermixing is favored by entropy and allows for a reduction of elastic energy in the layer and substrate.

For stable operation and to avoid interaction with surface states, the QDs need to be overgrown with a semiconductor layer. This step, usually consisting of GaAs overgrowth in the case of InGaAs QDs, generally brings in strong modifications in the structural properties of the S-K islands~\cite{Rastelli2001, Costantini2006}.  Specifically, the In-rich island top is driven away by the GaAs, leading to a reduction of QD height. This effect, which is due to the fact that Ga atoms tend to ``avoid'' the strained island top (regions of high chemical potential for Ga) and In atoms tend to wet the surrounding Ga-rich surface because of its lower surface energy, can be enhanced by interrupting the GaAs overgrowth after deposition of a layer with thickness $t_{\rm   cap}$, followed by an increase of substrate temperature to desorb In, resulting in QDs with height close to $t_{\rm  cap}$, as illustrated in Fig.~\ref{FIG:InGaAsQDs}(d). Through this method, referred to as ``In-flush'' or ``partial-capping-and annealing''~\cite{Fafard1999,Keizer2011,Wang2006}, the low-temperature ground-state emission wavelength of the InGaAs QDs can be controllably blue-shifted from $\sim$1200~nm to $\sim$890~nm. For proof-of-principle experiments, wavelengths below $\sim 950$~nm are favorable because of the availability of Si-based detectors and cameras and Ti:Sa lasers. An alternative method to blue-shift the emission wavelength of InGaAs QDs consists in post-growth {\em rapid thermal processing}, resulting in In-Ga interdiffusion~\cite{Fafard1999,Rastelli2004}. Since bulk-intermixing is involved in this process (with higher activation energies compared to surface processes), processing temperatures exceeding 800$^\circ$C are usually required and the resulting QDs have very different properties compared to InGaAs QDs obtained with the In-flush method: the confinement potential is much larger and the In fraction in the alloy is lower, resulting, e.g.\ in increased oscillator strengths~\cite{Langbein2004}.

On the other hand, to obtain light emission in the telecom O-band (about 1300~nm) at low temperatures, QD flattening during capping should be avoided. This can be achieved by capping the InGaAs QDs with an InGaAs {\em strain-reducing layer} instead of pure GaAs~\cite{yeh2000,Alloing2005}. The presence of In in this layer reduces the average lattice mismatch between deposited material and island top, facilitating overgrowth, and enriches the surface with In, limiting the out-diffusion of QD material to the surrounding surface. 

It should be noted that the maximum wavelength of InAs QDs is set by the energy bandgap of bulk InAs (415~meV at cryogenic temperatures, corresponding to a wavelength of almost 3~$\mu$m), so that emission in the C-band is in principle easy to reach. In addition to the above-mentioned material intermixing, which leads to alloyed InGaAs QDs with usually more than 40\% Ga fraction, the large lattice mismatch between InAs and GaAs limits the maximum QD size before {\em plastic relaxation} (i.e. defect formation)~\cite{Marzegalli2007} and -- at the same time -- the large compressive stress exerted by GaAs on InAs leads to a substantial bandgap increase of the QD material. Since strain is one of the main driving forces for alloying and for bandgap increase, the obvious solution is growing InAs QDs on substrates with $a_s$ closer to the lattice constant of InAs. The most common solutions are either InP substrates (with $\epsilon\simeq-3\%$), or InGaAs {\em virtual substrates} (or {\em metamorphic buffers}) grown on GaAs. High-quality emission in the C-band has been demonstrated for InAs QDs embedded in InAlGaAs barriers lattice-matched to InP(001) grown by MBE~\cite{Benyoucef2013,Yacob2014} (see Fig.~\ref{FIG:InGaAsQDs}(f)) as well as in InP barriers~\cite{Skiba-Szymanska2017} (see Fig.~\ref{FIG:InGaAsQDs}(e)). Metamorphic buffers are layers capable of accommodating the lattice mismatch between an In$_x$Ga$_{\rm 1-x}$As final layer acting as substrate and the GaAs substrate through {\em misfit dislocations}. Excellent results have been recently achieved by MOVPE~\cite{Sittig2022} and led to InAs/InAsGaAs QDs emitting in the C-band and grown on cost-effective GaAs(001). 

\begin{figure}[h!]
\centering\includegraphics[width=12cm]{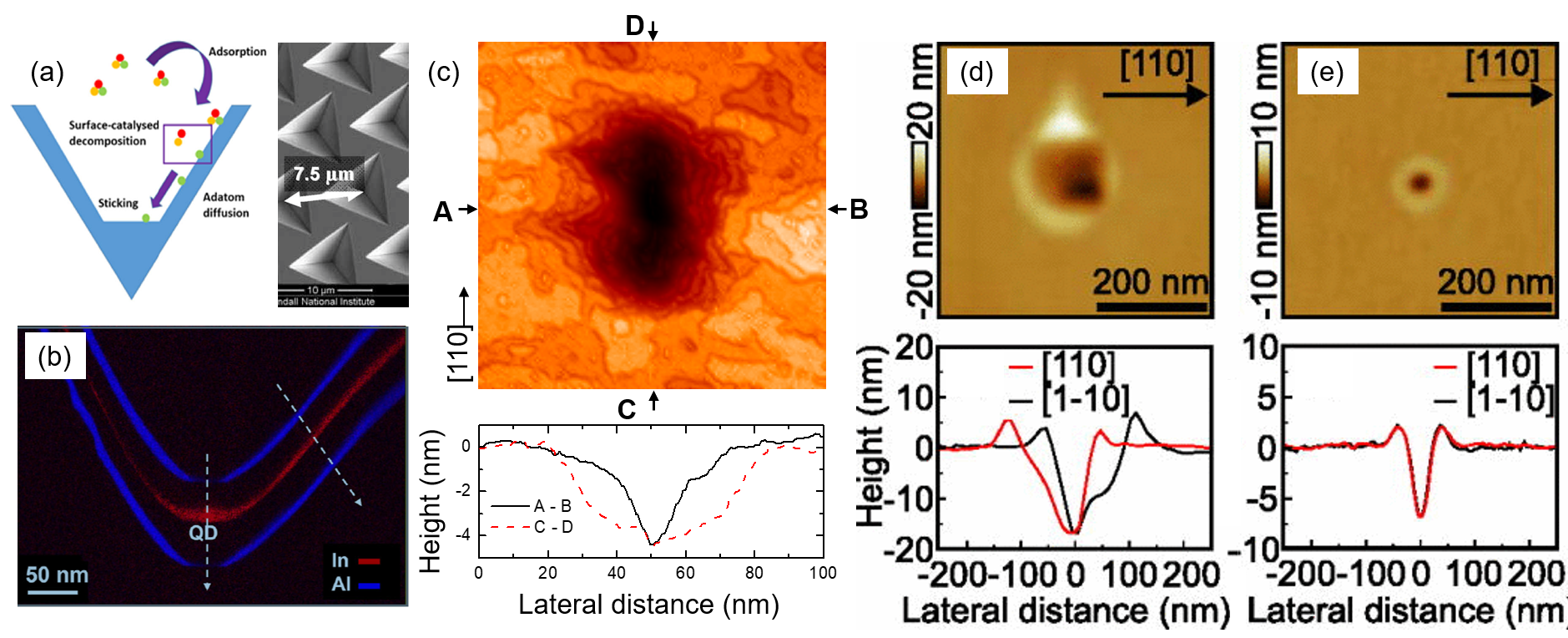}
\caption{QDs in nanoholes. (a) Scanning electron microscope (SEM) image of an array of pyramidal recesses in a GaAs(111)B substrate, filled with an (Al,In,Ga)As heterostructure; (b) Cross-sectional AFM image showing a heterostructure grown in such pyramids. (c) STM image of a nanohole on an Al$_{0.45}$Ga$_{0.55}$As surface obtained by MBE and {\em in situ} etching of a template of S-K /InGaAs/GaAs QDs followed by  overgrowth with a 7-nm thick Al$_{0.45}$Ga$_{0.55}$As layer; linescans of the elongated nanohole are shown in the bottom panel. (d) Similar to (c) but with nanoholes obtained on a GaAs surface by Ga droplet etching followed by overgrowth with 7-nm Al$_{0.45}$Ga$_{0.55}$As. (e) Symmetric nanoholes obtained on a 100~nm thick Al$_{0.4}$Ga$_{0.6}$As layer via optimized Al-droplet etching. %(f) Nanoholes obtained by electron beam lithography (EBL) and etching on GaAs(001) followed by AlGaAs overgrowth. The bottom panel shows a photoluminescence (PL) map of an array of GaAs QDs obtained by filling the nanoholes with GaAs followed by AlGaAs overgrowth. 
(a) Reproduced from~\cite{Dimastrodonato2010} with permission of the publisher, (b) reproduced from \href{https://nanoscalereslett.springeropen.com/articles/10.1186/1556-276X-6-567}{\textit{Juska et al. 2011}} \cite{Juska2011} under Creative Commons CC BY license, (c) Reprinted figure with permission from Ref.~\citeonline{Rastelli2004}  Copyright (2004) by the
American Physical Society. (d, e) Reprinted from \href{
https://doi.org/10.1063/1.4802088}{\textit{Huo et al. 2013}} \cite{Huo2013}, with the permission of AIP Publishing.}
\label{FIG:QDs_in_NHs}
\end{figure}
\subsubsection{Quantum dot fabrication via nanohole filling} \label{sec:QDsinNHs}
The main limitations of the S-K method are: (i) the difficulty of achieving suitable QD surface densities suitable for single-QD devices over large areas, making the selection of ``sweet spots'' on wafers necessary; (ii) the structural disorder and anisotropies due to inhomogeneous alloying, which limits the QD usability as hosts of coherent spins and as sources of entangled photons; (iii) the unsuitability for the creation of QDs out of almost lattice-matched material combinations. 

A method overcoming some or all of these limitations relies on the creation of dimples (or ``nanoholes'', Fig.~\ref{FIG:QDGROWTH}(d1)) on the bottom barrier material, followed by the growth of a ``planar’’ heterostructure with carefully chosen growth parameters to allow either quasi-conformal overgrowth of the dimple (for barrier material) or accumulation of QD material at its bottom (region of largest curvature and lowest chemical potential, see Eq.~\ref{eq:chempot}), leading to local thickening and QD formation (Fig.~\ref{FIG:QDGROWTH}(d2)). In turn, nanoholes can be created either by lithography and etching before growth or by {\em in situ} methods. An example of the former (see Fig.~\ref{FIG:QDs_in_NHs}(a)) is represented by tetragonal recesses obtained by anisotropic etching of lithographically-defined apertures on a GaAs(111)B surface~\cite{Sugiyama1995} and overgrown via MOVPE to obtain site-controlled arrays of QDs with very high ensemble homogeneity~\cite{Baier2004a}. A cross-sectional energy-dispersive-X-ray elemental map of an InGaAs QD at the center of an AlAs/GaAs/AlAs heterostructure is shown in Fig.~\ref{FIG:QDs_in_NHs}(b).

A notable material system combination for which the S-K mode cannot be used to fabricate QDs is represented by Al$_x$Ga$_{\rm 1-x}$As alloys, that have a lattice constant differing by at most 0.1\% compared to GaAs. The first attempts to create ``hierarchically self-assembled'' QDs via nanohole filling relied on the selective in situ etching of S-K InGaAs QDs during MBE, leading to dimples on a GaAs surface. Such nanoholes were then overgrown with AlGaAs at relatively low surface temperature~\cite{Rastelli2004} to allow for quasi-conformal overgrowth (see Fig.~\ref{FIG:QDs_in_NHs}(c)). Nanohole filling was then obtained by deposition of a thin GaAs layer and a growth interruption, allowing the QD material to diffuse and accumulate in the nanoholes. Since nanohole fabrication relied on a template of S-K InGaAs QDs, there was no improvement in the QD density control. In addition, the maximum thickness of the lower AlGaAs barrier was limited due to undesired nanohole filling with AlGaAs, making the resulting heterostructure poorly suited  for photonic integration. These QDs had however good optical quality and a well-defined elongated shape, enabling detailed studies on the relation between structural and optical properties of QDs~\cite{Rastelli2004, Plumhof2010,Wang2009c} as well as pioneering experiments on slow-light in Rb vapors with photons emitted by QDs~\cite{Akopian2011}. Attempts to replace the self-assembled nanoholes with ex situ etched nanoholes resulted in good site-control but poorer optical quality~\cite{Kiravittaya2006} due to interaction of the QD states with interface defects.

Later on, {\em local droplet etching} was discovered~\cite{Wang2007}, resulting in structures with improved control of density~\cite{Atkinson2012}. By focusing on III-V semiconductors, the process consists in the deposition of group-III elements (Ga, In, Al) in absence of group-V flux, leading to the formation of metal droplets on the surface (Fig.~\ref{FIG:QDGROWTH}(e1)). The group-V gradient at the interface between droplet and III-V semiconductor drives the diffusion of group-V elements (As, in the case of GaAs) into the droplet and consequent local liquefaction of the substrate. Under a reduced As flux, the etching process continues until a nanohole remains on the surface. For a recent review, see~\cite{Gurioli2019}. An example of GaAs nanoholes obtained by Ga-droplet etching followed by overgrowth with a thin AlGaAs layer is shown in Fig.~\ref{FIG:QDs_in_NHs}(d)). QDs with consistently low density across large areas are easily obtained via local droplet epitaxy. 

As in the case of GaAs nanoholes obtained by in situ etching of InGaAs QDs, the main limitation of Ga-assisted local-droplet-etching on GaAs is the limited maximum thickness of the AlGaAs barrier. This problem was finally solved by implementing local droplet etching directly on AlGaAs surfaces and using Al droplets~\cite{Heyn2009}. By optimizing the amount of Al used for droplet formation and other growth parameters, highly symmetric nanoholes were demonstrated~\cite{Huo2013} (Fig.~\ref{FIG:QDs_in_NHs}(e))), resulting in GaAs QDs with high in-plane symmetry and excellent optical properties, ideally suited as sources of polarization-entangled photons fully compatible with photonic integration~\cite{Liu2019}. For a recent review on these QDs, see~\cite{DaSilva2021} and for recent device achievements, see Section~\ref{sec:SPS_Performance_780}.

\subsubsection{Site-controlled quantum dots via guided self-assembly} \label{sec:SCQDs}
Ex situ patterned nanoholes have been successfully used to guide the formation of S-K QDs only at the desired substrate positions~\cite{Schmidt2007, Herranz2015} and almost perfect long-range ordering of InGaAs QDs has been reported~\cite{Kiravittaya2009}. The formation mechanism relies on the filling of the nanoholes with a diluted InGaAs alloy on top of which InGaAs preferentially form because of the locally reduced lattice mismatch compared to the surrounding GaAs surfaces, as sketched in Fig.~\ref{FIG:QDGROWTH}(f). 
Since nanoholes are energetically unfavorable, the growth of a thick buffer layer to bring the QDs away from the defective processed interface tends to flatten the surface, reducing the efficiency of site-control. Compromises have been found, leading to QDs with good optical properties~\cite{Joens2013}. To increase the interface-QD distance, diffusion  anisotropies have been used successfully used~\cite{Yakes2013}, leading to QDs with improved optical quality. 

\begin{figure}[h!]
\centering\includegraphics[width=12cm]{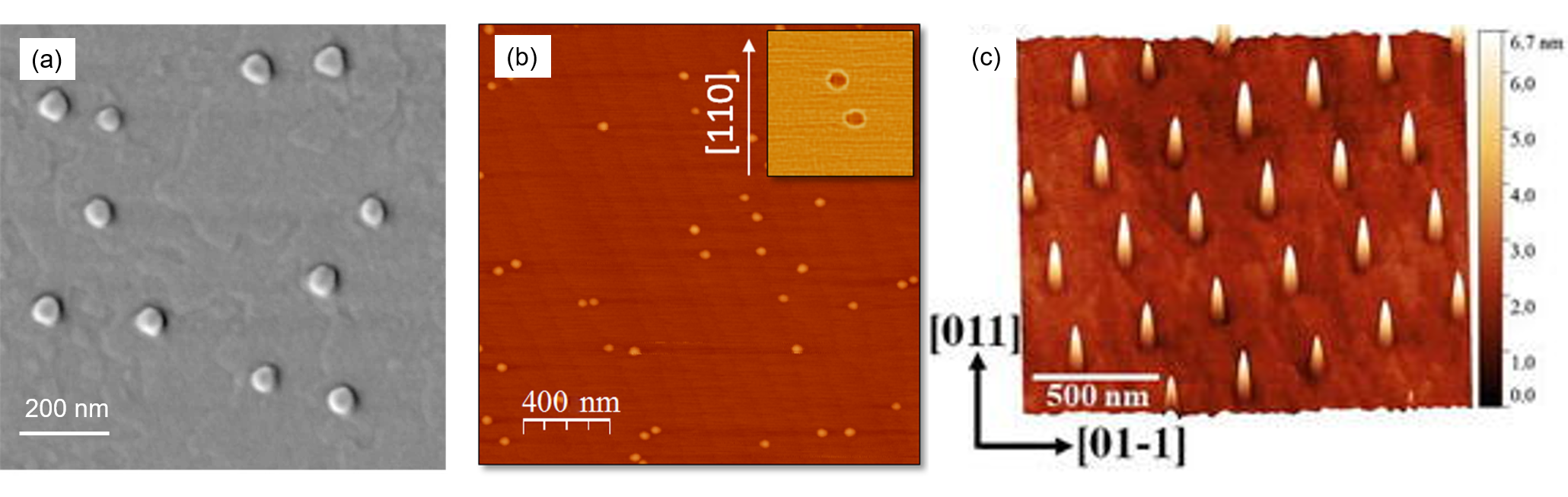}
\caption{Quantum dots obtained by droplet epitaxy. (a) InAs/InP(001) QDs with emission in the telecom C-band; (b) GaAs/AlGaAs QDs obtained on GaAs(111)A via ``high-temperature droplet epitaxy''; (c) site-controlled GaAs/AlGaAs QDs on GaAs(001) via laser interference.
(a) Reprinted figure with permission from \href{https://doi.org/10.1103/PhysRevApplied.8.014013}{\textit{Skiba-Szymanska et al. 2017}} ~\cite{Skiba-Szymanska2017} Copyright (2017) by the American Physical Society, (b) reprinted from \href{https://doi.org/10.1038/s41598-020-62248-9}{\textit{Bietti et al. 2020}} \cite{Bietti2020} under Creative Commons CC BY license. (c) reprinted from \href{https://aip.scitation.org/doi/10.1063/5.0045817}{\textit{Han et al. 2021}} \cite{Han2021} under Creative Commons CC BY license.}
\label{FIG:DE_QDs}
\end{figure}

Achieving the same quality as fully self-assembled QDs is still an open challenge and new in situ patterning methods, such as laser interference in MBE or growth through stencil masks are currently under investigation and have already shown promising results both for guiding the formation of InGaAs S-K QDs~\cite{Wang2023} and Ga droplets~\cite{zolatanosha2017} for GaAs QDs via droplet epitaxy~\cite{Han2021}. An example of such QDs is shown in Fig.~\ref{FIG:DE_QDs}(c).

\subsubsection{Quantum dots obtained by droplet epitaxy} \label{sec:DEQDs}
Historically, metal droplets were first used to directly obtain QDs rather than nanoholes in the so-called {\em droplet epitaxy} method, see review article~\cite{Gurioli2019}. To this aim, droplets of group-III elements (typically In or Ga) are exposed to a flux of group-V elements (typically As) to obtain their recrystallization, followed by barrier overgrowth (see Fig.~\ref{FIG:QDGROWTH}(e1-e3)). To prevent substrate etching, exposure must be performed at relatively low substrate temperatures, leading to the occurrence of point defects and difficulty to obtain material with quality as high as in QDs obtained with the S-K or local-droplet-etching methods, at least for what concerns GaAs/AlGaAs(001) QDs. High quality QDs have been recently achieved by performing the growth on GaAs(111)A substrates~\cite{Bietti2020}, see Fig.~\ref{FIG:DE_QDs}(a). Excellent results have been obtained also for InAs/InP QDs emitting in the C-band~\cite{Skiba-Szymanska2017} (Fig.~\ref{FIG:DE_QDs}(b)). As in the case of local droplet etching, the main advantage of droplet epitaxy over S-K growth is the improved QD-density control and also the higher in-plane symmetry of the obtained nanocrystals. This is exemplified by comparing pyramidal InAs QDs formed on InP by S-K growth (Fig.~\ref{FIG:InGaAsQDs}(e)) with the QDs obtained by droplet epitaxy in  Fig.~\ref{FIG:DE_QDs}(b).

\subsubsection{Quantum dot molecules} \label{sec:QDMs}
In addition to single QDs, there are experiments and applications that rely on interacting QDs. Among the different possible interactions, we mention here tunnel-coupled QDs, which can be obtained by vertical stacking of two or more QDs to form so called QD molecules~\cite{Bayer2001a}. For S-K QDs, the strain produced by a buried QD guides the formation of the next QD right on top of the buried one~\cite{Wang2009b}, providing a convenient way to achieve vertical stacking, as seen in Fig.~\ref{FIG:InGaAsQDs}(d). For QDs in nanoholes, vertically stacked QDs can be obtained by beginning with sufficiently deep nanholes and alternating QD material and barrier materials~\cite{Heyn2017}. Nanoholes or other surface features can also be used to guide the formation of closely spaced QDs in the growth plane~\cite{Wang2009b}.

\subsubsection{Quantum dots in nanowires} \label{sec:QDinwires}
While the methods discussed so far primarily lead to the formation of QDs in planar structure (an exception is represented by the QDs in inverted pyramids), QDs have been successfully fabricated also in nanowire structures~\cite{Francaviglia2016} following two possible routes: the vapor-liquid-solid growth, in which a metallic droplet (usually gold) acts as a catalyst for the vertical growth of nanowires on a substrate via the vapor-liquid-solid growth, and the catalist-free method, in which selective growth is achieved by opening nanometric apertures on an oxide layer (e.g. silicon oxide) on a substrate (e.g. silicon). Different from planar growth, nanowires allow strain due to lattice mismatch to be efficiently relaxed, thus enabling the growth of a richer set of material combinations, which would be incompatible for planar heterostructures. In addition, also the crystal structure is not necessarily imposed by the substrate and materials usually crystallizing in the zincblende structures are found to crystallize in the wurzite structure in nanowires. As in planar growth, heterostructures can be created along the wires. In particular, segments of QD material can be embedded in higher bandgap barriers. Lateral growth and also etching are also possible by proper tuning of the growth parameters, leading to a rich playground for nanostructure formation. For instance, {\em crystal phase} QDs have also been created by alternating segments of the same material but different lattice structure~\cite{Akopian2010}.

Site-controlled growth is easily achieved by patterning the precursor droplets or apertures. In addition, the wire geometry is favorable for improved light extraction both in free-space and integrated photonics via pick-and-place (see Section~\ref{sec:deterministic_fabrication}). 
Among the different material systems explored so far, InAsP QDs in InP barriers have demonstrated very good performance~\cite{Reimer2012,Dalacu2012}.

\section{Nanofabrication of single-quantum-dot devices}\label{sec:fab}

The nanofabrication of QLS-devices for applications in photonic quantum information technology is technologically very demanding. For instance, it requires a precise integration of individual QDs into nanophotonic structures and their spectral matching to the device's optical modes. In turn, the photonic structures with sub-$\mu$m feature sizes must precisely meet the design specifications of the numerical modeling. In addition, recent results show that targeted post-processing, e.g. via surface passivation, is often required to achieve optimal quantum-optical emission properties. Moreover, for practical applications, innovative concepts for applying electrical gates to or directly fiber-pigtailing respective quantum devices are beneficial. Not least, to gain the best out of multiple worlds, hybrid device-concepts are pursued, e.g. allowing for strain-induced spectral control of the QD emission in semiconductor-piezo integrated devices.

The epitaxial growth of QD heterostructures and numerical optimizations of device geometries is typically followed by several delicate processing steps during the device nanostructuring in a clean room environment. These mainly include the deposition of thin layers, optical lithography and electron beam lithography, as well as wet and dry chemical etching. In the context of this article, the lithography methods are of particular relevance. During lithography, the envisioned device geometry is transferred to a later etching mask. At the same time, it determines the position of the quantum emitter in the target structure. The latter point is particularly important in the case of single-QD devices. Due to the self-organized nature of the QD growth, the position of the emitter in the respective structure is completely undefined if standard lithography methods are used, which have been optimized, e.g. to produce semiconductor lasers and classical photonic circuits. Moreover, the spectral matching between emitter and nanophotonic structure is in general not guaranteed, being a major issue for resonator-based approaches. Conventional lithography usually results in a process yield for individual QD devices of below 1\%, rendering the scaling to e.g. complex IQPCs virtually impossible. Hence, to overcome these hurdles, deterministic process technologies are a crucial tool for the controlled and scalable fabrication of single-QD devices.

In the following, innovative nanotechnology methods are presented that have been developed and optimized in recent years for the deterministic fabrication of QD devices. Subsequently, concepts are presented that allow for the direct on-chip fiber coupling of QD devices, enabling quantum network integration. In addition, open challenges and future optimization approaches are discussed.

\subsection{Deterministic fabrication technologies}\label{sec:deterministic_fabrication}

A major challenge in the single-QD device fabrication is to integrate individual emitters with the desired optical and quantum-optical properties with high alignment accuracy into photonic nanostructures. This asks for deterministic nanofabrication technologies, which we introduce in the next subsections.  

\subsubsection{Pick-and-place technique}

The first deterministic process technology to be presented here is called pick-and-place technique. This approach is often used for the fabrication of heterogeneous IQPCs that contain emitter structures and waveguide structures made of different materials. For example, it can act as a powerful nanotechnology platform for the deterministic integration of III/V QDs into silicon-based IQPCs.

The aim of pick-and-place approaches is usually to process heterogeneous quantum devices in a scalable manner with high process yield. For this purpose, a multi-stage manufacturing process is used, where first a large number of optically active QD structures and the host structures (e.g. waveguide circuits) are independently processed on different chips. In a next step, QD devices suitable for the transfer to e.g. photonic waveguides are selected by spectroscopic means. Finally, using a micromanipulator or a rubber stamp technique~\cite{Katsumi2019}, the corresponding QD structures are detached from the host substrate and transferred to the target structure where they are integrated via van der Waals forces. While this transfer can be achieved with an accuracy of 10-100\,nm, it can hardly be automatized, thus limiting the practicality of this approach. Still, the pick-and-place technique is very suitable for efficient prototyping, as it allows for the independent optimization of the active QD structure and the passive counterpart, respectively, and a flexible integration of both.

\begin{figure}[h!]
\centering\includegraphics[width=12cm]{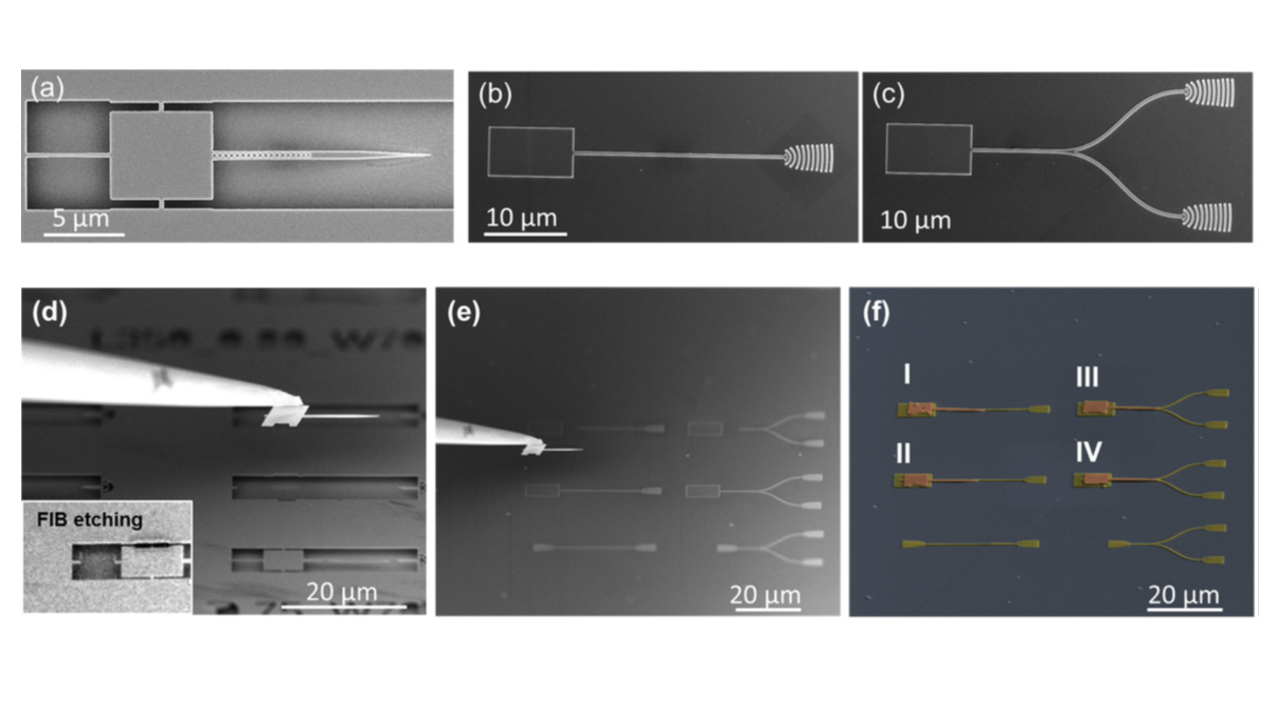}
\caption{SEM images illustrating deterministic nanofabrication via the pick-and-place technique. (a) EBL fabricated InGaAs/GaAs QD-nanobeam structure. (b, c) Straight waveguide and y-shaped 50/50 waveguide including grating outcouplers made of silicon. (d, e) Pick-and-place step transferring the QD-nanobeam to a y-shaped 50/50 waveguide using a microprobe tip combined with a focused ion beam and SEM. (f) False color SEM image of four fully processed hybrid QD-waveguide devices. Reprinted with permission from
Ref.~\cite{Kim2017}. Copyright 2017 American Chemical Society.}
\label{pickplace1}
\end{figure}

Figure~\ref{pickplace1} shows a concrete example in which the pick-and-place technique was used for the deterministic fabrication of QD quantum circuits~\cite{Kim2017}. The aim was to integrate an active QD structure into a silicon photonic chip. The prototype IQPC consists of an InAs/InP QD in a tapered nanobeam resonator (Fig.~\ref{pickplace1}(a)) whose emission is adiabatically coupled into the underlying silicon waveguide. This waveguide is in the simplest case linear, or branches off in an on-chip 50/50 beam splitter with two output waveguides, each ending in a grating coupler for vertical outcoupling of light (Fig.~\ref{pickplace1}(b, c)).

Due to the random position (and inhomogeneous broadening of the ensemble emission) of the self-assembled QDs in the growth plane, the spatial (and spectral) overlap of the QDs with the field maximum in the nanobeams can not be guaranteed in this conventional manufacturing process, resulting typically in a yield of suitable components of approximately 1\%. Hence, suitable QD-nanobeams, which fulfill the desired spectral and spatial properties, are selected before the transfer to waveguides. The corresponding QD-nanobeam membrane structures are then removed from the host chip by focus ion beam milling and transferred to the desired position on the waveguide using a micromanipulator (see Fig.~\ref{pickplace1}(d-f)). Van der Waals forces ensure the necessary adhesion of the QD-nanobeam structures to the micromanipulator and the waveguide structures. The transfer is reproducible and takes place with an accuracy better than 100\,nm. The placement routine of the nanobeam, however, is relatively time-consuming with a duration of about 1\,hour.

Pick-and-place techniques as described above have been applied for the fabrication of single QD quantum devices in various ways. For example, epitaxially grown InAsP nanowires with integrated QDs were transferred to silicon nitride waveguides via pick-and-place, where they can in turn act as single-photon emitters~\cite{Elshaari2017}. This work shows an attractive aspect of the nanowire technology: The host structure can be designed to include complex photonics structured in parallel to the emitter structure. Such structures can be fabricated  using methods of integrated photonics in order to obtain a functional IQPC after heterogeneous integration. For example, as shown in Ref.~\cite{Elshaari2017}, electrically controlled filter elements and single-photon on-chip multiplexers can be integrated.

Noteworthy, pick-and-place techniques are attractive alternatives to the complex epitaxial growth of hybrid heterostructures of III-V compound semiconductors on e.g. Si wafers or the fabrication of heterogeneous quantum devices based on wafer-bonded QD heterostructures.

\subsubsection{Marker-based lithography techniques}

Another strategy for the fabrication of single-QD devices is marker-based lithography. In this approach, the location of suitable quantum emitters is first identified relative to alignment markers, usually via optical imaging. In a second step, the intended nanophotonic structures are defined in the appropriate resist at the location of pre-selected QDs using electron beam lithography. Marker-based lithography is a flexible method that can be used for different quantum emitters, emission wavelengths and device concepts. It is currently very popular to fabricated CBG resonators with deterministically integrated QDs.

\begin{figure}[h!]
\centering\includegraphics[width=12cm]{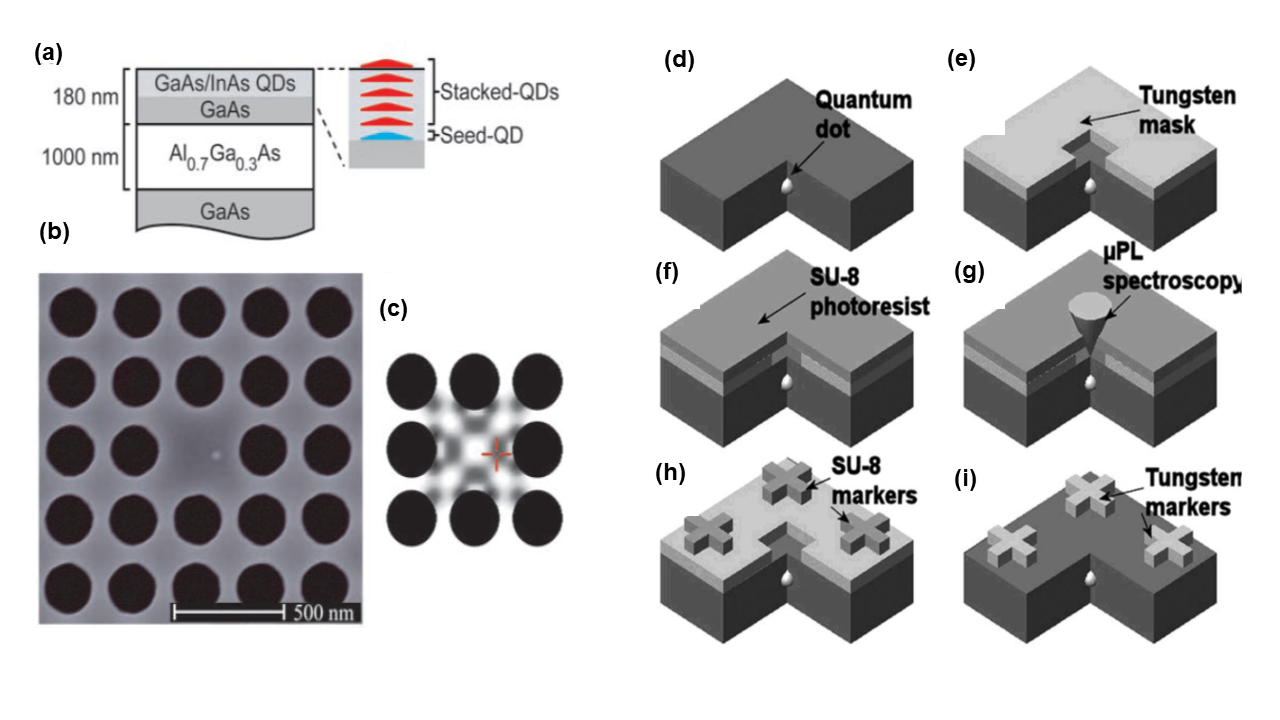}
\caption{Marker-based deterministic device processing. (a-c) Deterministic fabrication of a QD photonic crystal (PC) cavity device. Using the growth of vertically stacked QDs (a) the position of the target seed QD which is blue-shifted to be in resonance with the PC cavity can be retrieved by SEM imaging at the sample surface. Using marker-based EBL, the nanocavity can be aligned with the selected QD (b) to ensure optimum mode overlap (c). (d-f) Schematic illustration of cryogenic laser lithography to pre-select a QD and pattern alignment markers relative to its position. Using (a) a QD sample patterned with a tungsten mask (e) and SU-8 photoresist (f) $\mu$PL spectroscopy is performed at cryogenic temperatures to determine the position of a target QD (g) before markers spatially aligned to the QD are written into SU-8 by optical lithography in the cryogenic $\mu$PL. Finally, the markers are transferred by into Tungsten by reactive ion etching (i). In a following EBL process, the fabricated cross-markers could be used to align a photonic nanostructure to the selected QD using marker-based EBL. (a-c) From Ref.~\cite{Badolato2005}. Reprinted with permission from AAAS. (d-i) Reprinted from Ref.~\cite{Lee2006}, with the permission
of AIP Publishing.}
\label{Fig:markerbased1}
\end{figure}

The origins of marker-based lithography lie in technological developments aimed at producing QD nanoresonators for the study of cQED effects. Here, the spectral and spatial resonance between the emitter and the resonator mode is a crucial requirement that cannot be reproducibly achieved with conventional manufacturing technologies. To counteract this problem, in Ref.~\cite{Badolato2005} vertically strain-correlated stacked QDs were used as presented in Fig.\ref{Fig:markerbased1}(a-c). In a stack of 6 QDs, the bottom QD, which was intended for the cQED experiments, was blue-detuned and the position of the top QD could be determined on the sample surface using SEM images relative to alignment markers. Subsequently, PC-cavities were produced using EBL and etching techniques, the position of which was specifically aligned to the detected QDs. Overall, this advanced nanotechnology concept allowed the regime of weak~\cite{Badolato2005} and strong coupling~\cite{Hennessy2007} in QD-nanocavity systems to be implemented in a controlled manner for the first time.

The aforementioned marker-based manufacturing process first showed the potential of deterministic device fabrication technologies. However, it is limited in two ways. On the one hand, it is based on a stack of strain-coupled QDs, which limits the device compatibility to near-surface structures, and on the other hand, only the position of the QDs but not their spectral features (especially the emission wavelength) could be determined during device fabrication. In other words, only the spatial, but not the spectral resonance between the emitter and the resonator mode could be controlled. To overcome this problem, cryogenic laser photolithography was developed, which provides knowledge of the position and spectral properties of selected QDs~\cite{Lee2006}. In the first step of this nanotechnology concept, which is illustrated in Fig.~\ref{Fig:markerbased1}(d-i), the sample surface is scanned using $\mu$PL spectroscopy in order to determine the position and the spectral position of suitable QDs. Immediately afterward, the laser alignment marker is used to write in a photon-sensitive lacquer relative to the position of the pre-selected QDs. It is worth noting that both processes take place at cryogenic temperatures (4 K) to ensure a sufficiently high luminescence yield of the QDs. In the final step, the marker structures are transferred to the semiconductor material by dry etching. With this method, the QD positions can be determined with an accuracy of $\pm$50 nm and retrieved with an accuracy of $\pm$150 nm via marker detection, and the spectral accuracy is about 1 nm. The method described was not used in the following for the production of single-QD quantum devices, but it can be regarded as an important basis and trigger for the development of the deterministic fabrication methods described in the following.

\begin{figure}[h!]
\centering\includegraphics[width=12cm]{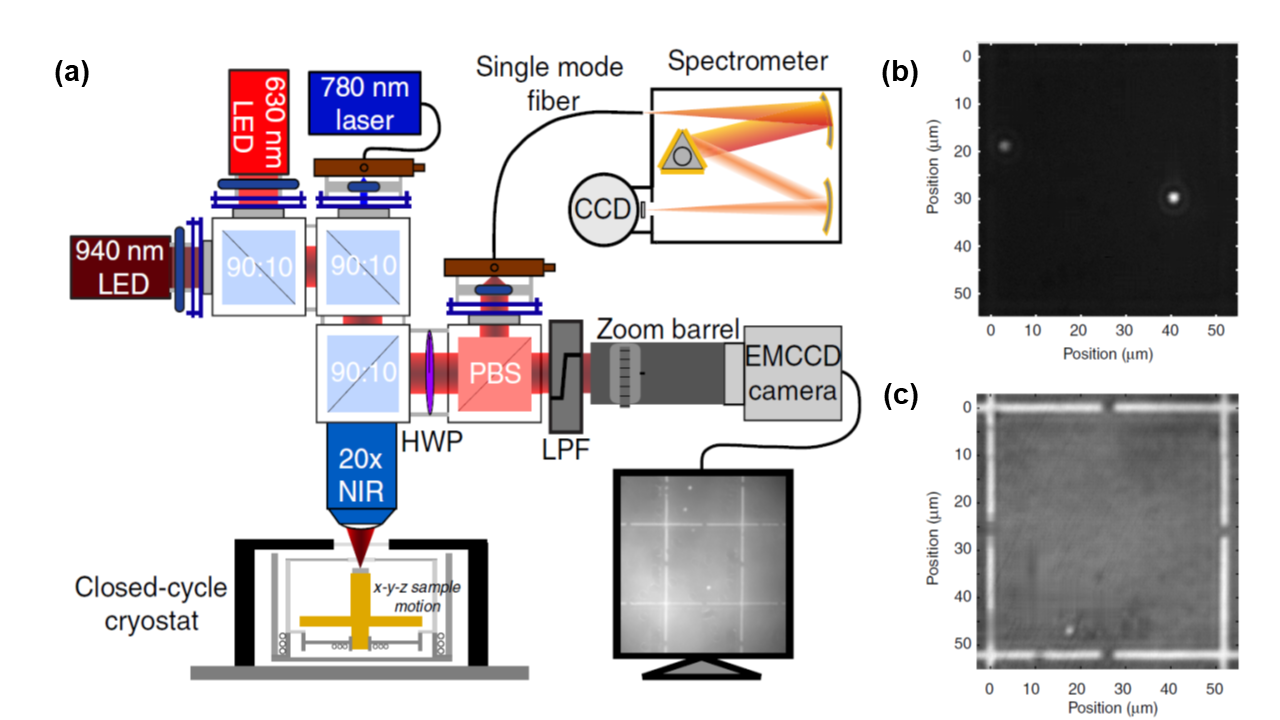}
\caption{Marker-based deterministic device processing using cryogenic optical imaging. (a) Experimental setup including two light emitting diodes (LEDs) (at 630 nm and 940 nm) and a laser (at 780), several beam-splitters and optical filters, a closed-cycle cryostat, an electron multiplied charged couple device (EMCCD) for optical imaging and a spectrometer for $\mu$PL investigations. (b) Fluorescence image under LED excitation at 630 nm and a 900 nm low-pass filter to suppress reflected light. Emission of two QDs and their positions (in the given coordinate system) can be identified. (c) Optical image of the sample under LED illumination at 630 nm. Metallic alignment markers are clearly seen. Overlying the two images allows one to determine the positions of pre-selected QDs with respect to the alignment markers with better than 30 nm accuracy. Reproduced from Ref.~\cite{Sapienza2015} under Creative Commons CC BY license.}
\label{Fig:markerbased2}
\end{figure}

The full potential of marker-based lithography was first demonstrated by Davanco et al. Ref.~\cite{Sapienza2015}. In this work, QD SPSs based on CGB resonators, also called bullseye resonators, were fabricated deterministically with spatial and spectral control. The precise fabrication process made it possible to produce sources with a photon extraction efficiency of (48 $\pm$ 5)\%, which agrees very well with the theoretically predicted value of 50\%. In the underlying multi-stage nanoscale optical positioning process illustrated in Fig.~\ref{Fig:markerbased2}, first, metallic alignment markers were structured on the sample surface using EBL. In the second step, performed at 6 K, the sample surface was illuminated over a large area (200 $\mu$m diameter) in a modified $\mu$PL setup with a 630 nm LED. In the detection path, either the reflected light or the fluorescence of the QD could be recorded with an electron multiplied charged couple device through a corresponding spectral filter. Alternatively, a laser was used to measure the $\mu$PL spectra of individual QDs. An additional 940 nm LED could also be used to record the reflected light of the markers and the luminescence of the QDs at the same time to improve the alignment accuracy. In this experimental configuration, by comparing the images and the $\mu$PL data, the positions and emission wavelengths of suitable QDs relative to the alignment marks could be determined with an average position uncertainty of <30 nm. In the final processing step, bullseye resonators, which are spatially and spectrally matched to selected QDs, were structured using EBL and dry chemical etching. The deterministically produced quantum devices are characterized by a high extraction efficiency ((48 $\pm$ 5)\%) and a very good single-photon purity of $g^{(2)}(0)= 0.009 \pm 0.005$. Furthermore, a Purcell effect of 4 (significantly below the expected value of 11) could be determined due to the increased light-matter interaction in the resonator structure.

The nanoscale optical positioning method presented has evolved since its first demonstration and has already been used many times for the deterministic production of various QLSs. These are usually based on QDs in CBG resonators~\cite{Liu2019, Wang2019a}, but micropillars~\cite{He2017b, Wang2019a} and PC-cavity-based devices~\cite{Chu2020} have also been manufactured using this process. Current work shows that it can also be used for quantum devices with emission in the telecom O-band~\cite{Xu2022}. 

It is interesting to note that recently a marker-based technology has been developed that uses cathodoluminescnce (CL) mapping instead of optical imaging. This method has the advantage that the QD selection and the EBL are carried out in the same system and therefore in principle also in the same coordinate system, which promises higher alignment accuracies in the future compared to optical imaging in combination with EBL. It was used to demonstrate the scalable integration of multiple QDs into an optical waveguide system~\cite{Li2022}.

Although the marker-based positioning methods are very powerful, they are also relatively complex in practical implementation due to the multistep process flow. Furthermore, the alignment markers can have a disruptive effect on the fabrication of larger structures, such as large-scale integrated quantum circuits. These limitations can be circumvented with in situ lithography concepts, which are presented and discussed in the following section.

\subsubsection{In situ lithography techniques}

Additional interesting deterministic nanostructuring technologies are in situ lithography techniques. In contrast to the approaches presented in the previous sections, these techniques do not require marker structures and are therefore comparatively simple in the process flow. The basic idea is to first determine the positions of suitable quantum emitters using optical spectroscopy or cathodoluminescence and then, in the same setup, to define the desired nanophotonic structure using optical lithography or EBL in the appropriate resist. 

\begin{figure}[h!]
\centering\includegraphics[width=12cm]{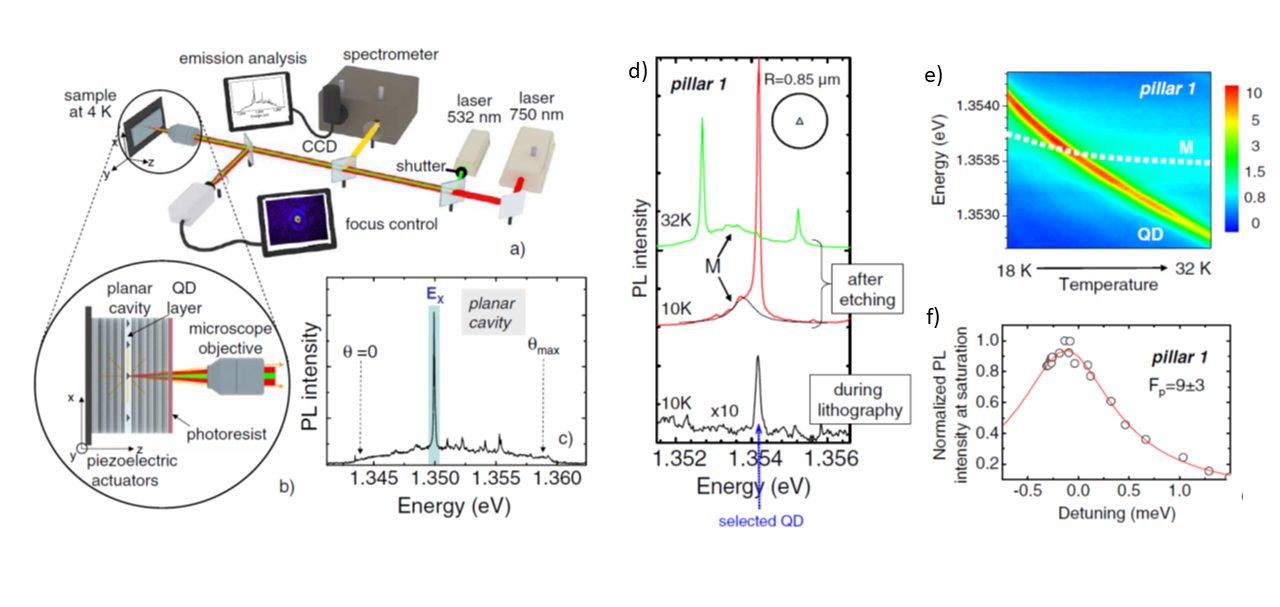}
\caption{In situ optical lithography of a spatially and spectrally resonant QD-micropillar system. a) Sketch of used experimental $\mu$PL setup extended by a second green laser for optical lithography. b) Layer design of the used QD-heterostructure with the photoresist layer on top and an illustration of the laser excitation scheme. c) $\mu$PL spectrum of the planar cavity during QD selection at 10 K. Emission of a single QD exciton $E_X$ is marked. d) $\mu$PL spectra of a selected QD during in situ lithography (black trace) and after etching of the micropillar with a radius of 0.85 $\mu$m at 10 K and 32 K (red and green traces). Emission of the fundamental cavity mode M is marked. e) Temperature tuning of the QD exciton X through resonance with the cavity mode M. The QD-micropillar system is in the weak coupling regime of cQED and shows enhanced emission at spectral resonance. f) Corresponding normalized PL intensity as function of detuning between X and M. The fit (red trace) yields a Purcell-factor of $9\mp3$. Reprinted from \href{https://doi.org/10.1103/PhysRevLett.101.267404}{\textit{Dousse et al. 2008}} \cite{Dousse2008}. Copyright (2008) by the American Physical Society.}
\label{Fig:iPL}
\end{figure}

Pioneering work in in situ lithography was performed by Dousse et al. in Ref.~\cite{Dousse2008}. In this work, in situ optical lithography was developed and used for the first time to produce a spectral and spatially resonant QD-micropillar system for the controlled study of cQED effects. The experimental setup consists of a low-temperature $\mu$PL unit, which has been expanded to include a second laser on the excitation side (see Fig.~\ref{Fig:iPL}a). While the first laser with a wavelength of 750 nm is used for optical excitation of the QD sample, the second laser with a wavelength of 532 nm is used for lithography. In the process sequence, the QD sample, which in Dousse et al. corresponded to a planar microresonator structure with a lower and an upper DBR with an intermediate cavity with an integrated QD layer (see Fig.~\ref{Fig:iPL}b), is first coated with a positive photoresist and then mounted onto the cold finger of a He-flow cryostat. After that, part of the sample surface (a few $\mu$m in x- and y-direction) is scanned for the QD selection by $\mu$PL mapping at typically 10 K, for which the 750 nm laser is used (while the 532 nm laser is blocked), which does not affect the photoresist. Here, QDs are selected specifically with regard to their emission wavelength (and $\mu$PL intensity) in order to later be brought into spectral resonance with the fundamental emission mode of a micropillar cavity. An exemplary $\mu$PL spectrum of the planar microcavity is shown in Fig.~\ref{Fig:iPL}c). Once a suitable QD has been found, the sample position is optimized for maximum $\mu$PL intensity before the long wavelength laser is blocked, and the photoresist is exposed with the unblocked short wavelength laser at the location of the QD. Here, the diameter of the exposure spot, which specifies the diameter of the later micropillar, can be adjusted within certain limits via the exposure time. Finally, reactive ion etching is applied to produce the micropillar. 

Fig.~\ref{Fig:iPL}d) compares $\mu$PL spectra of the planar cavity (black trace) with spectra of the processed QD-micropillar with a radius of 0.85 $\mu$m taken at two temperatures. The QD-micropillar is well-structured and spectral resonance between the single-QD exciton and the resonator mode can be achieved at about 20 K. A corresponding temperature tuning $\mu$PL map is presented in Fig.~\ref{Fig:iPL}e) and an evaluation of the normalized PL intensity yields a Purcell-effect of $9\pm3$. These results reflect the high potential of in situ optical lithography, which has been used very successfully since the first report, to demonstrate, for instance, bright sources of indistinguishable photons~\cite{Somaschi2016} and entangled photon pairs~\cite{Dousse2010}.

Despite the great success of in situ optical lithography, this technique also has disadvantages. Above all, it is based on the resist exposure using laser light, which limits the resolution and structure accuracy to a few 100 nm. Furthermore, no complex structures such as optical waveguides can be defined. In order to circumvent these limitations, the method of in situ electron beam lithography was developed~\cite{Rodt2021a}. This technique uses CL spectroscopy in combination with EBL to select QDs (mainly at cryogenic temperatures), and then to define the desired nanophotonic structure with high alignment accuracy relative to the selected QD. In this way, the in situ EBL combines the advantages of user-friendly CL mapping with the high flexibility and resolution of the EBL in a unique deterministic nanoprocessing technology.

\begin{figure}[h!]
\centering\includegraphics[width=12cm]{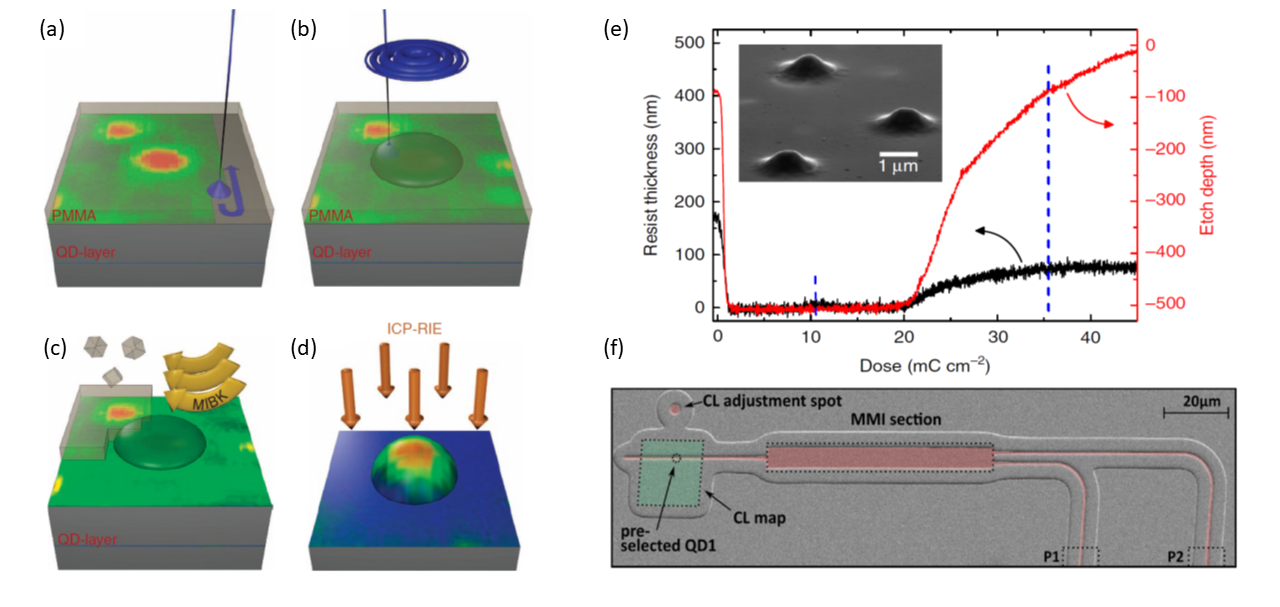}
\caption{In situ electron beam lithography of single-QD quantum devices. In the in situ EBL process flow, suitable resist (PMMA or CSAR) are first spin-coated on the sample surface before CL mapping is performed with low dose at cryogenic temperatures to select suitable QDs based on the luminescence yield and spectral properties. Then, still at cryogenic temperature, the desired nanophotonic is written in the resist with higher dose at the position of a selected QD, where gray-scale EBL can be applied to shape for instance 3D microlenses by locally inverting the resist (b). Afterward, the resist is developed in the clean room (c) leaving the defined structure as etch mask in the subsequent reactive ion etching step (d).  Contrast curve of PMMA resist at 5 K. Inversion of the positive to resist starts at a dose value of 20 mC cm$^{-2}$. Inset: SEM image of 3 deterministically fabricated QD-microlenses. (f) On-chip quantum circuit fabricated by in situ EBL. The circuit includes a deterministically integrated QD in the input waveguide of a multi-mode interference beam-splitter with two exit ports. (a-d) Reproduced from Ref.~\cite{Gschrey2015} under Creative Commons CC BY license. (e, f) Adapted with permission from Ref.~\cite{Schnauber2018}. Copyright 2018 American Chemical Society.}
\label{Fig:iEBL}
\end{figure}

The process sequence of the in situ EBL method is shown in Fig.~\ref{Fig:iEBL}. After the electron beam sensitive positive tone resist has been spin-coated onto the sample surface, it is mounted in a SEM with a CL extension and a He-flow cryostat. For the QD selection, sample areas of usually around 50 $\mu$m x 50 $\mu$m are scanned with a low dose below the onset dose for resist inversion (typically 20 mC cm$^{-2}$, see Figure~\ref{Fig:iEBL}(e)) in order to create a CL map as indicated in Fig.~\ref{Fig:iEBL}(a). Based on the data obtained, the positions of suitable QDs are determined, using the luminescence intensity and the spectral properties of the QD as criteria. At these QD positions, the photonic nanostructures are then exposed into the resist using EBL as illustrated in Fig.~\ref{Fig:iEBL}(b), with a dose above the onset dose for inversion being selected in order to locally invert the resist and reduce its solubility in the subsequent development step (Fig.~\ref{Fig:iEBL}(c). The CL mapping and the actual EBL are carried out at cryogenic temperature and allow an alignment accuracy between QD and nanophotonic structure of about 30-40 nm~\cite{Gschrey2015align}. This alignment accuracy is essentially limited by the mechanical drift of the cold finger (in the semi-professional SEM used). In the future, professional in situ EBL systems with an interferometer stage will probably be able to achieve significantly better alignment accuracies. With a suitable selection of the exposure dose above the onset value, i.e. in the range of 20-40 mC cm$^{-2}$ (see contrast curve in Fig~\ref{Fig:iEBL}(e)), three-dimensional nanostructures can be defined in the resist via gray-scale EBL, as used, for example, used for the deterministic fabrication of QD microlenses, as shown in Fig~\ref{Fig:iEBL}(b)). In the subsequent development process (Fig~\ref{Fig:iEBL}(c)), the non-inverted resist is removed so that the desired structure remains as an etching mask on the sample surface. In the final step, reactive ion etching is applied to transfer the patterned structure into the semiconductor material. As a result, microlenses with sub-$\mu$m feature sizes and deterministically integrated QDs are fabricated (see, Fig~\ref{Fig:iEBL}(e), inset). Such QD-microlenses act as bright sources of indistinguishable photons, as demonstrated in Ref.~\cite{Gschrey2015}. 

The great potential of in situ EBL technology for the fabrication of complex QD nanostructures with nanometer feature sizes was demonstrated in Ref.~\cite{Schnauber2018}, where the technique was applied for patterning a photonic quantum circuit with a deterministically integrated QD. The structure shown in Fig.~\ref{Fig:iEBL}(f) includes the deterministically integrated QD in a linear waveguide, via which the photons emitted by the QD are transmitted into the input port of a multi-mode interference beam splitter with a 50/50 splitting ratio. The latter was used in Ref.~\cite{Schnauber2018} as an on-chip-integrated beam splitter and enabled the authors to perform a quantum optical Hanbury Brown Twiss experiment on chip. 

Furthermore, hybrid waveguide systems~\cite{Schnauber2019} and structures for the controlled study of chiral light-matter interactions~\cite{Mrowiski2019} were fabricated using in situ EBL. An important milestone in the scalable fabrication of quantum circuits was recently achieved by deterministically integrating two QDs into the input waveguides of a multi-mode interference beam splitter~\cite{Li2022}. In the future, this approach, in combination with spectral fine-tuning using the quantum confined Stark effect~\cite{Schnauber2021}, can be used to develop highly functional IQPCs, for example for a fully integrated boson sampling chip.

% combination: marker based / in-situ: https://aip.scitation.org/doi/full/10.1063/1.4926995

% combination: Indistinguishable Photons from Deterministically Integrated Single Quantum Dots in Heterogeneous GaAs/Si3N4 Quantum Photonic Circuits: https://www.nature.com/articles/ncomms11183 

\newgeometry{left=1cm, right=1cm}
\begin{table}
\centering
\caption{Comparison of most relevant deterministic QD-device processing technologies. The table indicates the relative complexity (complex.), whether the technology is marker based (MB), if spectral selection (SS) (using a spectrometer) of QDs is possible, and which type of lithography is performed (optical lithography, EBL, at cryogenic temperatures (low-T) or at room temperature (RT)). It also provides information about position accuracy (PA), alignment accuracy (AA), lithography resolution (LR), and the related references. Here, PA refers to the accuracy with which the position of a QD can be determined. AA is the accuracy with which the QD is positioned in the nanophotonic structure.}
\label{Tab_fab}
\begin{threeparttable}
\begin{tabular}{cccccccccl}
\hline
\begin{tabular}[c]{@{}c@{}} Method \\ \end{tabular} & \begin{tabular}[c]{@{}c@{}} Complex. \\ \end{tabular} & \begin{tabular}[c]{@{}c@{}} MB \\\end{tabular} & \begin{tabular}[c]{@{}c@{}} SS\\ \end{tabular} & \begin{tabular}[c]{@{}c@{}} Litho. \\ \end{tabular} & \begin{tabular}[c]{@{}c@{}}  PA \\ \end{tabular} & \begin{tabular}[c]{@{}c@{}} AA \\ \end{tabular} & \begin{tabular}[c]{@{}c@{}} LR \\ \end{tabular} & \begin{tabular}[c]{@{}c@{}} Ref.\\ \end{tabular} \\ \hline

Pick-and-place & high & no & no & EBL, RT  & -- & $\approx 200$ nm & < 10 nm  & \cite{Kim2017}\\
 Optical imaging & medium & yes & no & EBL, RT  & $\approx 10$ nm & < 30 nm & < 10 nm  & \cite{Sapienza2015}\\
 CL imaging & medium & yes & yes & EBL, RT & 10 nm & 40 nm & < 10 nm  & ~\cite{Li2022}\\
 In situ opt. litho. & low & no & yes & optical, low-T  &  $\pm$ 50 nm & $\pm$ 50 nm & > 100 nm\ & \cite{Dousse2008}\\
In situ EBL & low & no & yes & EBL, low-T & 25 nm & 30-40 nm & < 10 nm & \cite{Gschrey2015align}\\
\hline
\end{tabular}
\end{threeparttable}
\end{table}
\restoregeometry

In summary, today there are a number of very powerful deterministic processing technologies available to integrate individual emitters with high accuracy into photonic nanostructures. The most relevant methods are given in Table~\ref{Tab_fab} together with the most important technology parameters. They differ in their complexity and in their alignment and structural accuracy. The relatively complex pick-and-place technique is based on a multi-stage process which, in addition to structure fabrication, includes structure selection and structure transfer. In addition to the actual structure fabrication, the marker-based technologies require the marker processing and a precise alignment of the mapping data with the marker coordinate system. In comparison, the two in situ lithography processes are technologically relatively simple because they do not require any marker structures. The achievable alignment and structure accuracies are similar with the techniques based on EBL. In this regard, in situ optical lithography has to accept compromises, especially with regard to structural accuracy.

\subsection{On-chip fiber coupling of quantum light sources}\label{sec:Nanofab_FC}

In the last two decades, enormous progress has been made in the development and fabrication of QLSs based on semiconductor QDs. As described in Sections~\ref{sec:epi} and~\ref{sec:fab}, innovative growth and fabrication technologies were developed and used, and almost ideal emission properties could be achieved (see Section~\ref{sec:optprop}). So far, however, these quantum devices have been operated and studied almost exclusively on a laboratory scale in proof-of-principle experiments. With regard to real applications, for example in photonic quantum technology, further development stages are necessary. In fact, fiber coupling of QLSs can enable the transmission of quantum information over long distances and the generation of remote entanglement between separated quantum systems to create quantum networks and to enable distributed quantum information processing. 

 Several different approaches have been proposed for achieving high coupling efficiency between a QD and a single-mode fiber. These approaches aim at connecting QLSs to optical fibers in a robust manner and with high coupling efficiency. Here, the coupling to single-mode optical fibers compatible with standard telecom components is particularly interesting and important to allow direct application, for example in QKD. On the source side, single-photon emitters with emission in the telecom O-band and C-band at 1.3 $\mu$m and 1.55 $\mu$m wavelengths are of particular relevance in order to transmit quantum information over medium and long distances. Additionally, sources with emission wavelengths below 1 $\mu$m are interesting for local interconnects, for example to generate photonic input states for quantum computers and simulators in a convenient manner via optical fibers.

 \begin{figure}[h!]
\centering\includegraphics[width=12cm]{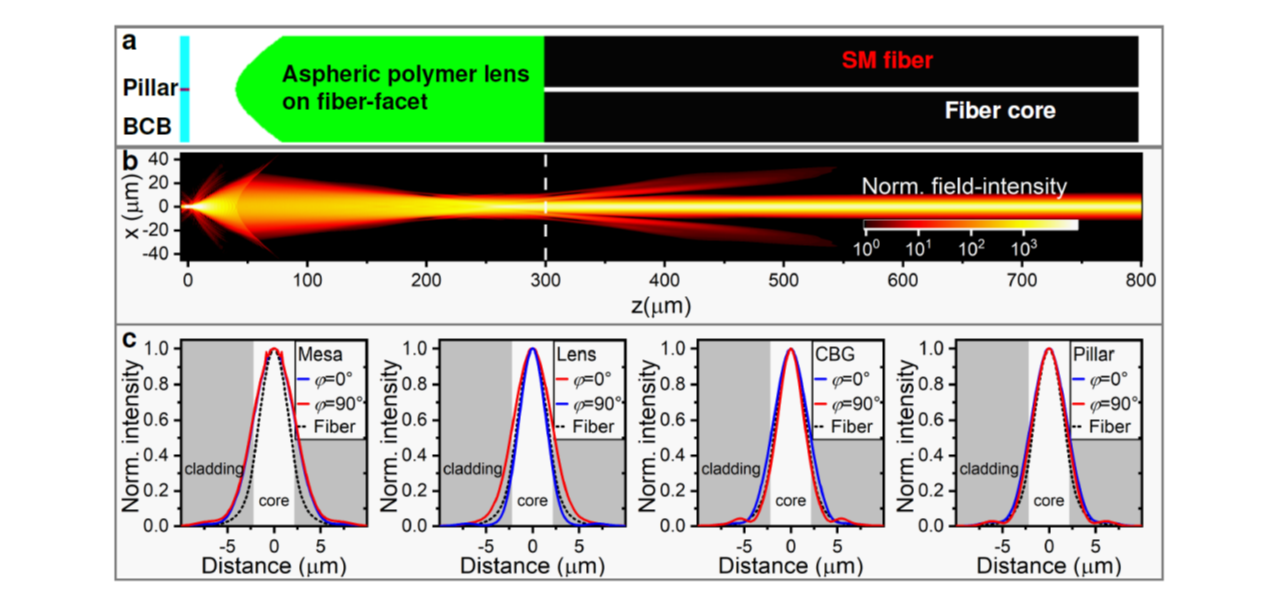}
\caption{Numeric simulation and optimization of fiber-coupled QD single-photon sources (SPSs). (a) Numerical setting, including the SPS (QD-micropillar planarized with benzocyclobutene in this case), mode-matching aspheric microlens and a single-mode optial fiber. (b) Corresponding light field intensity distribution calculated via the finite element method. The light is collected by the aspheric microlens and focused on the core of the fiber. (c) Calculated mode profile of the optical fiber (dashed black trace) and of the incident emission of a QD integrated into a micromesa, microlens, CBG resonator, and a micropillar for two orientations of the emitter dipole. The calculations yield excellent mode overlap of 89\%, 92\%, 90\% and 95\% for the four considered geometries. Reprinted from Ref.~\cite{Bremer2022}.}
\label{Fig:fiber0}
\end{figure}

One of the most important parameters of fiber-coupled SPS is the overall coupling efficiency $\eta_{\rm tot}$, i.e. the probability that a photon will be coupled into the core of the fiber after a trigger event. This parameter results from the product of the QD occupation probability $\eta_{\rm exc}$, the internal quantum efficiency of the QD $\eta_{\rm int}$, the photon extraction efficiency $\eta_{\rm ext}$ and the coupling efficiency between source and fiber $\eta_{\rm sf}$: $\eta_{\rm tot} = \eta_{\rm exc} \times \eta_{\rm int} \times \eta_{\rm ext} \times \eta_{\rm sf}$. $\eta_{\rm exc}$ and $\eta_{\rm int}$ are close to one when using suitable resonant excitation and high-quality QDs. $\eta_{\rm ext}$ depends on the device design, which is usually optimized using numerical methods, and in the case of CBG structures reaches values beyond 80\%~\cite{Liu2019} but for high NA of the collecting optics in the range of 0.6-0.8, as usually present in quantum-optical experiments.

Concerning fiber-coupling solutions, it is therefore a major challenge to achieve high $\eta_{\rm sf}$. On the one hand, conventional single-mode fibers with a small refractive index contrast between core and cladding have a large mode size, causing a small NA of around 0.1. Furthermore, simple settings suffer from poor mode matching between source and fiber. Thus, the optical fiber collects only a small fraction of photons emitted from the QDs.

In order to counter these problems, one interesting approach is the evanescent field coupling using tapered fibers. A fiber-pulling technique is able to form tapered micro-fibers, which couples the emission from QD devices with high efficiency of 23\%~\cite{Lee2015}. It is also possible to use a single-side tapered fiber and integrate it with tapered QD devices~\cite{Daveau2017}. Proper designs and alignments can lead to adiabatic mode transfer between tapered fiber and QD devices with near-unit coupling efficiency~\cite{Tiecke2015}. Although evanescent couplings via tapered fibers are very effective in improving coupling efficiency, the approach could be impractical due to the lack of mechanical stability and the need for continuous alignment to maintain high coupling efficiency. 

A further approach aims at far-field coupling between the sources and the fiber. Maximizing $\eta_{\rm tot}$ in this setting with a large parameter space requires high computational effort. A recent comprehensive numerical study maximized $\eta_{\rm tot}$ for micromesas, microlenses, micropillars and CBGs for emission wavelengths of 930 nm, 1.3 $\mu$m and 1.55 $\mu$m coupled to a single-mode fiber~\cite{Bremer2022}. Here, an intermediate achromatic microlens was considered to maximize the mode matching between source and fiber, as presented in Fig.~\ref{Fig:fiber0}(a). Fig.~\ref{Fig:fiber0}(b) shows the calculated field intensity when considering a QD-micropillar and panel (c) compares the mode profile of the light field at the fiber facet and the light field confined in the core of the fiber. Excellent mode overlap of up to 95\% and total efficiencies of up to 83\% were achieved by numeric optimization of the SPS-lens-fiber system. 

\begin{figure}[h!]
\centering\includegraphics[width=12cm]{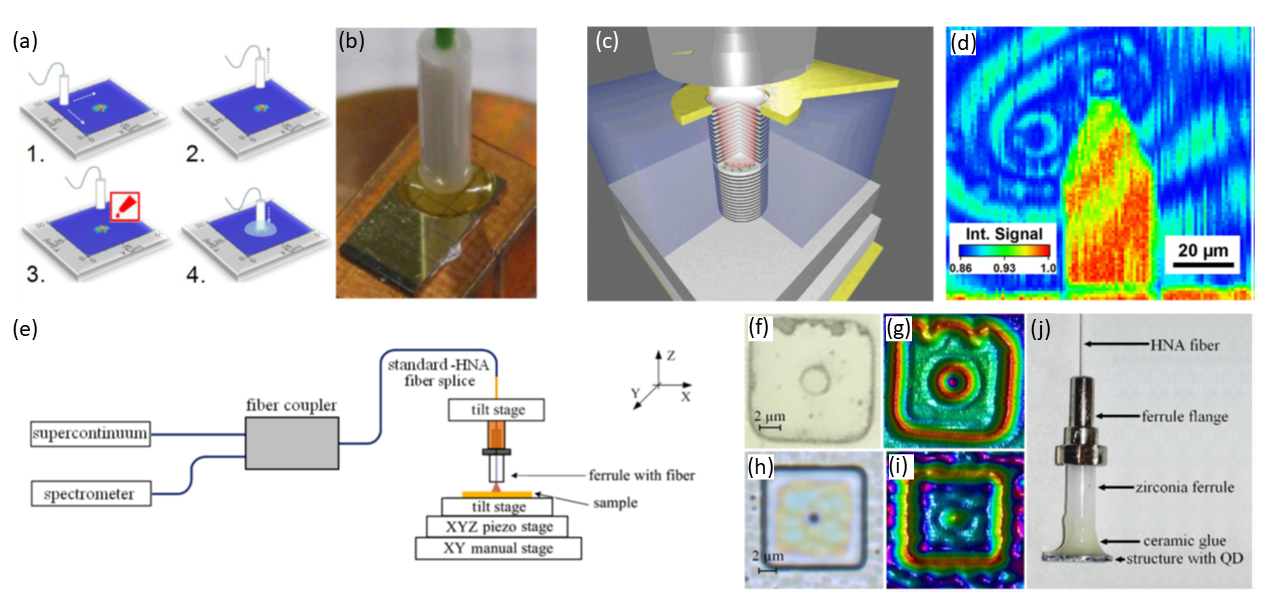}
\caption{Fiber-coupling techniques based on optical source-fiber alignment at room temperature. (a) Optical alignment and gluing of a single-QD micromesa to a multimode fiber. 1. Laser light is coupled into the fiber, which is scanned across the sample surface. 2. At the position of the QD-micromesa wetting layer, emission is generated and collected by the fiber as optical feedback. 3. After lifting the fiber at the position of the QD-micromesa epoxy glue is applied to the surface. 4. The fiber is brought into contact with the sample and the glue is cured. (b) Optical image of the fiber-coupled QD-micromesa. (c) Schematic view of a fiber-coupled, electrically driven QD-micropillar. (d) Corresponding reflected laser signal obtained by scanning a single-mode fiber across the sample surface. The electrical contact leading to high reflected intensity is clearly identified. Before gluing the fiber similar to (a) fine adjustment is performed by maximizing electroluminescence of the micropillar collected by the single-mode fiber. Then the fiber is glued similar to (a). (e) Alternative fiber alignment technique using the interference signal between a single-mode fiber and the sample surface as optical feedback while scanning the sample surface. Optical images of micromesas with 2 $\mu$m (f) and 0.5 $\mu$m (h) diameter in the center of an etched sample area. (g, h) Corresponding surface maps of the collected interference signal. The psoitions of the mesas can be determined with 50 nm (lateral) accuracy. (j) Optical image of a QD-micromesa with emission in the telecom O-band. The interference method was used for mesa-fiber alignment. (a, b) Reproduced from Ref.~\cite{Schlehahn2018} under Creative Commons CC BY license. (c-d)  Reprinted from Ref.~\cite{Rickert2021} with the permission
of AIP Publishing. (e-j) Reprinted from Ref.~\cite{Zolnacz2019}.}
\label{Fig:fiber1}
\end{figure}

The on-chip fiber coupling of QLSs based on semiconductor QDs is technologically challenging for several reasons. For example, QDs are typically embedded in intricate nanophotonic structures or cavities to maximize photon extraction efficiency. These have lateral dimensions in the micrometer range, so that a sub-$\mu$m alignment accuracy between the source and the single-mode fiber is required to achieve high photon coupling efficiency between the two. Furthermore, QDs must be operated at cryogenic temperatures in order to ensure a sufficiently high luminescence yield. This complicates the fiber source adjustment, which usually cannot be done in the cryostat at the operating temperature of the QD. It should be mentioned that in the case of evanescent microfiber-coupled QD structures, a low-temperature alignment using x-y-z stages is possible in principle, but permanent bonding of the fiber is not. Therefore, adjustment and coupling techniques must be developed and used that do not require optically active alignment to the QD signal at room temperature and ensure a robust and permanent source-fiber connection. It is important to note that these interconnect solutions must be capable of operating at a few 10 K and survive many cool-down cycles while maintaining micron-precise source-fiber coupling. In order to ensure this, it is important, for example, that the coefficients of thermal expansion of the materials used, and in particular the adhesive used to fix the fiber holder, differ only slightly. In practice, this could only be guaranteed to a limited extent, so that mechanical stresses usually arise, which lead to a temperature-dependent (strain-induced) spectral shift of the QD emission of up to a few nm~\cite{Bremer2020}.

A straight-forward fiber coupling solution is based on optical adjustment of the fiber and subsequent gluing of it at the position of the QD-SPS. The challenge here is optical adjustment at room temperature without a direct signal from the QD itself. In Ref.~\cite{Schlehahn2018}, the wetting layer signal of a QD-micromesa, which is strong enough even at room temperature, was used to align the fiber. As shown in Fig.~\ref{Fig:fiber1}(a), the fiber was scanned over the relevant sample area under optical excitation from the fiber by laser light. The spatially resolved luminescence was collected via the same fiber and analyzed by a spectrometer with regard to wetting layer emission. In this way, the position of the QD micromesa could be reliably determined via the wetting layer signal to then glue the fiber aligned to the position of the QD-micromesa. A correspondingly fabricated fiber-coupled QD-micromesa is shown in Fig.~\ref{Fig:fiber1}(b). This emits single photons with a wavelength of 930 nm directly into a multimode fiber. A similar approach was taken in Ref.~\cite{Rickert2021} and applied to couple an electrically driven QD-micropillar fiber (see Fig.~\ref{Fig:fiber1}(c)). In this case, the reflected signal from the electrical contact was used for the rough adjustment (see Fig.~\ref{Fig:fiber1}(d)) before the fine adjustment was made using electroluminescence from the QD-micropillar. Using this approach, the single-mode fiber coupling of the QD-micropillar was achieved. Another variant of this approach uses interference phenomena between the fiber facet and the sample surface to determine the position of the QD-structure~\cite{Zolnacz2019}. As shown in Fig.~\ref{Fig:fiber1}(e), the sample surface was scanned with light from a supercontinuum source and the reflected light was analyzed using a spectrometer. In this way, spatially resolved interference images can be recorded (see Fig.~\ref{Fig:fiber1}(g, i)) in order to determine the position of QD-microstructures with a sub-$\mu$m extension with a lateral resolution of 50 nm, before gluing to the fiber is performed. With this method, it was possible to couple a QD-micromesa with emission in the telecom O-band to a single-mode fiber (SMF) (see Fig.~\ref{Fig:fiber1}(j)), which was then integrated into a stand-alone SPS~\cite{Musial2020}.

\begin{figure}[h!]
\centering\includegraphics[width=12cm]{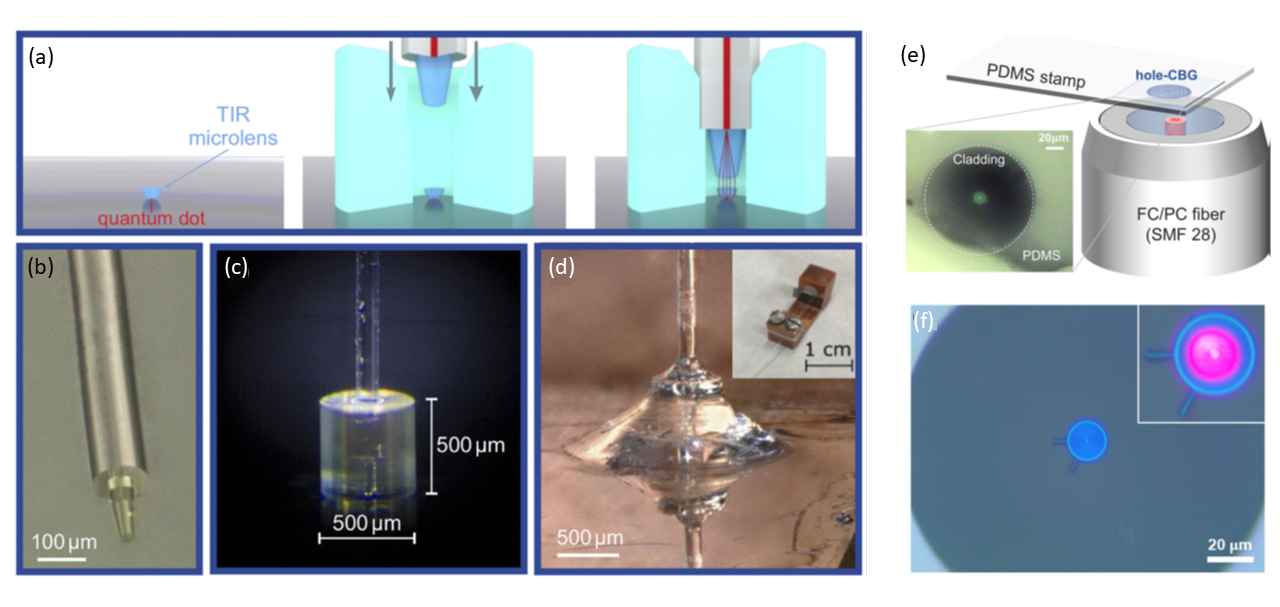}
\caption{SPS fiber coupling via a 3D printed holder ((a-d)) and a polydimethylsiloxane stamp ((e,f)), respectively. (a) Schematic of the coupling scheme. First, a total internal reflection  microlens is printed onto a QD-micromesa via 3D two-photon lithography to efficiently extract the photons from the SPS. Then the fiber holder, aligned with sub-$\mu$m to the source, is printed with 3D two-photon lithography. Finally, the fiber with a numerical aperture (NA) matched focus lens printed to its end facet is inserted into the holder to collect the photons from the total internal reflection lens. (b-d) Optical microscopy images of the lensed fiber, the holder with inserted fiber and the glued fiber. The inset in (d) shows the device mounted to a copper holder. (e) Schematic view of pick-and-place transferring a hole-CBG SPS to the core of a standard telecom fiber using a polydimethylsiloxane stamp. The optical image shows the facet of the fiber with the aligned hole-CBG (green color) in the center. (f) Optical microscopy image of the finished device and close-up view of the laser-illuminated hole-CBG. (a-d) Reproduced from Ref.~\cite{Bremer2020} under Creative Commons CC BY license. (e, f) Reproduced from Ref.~\cite{Jeon2022} with permission from John Wiley and Sons.}
\label{Fig:fiber2}
\end{figure}

An interesting alternative to the coupling techniques described is based on 3D two-photon lithography. This modern microstructuring process can be used to produce microlenses on the one hand and fiber holders on the other, very flexibly and with great precision. The application of this technique for the fabrication of fiber-coupled SPSs is illustrated in Fig.~\ref{Fig:fiber2}(a-d). As seen in panel (a), a high-NA total internal reflection microlens can be printed directly over the QD-SPS via 3D two-photon lithography to effectively collect the emitted photons and focus them onto the fiber core. The fiber itself is guided over a holder, also manufactured using 3D two-photon lithography, and aligned with the source. For a durable connection, the holder is also glued to the semiconductor chip together with a lensed fiber in the last process step. In this way, an SMF-coupled QD-SPS with excellent emission properties at 930 nm was realized~\cite{Bremer2020}.

Still another approach uses the pick-and-place technique for SPS fiber alignment. Fig.~\ref{Fig:fiber2}(e,f) shows a corresponding example in which a QD-SPS based on a hole-based circular Bragg grating (hole-CBG), i.e. a CBG resonator in which the trenches are replaced by hole arrays, was coupled to an SMF. For this purpose, the hole-CBG structure was transferred after fabrication onto a polydimethylsiloxane stamp, with the help of which it was transferred to the facet of the SMF with high precision. The excellent alignment accuracy was impressively demonstrated by the backside illumination of the hole-CBG structure (see Fig.~\ref{Fig:fiber2}(f)). In this way, a fiber-coupled telecom O-band SPS with a total efficiency of 4.6\% was fabricated~\cite{Jeon2022}. Similar results, with an total efficiency of 5.8\%, were achieved before for an nanowire-coupled InGaAs QD emitting at 970 nm~\cite{Cadeddu2016}.

The techniques and results presented show the enormous development in the field of fiber-coupled SPSs in recent years, and Table~\ref{Tab_fiber} summarizes the-state-of the-art in the field. Despite the enormous progress, great efforts are still necessary, in particular to further increase the source-fiber coupling efficiency so that the theoretically predicted values can be achieved. The points discussed reflect only a small part of the activities. For more details on fiber-coupled SPSs, we refer to a recently published review article~\cite{Bremer2022}.

\newgeometry{left=1cm,right=1cm}
\begin{table}
\centering
\caption{Comparison of fiber-coupled QD-SPSs. The table contains information about the SPS design, the fabrication method (deterministic nanofabrication yes/no), the alignment technique, near-field (NF) or far-field (FF) coupling, type of fiber;  multi-mode fiber (MMF), single mode fiber (SMF), micro-fiber (MF), permanent coupling (yes/no), wavelength, extraction and total efficiency and the corresponding reference. The table does not include solutions where the fiber coupling is performed outside the cryostat (using intermediate free-space optics).}
\label{Tab_fiber}
\begin{threeparttable}
\begin{tabular}{ccccccccccl}
\hline
\begin{tabular}[c]{@{}c@{}} Structure\\ \end{tabular} & \begin{tabular}[c]{@{}c@{}} Det. fab. \\ \end{tabular} & \begin{tabular}[c]{@{}c@{}} Alignment\\\end{tabular} & \begin{tabular}[c]{@{}c@{}} NF/FF\\ \end{tabular} & \begin{tabular}[c]{@{}c@{}} Fiber\\ \end{tabular} & \begin{tabular}[c]{@{}c@{}} Perma.\\ \end{tabular} & \begin{tabular}[c]{@{}c@{}} $\lambda$\\ \end{tabular} & \begin{tabular}[c]{@{}c@{}} $\eta_{\rm ext}$\\ \end{tabular} & \begin{tabular}[c]{@{}c@{}} $\eta_{\rm tot}$\\ \end{tabular} & \begin{tabular}[c]{@{}c@{}} Ref.\\ \end{tabular} \\ \hline
QD-PC-cavity & no & Optical & NF & MF & no & 914 nm & 41\% & 23\% & \cite{Lee2015}\\
 QD-micromesa & yes & PL wetting layer & NF & MMF & yes & 930 nm & 0.28\% & 0.28\% & \cite{Schlehahn2018}\\
QD-microlens & yes & 3D printing & FF & SMF & yes & 930 nm & -- & 0.56\% &  \cite{Bremer2020}\\
 QD-micropillar & no & Reflection, EL & NF & SMF & yes & 930 nm & -- & -- & \cite{Rickert2019}\\
 QD-micromesa & yes & Interference & NF & SMF & yes & 1.3 $\mu$m & -- & 1\%& \cite{Musial2020}\\
  QD-nanowire & no & Pick-and-place & NF & SMF & yes & 970 nm & -- & 5.8\%& \cite{Cadeddu2016}\\
  QD-hole-CBG & no & Pick-and-place & NF & SMF & yes & 1.3 $\mu$m & -- & 4.6\%& \cite{Jeon2022}\\ \hline
\end{tabular}
\end{threeparttable}
\end{table}
\restoregeometry
%Plug-and-Play Single-Photon Devices with EfficientFiber-Quantum Dot Interface DOI: 10.1002/qute.20220002

\section{Performance of quantum dots as stationary qubits and as sources of flying qubits}\label{sec:optprop}
%Todo Armando (GaAs QD based sources), Tobias (InGaAs based sources), Stephan (telecom WL sources)  \\

%\emph{This section discusses the properties of QDs acting as qubits and as sources of flying qubits in the relevant wavelength ranges (e.g. for fiber based quantum communication).}\\

%Topics to be included (including optical and quantum optical properties): 
%\begin{itemize}
%    \item GaAs QD based sources of flying qubits at 780 nm > Armando
%   \item InGaAs based sources flying qubits at 900+ nm > Tobias
%    \item Telecom WL sources flying qubits at 1300 nm and 1550 nm > Stephan
%    \item QDs as stationary qubits, spin properties
%    \item spin-photon interfaces based on QDs and QD molecules > Stephan
%    \item QDs and QD molecules for cluster state generation > Stephan
%    \item Generation of multi-photon states, GHZ states, obital angular momentum states
%\end{itemize}

This section is dedicated to the optical and quantum optical properties of QD-based QLSs, spin-photon interfaces and photonic cluster state sources. In this context, it is interesting to note that quantum nanophotonics was inspired by and benefited greatly from previous studies on atomic cQED and quantum optics. For example, S. Haroche and A. Aspect were the first to demonstrate vacuum Rabi oscillations in a strongly coupled atom-microcavity system ~\cite{Brune1996} and to use the radiation cascade of calcium for realizing the Einstein-Podolsky-Rosen-Bohm Gedankenexperiment~\cite{Aspect1982}, respectively. These and other concepts from quantum optics were adopted later to demonstrate cQED effects also in semiconductor systems~\cite{Weisbuch1992,Gerard1998,Reithmaier2004,Yoshie2004} and to apply the radiative biexciton-exciton cascade of QDs for the generation of time-correlated~\cite{Moreau2001a} and polarization entangled photon pairs~\cite{Akopian2006,Young2006}. Significant progress has been made since then regarding the application of QDs in photonic quantum technology, and in the following QLSs are first presented, with all relevant emission wavelengths in the range from 780 nm to 1.55 $\mu$m being discussed. They are compared with regard to important emission parameters such as photon extraction efficiency and quantum properties such as single-photon purity and indistinguishability. In the second part of this section, we turn to concepts for efficient spin-photon interfaces and entangled photon pairs as well as photonic cluster state generation.

%spin properties, quantum memories, quantum dot molecules, cluster state, spin-photon interfaces dorian gangloff?

\begin{figure}[h!]
\centering\includegraphics[width=12cm]{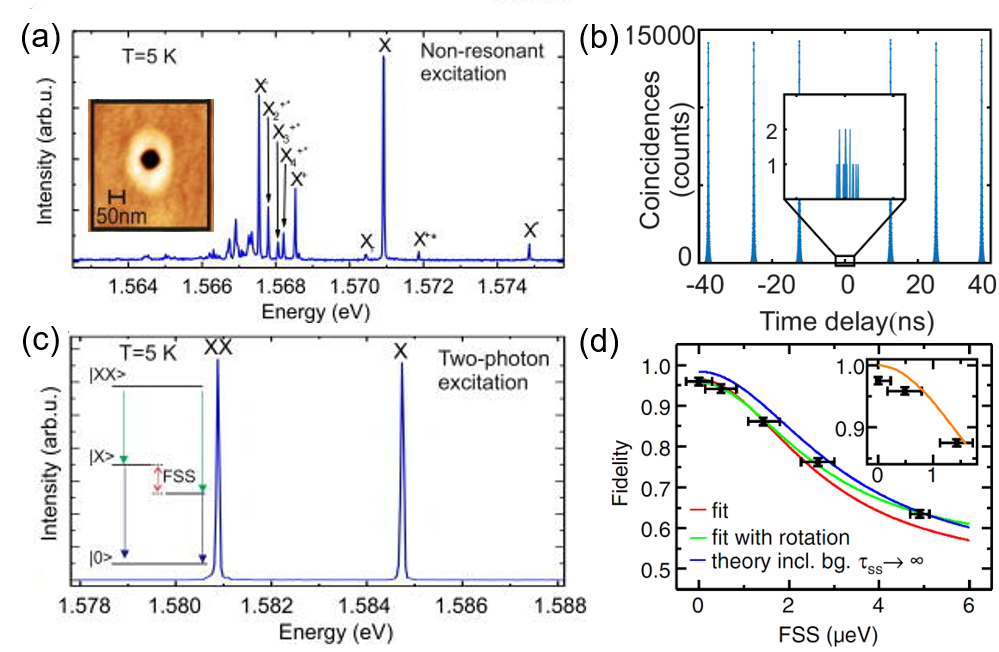}
\caption{Performance of GaAs QDs in AlGaAs nanoholes obtained by local Al-droplet etching. (a) Typical example of PL spectrum of a GaAs QD under non-resonant excitation, showing the isolated neutral exciton emission (X) and additional emission lines attributed to ground state and hot trions. (b) Second-order autocorrelation histogram for the X line under two-photon excitation of the XX state. (c) Example of PL spectrum under two-photon excitation (TPE). (d) XX-X polarization entanglement fidelity as a function of the excitonic fine-structure-splitting $E_{\rm FSS}$ -- see inset of (c) -- tuned via strain, induced by a microprocessed piezoelectric actuator. %(e-f) Sketch of a CBG resonator integrated onto a microprocessed piezoelectric actuator and XX and X lifetimes under two-photon excitation, from~\cite{Rota2023}.
(a -- c) Reprinted from \href{
https://doi.org/10.1063/5.0057070
}{\textit{Covre da Silva et al. 2021}} \cite{DaSilva2021} under Creative Commons CC BY license. (d) Reprinted from \href{https://journals.aps.org/prl/abstract/10.1103/PhysRevLett.121.033902}{\textit{Huber et al. 2018}} \cite{Huber2018}. Copyright (2018) by the American Physical Society.} 

\label{FIG:GaAsQDsperf}
\end{figure}
\subsection{Quantum dot single- and entangled photon sources emitting around 780\,nm}\label{sec:SPS_Performance_780}
QLSs with emission in the near-infrared are interesting for free-space quantum communication and integrated quantum photonics. Among different QD types, we focus here on GaAs QDs in nanoholes obtained via the local droplet etching (see Section~\ref{sec:QDsinNHs}) and with emission wavelength around 780~nm~\cite{DaSilva2021}. Besides the facility of obtaining QDs with low density ($<10^8$~cm$^{-2}$) for single QD devices, the negligible strain, limited alloy disorder, high ensemble homogeneity, high in-plane shape symmetry, and relatively large QD size compared to the free-exciton Bohr radius in GaAs~\cite{Reindl2019} lead to rather unique properties. In particular, the large QD size yields enhanced oscillator strengths, manifesting in spontaneous radiative decay times of the order of 200~ps for confined excitons and trions and 100~ps for biexcitons~\cite{Jahn2016,Huber2017,Keil2017}. This allows high-rate excitation and alleviates the effect of dephasing mechanisms. 

In spite of the dense excitonic levels resulting from the ``weak confinement'' regime (see PL spectrum in Fig.~\ref{FIG:GaAsQDsperf}(a)), driving QDs resonantly with the two-photon-excitation method results in PL spectra dominated by the XX and X emission (Fig.~\ref{FIG:GaAsQDsperf}(b)) and outstanding $g^{(2)}(0)$ values below 10$^{-4}$~\cite{Schweickert2018}, see Fig.~\ref{FIG:GaAsQDsperf}(c). 

The short lifetime, and hence relatively large Fourier-transform-limited (natural) linewidths of about 2-3~$\mu$eV, combined with high in-plane symmetry (see Fig.~\ref{FIG:QDs_in_NHs}(e) and inset of Fig.~\ref{FIG:GaAsQDsperf}(a)), which result in ensemble-averaged $E_{\rm FSS}$ of $<3~\mu$eV~\cite{DaSilva2021}, make these QDs ideally suited as sources of polarization-entangled photon pairs~\cite{Keil2017,Huber2017,Liu2019}. To ensure full cancellation of the fine-structure-splitting and simultaneous tuning of the emission energy, multiaxial strain actuators based on laser-microprocessed piezoelectric substrates have been introduced~\cite{Trotta2015,Martin-Sanchez2016}. By employing them to tune the $E_{\rm FSS}$ of a GaAs QD, entanglement fidelities of up to about 98\% have been observed~\cite{Huber2018} (see Fig.~\ref{FIG:GaAsQDsperf}(d)). Recent investigations have pointed out that the residual deviation from perfect entanglement is due to a combination of several phenomena, which we discuss in more detail in Sec.~\ref{sec:entangledpair}. Also for entangled photon generation the short intrinsic QD lifetimes contribute to alleviate the effect of inhomogeneous dephasing. At the same time, the large QD size and consequently small spacing between ground-state and excited states make state-of-the-art GaAs QDs vulnerable to thermally activated decoherence due to interactions with acoustic phonons~\cite{Lehner2023}.

The high ensemble homogeneity of GaAs QDs (wavelength spread of a few nanometers in an ensemble~\cite{Rastelli2004b,Heyn2009,Keil2017,DaSilva2021}) facilitates experiments and applications relying on TPI among photons emitted by independent QDs~\cite{Benyoucef2009,Reindl2017,Zhai2022} and have led to the highest TPI visibility to date~\cite{Zhai2022}, as discussed in Section~\ref{sec:applications_advanced_QKD} (see Fig.~\ref{fig:Fig_Advanced_QKD}(c)). 
High ensemble homogeneity combined with post-growth fine-tuning provided by strain and electric fields may lead to scalable QD hardware. %The device sketched in Fig.~\ref{FIG:GaAsQDsperf}(e) represents an important step in this direction: 
%A QD deterministically embedded in a CBG resonator (see Section~\ref{sec:deterministic_fabrication}) integrated onto a multiaxial strain actuator~\cite{Trotta2015}. With this kind of device, Rota et al.~\cite{Rota2023} have recently demonstrated single-photon extraction efficiencies of 0.77 as well as Purcell factors of about 10, resulting in ultrashort radiative decay times for the XX-X cascade in the embedded QDs, see Fig.~\ref{FIG:GaAsQDsperf}(f). 

\subsection{Quantum dot quantum light sources emitting at around 900\,nm}\label{sec:SPS_Performance_NIR}

As the pioneering work on S-K QDs was carried out on the InGaAs/GaAs material system, typically resulting in QD emission wavelengths between 890\,nm to 970\,nm (see Section~\ref{sec:fab_SK}), QD-based QLSs emitting in this wavelength range have the longest history. Since the first-time demonstration of single-photon emission from epitaxial QDs by Michler et al. in 2000 \cite{Michler2000}, these QLSs developed to a mature quantum technology enabling high performance quantum light generation. To achieve high photon extraction efficiencies, QDs can be integrated into different types of photonic structures \cite{Barnes2002}, including microlenses, micropillars, CBG resonators, PC cavities and photonic wires as discussed in previous sections.

\begin{figure}
  \includegraphics[width= \linewidth]{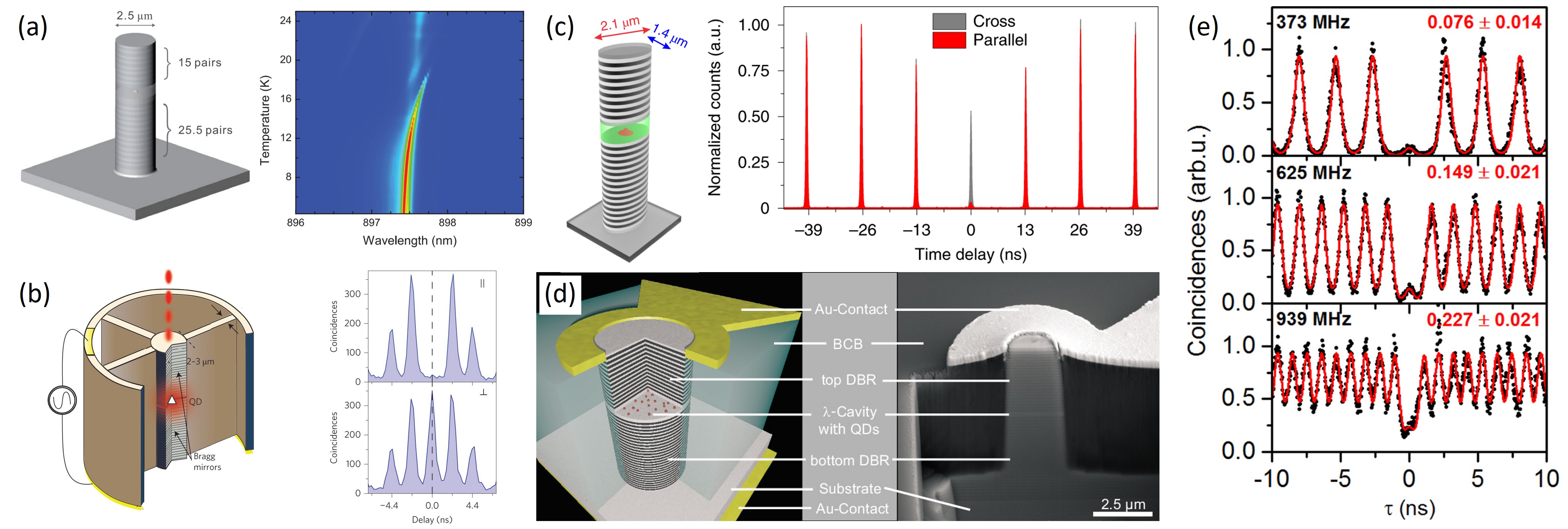}
  \caption{(a) Schematic (left) and spectral detuning vs. temperature map (right) of a QD micropillar cavity used for the generation of highly indistinguishable photons in Ref.~\cite{Ding2016}. (b) Illustration of a deterministically-fabricated electrically-gated QD micropillar cavity (left) and measurement data of HOM two-photon interference (right) from Ref.~\cite{Somaschi2016}. (c) Schematic of an elliptical QD micropillar cavity (left) enabling the highest photon extraction efficiencies and highly indistinguishable photons (right). (d) Schematic and cross-sectional SEM image of a single-photon light-emitting diode based on electrically contacted p-i-n doped micropillar cavities with self-organized InAs/GaAs-QDs \cite{Heindel2010}. (e) $g^{(2)}(\tau)$ measurements on a single-photon LED from (d) as a function of the excitation repetition rate up to the GHz range \cite{Schlehahn2016}. 
  (a) reprinted by permission from \href{https://doi.org/10.1103/PhysRevLett.116.020401}{\textit{Ding et al. 2016}} \cite{Ding2016} Copyright 2016 by the American Physical Society, (b) adapted from \href{https://doi.org/10.1038/nphoton.2016.23}{\textit{Somaschi et al. 2016}} \cite{Somaschi2016} with permission of Springer Nature: Copyright 2016 Springer Nature, (c) reprinted from \href{https://doi.org/10.1038/s41566-019-0494-3}{\textit{Wang et al. 2019}} \cite{Wang2019a} with permission of Springer Nature: Copyright 2019 Springer Nature, (d) reprinted from \href{http://aip.scitation.org/doi/10.1063/1.3284514}{\textit{Heindel et al. 2010}} \cite{Heindel2010} with the permission of AIP Publishing, (e) reprinted from \href{https://doi.org/10.1063/1.4939831}{\textit{Schlehahn et al. 2016}} \cite{Schlehahn2016} under Creative Commons CC BY license.}
  \label{fig:Fig_SPS_Performance_NIR}
\end{figure}

A frequently used type of photonic structure are micropillar cavities \cite{Gerard1998,Solomon2001}, which enable large Purcell enhancements and high photon extraction efficiencies in a narrow spectral range along with directional emission normal to the sample surface~\cite{Pelton2002}. Using resonant excitation, single photons can be generated on-demand with near-unity generation probability while keeping dephasing low. In 2016, Ding et al. \cite{Ding2016} reported an SPS simultaneously achieving a high photon extraction efficiency of 66\%, a single-photon purity of 99.1\%, and photon indistinguishability of 98.5\% (see Fig.~\ref{fig:Fig_SPS_Performance_NIR}(a)) using QD-micropillars with Q-factor around 6000 \cite{Unsleber2016a}. While this high performance level has been reached with non-deterministic device approaches here, deterministic fabrication technologies are useful to increase the device yield by spatio-spectrally matching the emitter-mode coupling. Using in situ photolithography \cite{Dousse2008}, Gazzano et al. fabricated deterministic micropillar cavities containing single pre-selected QDs \cite{Gazzano2013} with high photon extraction efficiency and high degree of photon indistinguishability of 0.79$\pm$0.08 and (82$\pm$10)\%, respectively. Implementing electrical gates in p-i-n doped micropillars, this technology has been further developed to enable a spectral fine-tuning of the quantum emitters \cite{Nowak2014}. This enabled the fabrication of a near-optimal SPS with a photon indistinguishability of up to $(99.56\pm0.45)$\% in 2016 \cite{Somaschi2016}. In a similar work by Unsleber et al., extraction efficiencies of up to $(74\pm4)$\% have been achieved using a deterministically fabricated QD-micropillar device \cite{Unsleber2016}. 

Experimental realizations of coherent resonant excitation schemes often rely on polarization filtering for suppressing the excitation laser at the wavelength of the quantum emitter, which in-turn reduces the photon extraction efficiency by 50\%. In 2019, this limitation has been overcome by Wang et al. using polarization-selective Purcell microcavities~\cite{Wang2019a}. Here, narrow-band elliptical micropillars, as previously used to achieve linearly polarized emission with Purcell-enhanced photon extraction efficiency under non-resonant excitation~\cite{Moreau2001}, (see Fig.~\ref{fig:Fig_SPS_Performance_NIR}(c)) and broad-band elliptical Bragg gratings (cf. discussion on CBGs in Section~\ref{sec:SPS_Performance_780}) were employed to realize a polarization-orthogonal excitation–collection scheme minimizing the polarization filtering loss under resonant excitation. The authors demonstrated a polarized single-photon efficiency of 0.60$\pm$0.02, a single-photon purity of 0.975$\pm$0.005 and an indistinguishability of 0.975$\pm$0.006 for their micropillar device.

Another route to combine coherent pumping with high photon extraction efficiencies are advanced excitation schemes using driving laser fields which are spectrally detuned with respect to the transitions of the quantum emitter. Examples are two-photon resonant excitation of the XX-X cascade without \cite{Brunner1994,Stufler2006,Jayakumar2013,Mueller2014} and with \cite{Sbresny2022} stimulation pulse, dichromatic excitation of a single transition \cite{He2019,Koong2021}, or recent theory proposals for swing-up schemes using spectrally far-detuned frequency-modulated laser pulses~\cite{Bracht2021}. The latter is also referred to as SUPER (for Swing UP of quantum emittER population) scheme and has recently been demonstrated for the first time experimentally by Karli et al.~\cite{Karli2022}.

For some applications, including those relying on the XX-X cascade (e.g. for the generation of polarization entangled photon pairs), it is beneficial to have a high photon extraction efficiency in a wider spectral range. Examples of such photonic structures offering broad-band capability are photonic (nano)wires \cite{Claudon2010}, lens structures, and the aforementioned CBGs \cite{Ates2012}. Following the pioneering top-down QD-photonic-nanowire approach of Claudon et al.~\cite{Claudon2010} with a photon extraction efficeincy of 72\% and $g^{(2)}(0) \approx 0.01$, a bright SPS based on bottom-up grown tapered InP nanowires with integrated positioned InAsP QDs was demonstrated by Reimer et al. in 2012 \cite{Reimer2012}, with a reported photon extraction efficiency of 42\% and a measured antibunching value of $g^{(2)}(0)<0.5$ under continuous wave excitation. While the spatial distribution of the nanowires was statistically random in this work, pre-patterned substrates can be used to achieve site-controlled growth of photonic nanowires with integrated single quantum emitters \cite{Heinrich2010}. Employing deterministically fabricated microlenses with embedded pre-selected QDs, Gschrey et al.\ demonstrated in 2015 an SPS~\cite{Gschrey2015} with a broadband photon extraction efficiency of $(23\pm3)$\% into an NA of 0.4, low multi-photon emission probabilities $g^{(2)}(0)<0.01$, and high photon indistinguishability of (80$\pm$7)\%, even beyond saturation of the quantum emitter. In follow-up work, these QD-microlenses were used to explore dephasing mechanisms limiting the photon indistinguishability \cite{Thoma2016}, revealing photon indistinguishability of up to $(96\pm4)$\% under quasi-resonant excitation of the quantum emitter at short temporal separations (2\,ns) and low temperatures (10\,K). As revealed in this study, the semiconductor environment in QD samples results in non-Markovian noise correlations, which can lead to reduced photon-indistinguishability at larger temporal separation (see Fig.~\ref{fig:Fig_SPS_Performance_NIR}(d)). 
Also, to further push the achievable single-photon flux at a given extraction efficiency, Schlehahn et al. demonstrated an innovative approach using a mode-locked vertical-external-cavity surface-emitting laser at 500\,MHz repetition rate \cite{Schlehahn2015a}.

A major advantage of semiconductor based QLSs, which is not exploited in experiments using optical excitation, is the possibility to realize complex engineered devices including diode structures for electrical charge carriers injection. This is highly beneficial for applications, not only because higher degrees of device integration become possible, as bulky laser systems become obsolete, but also the clock rate of quantum cryptographic implementations can easily be adjusted and pushed to their limits (see also section \ref{sec:application_SPS_QKD}). The first electrically injected QD-based SPS was reported in pioneering work by Yuan et al. \cite{Yuan2002}. Later, the photon extraction efficiency of electrically triggered SPSs could be significantly increased to 34\%, by embedding QDs in p-i-n doped micropillar cavities with ring-shaped top-contacts \cite{Heindel2010} (see Fig.~\ref{fig:Fig_SPS_Performance_NIR}(d)). In follow-up work by the authors, the overall efficiency could be pushed further to values exceeding $60\%$ (including electrical losses), while excitation repetition, or clock, rates of up to the GHz-range were achieved for this type of device \cite{Schlehahn2016} (cf. Fig.~\ref{fig:Fig_SPS_Performance_NIR}(e)). As discussed in Section~\ref{sec:application_SPS_QKD}, these efficient single-photon emitting diodes have in turn also be employed for the first QKD experiments using electrically injected QD-devices \cite{Heindel2012,Rau2014}. In another approach, QDs were embedded in diode structures to electrically generate polarization-entangled photon pairs via the XX-X radiative cascade \cite{Salter2010,Muller2018}. These so-called entangled light-emitting diodes have later been employed for the first entanglement-based QKD experiments using QD-devices \cite{Dzurnak2015} (see Section~\ref{sec:application_EPS_QKD} for details).
%
%Other types of photonic structures used to enhance the performance of QD-based quantum light sources include photonic crystal cavities \cite{Madsen2014,Kim2016}, circular Bragg gratings \cite{Davanco2011,Liu2019,Wang2019}, open cavity systems \cite{Tomm2020}, on-chip waveguide based structures \cite{Uppu2020}.

\subsection{Quantum dot quantum light sources emitting in the telecom O- and C-band}\label{sec:SPS_performance_telecom}

With regard to fiber-based quantum networks, QLSs with emission in telecom O-band and C-band at 1.3 $\mu$m and 1.55 $\mu$m wavelength form important building blocks. In these transmission bands, glass fibers have a local minimum attenuation of 0.31 dB/km at 1.3 $\mu$m and an absolute minimum attenuation of 0.15 dB/km at 1.55 $\mu$m wavelengths, which makes them ideal for optical data transmission over long distances. Noteworthy, the O-band is relevant due to a material dispersion that is close to zero, above all for high-bit-rate quantum communication over medium distances of up to around 50 km.

Compared to many other QLSs, which are based, for example, on nitrogen vacancy centers in diamond with fixed spectral properties, semiconductor properties have the great advantage that their emission wavelength can be flexibly nanoengineered through the choice of material and the growth conditions. InGaAs QDs on a GaAs substrate or on an InP substrate are particularly relevant for emissions in the telecom O- and C-band as discussed in section~\ref{sec:epi}. Due to the described extensive optimization in the epitaxial growth of telecom QDs, enormous progress has been made in the field of O-band and C-band QD-SPS in recent years.

Regarding the development of telecom-wavelength QD-SPSs, epitaxially grown quantum emitters were integrated into various nanophotonic structures to increase photon extraction efficiency.  Moreover, cavity-based concepts, in analogy to NIR SPSs discussed in Sections \ref{sec:SPS_Performance_NIR}, use cQED effects for performance optimization, for instance in terms of high photon indistinguishability. 

\begin{figure}[h!]
\centering\includegraphics[width=14cm]{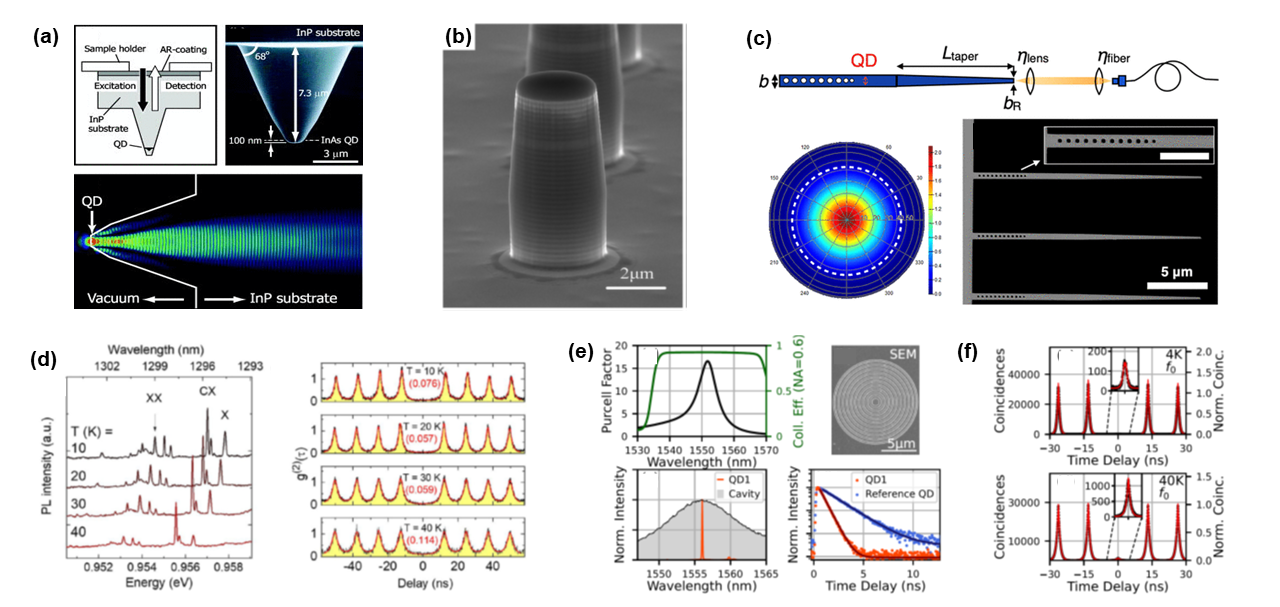}
\caption{Telecom wavelength QD-SPSs. (a) Photonic horn based QD-SPS emitting in the C-band~\cite{Takemoto2007}. (b) Telecom O-band micropillar SPS based on InAs/GaAs QDs in a Al$_{0.9}$Ga$_{0.1}$As/GaAs
DBR cavity~\cite{Chen2017}. (c) Tapered nanobeam based SPS with emission in the O-band Courtesy of~\cite{Lee2020}. (d) Emission spectrum and temperature dependent $g^{(2)}(\tau)$ functions of a deterministically fabricated O-band micromesa SPS~\cite{Srocka2020a}. The device shows high temperature stability and strong multi-photon suppression up to 40 K with $g^{(2)}(0) = 0.076$ at 4 K. (e) Optical characteristics and SEM image of a C-band CBG-SPS and (f) temperature dependent $g^{(2)}(\tau)$ functions demonstrating also high stability up to 40 K, and $g^{(2)}(0) = 0.0052$ at 4 K~\cite{Nawrath2022}. (a) Reproduced from Ref.~\cite{Takemoto2007} under Creative Commons CC BY license. (b) Reproduced from Ref.~\cite{Chen2017} under Creative Commons CC BY license. (c) Reprinted with permission from
Ref.~\cite{Lee2020} 2020 American Chemical Society. (d) Reprinted from Ref.~\cite{Srocka2020a}, with the permission of AIP Publishing. (e, f) Reprinted with permission from Ref.~\cite{Nawrath2022}. Copyright {2022} American Chemical Society.}
\label{Fig:telecomSPS}
\end{figure}

On the one hand, the implemented concepts include simple QD micromesas, QD microlenses and QD solid immersion lenses, some of which were manufactured deterministically~\cite{Sartison2018,Srocka2018,Srocka2020}. Using this broadband approach, O-band (C-band) SPSs with extraction efficiencies $\eta_{\rm ext}$ of up to 17\%~\cite{Yang2020} (13\%~\cite{Musial2021}) could be demonstrated. Temperature emission spectra and $g^{(2)}(0)$ functions of a O-band micromesa SPS are presented in Fig.~\ref{Fig:telecomSPS}(d)~\cite{Srocka2020a}. The high temperature stability of this device makes it compatible with cooling via a stand-alone stirling cryocooler with a base temperature of about 30 K~\cite{Schlehahn2015}. Single-photon emission was achieved with a high multi-photon suppression with  $g^{(2)}(0)$ as low as 0.027 $\pm$ 0.005~\cite{Srocka2020}. Further broadband telecom QD-SPSs include photonic wires~\cite{Jaffal2019,Haffouz2020} and photonic horn devices~\cite{Takemoto2007} (see Fig.~\ref{Fig:telecomSPS}(a)) with $\eta_{\rm ext} = $ 11\% (C-band) and $g^{(2)}(0)$ in the few percent range. Noteworthy, photonic wires based on epitaxially grown QDs can cover both the O-band and C-band depending on the height and diameter of the dot-in-a-rod structure~\cite{Jaffal2019}. 

Resonator-based telecom O-band QD-SPSs include vertically emitting micropillars as presented in Fig.~\ref{Fig:telecomSPS}(b) with $\eta_{\rm ext} = 3.3\%$~\cite{Chen2017} and PC-based devices with $\eta_{\rm ext} = 36\%$~\cite{Kim2016} and $g^{(2)}(0)$ values of 0.14 and 0.085, respectively. Moreover, laterally emitted tapered nanobeam devices were developed which feature $\eta_{\rm ext} = 27\%$ and $g^{(2)}(0) < 0.1$ in the O-band~\cite{Lee2020} (see Fig.~\ref{Fig:telecomSPS}(c)). Recently there have also been interesting developments regarding CBG SPS, which, thanks to their wavelength and material flexibility, also promise very good emission properties such as $\eta_{\rm ext}$ exceeding 90\% in the telecom wavelength range~\cite{Rickert2019, Bremer2022}. In experiment, CBG-SPS have been demonstrated in both O-band~\cite{Xu2022} and C-band~\cite{Kolatschek2021,Nawrath2022} with $\eta_{\rm ext}$ 23\% ~\cite{Xu2022} and $g^{(2)}(0) = 0.01$, and 17\% combined with very good single-photon purity of $g^{(2)}(0) = 0.0052$~\cite{Nawrath2022} (see Fig.~\ref{Fig:telecomSPS}(e,f)). 

Overall, these results demonstrate the tremendous advances in telecom QD-SPSs that have been made in recent years. It is noticeable that achieved $\eta_{\rm ext}$ with values of about 10-40\% are well below the theoretically predicted values of over 90\%~\cite{Rickert2019, Bremer2022} for these wavelengths and also behind the experimental $\eta_{\rm ext}$ values that were achieved for comparable NIR QD-SPSs (see Sections \ref{sec:SPS_Performance_780} and~\ref{sec:SPS_Performance_NIR}). On the one hand, this issue can be related to a non-ideal position of the QD in the nanophotonic structure. On the other hand, the systematic and clear deviation of the $\eta_{\rm ext}$ strongly indicates that the optical quality of the telecom QDs, in terms of internal quantum efficiency, can be a problematic factor that directly affects brightness, and $\eta_{\rm ext}$ if it includes the internal quantum efficiency as factor as it is often the case in experimental evaluations. Determining the internal quantum efficiency for quantum emitters is a nontrivial task. One way to extract this important parameter is to perform time resolved PL studies under variation of the optical density of state at the position of the QDs. Experimentally, this is done by systematically reducing the capping layer thickness of the QD sample~\cite{Johansen2008}. A first study of this kind for O-band InGaAs QDs revealed an internal quantum efficiency of (85$\pm$10)\%~\cite{Groe2021}, which is a promising, but also non-ideal, value. In the future, it will be interesting to determine the internal quantum efficiency also for C-band QDs, and to include this parameter in the evaluation of QD-SPSs in order to explain a possible mismatch between theoretical predicted and experimentally obtained $\eta_{\rm ext}$. 

\begin{figure}[h!]
\centering\includegraphics[width=12cm]{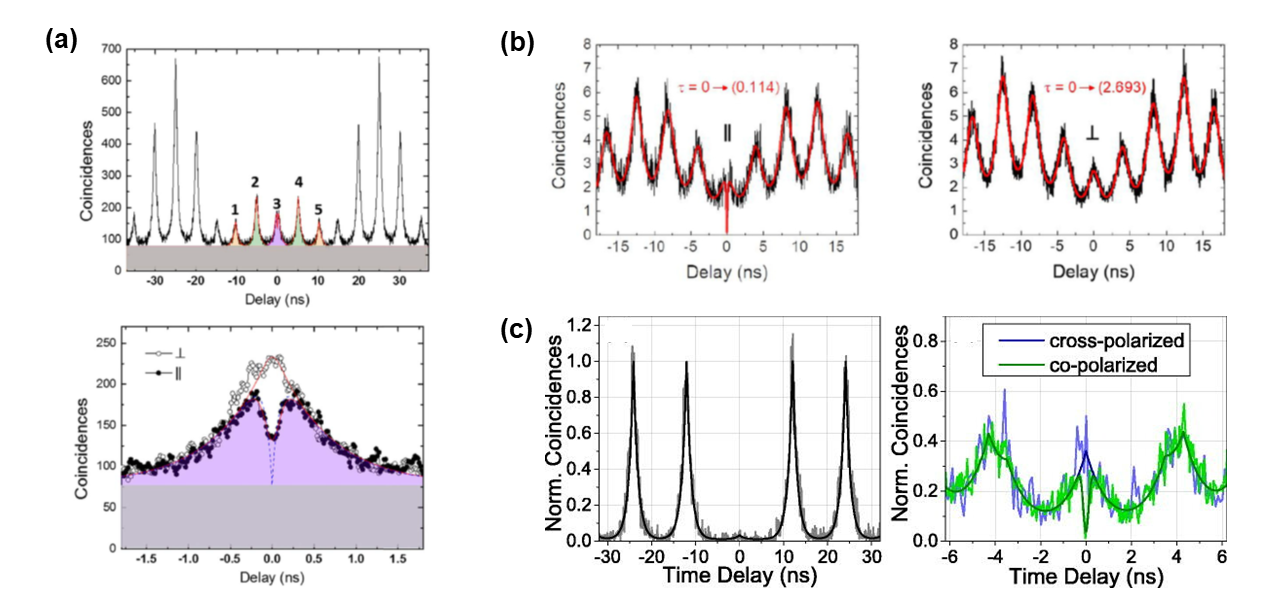}
\caption{Photon indistinguishability of telecom wavelength QD-SPSs determined via two-photon interference measurements in HOM configuration. (a) HOM-correlation histogram of an O-band PC-based QD-SPS for parallel polarization and zoom-in (lower panel) of the center peak for parallel (solid dots) and orthogonal polarizations (open dots). Comparing the data under parallel and orthogonal polarization yields $V_{\rm TPI} = 18\%$. (b) HOM-correlation histogram of a deterministically fabricated O-band QD-micromesa for parallel (left) and orthogonal (right) polarization. Evaluation of the data results in $V_{\rm TPI} = 12\%$. (c) Photon autocorrelation histogram (left) and HOM-correlation histogram (right) of a C-band CBG-based QD-SPS under strict resonant excitation. Fitting the data yields $g^{(2)}(0) = 0.0236\pm0.019$ and $V_{\rm TPI} = 0.1446\pm0.015\%$. (a) Reprinted  from Ref.~\cite{Kim2016}. (b) Reprinted from Ref.~\cite{Srocka2020a}, with the permission of AIP Publishing. (c) Reprinted from Ref.~\cite{Nawrath2021}, with the permission of AIP Publishing.}
\label{Fig:telecomHOM}
\end{figure}

Another important aspect is the photon indistinguishability of the telecom QD-SPSs. While for NIR SPSs, values close to one are obtained almost routinely (see section \ref{sec:SPS_Performance_NIR}), in the case of telecom sources, it is still a major challenge to achieve significant photon indistinguishability. In fact, so far, results from HOM experiments show (non post-selected) two-photon interference visibility $V_{\rm TPI}$ of a maximum of about 20\% in the O-band~\cite{Kim2016, Srocka2020a,Lee2020} (see Fig.~\ref{Fig:telecomHOM}(a,b)) and 15\% in the C-band~\cite{Nawrath2022}, even under resonant excitation~\cite{Nawrath2021} (Fig.~\ref{Fig:telecomSPS}(c)). It is noteworthy that a significantly higher post-selected TPI visibility is often mentioned, but this visibility is mainly limited by the temporal resolution of the HOM setup compared to the coherence time of the photons and is not relevant for typical applications in photonic quantum technology. For instance, in Ref.~\cite{Kim2016} a post-selected TPI visibility of $0.97\pm0.04$ (compared to the non-post-selected value of $V= 0.18\pm0.01$) was determined, considering a HOM resolution of 200 ps and a coherence time of ($150\pm29$) ps.  In two-photon interference measurements, spectral diffusion of the emitter significantly reduces the achievable visibility. Thus, the moderately high photon indistinguishability of the telecom QD-SPSs indicates electronic and magnetic fluctuations in proximity to the QDs, which lead to increased decoherence and spectral fluctuations and thereby limit the TPI visibility~\cite{Kuhlmann2015,Malein2016}. Defect states in the strain reducing layer are possible causes of electronic fluctuations in O-band InGaAs QDs~\cite{Srocka2020a}. Future growth optimizations should therefore aim at optimizing the strain reducing layer to ensure stable electrostatic conditions around the QDs. Furthermore, externally applied electric fields, as successfully practiced in the case of NIR QD-SPS~\cite{Somaschi2016}, could be used for charge noise in order to maximize $V_{\rm TPI}$.\\

Before me move on to the discussion of spin-photon interfaces and photonic-cluster-state sources, we summarize in Table~\ref{Tab_QLS} the state-of-the-art of QD-based QLSs in the presented wavelength ranges from 780 nm to 1550 nm. As can be seen, a variety of device structures have been used to achieve high-performance QLS with close-to-ideal values in terms of single-photon purity, indistinguishability and entanglement fidelity, as well as photon extraction efficiencies exceeding 80\% at emission wavelengths below 1$\mu$m. In contrast, despite enormous progress has been achieved for QLSs emitting in the telecom O- and C-band, there is still a lot of room for improvement especially regarding the indistinguishability which still below about 20\% in the best case.    

\newgeometry{left=0.5cm,right=0.5cm}
\begin{table}
\centering
\caption{Comparison of state-of-the-art QLSs based on semiconductor QDs. (abbreviations: det. fab.: deterministic fabrication, exc. scheme: excitation scheme, NB/BB: narrowband/broadband enhancement of emission, $F_P$: Purcell-factor, $F$: entanglement fidelity, TPR: two-photon-resonant excitation, RF: resonance fluorescence, non-res.: non-resonant excitation.}
\label{Tab_QLS}
\begin{threeparttable}
\begin{tabular}{ccccccccccccl}
\hline
\begin{tabular}[c]{@{}c@{}} Structure\\ \end{tabular} & \begin{tabular}[c]{@{}c@{}} Det. fab. \\ \end{tabular} & \begin{tabular}[c]{@{}c@{}} Exc. scheme \\\end{tabular} & \begin{tabular}[c]{@{}c@{}} $\lambda$\\ {[}nm{]}\\ \end{tabular} & \begin{tabular}[c]{@{}c@{}} NB/BB \\ \end{tabular} & \begin{tabular}[c]{@{}c@{}} $F_P$ \\ \end{tabular} & \begin{tabular}[c]{@{}c@{}} $\eta_{\rm ext}$ \\ \end{tabular} & \begin{tabular}[c]{@{}c@{}} $g^{(2)}(0)$ \\(pulsed) \\ \end{tabular} & \begin{tabular}[c]{@{}c@{}} $V_{\rm TPI}$\\ \end{tabular} & \begin{tabular}[c]{@{}c@{}} $F$\\ \end{tabular} & \begin{tabular}[c]{@{}c@{}} Ref.\\ \end{tabular} \\ \hline

%Structure, Det. fab., Excitation scheme, lambda, NB/BB, Purcell, eta, g2 (pulsed), V, Fidelity, Ref.  

CBG & yes & TPR & 780  & BB & $\approx$ 3 & (85$\pm$3)\% & <1\% & (90.3$\pm$0.3)\% & (88$\pm$2)\% & \cite{Liu2019}\\

Planar DBR & no & RF & 780  & NB & $\approx$ 10 & 2.3\% &  1\%& (98.2$\pm$1.3)\% & (85.0$\pm$1.0)\% & \cite{Zhai2022}\\

Micropillar & yes & non-res. & 930  & NB &  $\approx$ 3-4 & 34\% &  17-40\% & -- & 67\% & \cite{Dousse2010}\\

Photonic wire & no & non-res. & 930  & BB & -- & 72\% &  <1\% & -- & -- & \cite{Claudon2010}\\

Micropillar & yes & RF & 930  & NB &  $\approx$ 8 & 65\% &  (0.28$\pm$0.12)\% & (99.56$\pm$0.45)\% & -- & \cite{Somaschi2016}\\

Micropillar & yes & RF & 930  & NB &  $\approx$ 6 & 66\% &  <1\% & 98.5\% & -- & \cite{Ding2016}\\

Micropillar & no & el. & 930  & NB &  $\approx$ 3 & 61\% &  (0.076$\pm$0.014)\%  & (41.1$\pm$9.5)\%  & -- & \cite{Schlehahn2016}\\

Open cavity & no & RF & 930  & NB & $\approx$ 11 & 82\% &  2.1 & 96.7\% & -- & \cite{Tomm2021}\\

PC cavity & no & non-res. & 1300 & NB & $\approx$ 4 & 36\% &  (8.5$\pm$2.2)\% & 18\% & -- & \cite{Kim2016}\\

Micromesa & yes & p-shell & 1300  & BB & -- & 5-10\% &  2-4\% & 12\% & -- & \cite{Srocka2020a}\\

CBG & no & p-shell & 1550  & BB & 3 & 17\% &  (0.52$\pm$0.10)\% & 8\% & -- & \cite{Nawrath2022}\\\hline

\end{tabular}
\end{threeparttable}
\end{table}
\restoregeometry

\subsection{Quantum dot spin-photon interfaces}\label{sec:spin_photon}

Spin-photon interfaces are important links between stationary qubits and flying qubits. Due to this functionality, they have diverse uses in quantum networks. In quantum repeater networks, for example, in connection with quantum memories, they can temporarily store the information to be transmitted locally. They are also needed to exchange quantum information between quantum processor nodes of a future quantum internet via quantum channels~\cite{Cirac1999, Kimble2008}, which requires the coherent coupling of distant (stationary) qubits. 

\begin{figure}[h!]
\centering\includegraphics[width=14cm]{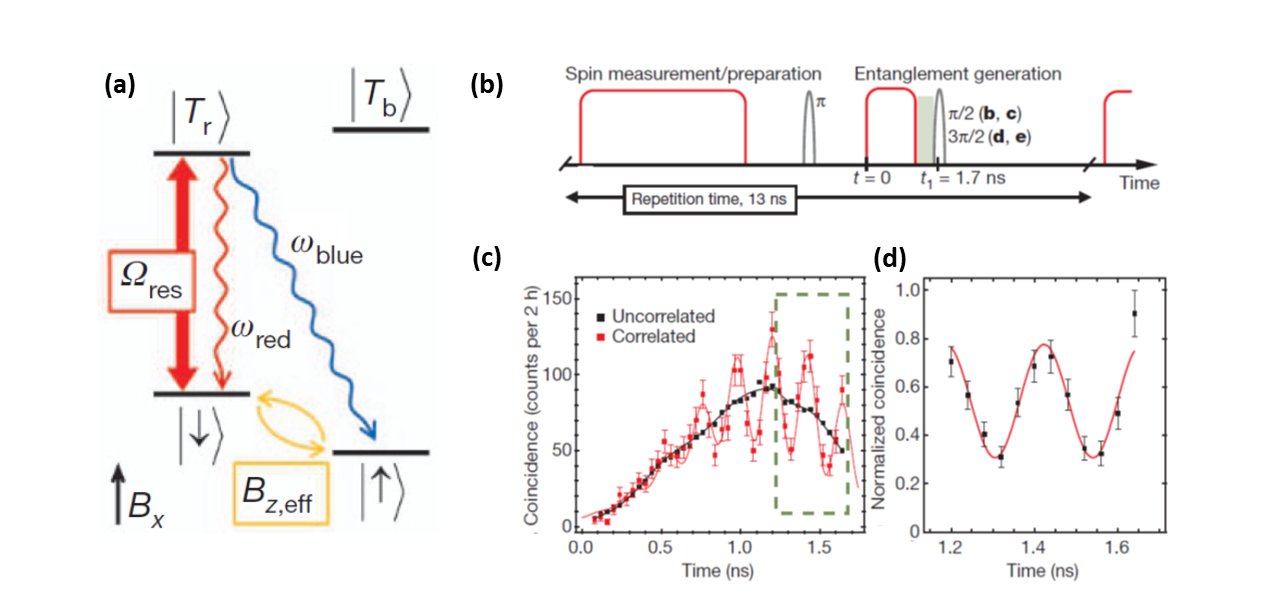}
\caption{Spin-photon entanglement using a charged QD. (a) Energy-level
diagram of a single-electron-charged InGaAs QD under application of a magnetic field $B_x$ in Voigt geometry. (b) Pulse
sequence used to generate and verify spin-photon entanglement in the charged QD system. (c) Corresponding time correlogram between single-photon detection following the entanglement pulse and the detection of a photon during the first measurement/preparation pulse after a $\pi/2$-pulse (red squares). Black squares correspond to reference measurements between spin and photon detection events, taking different excitation/preparation cycles. (d) Normalized coincidences obtained by normalization by counts from correlated spin–photon pairs. Reprinted from \href{https://doi.org/10.1038/nature11573}{\textit{Gao et al. 2002}}~\cite{Gao2012} with permission of Springer Nature: Copyright 2012 Springer Nature.}
\label{Fig:spinphoton1}
\end{figure}

The basic idea behind spin-photon interfaces based on QDs is to entangle the spin degree of freedom of a confined electron or hole with the energy or polarization of an emitted photon. A milestone in this context was reached by W. B. Gao et al. with a sophisticated quantum optical experiment~\cite{Gao2012} using ultra-fast optical quantum control of a single QD spin which had been demonstrated before by D. Press et al.~\cite{Press2008}. The concept by W.B. Gao et al. is based on previous experiments on spin-state-dependent resonance fluorescence from a single-electron
charged QD~\cite{Ylmaz2010} and aims at realizing the entangled spin-photon state $\ket{\Psi}= \frac{1}{\sqrt{2}}(\ket{\uparrow}\ket{\omega_{\rm red};H} + i \ket{\downarrow}\ket{\omega_{\rm blue};V})$. It uses the optical control of a QD trion $T_r$ (negatively charged QD) as depicted in Fig.~\ref{Fig:spinphoton1}(a). The depicted level scheme is based on a magnetic-field-induced splitting of the electron's spin states (magnetic field in Voigt configuration), with $\ket{\uparrow}$ and $\ket{\downarrow}$ respectively denoting spins parallel
and antiparallel to the magnetic field direction. In the applied pulse sequence (see Fig.~\ref{Fig:spinphoton1}(b), first a 5 ns resonant laser pulse $\Omega_{\rm res }$, drives the transition $\ket{\downarrow} \leftrightarrow \ket{T_r}$ and prepares the QD with high probability of 87\% in the $\ket{\downarrow}$ state. Then a 4 ps $\pi$-pulse is applied to transfer the QD into the $\ket{\downarrow}$ state. Subsequently, the entangled spin-photon pair is generated with a 1.2 ns resonant ''entangler pulse'', and a $\pi/2$ or $3\pi/2$ ''measurement/preparation'' pulse rotates the electron spin and project it into the $(\ket{\downarrow} - i\ket{\uparrow})/\sqrt{2}$ and $(\ket{\downarrow} + i\ket{\uparrow})/\sqrt{2}$ state, respectively, after a photon detection event. The whole sequence is repeated every 13 ns according to the pulsed laser repetition frequency. The spin-photon entanglement process is verified by photon correlation measurements between the single-photon detection
events induced by the entanglement pulse and the detection
of a photon during the measurement/preparation pulse. In the corresponding coincidence diagrams depicted in Fig.~\ref{Fig:spinphoton1}(c,d) (for a $\pi/2$ measurement/preparation pulse) starting at the onset of the entanglement pulse (t = 0) presented in Fig.~\ref{Fig:spinphoton1}b), the oscillations appearing for the correlated trace (red data points) clearly reflect the superposition of the photonic state in its two frequency components. In fact. the appearance of these oscillations constitutes a remarkable manifestation of the quantum coherence of the entangle spin-photon state. Comparing the result with correlations measured in different excitation/preparation cycles (black data points) yields an entanglement fidelity of $F = ~0.46\pm0.04$. For more details on the experimental scheme, we refer to Ref.~\cite{Gao2012}.

Since the first demonstration of spin-photon entanglement in the QD system, there has been a number of interesting follow-up work in this area. In the same year, J. R. Schaibley et al. also spin-photon entanglement in a QD with electric charge carrier control~\cite{Schaibley2013}. Also, resonant optical excitation with a similar pulse scheme as described above was used to generate spin-photon entanglement in a four-level QD system generated by a magnetic field in Faraday configuration. In their case, an entanglement fidelity of $0.59\pm0.04$ was achieved. Also in 2012, again using an effective four-level QD system, K. De Greve et al. succeeded in demonstrating spin-photon entanglement with a fidelity of $0.80\pm0.085$, whereby the emitted photon was transmitted to the telecom C-band via frequency conversion~\cite{DeGreve2012}. This is an important step to enable long-distance quantum networks in which photons are transmitted in optical fibers with low loss.

\begin{figure}[h!]
\centering\includegraphics[width=14cm]{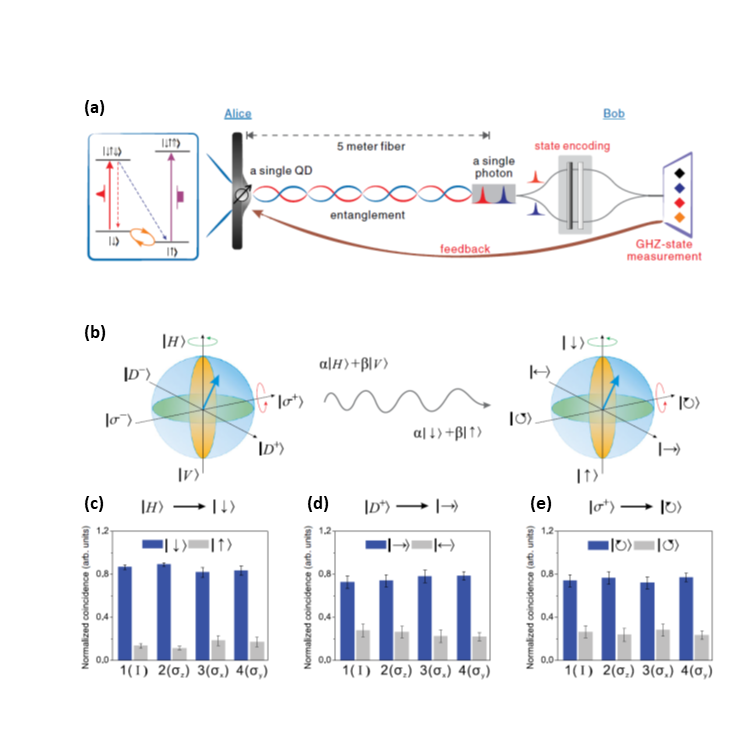}
\caption{Quantum state transfer via spin-photon entanglement. (a) Level scheme of the used QD and illustration of the quantum state transfer concept between Alice and Bob separated by 5 m. After generating spin-photon entanglement, Alice sends the frequency-encoded photon qubit to Bob, where the to-be-teleported state is prepared in the photon’s polarization. Measuring the polarization, frequency, and path degrees of freedom of the photon jointly on the four Greenberger–Horne–Zeilinger (GHZ)-state basis and using the obtained results for feedback, the photon polarization is deterministically
transferred to the QD spin at Alice. (b)  Schematic illustration of the photon-to-spin remote state mapping process from the photon’s to the spin’s Bloch sphere. (c) Coincidence diagprams comparing the experimental results of quantum state transfer with intended outcomes (blue bars) with undesired outcomes (gray bars) of the target photon states ($\ket{H}$,$\ket{D^+}$, $\ket{\sigma^+}$) and correlated spin states ($\ket{\downarrow}$,$\ket{\rightarrow}$, $\ket{\circlearrowright}$). Reprinted from \href{https://doi.org/10.1103/physrevlett.119.060501}{\textit{He et al. 2017}} \cite{He2017a}. Copyright (2017) by the American Physical Society.}
\label{Fig:spinphoton2}
\end{figure}

In another groundbreaking work, Y. He et al. demonstrated the quantum state transfer from a single photon to a distant QD electron spin using spin-photon entanglement~\cite{He2017}. This type of quantum information transmission is another important resource in the development of quantum networks that can be used for distributed quantum processing in the future.  The concept developed by Y. He et al. is shown schematically in Fig.~\ref{Fig:spinphoton2}(a). Again, a four-level system with a $\lambda$ scheme between the QD trion and the two spin states of the ground state is used to generate spin-photon entanglement at Alice via a pulse scheme as described above. The frequency-encoded photon qubit generated is then transmitted to Bob at a distance of 5 m in order to carry out state encoding of the qubits' polarization there. Finally, polarization, frequency, and path degrees of freedom of the photon are measured jointly on the four GHZ state basis, and via a feedback signal the photon polarization detected at Bob is deterministically transferred to the QD spin at Alice. It is interesting to note that ultrafast optical spin echo was applied for prolonging the QD's spin coherence to enable the remote state transfer experiment. Fig.~\ref{Fig:spinphoton2}(b) illustrates the quantum state transfer from Bob to Alice, where a state vector on the photon's Bloch sphere is transferred to the state vector on the spin's Bloch sphere via a flying qubit. The experimental results obtained in this way are shown in Fig.~\ref{Fig:spinphoton2}(c-e). Normalized coincidence counts are plotted, which show the probability that Bob's target photon state was successfully transferred to Alice's spin state. The coincidences of the desired state are shown in blue and those of the undesired state in gray, and analyzing these results yields quantum state transfer fidelities of $F_{\ket{H}}=0.851\pm0.017$, $_{\ket{D^+}}=0.756\pm0.027$, and $_{\ket{\sigma^+}}=0.747\pm0.027$. Based on the results obtained, it will be interesting to extend the distance between Alice and Bob in the future and also to perform quantum state transmission over longer distances via optical fibers.

Further work on QD-based spin-photon interfaces includes waveguide and resonator systems with and without electrical control of the QD states. In such quantum devices, effects of light-matter interaction are used, for example to increase the spin initialization efficiency and qubit gate fidelity and thus to enable the generation of deterministic spin-photon entanglement in the future. In this context, Z. Luo et al. demonstrated that an electrically contacted QD photonic crystal nanocavity exhibits spin-dependent cavity reflectivity in the strong coupling regime~\cite{Luo2019}. Here, the reflectivity can also be controlled electrically, to deterministically load and stabilize electron spin inside QDs. Waveguide-based spin-photon interfaces include QD-nanobeam structures for the coherent optical control of a QD spin-qubit\cite{Ding2019}, and crossed suspended waveguides for interfacing
an optically addressed spin qubit to a path-encoded photon~\cite{Luxmoore2013}. 

Overall, enormous progress has been made in the development of spin-photon interfaces and in the realization of spin-photon entanglement in the last decade. This was made possible on the one hand by the high optical quality of the QDs, but above all by innovative ideas and very sophisticated quantum optical experiments. Further advances in the field can be achieved through increased in- and out-coupling efficiency, in which QDs are integrated into CBG resonators, for example. Furthermore, it will be interesting to increase the spin coherence time, e.g. for quantum state transfer over large distances. As shown in Ref.~\cite{Stockill2016}, all-optical Hahn echo decoupling can be used for this purpose, through which the electron spin coherence times from few tens of nanoseconds to the microsecond regime could be increased.

\subsection{Quantum dots for entangled photon pair generation}\label{sec:entangledpair}

Quantum entanglement is not only an intriguing physical effect, but also a key resource in photonic quantum technology. An application example is the quantum repeater concept, which is based on entanglement distribution between distant nodes of a quantum network, see Sections\ref{sec:application_scenarios_QKD} and~\ref{sec:applications_advanced_QKD}. In order to implement corresponding applications, quantum light sources are required, which in the best case emit entangled photon pairs on demand. Widespread sources of entanglement pairs, which, however, generally do not meet the requirement of making photons available to the user at the touch of a button, are based on spontaneous parametric down-conversion processes~\cite{Zhang2021}. Sources of this type with a non-deterministic emission process have already been widely used in photonic quantum technology. However, the classical emission statistics intrinsically limits their brightness in terms of the average photon pair generation probability per pulse to a rate that is typically < 11\%~\cite{Pan2012}. This imposes a great challenge in advancing efficiency-demanding photonic quantum information technologies. In principle, this limitation can be tackled by heralding of photons, but only at the cost of a large experimental overhead~\cite{GPuigibert2017}.

\begin{figure}[h!]
\centering\includegraphics[width=14cm]{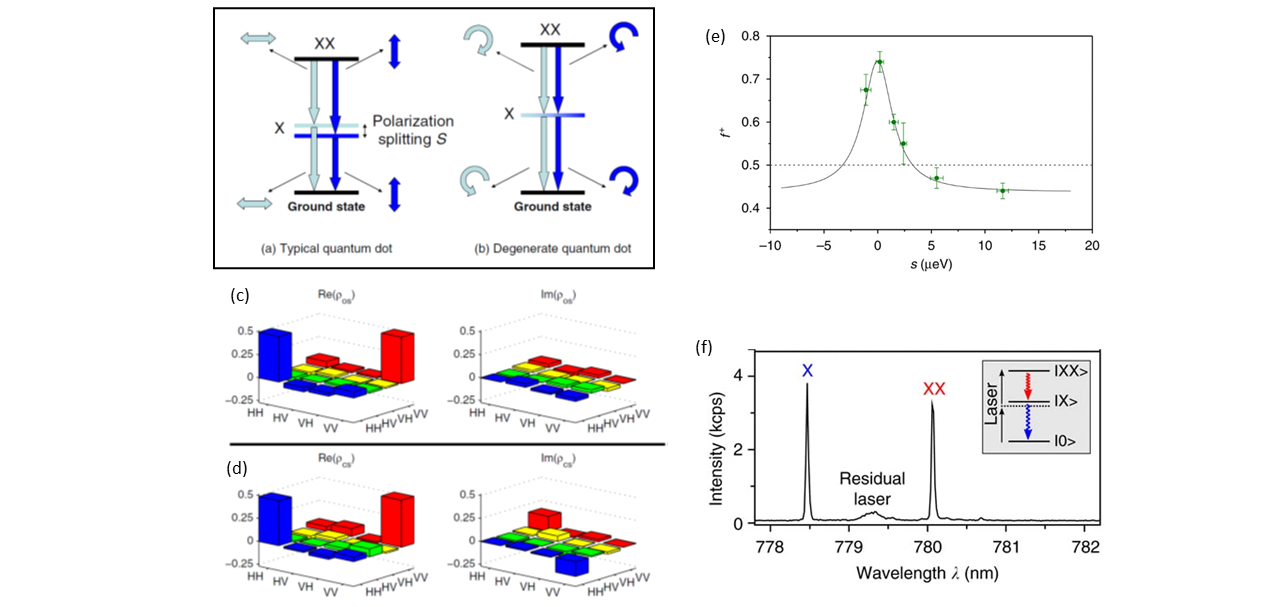}
\caption{Generation of entangled photon pairs from semiconductor QDs. Energy scheme illustrating the radiative decay of the biexciton state (XX) in (a) a typical QD and (b) a QD zero fine-structure splitting. In (a) the biexciton-exciton cascade generates a pair vertically or horizontally colinearly polarized photons. In the ideal case presented in (b) leads to a super-positions of cross-circularly polarized photon pairs which are polarization entangled. Measured two-photon density matrix of photon pairs emitted by the biexciton-exciton cascade using spectral filtering of (c) 200 $\mu$eV and (d) 25 $\mu$eV, respectively. (e) Entanglement fidelity ($f^+$) as a function of the strain-controlled fine structure splitting $s$. The dashed line indicates the classical value of 0.5. Quantum mechanical entanglement is achieved for $|s|\lessapprox 2.5$ $\mu$eV. (f) QD emission spectrum under resonant TPE of the biexciton state. The underlying excitation scheme is schematically shown in the inset. The excitation laser signal is strongly suppressed by using notch filters. (a, b) Figure reproduced with permission from Ref.~\cite{Young2006} © IOP Publishing and Deutsche Physikalische Gesellschaft. Reproduced by permission of IOP Publishing, CC BY-NC-SA. (c, d)  Reprinted from \href{https://doi.org/10.1103/PhysRevLett.96.130501}{\textit{Akopian et al. 2006}} \cite{Huber2018}. Copyright (2006) by the American Physical Society. (e) Reproduced from Ref.~\cite{Zhang2015} under Creative Commons CC BY license. (f) Reproduced from Ref.~\cite{Keil2017} under Creative Commons CC BY license.}
\label{Fig:entpairs}
\end{figure}

In contrast to photon sources which are based on parametric downconversion QDs basically offer the possibility to develop on-demand sources of entangled photon pairs. For this purpose, the biexciton-exciton cascade of QDs can be used in an excellent way, as suggested in the seminal work by Benson et al.~\cite{Benson2000}. In fact, temporally correlated photon pairs are created in the emission cascade of QDs~\cite{Moreau2001a}, with their polarization in one of the maximally entangled Bell states provided the QD has vanishingly small (on the scale of the homogeneous linewidth) fine-structure splitting, as mentioned in Sec.~\ref{sec:theory_qd_states}. This ideal degenerate case is compared in Fig.~\ref{Fig:entpairs}(b) with the typical QD situation with finite fine-structure splitting $S$ in panel (a). In the latter case, the ''which path'' information is maintained so that only classical time-correlations can be observed between the biexciton and exciton photons.    

The generation of polarization-entangled photons via the biexciton-exciton cascade was first achieved in 2006 by Akopian et al.~\cite{Akopian2006} and Young et al.~\cite{Young2006}. In the first case, the studied QD exhibited a fine-structure splitting of more than a factor of 10 larger than the homogeneous linewidth, so spectral post-selection was necessary (at the expense of photon flux) to detect quantum mechanical entanglement. To prove polarization entanglement, quantum tomography measurements were carried out in both experiments, which are based on a total of 16 different polarization-resolved photon correlation measurements. Corresponding two-photon density matrices are shown in Fig.~\ref{Fig:entpairs} (c, d) for a QD with a fine structure splitting of ($27 \pm 3$) $\mu$eV for a spectral selection of 200 $\mu$eV (c) and 25 $\mu$eV (d), respectively. While an evaluation of the matrices shown in panel (a) without spectral post-selection and with vanishingly small imaginary entries does not result in any quantum mechanical entanglement, the density matrices of the photon pairs in panel (b) meet the Peres criterion for entanglement by more than 3 standard deviations~\cite{Akopian2006}. Similarly, in Ref.~\cite{Young2006} it was shown that $>70$\% of the detected photon show polarization entanglement. 

Based on these milestone results, many other important results related to QDs as sources of entangled photon pairs have since been obtained. An important direction of development was to minimize the excitonic fine structure splitting of QDs. This was achieved on the one hand through optimized growth methods, and on the other hand through the post-growth manipulation of this quantity~\cite{Plumhof2012}, above all via strain tuning~\cite{Seidl2006,Trotta2016}. 

In the area of epitaxial growth, using substrates with highly symmetrical crystal orientations such as (111)-oriented GaAs and the growth of strain-free GaAs/AlGaAs QDs are of particular interest in order to achieve low fine structure splittings for efficient generation of polarization-entangled photon pairs~\cite{Huo2013,Chung2016} (see Section~\ref{sec:epi} for details on the epitaxial growth of such QDs). 

On the other hand, strain tuning is an attractive method to control the electronic properties of QDs and in particular the fine structure splitting. In the context of polarization-entangled photon pairs, Ref.\cite{Zhang2015} impressively shows that strain tuning can also influence entanglement fidelity and that it can be maximized for $E_{\rm FSS}=0$, as presented in Fig.~\ref{Fig:entpairs}(e) in accordance with theoretical expectations. Moreover, as presented in Ref.~\cite{Trotta2016}, three-directional strain engineering can be used to generate polarization-entangled photons whose energy can be tuned, in this case to the two D$_1$ lines of Cs, without degrading their degree of entanglement, which is highly interesting for hybrid quantum systems aiming for instance at combining efficient QD quantum emitters with atomic based quantum memories. 

Further important work in the field of polarization-entangled photon pairs from QDs aims at the coherent preparation of the biexciton state. For this purpose, resonant two-photon excitation (TPE) can be used~\cite{Brunner1994,Stufler2006,Jayakumar2013}, which is shown schematically in Fig.~\ref{Fig:entpairs}(f). With this method, the energy of the exciting laser is chosen in such a way that the two-photon energy is sufficient to directly prepare the biexciton state. In practice, this means that the laser energy corresponds $(E_X+E_{XX})/2$ with the energies of the exciton $E_X$ and the energy of the biexciton $E_{XX}$. The corresponding laser straylight (after efficient suppression by notch filters) is observable in Fig.~\ref{Fig:entpairs}(f) together with the generated bexciton (XX) and exciton (X) emission lines. The TPE excitation scheme was first used in Ref.~\cite{Mueller2014} for the on-demand generation of indistinguishable polarization-entangled photon pairs, in which the biexciton population was deterministically prepared by a $\pi$-pulse with high efficiency. It was possible to simultaneously show ultrahigh purity ($g^{(2)}(0) < 0.004$), high entanglement fidelity (0.81 $\pm$ 0.02), high two-photon interference with non-post selective visibilities of 0.86 $\pm$ 0.03 for the biexciton and 0.71 $\pm$ 0.04 for the exciton. Since then, the TPE excitation scheme has been used in many works aiming at the coherent control of the QD biexciton-exciton cascade~\cite{Ardelt2016,Hargart2016,Bounouar2017} and on the on-demand generation of polarization entangled photon pairs~\cite{Jayakumar2013,Keil2017,Huber2017,Bounouar2018},
see also Fig.~\ref{FIG:GaAsQDsperf}. 

Interestingly, advanced applications in photonic quantum information technology such as quantum repeater networks based on entanglement distribution via Bell-state measurements require both high entanglement fidelity and high photon indistinguishability. In this context, it can be shown that due to temporal jitter induced by the biexciton-exciton cascade the maximum indistinguishability of photons generated is $\gamma_{XX}/(\gamma_{XX}+\gamma_{X})$, with the decay rates $\gamma_{XX}$ and $\gamma_{X}$ of the biexciton and exciton state~\cite{Schoell2020}. Thus, a typical ratio of the decay rates $\gamma_{XX}/\gamma_{X} = 0.5$ limits the achievable photon indistinguishability to 66\%. This limit can possibly be overcome by engineering the lifetime ratio using the Purcell effect in suitable resonator structures such as CBG cavities~\cite{Schoell2020}, by spectral filtering~\cite{BassoBasset2021,Schimpf2021_APL}, or by more advanced excitation schemes then TPE~\cite{Bracht2021,Liu2023}.

For sources of entangled photon pairs, the source brightness in terms of photon extraction efficiency is an important parameter, especially with respect to future applications. As with single-photon sources, this property can also be enhanced for photon pair sources by integrating the QD into an appropriate nanophotonic structure. However, here the situation is more complex due to the fact that extraction of both the biexciton and the exciton photons must be increased efficiently. Narrow-band photon extraction, as in the case of simple micropillar structures, is not suitable for this purpose, and broadband concepts are generally used. Experimental results on entangled photon pair sources with enhanced brightness include laterally coupled micropillars whose resonance frequencies were engineered to the biexciton and exciton energies of a deterministic integrated QD. This approach yielded entangled photon pair generation 12\% per excitation pulse~\cite{Dousse2010}. However, the coupled micropillar approach is comparatively complex, and recent work focuses mainly on broadband photon extraction concepts such as microlenses~\cite{Bounouar2018}, optical antennas~\cite{Chen2018} and CBG resonators~\cite{Liu2019, Wang2019} to increase the brightness of paired sources in which entanglement fidelities of about 0.9 and pair extraction efficiencies exceeding 0.6 were reported. 

Obtaining a QD entangled-photon pair source with ideal performance in terms of entanglement fidelity,  photon indistinguishability, and brightness is still an open challenge. In addition to the mentioned time-correlation inherent to the cascaded decay and affecting the photon indistinguishability, also the interaction of the laser excitation with the QD electronic states can deteriorate the ultimate performance of the sources via the AC-Stark effect. In fact, the commonly used TPE method relies on linearly polarized laser pulses with finite duration (typically >2-5~ps), which induce a temporary symmetry breaking even in the case of a QD with $E_{\rm FSS}=0$. Specifically, the AC-Stark effect induces an energy splitting of the excitonic levels and thus a drop in the entanglement fidelity for a fraction of the photon pairs characterized by a biexciton decay occurring while the laser field is still present. This effect becomes particularly pronounced for Purcell-enhanced QDs in photonic structures, in which the lifetime of the biexciton state approaches the duration of the laser pulses, as experimentally demonstrated by Basso~Basset et al.~\cite{BassoBasset2022AC}. These recent findings make it clear that the QD excitation method must be taken into account during the source optimization process and that the Purcell enhancement must be used with caution for increasing the source brightness and for alleviating other dephasing effects such as spin noise~\cite{Schimpf2023} and possibly time-correlation effects in the biexciton-exciton cascade. 

In addition to polarization entanglement, also time-bin entanglement, hyper-entanglement, and energy-time entanglement have been demonstrated with photon pairs emitted by QDs~\cite{Jayakumar2014,Prilmuller2017,Aumann2022,Hohn2023}. Time-bin entanglement is particularly attractive for fiber-based applications but up to now its efficiency remains limited compared to polarization entanglement because of the probabilistic nature of the used excitation schemes. The creation of deterministic time-bin entanglement is therefore highly desired. Also in this case a concerted design of source hardware and excitation method will be required.

\subsection{Quantum dots for photonic cluster state generation}\label{sec:cluster}

Entangled photonic states are the basis for advanced quantum communication schemes, photonic quantum computing and eventually the quantum internet. While polarization-entangled photon pairs can be generated e.g. via the QD XX-X radiative cascade~\cite{Benson2000}, as discussed in the previous section, it is a major challenge to generate entangled photonic states in a scalable manner. In this context, photonic cluster states play a large and important role and can enable for instance a one-way quantum computer~\cite{Raussendorf2001}. In addition to approaches generating photonic cluster states optically~\cite{Lu2007} or using ions~\cite{Lanyon2013} and nitrogen vacancy centers~\cite{Rao2015}, there are also very attractive concepts for their deterministic generation using QDs as we discuss in the following. In general, and related to the results presented in the previous section, corresponding concepts are based on the fact that sequentially emitted photons are entangled via a common stationary qubit. 

\begin{figure}[h!]
\centering\includegraphics[width=14cm]{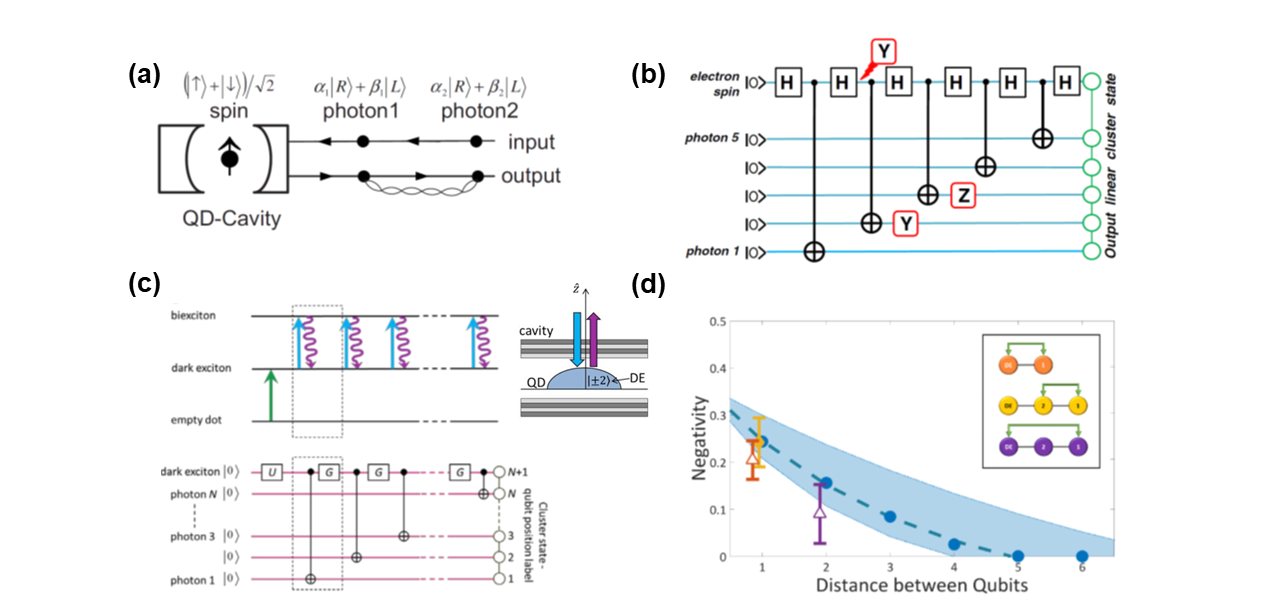}
\caption{QD-based photonic cluster state generation. (a) Photon entangling scheme based on electron spin interaction in a strongly coupled QD-microcavity system. (b) Circuit diagram for the on-demand photonic cluster state generation based on the repeated Hadamard (H) gate and CNOT gate operation. (c) Corresponding experimental level scheme (upper part), and associated circuit diagram (lower part) based on the coherent control of the dark exciton and biexciton of a QD integrated into a planar microcavity (inset). (DE: dark exciton) (d) Localizable entanglement in the generated photon cluster state as function of the distance $d$ between two qubits in the string. (a) Reprinted from \href{https://doi.org/10.1103/physrevb.78.125318}{\textit{Hu et al. 2008}} \cite{Hu2008}. Copyright (2008) by the American Physical Society. (b) Reprinted from \href{https://doi.org/10.1103/physrevlett.103.113602}{\textit{Lindner et al. 2009}} \cite{Lindner2009}. Copyright (2009) by the American Physical Society. (c, d) From Ref.~\cite{Schwartz2016}. Reprinted with permission from AAAS.}
\label{Fig:cluster}
\end{figure}

In Ref.~\cite{Hu2008}, C.Y. Hu et al. proposed to use a charged QD inside a microcavity to generate polarization photon entanglement. Here, the authors propose to use a giant circular birefringence induced by the strong coupling between the integrated QD and the resonator mode to make a photon-spin entangling gate. In their approach, independent photons interacting with a single QD electron spin in the superposition state $(\ket{\uparrow}+ \ket{\downarrow})/\sqrt{2} $ are entangled as soon as the initial state of a third incident photon is measured (see Fig.~\ref{Fig:cluster}(a)). In this way, tripartite GHZ states and, in principle, 1D photonic cluster states can also be generated by sequential application of the scheme. However, the entanglement fidelity of these states is severely limited by the spin-coherence time and the optical losses of the resonator. While a related QD-induced phase shift in a pillar microcavity was reported~\cite{Young2011}, the experimental implementation of the proposed scheme is still pending.
%Deterministic photon entangler using a charged quantum dot inside a microcavity, C. Y. Hu, W. J. Munro, and J. G. Rarity, Phys. Rev. B 78, 125318 – Published 16 September 2008
% Quantum-dot-induced phase shift in a pillar microcavity, A. B. Young, R. Oulton, C. Y. Hu, A. C. T. Thijssen, C. Schneider, S. Reitzenstein, M. Kamp, S. Höfling, L. Worschech, A. Forchel, and J. G. Rarity, Phys. Rev. A 84, 011803(R) – Published 13 July 2011

An interesting alternative for the on-demand generation of a continuous stream of 1D-dimensional cluster states was proposed by N. H. Lindner and T. Rudolph~\cite{Lindner2009}. It does not require strong light-matter interaction and is also based on the photon polarization entanglement via a common electron spin, which in principle can be repeated any number of times in order to generate scalable photonic cluster states. As shown schematically in Fig.~\ref{Fig:cluster}(b), the scheme is based on the Hadamard (H) gate operation on the electron spin followed by a CNOT operation with the nth photon of the photonic cluster state. Repeated execution of this operation generates and entangles all photons of the cluster state across the common electron spin qubit. This concept is largely immune to decoherence, and it can be shown that standard spin errors affect only 1 or 2 of the emitted photons at a time. 

I. Schwartz et al. succeeded in experimentally implementing the scheme proposed by Lindner and Rudoplh for the first time, with the resulting qubit corresponding to a QD confined dark exciton with a long spontaneous lifetime~\cite{Schwartz2016}. In their experimental approach, as shown schematically in Fig.~\ref{Fig:cluster}(c), first a dark exciton is deterministically initialized in its higher energy spin eigenstate (green arrow). Then the dark exciton is repeatably excited to the biexciton state (blue arrows) which results in the subsequent emission of single photons (magenta arrows) forming the 1D cluster state. The corresponding circuit diagram is presented in the lower part, where initialization of the dark exciton is performed by the U gate operation, followed by the excitation-emission represented by a CNOT gate operation and a timed dark exciton precision to prepare for the next entanglement step via a single qubit gate operation G (we refer to Ref.~\cite{Schwartz2016} for more details). The experimental implementation of these qubit operations is very demanding and requires the synchronized resonant excitation of the QD system via several lasers of different wavelengths, and sophisticated correlation measurements between the emitted photons in order to prove the entanglement and to determine the fidelity. The result of corresponding quantum-optical measurements is presented in Fig.~\ref{Fig:cluster}(d). It shows the negativity as a measure of the achieved localizable entanglement in the generated photonic cluster state over the photon distance $d$. The experimental data shows the measured negativity of localizable entanglement between the dark exciton and the emitted photon after one application of the cycle (orange data point), and in a two and a three qubit string (orange and purple data points), respectively. An extrapolation of the achieved fidelity to larger distances between qubits indicated the robustness of the multipartite entanglement in the state produced by our device, and promises entanglement to persist up to 5 qubits.

The discussed results show the possibility to generate photonic cluster states with the help of QDs. However, further technological and experimental improvements are necessary for use in quantum information processing. On the one hand, the performance in terms of the generation rate can be significantly increased by using, for example, a CBG resonator with efficiencies beyond 70\%~\cite{Liu2019} instead of a planar resonator with a photon extraction efficiency of < 20\%, or by using deterministically manufactured QD microlenses~\cite{Heindel2017}. Furthermore, through a targeted optical~\cite{Schmidgall2015} or electrical~\cite{Schlehahn2016} occupation of the QD, a higher repetition rate of the entanglement cycle, and thus in turn an increased generation rate of the cluster states, could be achieved. With regard to large-scale quantum networks based on entanglement distribution using BSM, photon indistinguishability is also an important parameter. In fact, by using photonic cluster states or graph states, which provide redundancy against photon loss and the probabilistic nature of photonic BSMs, all photonic quantum repeaters can be realized that do not require complex quantum memories~\cite{Azuma2015}. Recently, an important step was taken in this direction, in which cluster states with a characteristic entanglement decay length of about ten photons, and a photon indistinguishability of about 80\% were generated with a GHz rate using the QD heavy hole as an entangler~\cite{Cogan2023}. Based on these results, efficient fusion could be made in the future of cluster states to get more complicated graph states for demonstrating all-photonic quantum repeaters. Beyond that, 2D photonic cluster states are of even higher interest for photonic quantum computing, and proposals exist, which promise their efficient generation by coupled QDs, i.e. quantum dot molecules~\cite{Economou2010,Vezvaee2022}. Such 2D photonic cluster generators can strongly benefit from technological advances in the deterministic fabrication of bright electrically tunable QD molecule devices, as recently reported by J. Schall et al.~\cite{Schall2021}.

\section{Integrated quantum photonics with QDs}\label{sec:integration}

\begin{figure}[b!]
\centering\includegraphics[width= 13.35cm]{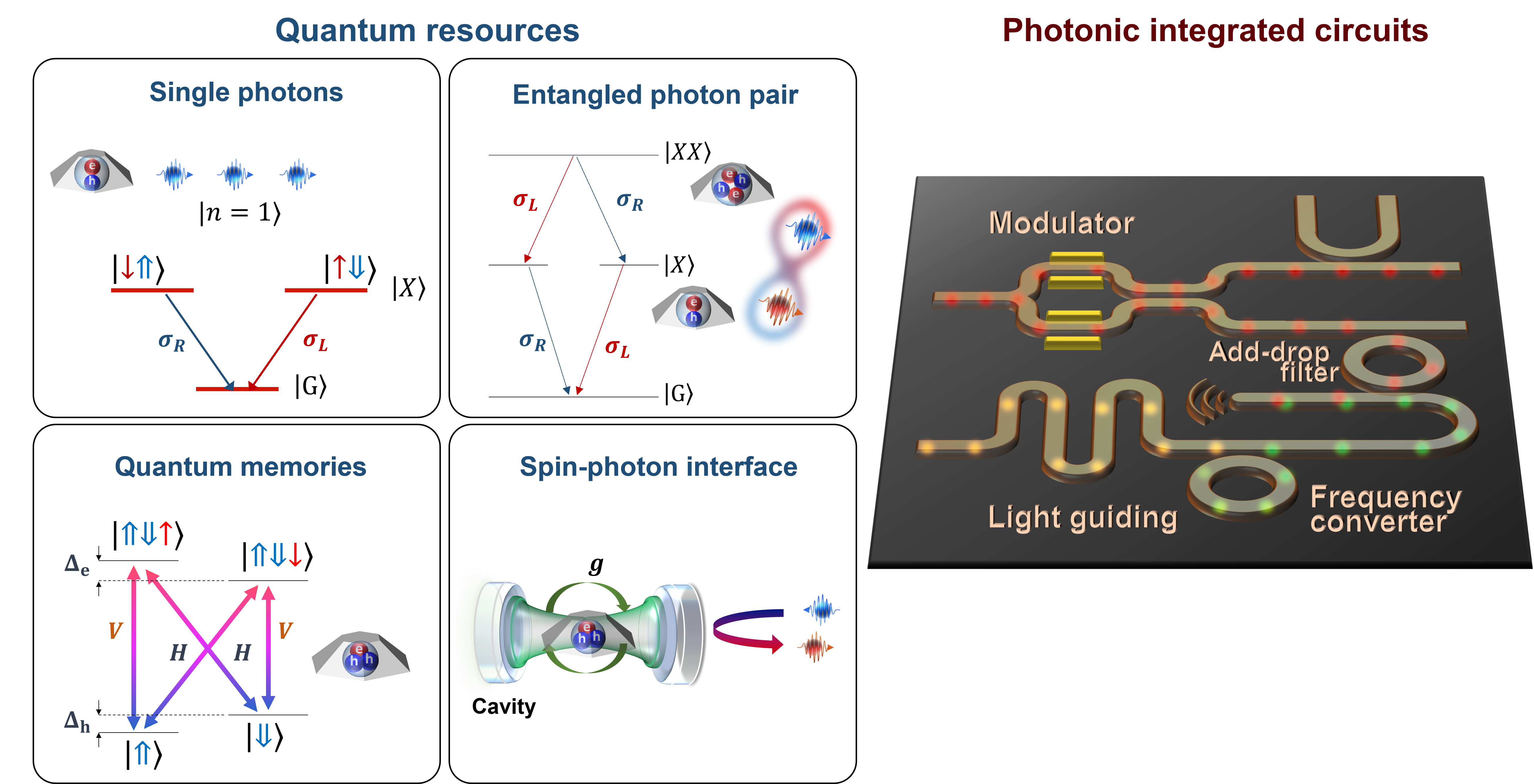}
\caption{Integration of quantum resources %- with a PIC
in an integrated quantum photonic circuit (IQPC). A variety of exciton complexes in QDs serves single photons, polarization-entangled photon pairs, spin quantum memories, and spin-photon interfaces. IQPCs %-PIC 
offers low-loss, functional platforms for manipulating the path, phase, and frequency of photons.}
\label{Fig7_1}
\end{figure}

Integration of quantum emitters into photonic integrated circuits will play a crucial role in future quantum information technologies as %-they 
these circuits will  advance the performance, scalability, and functionality of quantum systems~\cite{Wang2020}. Much progress has already been made in the field of classical photonic integrated circuits. 
Matured growth and fabrication techniques in 
photonic integration enable the integration of a few hundred phase shifters and directional couplers that rapidly control the flow of light and optically map unitary operations on a miniaturized chip \cite{Bogaerts2020}. To bring the advantages of these low-loss and functional photonic platforms to quantum photonics, it is essential to combine them with QLSs. In a simple approach, quantum light can be employed with photonic platforms by external coupling or internal generation, usually based on nonlinear effects such as spontaneous parametric down-conversion or spontaneous four-wave mixing \cite{Signorini2020}. However, such sources are inherently probabilistic, so they have an unfavorable trade-off between achievable single-photon purity and generation rate. Besides that, additional detectors are required for heralding. Therefore, these types of QLSs pose fundamental limitations to the scalability and efficiency of integrated quantum photonics.

To address these challenges, new approaches to integrating solid-state quantum emitters are arising. In particular, incorporating QDs as active sources and hosts of quantum information encoded in photons and spins into low-loss and programmable photonic platforms offers several potential advantages: deterministic single photons and entangled photon pairs, quantum memories, and quantum light-matter interactions \cite{Hepp2019}. Thus, as shown in Fig.~\ref{Fig7_1}, combining these quantum resources into compact, integrated photonic platforms enables the generation, manipulation, storage, and detection of quantum states in a more efficient and functional way. Here, we introduce recent advances in semiconductor QDs integrated into scalable and functional photonic chips.

\subsection{Homogeneous integrated quantum photonic systems}

\begin{figure}[b!]
\centering\includegraphics[width= 13.35cm]{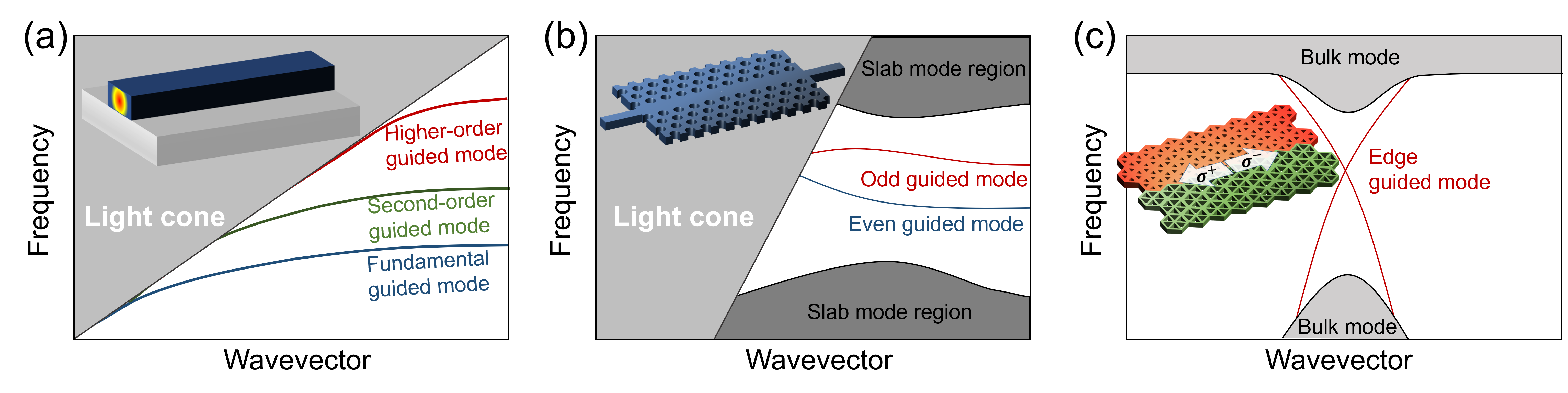}
\caption{A variety of nanophotonic waveguides and their photonic dispersion curves. (a) A ridge waveguide forms optical modes depending on its size and dimensions. (b) A photonic crystal waveguide creates slow light modes within photonic bandgaps. (c) A topological photonic waveguide uses helical topological edge states formed at the boundary of two photonic systems with different band topologies.}
\label{Fig7_2}
\end{figure}

The capability of wafer-scale growth of QDs in a thin film makes Group III-As materials as an ideal platform for integrated quantum photonics \cite{Dietrich2016}. In particular, Group III-As can host QDs with light emission in a wide spectral range, from  visible to telecom wavelengths and can be used to form the photonic structures of low-loss (<0.5 dB/cm) waveguides and high \textit{Q} (>100,000) resonators. Also, from its high refractive index and large electro-optic effect, a Mach-Zehnder interferometer with 50 GHz modulation speed has been demonstrated
\cite{Walker2019}. 

Generation, manipulation, and detection of single photons in an IQPC rely on the efficient interconnection between quantum resources and photonic elements via low-loss channels. Therefore, coupling QDs to linear waveguides is of paramount importance. Figure~\ref{Fig7_2}(a-c) displays different types of waveguides. A ridge waveguide provides a basic unit of photonic circuits and simply holds a single TE mode field that couples the in-plane dipoles of QDs with minimal optical loss. More functional waveguides can be made by photonic crystal structures. Although photonic crystal waveguides require more sophisticated nanofabrication processes, they have the ability to engineer the photonic density of states \cite{Baba2008}. An important feature of slow-light effects in photonic crystal waveguides is that it significantly enhances the emitter-waveguide coupling efficiency ($\beta$) and cooperativity ($\eta$). %- 
Near-unity $\beta$ and high $\eta$ over 60 have been reported \cite{Arcari2014}. Topological photonic waveguides are another important platform that supports topological edge states formed at the boundary of two photonic crystal waveguides with different topologies. Topological waveguides are robust against structural imperfections and allow for unusually narrow bending angles, reflectionless propagation, and unidirectional transport and are thus highly attractive in IQPCs.  

\begin{figure}[b]
\centering\includegraphics[width= 13.35cm]{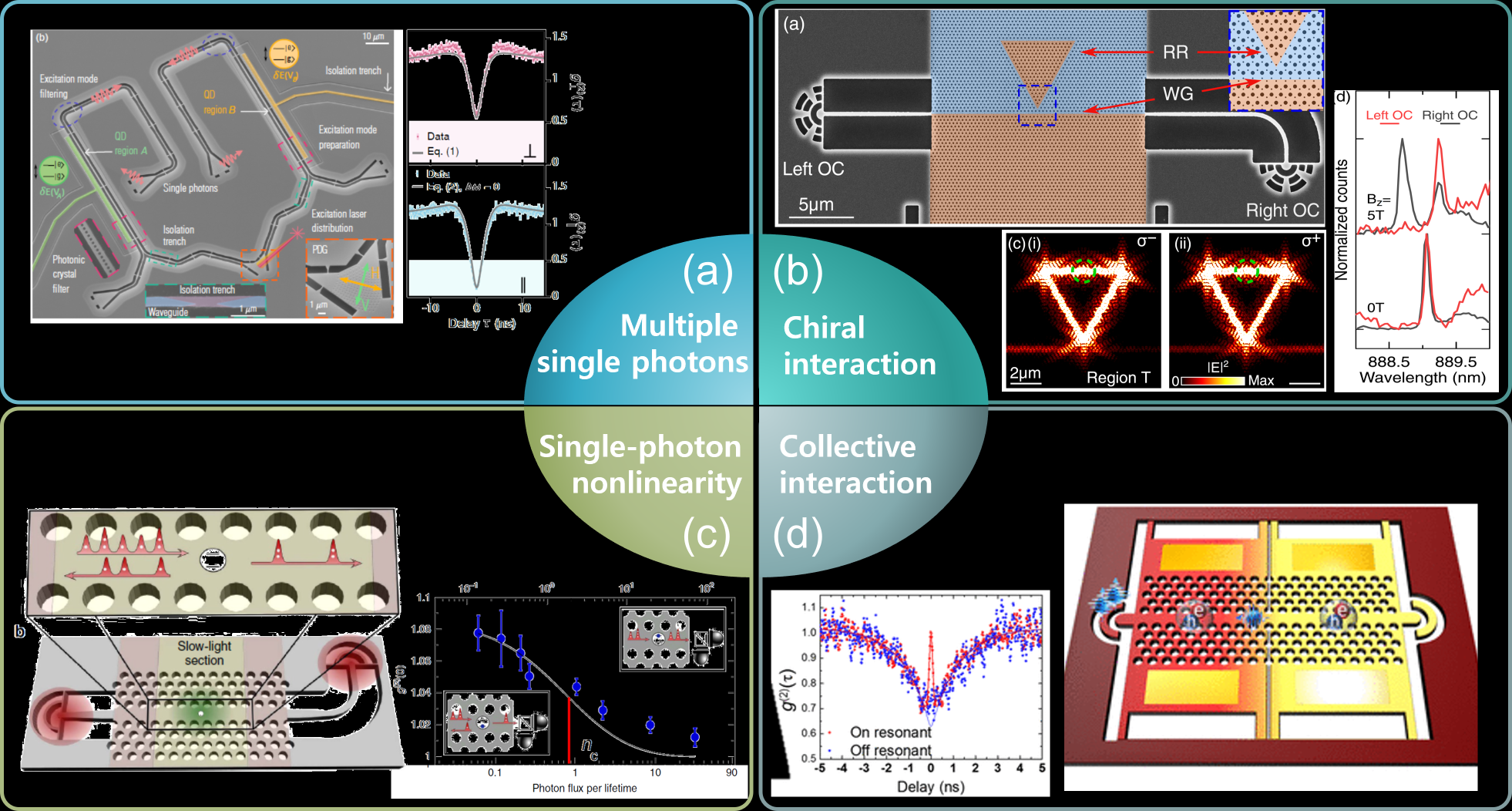}
\caption{Generation of on-chip coupling of single photons, chiral light-matter interaction, single-photon nonlinearity, and scalable interactions in a variety of QD-coupled waveguide systems. (a) Two separated QDs in a waveguide with independent frequency tuners produce indistinguishable photons. (b) QDs in a symmetry-broken waveguide show directional chiral coupling depending on their spin states. (c) Single-photon nonlinearity based on a single QD in a waveguide demonstrates deterministic few-photon scattering of a weak resonant laser, deforming photon statistics. (d) Two separated QDs in a waveguide are independently tuned to resonant, which leads to cooperative emissions. (a) Reproduced from Ref.~\cite{Papon2022}  under Creative Commons CC BY license. (b) Reproduced from Ref.~\cite{Mehrabad20}  under Creative Commons CC BY license. (c) Reproduced from Ref.~\cite{Javadi2015}  under Creative Commons CC BY license. (d) Adapted with permission from Ref.~\cite{Kim2018}. Copyright 2018 American Chemical Society.}.
\label{Fig7_3}
\end{figure}

Homogeneously integrated QD-waveguide systems have led to new opportunities for exploiting quantum optics in compact photonic platforms. Major progress has been made recently toward bright, coherent single photons in a variety of waveguide platforms \cite{Arcari2014, Uppu2020, Barik2018}. Even though, nowadays, a single QD itself can emit indistinguishable single photons with (quasi) resonant excitation techniques, the spectral randomness of each QD significantly limits the scalability of quantum systems. Scalable integration of multiple, ``identical'' quantum emitters has been realized using independent frequency tuning methods based on temperature, bias voltage, or strain \cite{Papon2022, Ellis2018, Grim2019}. Figure~\ref{Fig7_3}(a) shows a TPI experiment of resonant single photons from separate QDs. The frequency mismatch between remotely located QDs is eliminated via the quantum confined Stark effect, and TPI is performed off-chip. 

An important task with integrated quantum emitters in photonic waveguides is creating light-matter interfaces at the level of single photons. Introducing light-matter interactions provides new capabilities, such as deterministic spin-photon interfaces and quantum gates, which are difficult in linear quantum optics. 
Chiral light-matter interaction is an interesting example that creates spin-photon interfaces and quantum state-controlled directionalities \cite{Mitsch2014}. Chiral light-matter interaction has been investigated in various QD-coupled photonic waveguide platforms \cite{Mrowinski2019,Javadi2018,Söllner2015,Barik2018}. Chirality in photonic waveguides can arise from position-dependent spin-momentum locking effects of QDs \cite{Javadi2018, Mrowinski2019}, symmetry breaking in waveguide structures \cite{Söllner2015}, and helical edge modes of topological waveguides \cite{Barik2018, Mehrabad20}. Figure~\ref{Fig7_3}(b) illustrates a chiral quantum optical interface that enables nonreciprocal systems implementing directional conversion of spin information of QD excitons into path information of photons. Also, more recently, the chiral coupling of excitons and biexcitons to a waveguide demonstrated the deterministic generation of spatial path-entangled photon pairs \cite{Ostfeldt2022}. 

Another essential feature of light-matter interaction associated with QD-coupled waveguides is their strong single-photon nonlinearity. %- beyond the linear optics regime. 
For example, the optical transparency of QD-coupled waveguides can be controlled by the states of a single QD, known as dipole-induced transparency \cite{Faraon2008,Singh2022}, similar to electromagnetically induced transparency in three-level atoms.  When a single dipole couples to optical modes, it induces an abrupt change in the phase shift around the transparency window. This change in transmission from transparent to opaque can be controlled by weakly coupled QDs with a large Purcell effect without strong coupling, making it much easier to achieve in waveguides. Another example is a single-photon nonlinear device that induces deterministic few-photon scattering for a weak resonant laser on a single QD. Figure~\ref{Fig7_3}(c) shows the modification of transmitted photon number statistics depending on incoming photon numbers \cite{Javadi2015}. The strong single-photon nonlinearity of the QD-coupled waveguides is responsible for selectively filtering single photons, which can be used to implement single-photon transistors and deterministic Bell-state measurements~\cite{Ralph2015}. Also, advanced schemes for multiphoton-entangled states based on spin-photon interface, such as GHZ states or photonic cluster states prepared by waveguide-coupled single QDs, have been proposed \cite{Tiurev2022}. 

One important aspect for the QD-waveguide platforms compared to QDs coupled to high \textit{Q}, small mode-volume cavities is that the light-matter interaction in a waveguide requires less strict spectral and spatial matching conditions between QDs and modes, and thus developing integrated quantum photonic architectures is more feasible. Moreover, it is also possible to create photon-mediated long-range interaction between coupled multiple QDs. 
To achieve collective interaction between emitters in free space or in a homogenous medium, the emitters need to be placed at a short distance, comparable to the wavelength~\cite{Koong2022}, 
making independent spectral control difficult. 
In a single-mode waveguide, far-separated emitters can couple to the same optical mode, extending the interaction distance. With recent efforts, %-separated
multiple QDs in a waveguide have been successfully tuned into resonance and showed quantum interactions via cooperative emission (Fig.~\ref{Fig7_3}(d)) \cite{Kim2018, Grim2019}. Multiple quantum emitters in a waveguide with tunable long-range interactions open new perspectives for exploring complex ``multi-atomic'' systems. 

\subsection{Heterogeneous integrated quantum photonic systems}

A fully functional quantum photonic architecture requires on-chip integration of highly reliable quantum emitters, low-loss waveguides, fast phase-shifters, and highly efficient single-photon detectors. Also, the integration of other quantum and photonic components, such as quantum memories, spectral filters, and frequency converters, would be desirable for storing quantum states and manipulating photons. The primary limitation of monolithic integration approaches discussed above is that no single material can meet all of these functionalities. In recent years, hybrid quantum photonic architectures heterogeneously integrating multiple components from different material platforms have emerged as an alternative solution \cite{Kim2020,Elshaari2020}. 

Heterogeneous integration can be employed at different levels with different assembly techniques. For example, a GaAs thin film can be epitaxially grown on a silicon-on-insulator wafer \cite{Wei2020a}. To prevent crystal quality degradation due to the different lattice structure and lattice constant of GaAs and Si, active and passive materials for QDs and photonic circuits can be grown individually at wafer scale with high quality and then integrated using wafer bonding techniques \cite{Davanco2017}. 

Alternatively, integration can take place at the functional device level, such as single QD devices placed on prefabricated 
IQPCs~\cite{Katsumi2021, Kim2017, Zadeh2016}. 
In this approach, devices can be %-simply 
assembled by Van der Waals forces or direct bonding techniques, providing freedom in the choice of materials and device design. Furthermore, the device-to-device integration allows pre-characterization and post-selection of each component. Given the difficulties of spatial and spectral control of QDs during their growth process, selective integration features are crucial to increasing system yield in large-scale chips. 

\begin{figure}[t!]
\centering\includegraphics[width= 13.35cm]{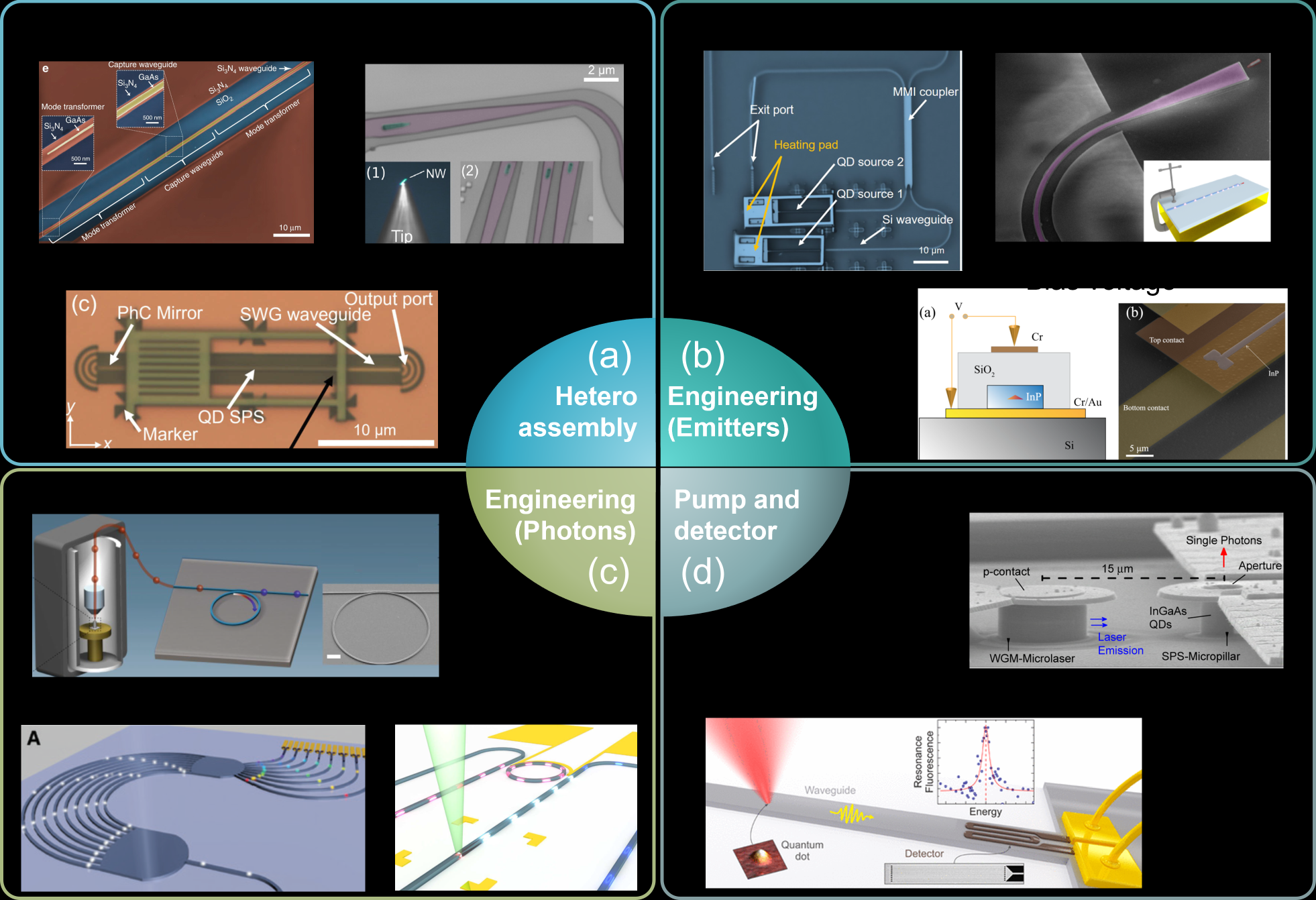}
\caption{Heterogeneous integration and manipulation for fully integrated quantum photonics. (a) Three different methods for hetero assembly of QDs: Wafer bonding (Reproduced from Ref.\cite{Davanco2017} under Creative Commons CC BY license.), Pick-and-place (Adapted with permission from Ref.~\cite{Zadeh2016}. Copyright 20016 American Chemical Society.), and Transfer printing (Adapted] with permission from Ref.~\cite{Katsumi2021}. \textcopyright The Optical Society) (b) Engineering emitter’s frequency by local temperature control (Reprinted from Ref.~\cite{Katsumi2020}, with the permission of AIP Publishing.), strain (
Adapted with permission from Ref.~\cite{Elshaari2018}. Copyright {2018} American Chemical Society.), and applying voltage bias (Reprinted from Ref.~\cite{Aghaeimeibodi2019}, with the permission of AIP Publishing.). (c) Various techniques can engineer the frequencies of single photons in integrated photonic circuits: (Top) four-wave mixing Bragg scattering in a microring resonator for frequency conversion (Adapted with permission from Ref.~\cite{Singh2019}. \textcopyright The Optical Society), (Bottom left) frequency multiplexing for spectral distribution (Adapted with permission from Ref.~\cite{Kahl2017}. \textcopyright The Optical Society), and (Bottom right) frequency filtering using a tunable add-drop filter (Adapted with permission from Ref.~\cite{Elshaari2017}. \textcopyright The Optical Society.). (d) Hybrid integration of micro pump lasers (Reprinted with permission from
Ref.~\cite{Munnelly2017}. Copyright 2017 American Chemical Society.) and superconducting nanowire single-photon detectors (Reprinted with permission from
Ref.~\cite{Reithmaier2015}. Copyright 2015 American Chemical Society.) in photonic circuits.}
\label{Fig7_4}
\end{figure}

As illustrated in Fig.~\ref{Fig7_4}(a), a number of groups have successfully demonstrated heterogeneous integration of a variety of QD structures and photonic material platforms using wafer bonding \cite{Davanco2017}, transfer printing \cite{Katsumi2020,Katsumi2021}, and pick-and-place techniques \cite{Zadeh2016,Kim2017}. As losses are a major source of error in photon-based quantum information processing, the high coupling efficiency of single photons from QDs to waveguides is the most important feature for the heterogeneous assembly of dissimilar platforms via post-processing.
The evanescent coupling between the QD structure and the photonic waveguide enables efficient transmission of single photons into photonic circuits. To further increase the coupling efficiency and directionality of single photons, QDs in tapered nanostructures and Bragg mirrors can be added \cite{Aghaeimeibodi2018}. In addition, to guarantee precise spatial alignment between QDs and photonic circuits, the site-controlled growth (see section~\ref{sec:SCQDs}) or fabrication techniques introduced in section~\ref{sec:deterministic_fabrication} can be adopted. Instead of fabricating QD devices for heterogeneous integration, QDs in a nanowire (see section~\ref{sec:QDinwires}) may also be useful as heterogeneously transferable SPSs. Nanowires can be easily picked and placed into photonic circuits using a micro tip, and they typically have a tapered shape, which is desirable for single-photon coupling with waveguides \cite{Zadeh2016,Mnaymneh2020}. Furthermore, excitons in epitaxial QDs generally couple to light with polarization in the growth plane, so laterally integrated QD devices on a waveguide predominantly couple with TE modes of a waveguide. On the other hand, nanowires can be placed in waveguides with orthogonal growth directions. As both TE and TM modes of the waveguide can be exploited in this configuration \cite{Zadeh2016}, a polarization-insensitive coupling is possible, which is important for supporting polarization-entangled photons from exciton and biexcitons. In the near future, to increase the processing speed and yield for scalable quantum photonic systems, characterization and integration processes could be fully automated \cite{Masubuchi2018}.

Along with the efficient integration of quantum emitters and waveguides, spectrally synchronizing multiple emitters and photons in a photonic chip is essential for achieving quantum interactions and interferences in all-integrated quantum photonic chips. A limited TPI visibility of transmitted photons increases the error rate and lowers the fidelity in the quantum gate operation or Bell-state analysis. In addition, the interaction between single photons and quantum memories also becomes inefficient when their spectra are detuned. Although it would be possible to pre-characterize and post-select proper QD devices with heterogenous integration, fine frequency tuning of emitters is still required for removing residual spectral mismatch. Similar techniques of frequency tuning used in monolithic integration can be applied to hetero-systems. Figure~\ref{Fig7_4}(b) shows the local engineering of the emitter’s frequency via heat, electric gates, or strain controls \cite{Elshaari2018,Aghaeimeibodi2019,Katsumi2020}. Applying elastic stress~\cite{Martin-Sanchez2018} is particularly useful because it not only tunes the emission frequency \cite{Seidl2006,Zander2009,Flagg2010} but also controls the fine structure splitting of excitons, which is essential for generating entangled photon pairs \cite{Seidl2006,Trotta2012,Chen2016,Trotta2016}. As fine-tuning techniques allow for spectral matching, TPI has been demonstrated between separated QDs on a hetero-waveguide platform \cite{Ellis2018}. For long-term measurement, the frequencies could be precisely controlled and monitored in real-time to be resonant during the operation \cite{Zopf2018}. Combining cavity structures in the QD devices can further engineer the optical properties of QDs in the weak and strong coupling regimes \cite{Osada2019}. 

Frequency control often requires conversion in a wide spectral range up to a few hundred nm. For example, most solid-state quantum emitters emit photons in the visible to near-infrared range, while important photonic platforms of Si photonic integrated circuits and optical fibers demand longer wavelengths, such as telecom wavelengths. However, the frequency tuning by engineering the emitters is generally limited to at most a few tens of nm~\cite{Bennett2010,Yuan2018}. Wider-range frequency conversion can be employed using nonlinearity in photonic waveguides or resonators. Figure~\ref{Fig7_4}(c) shows the on-chip frequency conversion of single photons from a single QD with a conversion efficiency of 12\% \cite{Singh2019}. A much higher conversion efficiency of 74\% has been recently reported on a lithium niobate on an insulator platform \cite{Wang2022}. Furthermore, frequency conversion allows connecting dissimilar quantum emitters, such as InAs QDs and nitrogen vacancies in diamonds. They are advantageous for QLSs and quantum memories, respectively. Therefore, interfacing different types of quantum emitters could open new pathways for hybrid quantum systems.

When using QDs in IQPCs, one remaining issue is the spectral separation of a single QD from the background fluorescence, including pumping laser, multi-exciton processes, and the emissions from other QDs. In free space, spatial and spectral isolation of single photons from such background noises can be easily done with confocal microscopy and spectral filters or monochromators. To implement such isolation of single photons in a photonic chip, on-chip single-photon spectrometers performing spectral demultiplexing can be employed using arrayed waveguide gratings \cite{Kahl2017}. Tunable add-drop filters can also serve as a spectral filter at a single frequency matched with a target QD \cite{Elshaari2017} (See Fig.~\ref{Fig7_4}(c)). 

In  Table~\ref{Tab_PIC}, we summarize key demonstrations of integrated quantum photonic systems incorporating QDs with different integration approaches and functionalities.

\begin{table}[]
\caption{Representative demonstrations of integrated quantum photonic systems with QDs. $V_{\rm TPI}$: TPI visibility, $T/T_0$: modulated transmission of a weak resonant laser by a single QD. The value in parentheses is a correction after deconvolution. $g^{(2)}(0)$: Change in the photon statistics of a transmitted laser by a single QD. The value in parentheses is a correction after deconvolution. $\beta$: QD-waveguide coupling efficiency, $g_0$: light-matter coupling strength of a QD-cavity system} 
\label{Tab_PIC}
\renewcommand{\arraystretch}{1.3}
\resizebox{\columnwidth}{!}{%
\begin{tabular}{cccccc}
\hline
\textbf{Integration} &
  \begin{tabular}[c]{@{}c@{}} \textbf{Types of QD}\\    \textbf{(Wavelength)}\end{tabular} &
  \textbf{Photonic  platforms} &
  \textbf{Functionality} &
  \textbf{Figure of merit} &
  \textbf{Ref.} \\
  \hline
Monolithic &
  \begin{tabular}[c]{@{}c@{}}InAs QD\\    (Near IR)\end{tabular} &
  \begin{tabular}[c]{@{}c@{}}GaAs Photonic \\ crystal waveguide\end{tabular} &
  \begin{tabular}[c]{@{}c@{}} Indistinguishable \\ single photons\end{tabular} &
  \begin{tabular}[c]{@{}c@{}}  $V_{\rm TPI}$=96\%\\ (up to 115 photons)  \end{tabular} & \cite{Uppu2020}   \\
Monolithic &
  \begin{tabular}[c]{@{}c@{}}InAs QD\\    (Near IR)\end{tabular} &
  \begin{tabular}[c]{@{}c@{}}GaAs   Photonic \\ crystal waveguide\end{tabular} &
  \begin{tabular}[c]{@{}c@{}}TPI\end{tabular} &
  \begin{tabular}[c]{@{}c@{}}$V_{\rm TPI}$=79\%\\    (two independent QDs)\end{tabular} & \cite{Papon2022} \\
Monolithic &
  \begin{tabular}[c]{@{}c@{}}InAs QD\\    (Near IR)\end{tabular} &
  \begin{tabular}[c]{@{}c@{}}GaAs   Photonic \\ crystal waveguide\end{tabular} &
  \begin{tabular}[c]{@{}c@{}}Single-photon   \\ nonlinearity\end{tabular} &
  \begin{tabular}[c]{@{}c@{}} $T/T_0$=8\%($\sim$35\%)  \\  $g^{(2)}(0)$ =1.08($\sim$2.1)\end{tabular} & \cite{Javadi2015} \\
Monolithic &
  \begin{tabular}[c]{@{}c@{}}InAs QD\\    (Near IR)\end{tabular} &
  \begin{tabular}[c]{@{}c@{}}GaAs   \\  waveguide\end{tabular} &
  Superradiance    &
  \begin{tabular}[c]{@{}c@{}} Two coupled QDs   \\Three coupled QDs  \end{tabular}
   & \begin{tabular}[c]{@{}c@{}} \cite{Kim2018}   \\\cite{Grim2019}  \end{tabular} \\
Monolithic &
  \begin{tabular}[c]{@{}c@{}}InAs QD\\    (Near IR)\end{tabular} &
  \begin{tabular}[c]{@{}c@{}}GaAs   topological \\ photonic crystal\end{tabular} &
  \begin{tabular}[c]{@{}c@{}}Chiral   \\ spin-photon interface\end{tabular} &
  $\eta_{\rm chiral}$=68\% & \cite{Barik2018}  \\

\begin{tabular}[c]{@{}c@{}}Heterogeneous\\    (Wafer bonding)\end{tabular} &
  \begin{tabular}[c]{@{}c@{}}InAs QD in \\ GaAs nanobeam\\    (Near IR)\end{tabular} &
   $\mathrm{Si}_3\mathrm{N}_4$ waveguide &
  \begin{tabular}[c]{@{}c@{}}Adiabatic   coupling \\ in a hybrid system\end{tabular} &
  $\beta$=20\% & \cite{Davanco2017} 
   \\
\begin{tabular}[c]{@{}c@{}}Heterogeneous\\    (Pick-and-place)\end{tabular} &
  \begin{tabular}[c]{@{}c@{}}InAs QD in \\ InP nanobeam\\    (Telecom)\end{tabular} &
  Si waveguide &
  \begin{tabular}[c]{@{}c@{}}Adiabatic   coupling \\ in a hybrid system\end{tabular} &
  $\beta$=32\% & \cite{Kim2017}
   \\
\begin{tabular}[c]{@{}c@{}}Heterogeneous\\    (Pick-and-place)\end{tabular} &
  \begin{tabular}[c]{@{}c@{}}InAsP QD in \\ InP nanowire \\ (Near IR) \end{tabular} &
   $\mathrm{Si}_3\mathrm{N}_4$ waveguide &
  \begin{tabular}[c]{@{}c@{}}Strain   tuning \\ in a hybrid system\end{tabular} &
  \begin{tabular}[c]{@{}c@{}} $\beta$=1\% \\ $\Delta_{\rm nm}$ =1.6 nm \end{tabular} &  \cite{Elshaari2018}
   \\ 
\begin{tabular}[c]{@{}c@{}}Heterogeneous\\    (Pick-and-place)\end{tabular} &
  \begin{tabular}[c]{@{}c@{}}InAsP QD in \\ InP nanowire \\ (Near IR) \end{tabular} &
  $\mathrm{Si}_3\mathrm{N}_4$ waveguide &
  \begin{tabular}[c]{@{}c@{}}Tunable   \\single-photon routing \\with a ring resonator\end{tabular} &
  \begin{tabular}[c]{@{}c@{}} $\beta$=24\%\\ Bandwidth = 40 nm \\ Selectivity = 15 dB\end{tabular} & \cite{Elshaari2017}
   \\
\begin{tabular}[c]{@{}c@{}}Heterogeneous\\    (Pattern transfer)\end{tabular} &
  \begin{tabular}[c]{@{}c@{}}InAs QD in \\ GaAs nanobeam\\    ($\sim$1.2$\mu$m)\end{tabular} &
  Si waveguide &
  \begin{tabular}[c]{@{}c@{}}Independently tunable \\ two QDs devices \end{tabular} &
  \begin{tabular}[c]{@{}c@{}} $\beta\sim$80\% \\ $\Delta_{\rm nm}$ =0.9 nm\end{tabular} & \cite{Katsumi2020}
   \\
\begin{tabular}[c]{@{}c@{}}Heterogeneous\\    (Pattern transfer)\end{tabular} &
  \begin{tabular}[c]{@{}c@{}}InAs QD in \\ GaAs nanobeam\\    ($\sim$1.2$\mu$m)\end{tabular} &
  Si waveguide &
  \begin{tabular}[c]{@{}c@{}}Strongly coupled \\ QD-cavity \end{tabular} &
  \begin{tabular}[c]{@{}c@{}} \textit{Q}=8,000 \\    $g_0$ = 69 $\mu$eV\end{tabular} &  \cite{Osada2019}

  \end{tabular}% 
}
%\hline

\end{table}

Optical quantum information processing starts with preparing photonic quantum states and ends with measuring single photons. Therefore, pumping lasers and single-photon detectors should be efficiently interfaced with the QD-containing IQPCs.  
One representative method is a fiber-optic interface that in-/out-couples external pump lasers and single-photon detectors with photonic circuits. Alternatively, hybrid integration can also bring these pump lasers and single-photon detectors directly into photonic circuits and enables all-on-chip configurations without free-space alignment. Figure~\ref{Fig7_4}(d) shows an integrated tunable microscale laser next to the emitter. The approach successfully demonstrated on-chip resonant optical excitation of a single QD \cite{Munnelly2017}. On-chip integrated single-photon detectors have also been demonstrated by several groups in the Si material system~\cite{Pernice2012,Marsili2013} and in the GaAs material system~\cite{Reithmaier2015, Schwartz2018, Gyger2021}. In particular, superconducting nanowire detectors can be easily integrated onto waveguides and can detect single photons with high efficiency and low timing jitter of about 50\% and 200 ps, respectively \cite{Schwartz2018}. 

Advances in hybrid integration enable us to utilize several functional building blocks from different platforms in a compact photonic circuit. The approach does not just leverage the strengths of multiple platforms of quantum sources, photonic chips, and detectors, but also brings new capabilities beyond linear quantum optics for a range of quantum applications. However, integrating emitters and detectors on a chip also imposes new constraints. Along with the increased complexity of manipulating quantum emitters and filtering their emissions, cryogenics temperature is vital for operating both emitters and detectors. Although the maturity of integrated photonic technology can establish complex linear transformations using multiple directional couplers with programable phases, they are mostly tested at room temperature. At cryogenic temperatures below 10 K, carriers freeze out and $\chi^{(2)}$ and $\chi^{(3)}$ nonlinear optical susceptibilities of materials are significantly altered. Therefore, additional technical development of photonic circuits may be needed for fully integrated quantum photonic architectures \cite{Dong2022}. In addition, even though combining multiple technologies have spurred the implementation of several protocols of quantum optics in a single photonic chip, addressing all issues from single-photon purity, Fourier-transform linewidth, coupling efficiency, and spectral match of solid-state quantum emitters to fabrication yield, reproducibility, and coupling efficiency of photonic devices at the same time is still a formidable challenge. Such large-scale integrated quantum photonics with solid-state quantum emitters was recently demonstrated with defect centers in diamonds \cite{Wan2020}. The 128-channel photonic chip integrates more than 70 germanium and silicon vacancies, generating spectrally identical single photons with near Fourier-transformed linewidths. Adopting a similar approach for QDs would require more effort to compensate for larger spectral randomness, but higher oscillator strength and reduced coupling with high-frequency phonons are expected for QDs. These features enable QDs to generate much brighter single photons at the zero-phonon line, which is crucial for speeding up large-scale quantum systems.

\section{Applications of single quantum dot devices in photonic quantum technology}\label{sec:applications}

So far, we introduced QDs as promising candidates for quantum information technologies and reviewed various aspects ranging from the fabrication of photonic devices to the evaluation of their quantum optical properties. Representing one of the most promising quantum emitter platforms, the research in the field soon also considered demonstrations of applications using QDs. In this section, we present an overview on applications which have already been implemented to date or are currently tackled in by the community using QD-based devices. We start with QD-based implementations of QKD and proof-of-concept experiments on quantum teleportation and entanglement swapping, as important steps towards larger quantum networks, and move on to boson sampling and photonic computing.

\subsection{Quantum key distribution}
As discussed in section \ref{sec:application_scenarios_QKD}, QDs can either be used in prepare-and-measure type settings (cf. BB84 protocol) or entanglement-based settings (cf. E91 protocol) of QKD. While single photons are sufficient for the first type of implementation, entangled photon pairs are required as a key resource for the latter. In addition, advanced protocols exploiting the concept of device-independence have been proposed, which remove security risks associated with various loopholes otherwise possible to exist in practical applications. In the following, we begin with discussing BB84-type implementations using optimized QD-SPSs.

\subsubsection{Single-photon quantum key distribution}\label{sec:application_SPS_QKD}

\begin{figure}[htbp]
  \includegraphics[width=\linewidth]{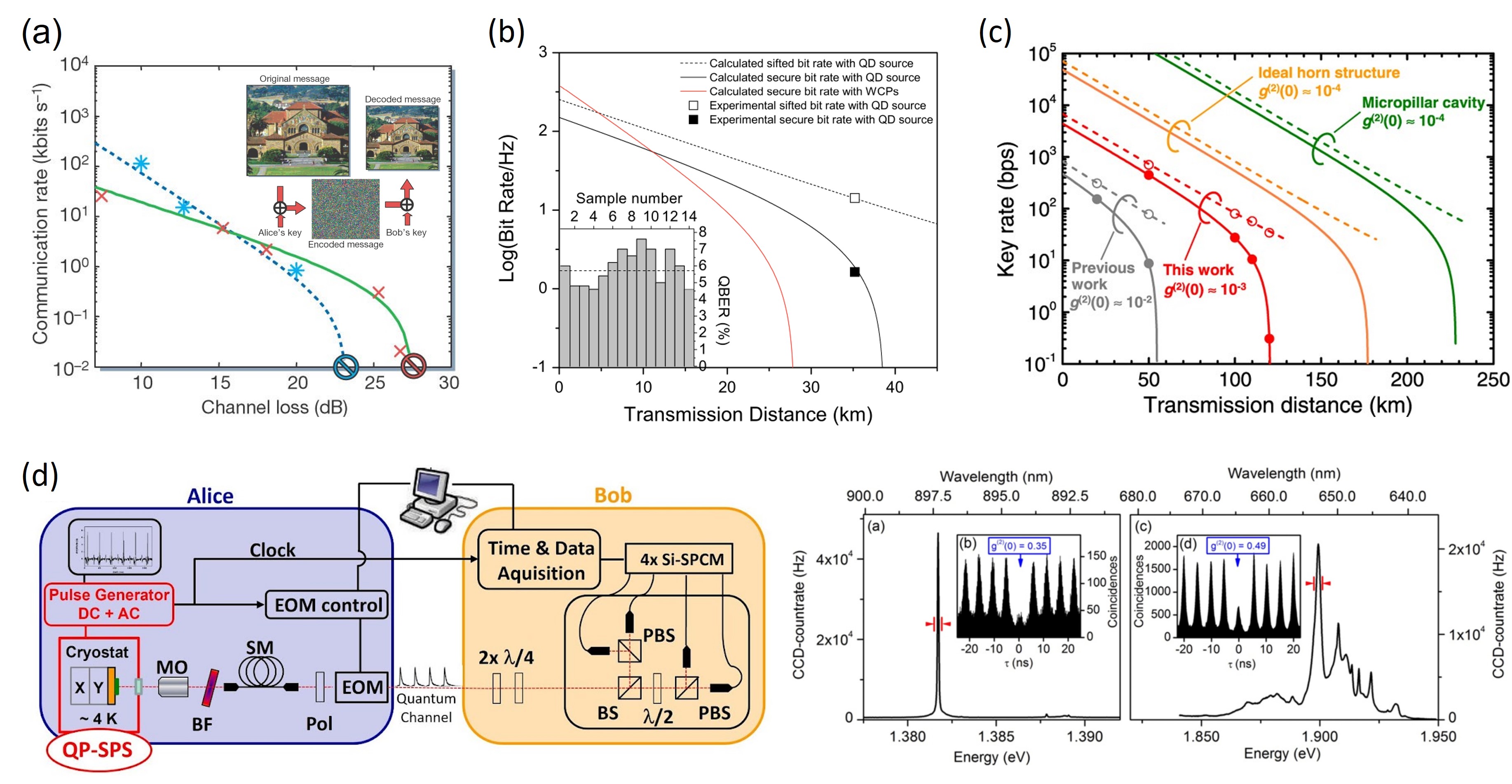}
  \caption{Selected single-photon QKD experiments using QD sources: (a) First QD-SPS QKD experiment by Waks et al. in 2002 \cite{Waks2002a}. (b) QD-based QKD at 1300\,nm wavelengths by Intallura et al. in 2009 using a phase-encoding over a 35\,km long optical fiber \cite{Intallura2009}. (c) Implementations of BB84 QKD by Takemoto et al. using QD single photons at telecom C-band wavelengths (1550\,nm) over 50\,km (gray line, \cite{Takemoto2010}) and 120\,km (red line, \cite{Takemoto2015}) of SFM-28 optical fiber (cf. gray line). Extrapolations by the authors (yellow and green lines) still show prospects for substantial future improvements. (d) In 2012 Heindel et al. demonstrated lab-scale single-photon QKD using two different types of electrically triggered QD-SPSs based on InAs QDs (900\,nm) and InP QDs (650\,nm), respectively \cite{Heindel2012}. (a) Reprinted from \href{https://www.nature.com/articles/420762a}{\textit{Waks et al. 2002}} \cite{Waks2002a} with permission of Springer Nature: Copyright 2002 Springer Nature. (b) Figure reproduced with permission from \href{https://iopscience.iop.org/article/10.1088/1464-4258/11/5/054005}{\textit{Intallura et al. 2009}}  \cite{Intallura2009}. IOP Publishing. All rights reserved. (c) Figures reproduced from \href{http://www.nature.com/articles/srep14383}{\textit{Takemoto et al. 2015}} \cite{Takemoto2015} under Creative Commons CC BY license. (d) reproduced with permission from \href{https://iopscience.iop.org/article/10.1088/1367-2630/14/8/083001}{\textit{Heindel et al. 2012}} \cite{Heindel2012} © IOP Publishing and Deutsche Physikalische Gesellschaft. Reproduced by permission of IOP Publishing. CC BY-NC-SA}
  \label{fig:Fig_SPS_QKD}
\end{figure}

The first implementation of single-photon QKD was reported by Waks et al. in 2002 \cite{Waks2002a}. Here, the authors used single-photons emitted by an optically triggered InAs QD integrated into a micropillar cavity \cite{Santori2002} to implement the BB84 protocol with polarization-encoded single-photon states. The non-resonant optical excitation at a rate of 76\,MHz resulted in a mean photon number of $\mu = 0.007$ injected into the quantum channel, as deduced from a measurement using a single-photon detector on Alice's side. Using a short free-space link with a variable attenuator, the photons were sent to Bob for polarization-state discrimination, photon detection, and post-processing. 
The experiments revealed an asymptotic secure key rate, calculated according to \cite{Waks2002}, of 25\,kbit/s with a QBER of $2.5\%$ in back-to-back configuration, i.e. vanishing losses in the quantum channel. Using the variable attenuator, a maximum tolerable channel loss up to which communication is possible (given by non-zero rate) of 28\,dB was observed. A comparison with attenuated laser pulses, not yet implementing decoy states, revealed, that the QD-SPS was able to outperform the attenuated laser at link losses exceeding 16\,dB (see \ref{fig:Fig_SPS_QKD}(a)). Overall, the SPS could tolerate about 4\,dB higher losses than the laser in this experiment. Noteworthy, decoy-state protocols nowadays allow for the in situ estimation of the multi-photon contribution to mitigate photon number splitting attacks and hence much higher average photon numbers in the laser pulses \cite{Wang2005}. As a result, the asymptotic key rate achieved by Waks et al. would not beat a decoy-state implementation using WCPs. 

To improve the secure key rates achievable in BB84-QKD for a given QD-QLS, Aichele et al. presented an elegant approach in 2004 \cite{Aichele2004}. In their work, the authors used the XX-X radiative cascade of a QD to generate two single-photons at slightly different energies with each excitation pulse, effectively doubling the achievable key rate. Using this scheme, a rate of secure bits per pulse of $5 \cdot 10^{-4}$ was demonstrated, which results in a communication rate of 38 kbit/s at a laser repetition rate of 76\,MHz.

While these first implementations used short laboratory-scale free-space optical (FSO) links as quantum channels, the use of optical fibers in ground-based communication scenarios has the practical advantages of being less susceptible to environmental fluctuations and being compatible with existing deployed fiber-networks. 
The first among several QD-based QKD experiments using optical fibers as quantum channels was conducted by Collins et al. using QD-generated single-photon pulses at a wavelength of 900\,nm sent through 2\,km of optical fiber \cite{Collins2010}.

To benefit from the minimal transmission losses possible in optical fibers, the community soon also tackled the fabrication of QDs operating at wavelengths in the second and third telecom window (O- and C-band) - work that has been pioneered by Ward et al.\cite{Ward2005}. Single-photon QKD using QDs at 1300\,nm was first demonstrated by Intallura et al. in 2009 using a QD-micropillar cavity optically excited above the bandgap \cite{Intallura2009}. As the quantum channel, the authors used 35\,km of standard SMF-28 optical fiber. To avoid polarization-state distortions, possible at long transmission distances in optical fibers, phase encoding has been implemented using path-length matched Mach-Zehnder-Interferometers at Alice and Bob. Using their QKD demonstration testbed operated at a clock rate of 1\,MHz, the authors reported a calculated (according to the so-called asymptotic GLLP~\footnote{Named after the authors D. Gottesman, H.-K. Lo, N. Lütkenhaus, and J. Preskill of Ref. \cite{Gottesman2004}.} rate equations) maximum secure key rate of about 160\,bit/s at a measured QBER of $5.9\%$ and achieved a non-zero key rate at a distance of 35\,km. This performance surpassed the distance limit of a WCP-source (without decoy states) in their setup (see Fig. \ref{fig:Fig_SPS_QKD}(b)). 

The first implementation of single-photon QKD in the telecom C-band (1560\,nm), i.e. at the lowest transmission loss possible in optical fibers, was reported only one year later by Takemoto et al. \cite{Takemoto2010}. In their phase-encoding setup, the SPS comprised a QD integrated into a horn-structure (cf. Ref.~\cite{Takemoto2007}). The authors achieved a maximum secure communication distance of 50\,km based on the asymptotic GLLP rate equations. Further, improving their QKD implementation, by employing low-noise single-photon detectors based on superconducting nanowires \cite{Hadfield2005} and a QD source with better single-photon purity ($g^{(2)}(0) = 0.005$), the same group presented an improved version of their QKD implementation in Ref.~\cite{Takemoto2015}. 
Here, the improvements resulted in a maximal communication distance of 120\,km - the longest transmission distance achieved in fiber-based single-photon QKD to date  (cf. Fig. \ref{fig:Fig_SPS_QKD}(c)).

While all aforementioned QKD experiments used optically excited QD devices, relying on pulsed laser systems, a major advantage of semiconductor based QLSs is the possibility to realize complex engineered devices including diode structures for electrical charge carrier injection. The electrical triggering of QD emission \cite{Yuan2002} enables both, higher degrees of device integration and flexibly adjustable clock rates in protocol implementations.

In 2012 Heindel et al. demonstrated lab-scale BB84-QKD experiments using two different types of single-photon emitting diodes operating in the near-infrared and visible spectral range, at 897\,nm, and 653\,nm, respectively (see Fig. \ref{fig:Fig_SPS_QKD}(d)) \cite{Heindel2012}. Employing engineered QD devices based on different material systems and growth techniques, their work highlighted the flexibility semiconductor based QLSs offer for quantum information technologies. The near-infrared SPS was based on an electrically contacted micropillar cavity, exploiting the Purcell effect to enhance the photon extraction efficiency \cite{Heindel2010}. For the shorter wavelength SPS, QDs were integrated into a quasi-planar DBR cavity structure \cite{Reischle2010}. Using the Purcell-enhanced SPS at 897\,nm under pulsed current injection at a clock-rate of 182.6\,MHz, the authors achieved sifted key rates of 27.2\,kbit/s at a QBER of $3.9\%$ and a $g^{(2)}(0)$ value of 0.35 at moderate excitation. The 653\,nm SPS was triggered at 200\,MHz, resulting in a sifted key rate of 95.0\,kbit/s at a QBER of $4.1\%$ and a $g^{(2)}(0)$ value of 0.49. These first proof-of-principle QKD experiments using electrically operated semiconductor SPSs were considered as a major step forward in photonic quantum technologies. Shortly after the lab-scale QKD experiments reported in 2012, the authors integrated the near-infrared-emitting SPS in a rather compact quantum transmitter setup to be employed for QKD field experiments in downtown Munich. These QKD experiments by Rau et al. \cite{Rau2014} comprised a 500\,m FSO link between two buildings of the Ludwigs-Maximilians-Universität Munich, with the transmitter and receiver units synchronized via GPS-disciplined oscillators. Using a single-photon LED modulated at a clock-rate of 125\,MHz, the authors achieved sifted key rates of 7.4\,kbit/s (11.6\,kbit/s) at a quantum bit error ratio of $7.2\%$ ($6.3\%$) and a $g^{(2)}(0)$ value of 0.39 (0.46) at low (moderate) excitation.

\begin{figure}[htbp]
  \includegraphics[width=\linewidth]{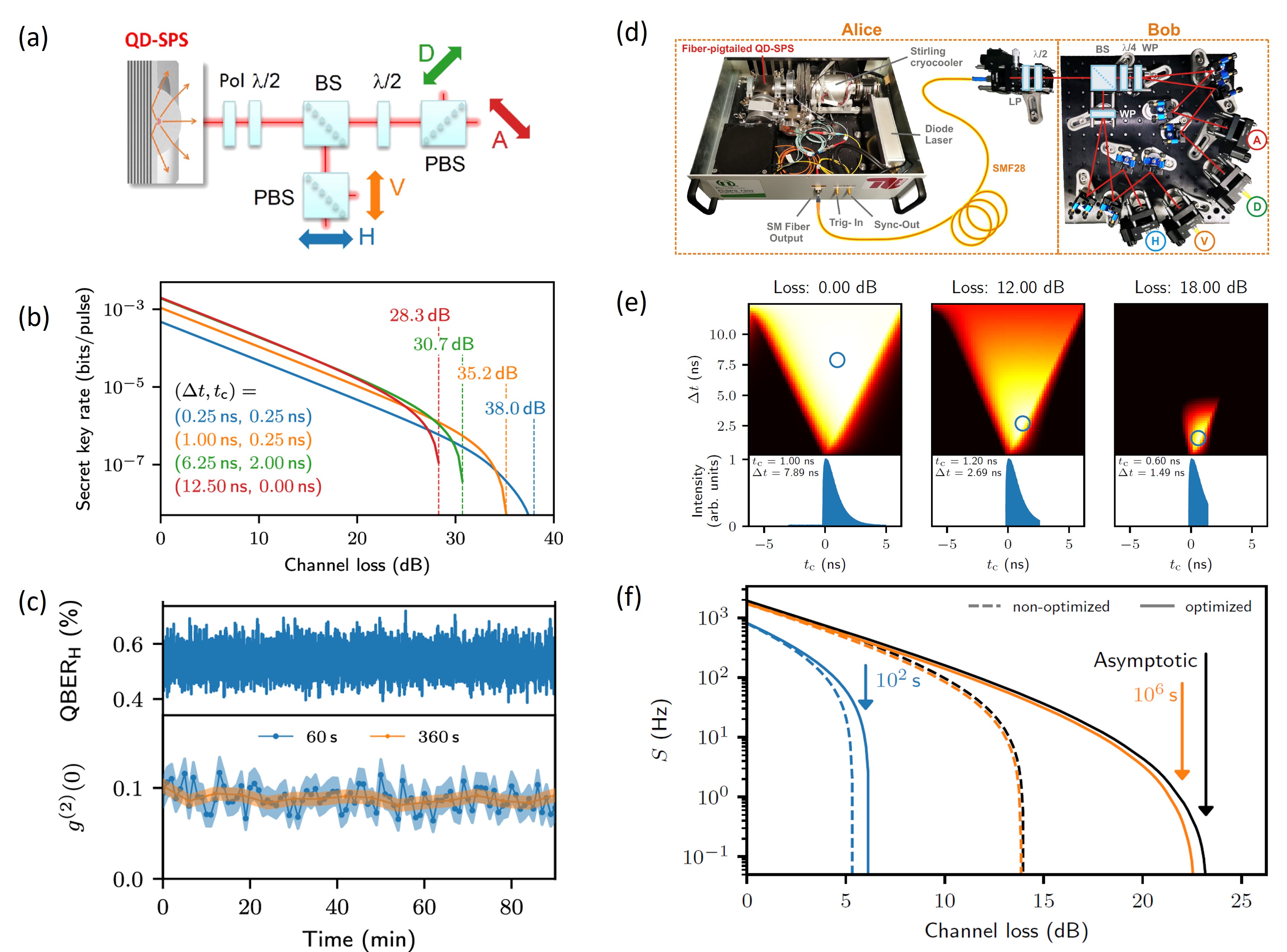}
  \caption{(a) BB84-QKD testbed using a triggered QD-SPS for the development of tools for the performance optimization of single-photon QKD, reported by Kupko et al. \cite{Kupko2020}. (b) Exploiting temporal filtering in this testbed, the maximal tolerable loss inside the quantum channel can be enhanced by 24\%. (c) QBER and $g^{(2)}(0)$ as a function of time for the QD-SPSs. The parameter can be directly used for the key distillation process, enabling a security monitoring in real time. (d) QKD testbed using a benchtop plug\&play telecom-wavelength QD-SPS providing single-photon pulses via an SMF28 optical fiber for polarization-coding. The 19-inch rack module houses a compact Stirling cryocooler including the fiber-pigtailed QD-device, a pulsed diode laser, and a fiber-based bandpass filter. (e) 2D temporal filtering for optimization of the expected secret key rate fraction $S$ as a function of the temporal width $\Delta t$ and the center $t_\textup{c}$ of the acceptance time window for different losses in the quantum channel. Blue circles mark the optimal parameter sets indicated in the time-resolved measurements in the lower panels. (f) Rate-loss diagrams considering the experimental data from (d) showing the asymptotic (black) and finite (blue/orange) key rate with (solid lines) and without (dashed lines) optimization of the temporal acceptance time window, respectively. In the finite-key scenario, accumulation times of 100\,seconds (blue) and 1 million seconds (orange) are considered. (a-c) reprinted from \href{http://dx.doi.org/10.1038/s41534-020-0262-8}{\textit{Kupko et al. 2020}} \cite{Kupko2020} under Creative Commons Attribution 4.0 International License, (d-f) reprinted from \href{https://doi.org/10.1063/5.0070966}{\textit{Gao et al. 2022}} \cite{Gao2022} with the permission of AIP Publishing.}
  \label{fig:Fig_SPS_QKD_advances}
\end{figure}

Table~\ref{tab:SPS_QKD} summarizes the QD-based single-photon QKD experiments discussed above. 
\newgeometry{left=1cm,right=1cm}
\begin{table}
\centering
\caption{Implementations of single-photon QKD based on the BB84 protocol and QD sources (abbreviations: light emitting diode (LED), polarization (Pol), free space optical (FSO), fiber-coupled (FC))}
\label{tab:SPS_QKD}
\begin{threeparttable}
\begin{tabular}{ccccccccccl}
\hline
\begin{tabular}[c]{@{}c@{}} Photonic Device\\ \end{tabular} & \begin{tabular}[c]{@{}c@{}} QD Material\\ \end{tabular} & \begin{tabular}[c]{@{}c@{}} $\lambda$\\ {[}nm{]}\end{tabular} & \begin{tabular}[c]{@{}c@{}} Pump\\ \end{tabular} & \begin{tabular}[c]{@{}c@{}} Coding\\ \end{tabular} & \begin{tabular}[c]{@{}c@{}} Clock\\ {[}MHz{]} \end{tabular} & \begin{tabular}[c]{@{}c@{}} FSO/FC\\ \end{tabular} & \begin{tabular}[c]{@{}c@{}} Sifted/Secure\\ Key Rate\end{tabular} & \begin{tabular}[c]{@{}c@{}} QBER\\ {[}\%{]}\end{tabular} & \begin{tabular}[c]{@{}c@{}} Ref.\\ \end{tabular} \\ \hline
Micropillar & InAs/GaAs & 880 & optic. & Pol & 76 & FSO (In-Lab) & - / 25 kbps & 2.5 & \cite{Waks2002a}  \\
Planar & InP/GaInP & 635 & optic. & Pol & 0.01 & FSO (In-Lab) & 15 bps / 5 bps & 6.8 & \cite{Aichele2004} \\
Micropillar & InAs/GaAs & 1300 & optic. & Phase & 1 & FC (35\,km)  & 10 bps / 1 bps & 5.9  & \cite{Intallura2009} \\
Planar Microcavity & InAs/GaAs & 895 & optic. & Pol & 40 & FC (2\,km) & - / 8-600 bps & 1.2-21.9 & \cite{Collins2010}\\
Optical horn & InAs/InP & 1580 & optic. & Phase & 20 & FC (50\,km) & 15-386 bps / 3-9 bps & 3.4-6  & \cite{Takemoto2010} \\
Micropillar LED & InGaAs/GaAs & 898 & elect. & Pol  & 182.6  & FSO (In-Lab) & 8-35 kbps / - & 3.8-6.7 & \cite{Heindel2012} \\
Resonant-cavity LED \tnote{ a)} & InP/GaInP & 653 & elect. & Pol  & 200 & FSO (In-Lab) & 9-117 kbps / - & 4.1-6.0 & \cite{Heindel2012}  \\
Micropillar LED & InGaAs/GaAs & 910 & elect. & Pol  & 125 & FSO (500\,m)  & 5-17 kbps / - & 6-9 & \cite{Rau2014} \\
Optical horn & InAs/InP & 1500 & optic.  & Phase & 62.5 & FC (120\,km) & 34 bps / 0.307 bps & 2-9 & \cite{Takemoto2015} \\ \hline
\end{tabular}
%\begin{tablenotes}
a) Same publication as above but QKD experiment performed by a different group;
%\end{tablenotes}
\end{threeparttable}
\end{table}
\restoregeometry

Despite the enormous progress seen in the development of telecom-wavelength SPSs (cf. Section~\ref{sec:SPS_performance_telecom}), the fabrication of devices offering high performance remains challenging. Therefore, recent work also considers quantum frequency conversion to transfer the emission of high-performance NIR QD-SPSs emitting to C-band wavelenghts \cite{Morrison2021}. Respective sources were first employed in proof-of-concept QKD experiments by Morrison et al. in 2022 \cite{Morrison2022}. Along this route, Zahidy et al. recently demonstrated QKD in an 18-km-long field-installed fiber link, generating a secret key at 2\,kbits/s at 9.6\,dB channel loss \cite{Zahidy2023}.

Other recent developments in the implementation of QD-based single-photon QKD aim at the performance optimization of single-photon QKD systems, as well as the development of compact devices for applications in practical scenarios. In this context, Kupko et al. studied the impact of temporal filtering on the performance of single-photon QKD and showed how the secret key rate and the achievable tolerable loss can be optimized using two-dimensional temporal filtering as presented in Fig.~\ref{fig:Fig_SPS_QKD_advances}(a-c)~\cite{Kupko2020}. In addition, the authors demonstrated real-time security monitoring by evaluating $g^{(2)}(0)$ in situ during key generation. Two years later, the same group reported on a benchtop QKD testbed using a stand-alone  fiber-coupled QD-SPS emitting at telecom O-band wavelengths \cite{Gao2022}. The plug\&play device emitted single-photon pulses at 1321\,nm and was based on a directly fiber-pigtailed deterministically fabricated QD-device integrated into a compact Stirling cryocooler housed in a 19-inch rack module (see Fig.~\ref{fig:Fig_SPS_QKD_advances}(d)). Emulating the BB84 protocol in their testbed, the authors achieved $g^{(2)}(0)=0.10\pm0.01$, a raw key rate of up to $(4.72\pm0.13)\,$kHz, and predicted tolerable losses of up to 23.19\,dB applying the 2D temporal filtering approach introduced in their previous work as presented in Fig.~\ref{fig:Fig_SPS_QKD_advances}(e, f). While Stirling-type refrigerators are the most compact solution to date, the achievable base temperatures are presently limited to about 27\,K. For applications which rely on the excellent coherence properties of QDs, e.g. for the generation of highly indistinguishable photons, small-footprint Gifford-McMahon cryocoolers in combination with compact compressors are an alternative.

It should be noted that, while this review focuses on QDs, several emerging quantum emitter platforms recently attracted significant attention in the context of quantum information technologies and QKD in particular. In proof-of-concept studies, confined excitons in hexagonal boron nitride (hBN) \cite{Zeng2022}, molecules of polyaromatic hydrocarbons \cite{Murtaza2022}, and monolayers of transition metal dichalcogenides \cite{Gao2023} were considered and evaluated for their application in QKD, including an implementation of the B92 protocol \cite{Bennett1992b} using a hBN-based SPS \cite{Samaner2022}.

While the QKD implementations discussed in this section were performed in prepare-and-measure configuration, we review QKD experiments using QD-based entangled photon pair sources in the next section.

\subsubsection{Entangled-photon quantum key distribution}\label{sec:application_EPS_QKD}
As discussed in section \ref{sec:application_scenarios_QKD}, QD-based QLSs can also be employed in entanglement-based QKD protocols. For this purpose, polarization-entangled photon pairs can be generated via the XX-X emission cascade (cf. Fig. \ref{fig_QD_Conf}(a)). Since the photons obey single-photon statistics, higher generation rates of entangled photons are possible compared to spontaneous parametric down-conversion sources \cite{Chen2018,Wang2019,Liu2019}. Thus, QDs can be used for entanglement-based implementations of QKD, where the entangled photons can either be distributed via FSO or fiber-optical links.

\begin{figure}[htbp]
  \includegraphics[width=\linewidth]{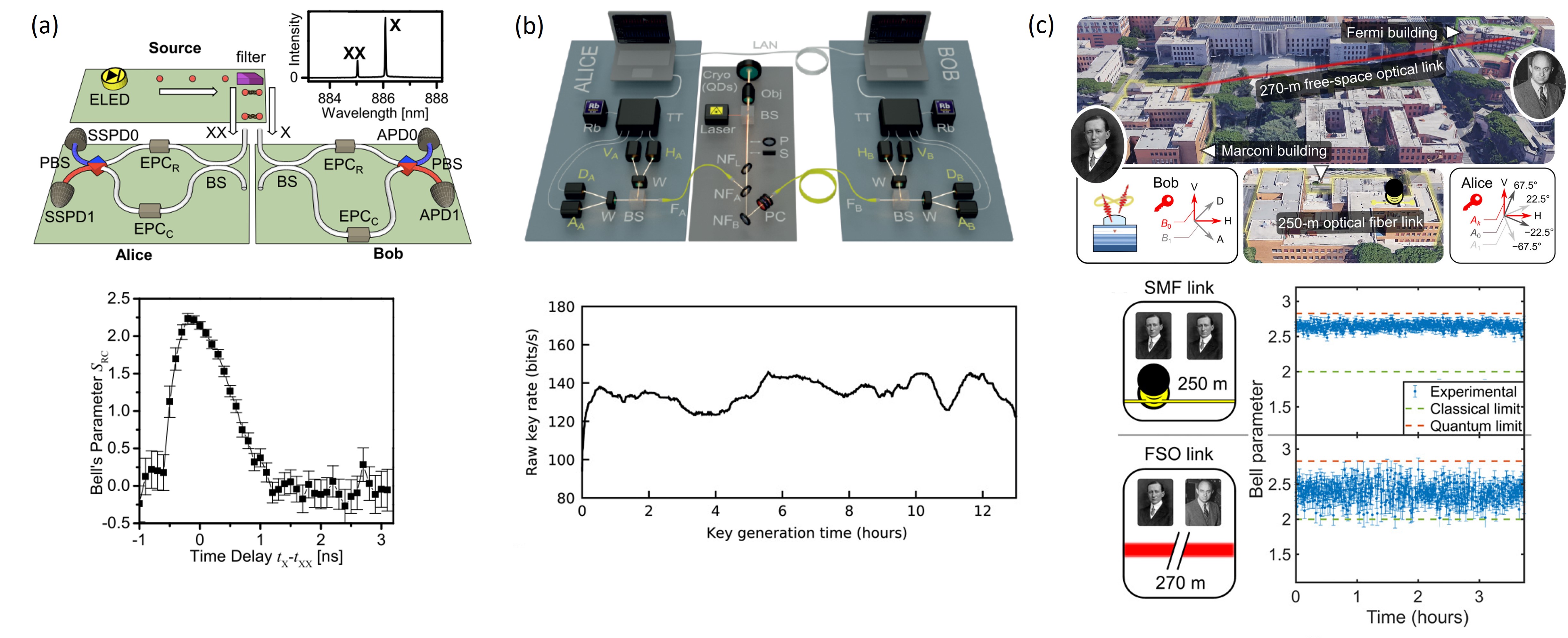}
  \caption{Entanglement-based implementations of QKD using QD-generated entangled photon pairs: (a) Experimental setup of the first demonstration of entanglement-based QKD by Dzurnak et al. \cite{Dzurnak2015} using an electrically triggered QD-device. Entanglement was verified by violating the CHSH inequality with an $S$-parameter $>\,2$. (b) Schimpf et al. \cite{Schimpf2021a} and (c) Basso Basset et al. \cite{Basset2021} realized entanglement-based QKD experiments of the BBM92 QKD protocol and an asymmetric Ekert protocol, respectively. Both groups used QD entangled photon pair sources coherently driven via two-photon excitation. (a,b,c) Reprinted from \href{http://aip.scitation.org/doi/10.1063/1.4938502}{\textit{Dzurnak et al. 2015}} \cite{Dzurnak2015}, with the permission of AIP Publishing. (d) reproduced from \href{https://advances.sciencemag.org/lookup/doi/10.1126/sciadv.abe8905}{\textit{Schimpf et al. 2021}} \cite{Schimpf2021a} under Creative Commons Attribution License 4.0, (e) reprinted from \href{https://advances.sciencemag.org/lookup/doi/10.1126/sciadv.abe6379}{\textit{Basso Basset et al. 2021}} \cite{Basset2021} © The Authors, some rights reserved; exclusive licensee AAAS. Distributed under a \href{http://creativecommons.org/licenses/by-nc/4.0/}{CC BY-NC 4.0 license.}}
  \label{fig:Fig_EPS_QKD}
\end{figure}
The first proof-of-concept demonstration of QD-based entanglement QKD was reported by Dzurnak et al. in 2015 \cite{Dzurnak2015}. Here, the authors implemented the BBM92 protocol using entangled photons generated via an entangled-light emitting diode, an electrically triggered QD-device introduced earlier by Salter et al. \cite{Salter2010}. Using the experimental setup depicted in Fig. \ref{fig:Fig_EPS_QKD}(a), the entangled photon pairs were distributed via optical fibers connected to the receiver stations of Alice and Bob, analyzing the polarization state of the XX- and X-photon, respectively. The photons emitted by the biexciton- and exciton-state were spatially separated using a spectral filter (cf. Fig. \ref{fig:Fig_EPS_QKD}(b)), before coupling them to the individual fiber links connected to Alice and Bob. Polarization-entanglement of the photon pairs was verified by violating the CHSH inequality, yielding a $S$-parameter $>\,2$ for vanishing time delays (cf. Fig. \ref{fig:Fig_EPS_QKD}c). In the QKD experiment a sifted key of 2053\,bits was transferred with a bit error rate of 9.8\%, below the threshold of 11\% required for the security of the BBM92 protocol, resulting in a final secret key of 949\,bits shared between Alice and Bob after error correction. A sifted key rate of about $10\,$bits/min was achieved in the experiment.

All QD-based implementations of QKD discussed so far, including those summarized in \ref{sec:application_SPS_QKD}, used in-coherent optical or electrical excitation. In 2021, two research groups independently reported the first entanglement based-QKD experiments employing coherently driven QD sources \cite{Schimpf2021a,Basset2021}. In both experiments, a GaAs QD fabricated by the droplet-etching technique (cf. Fig.~\ref{fig:Fig_Advanced_QKD}) as coherently excited via two-photon resonant excitation to generate entangled photon pairs. This enabled significant improvements in the single-photon purity and entanglement fidelity compared to the firs-time demonstration discussed above.

Schimpf et al. \cite{Schimpf2021a} distributed the entanglement of their QD-QLS using a 350\,m optical fiber, resulting in an asymptotic secure key rate of 86\,bits/s (cf. Fig. \ref{fig:Fig_EPS_QKD}(b)). Basso-Basset et al. \cite{Basset2021} realized a fiber link (250\,m) as well as a free-space link (270\,m) with comparable transmission lengths, allowing for a direct comparison of both channel types operated with the same QD-QLS (see Fig. \ref{fig:Fig_EPS_QKD}(c)). The authors observed larger Bell parameters and fewer fluctuations for the entanglement distribution via the fiber-based link, which used active feedback to compensate for polarization-state distortions. Moreover, the average raw key rate of 486\,bits/s achieved in the fiber-link was higher compared to the FSO link (60\,bits/s), due to environmental fluctuations in the FSO channel.

Noteworthy, the two groups implemented slightly different protocols in their entanglement-based QKD experiments. In the demonstration by Basso-Basset et al. an asymmetric version of the original E91 protocol was applied. Here, the violation of the CHSH inequality was evaluated using a subset of the transmitted bits, to quantify the degree of entanglement left after the photons propagated through the quantum channel. This reveals the amount of potential eavesdropping, in turn determining the amount of privacy amplification required for distillation of the secret key. On Alice's side, the authors measured in the basis set known to maximally violate the CHSH inequality, while the conventional BB84 basis was used for Bob's measurement. Compared to the original E91 protocol, this asymmetric approach reduces the number of detectors required. In the work by Schimpf et al. \cite{Schimpf2021a}, the BBM92 protocol \cite{Bennett1992} was implemented, which represents the entanglement-based analog of the BB84 protocol. Here, Alice and Bob both measure their respective halves of the entangled state in two conjugate bases and evaluate deviations in a subset of their data. The result determines the amount of privacy amplification required for distilling the secret key. Hence, in contrast to the work by Basso-Basset et al., the degree of entanglement (i.e. the $S$-parameter) is not monitored during the key generation.

Interestingly, the QD-sources used in the experiments by Schimpf et al. and Basso-Basset et al. exhibited a non-negligible blinking effect, which limited the photon pair extraction efficiency. This has been improved in a follow-up work by Schimpf et al., where the authors demonstrated entanglement QKD using a p-i-n doped QD diode delivering blinking-free entangled photon pairs under pulsed optical excitation \cite{Schimpf2021c}.

More recently, the group in Rome also achieved daylight operation in their urban 270-m-long FSO QKD-link \cite{Basset2023}. Here, Basso-Basset et al. used narrower spectral filtering and better stray-light suppression in combination with improved beam tracking to enable the continuous operation of their entanglement-based QKD-link over 3.5\,days under different light and weather conditions.

Table~\ref{tab:EPS_QKD} summarizes the entanglement-based QKD experiments employing QD-sources as discussed above. 

\newgeometry{left=1cm,right=1cm}
\begin{table}
\centering
\caption{Implementations of entanglement-based QKD using QD sources (abbreviations: light emitting diode (LED), free space optical (FSO), fiber-coupled (FC))}
\label{tab:EPS_QKD}
\begin{threeparttable}
\begin{tabular}{ccccccccccl}
\hline
\begin{tabular}[c]{@{}c@{}} Photonic Device\\ \end{tabular} & \begin{tabular}[c]{@{}c@{}} QD Material\\ \end{tabular} & \begin{tabular}[c]{@{}c@{}} $\lambda$\\ {[}nm{]}\end{tabular} & \begin{tabular}[c]{@{}c@{}} Pump\\ \end{tabular} & \begin{tabular}[c]{@{}c@{}} Protocol\\ \end{tabular} & \begin{tabular}[c]{@{}c@{}} Clock\\ {[}MHz{]} \end{tabular} & \begin{tabular}[c]{@{}c@{}} FSO/FC\\ \end{tabular} & \begin{tabular}[c]{@{}c@{}} Sifted/Secure\\ Key Rate\end{tabular} & \begin{tabular}[c]{@{}c@{}} QBER\\ {[}\%{]}\end{tabular} & \begin{tabular}[c]{@{}c@{}} Ref.\\ \end{tabular} \\ \hline
Planar Microcavity LED & InAs/GaAs & 885 & elect. & BBM92 & 50\tnote{ a)} & FC (In-Lab) & 0.2 bps & 0.1 bps & - & \cite{Dzurnak2015} \\
Planar Microcavity & GaAs/AlGaAs & 785 & optic. & E91\tnote{ c)} & 320 & FC (250\,m) & 243 bps & - & 3.4 & \cite{Basset2021} \\
" & " & " & " & " & " & FSO (270\,m) & 30 bps & 9 bps  & 4.0 &  " \\
Planar Microcavity & GaAs/AlGaAs & 785 & optic. & BBM92 & 80 & FC (350\,m) & 135 bps & 86 bps  & 1.9 & \cite{Schimpf2021a} \\
Planar Microcavity Diode\tnote{ b)} & GaAs/AlGaAs & 785 & optic. & BBM92 & 80 & FC (350\,m) & 55 bps (raw) & - & 8.4 & \cite{Schimpf2021c} \\
Planar Microcavity & GaAs/AlGaAs & 785 & optic.  & E91\tnote{ c)} & 320 & FSO (270\,m) & 106 bps  & 11.5 & 7.16 & \cite{Basset2023} \\ \hline
\end{tabular}
%\begin{tablenotes}
a) Time-multiplexed detector effectively reduced clock rate to below 1 MHz; b) Diode structure used for spectral tuning; c) Modified asymmetric E91 protocol;
%\end{tablenotes}
\end{threeparttable}
\end{table}
\restoregeometry

As discussed in section \ref{sec:application_scenarios_QKD}, the quantum cryptographic protocols of the type BB84, E91, or BBM92 are provably secure in an information theoretical sense. But if implemented with imperfect devices, security risks may arise from potential side-channel attacks. This motivates the exploration of DI-QKD protocols, as discussed in the following section.

\subsubsection{Towards Device-independent quantum key distribution}\label{sec:applications_advanced_QKD}
Fully or partially DI-QKD protocols, are constructed such that they close all, or some specific, loopholes resulting from device imperfections in practical implementations. An important example for such protocols is MDI-QKD, which relies on the quantum interference of remote QLSs. Implementing triggered QLSs in an MDI-QKD setting, by using spatially separated QDs, TPI visibility exceeding the fundamental limit of 50\% for WCPs becomes possible. As a result, the Bell-state measurements become more efficient \cite{Lee2021}, which is why substantial quantum advantages can be expected for MDI-QKD with QD-QLSs. The crucial prerequisite for implementations of MDI-QKD is the TPI at a beam splitter. While indistinguishable photons from the same QD-QLS have been reported several times with high TPI visibility \cite{Santori2002,Gazzano2013,Thoma2016,Somaschi2016,Wang2016,Ding2016}, this is much more challenging to achieve for photons emitted from remote, i.e. spatially separated, QD-QLSs. 
 
Here, the spectral properties are of particular importance, due to the self-organized nature and the semiconductor environment of the quantum emitters. This requires firstly a coarse spectral matching of quantum emitters using pre-selection of suitable QDs. Using deterministic fabrication technologies (cf. section ~\ref{sec:deterministic_fabrication}), the yield for identifying suitable candidates can be increased significantly. Additionally, a spectral fine-tuning of one of the QDs is typically required, for which several techniques can be used, such as the tuning via the temperature \cite{Giesz2015,Thoma2017}, strain \cite{Flagg2010,Reindl2017}, or electrical gates by exploiting the quantum-confined Stark effect \cite{Patel2010}. Noteworthy, these schemes are also directly compatible with fully integrated device concepts (see Fig. \ref{Fig7_4} and corresponding discussions).

\begin{figure}[htbp]
  \includegraphics[width=\linewidth]{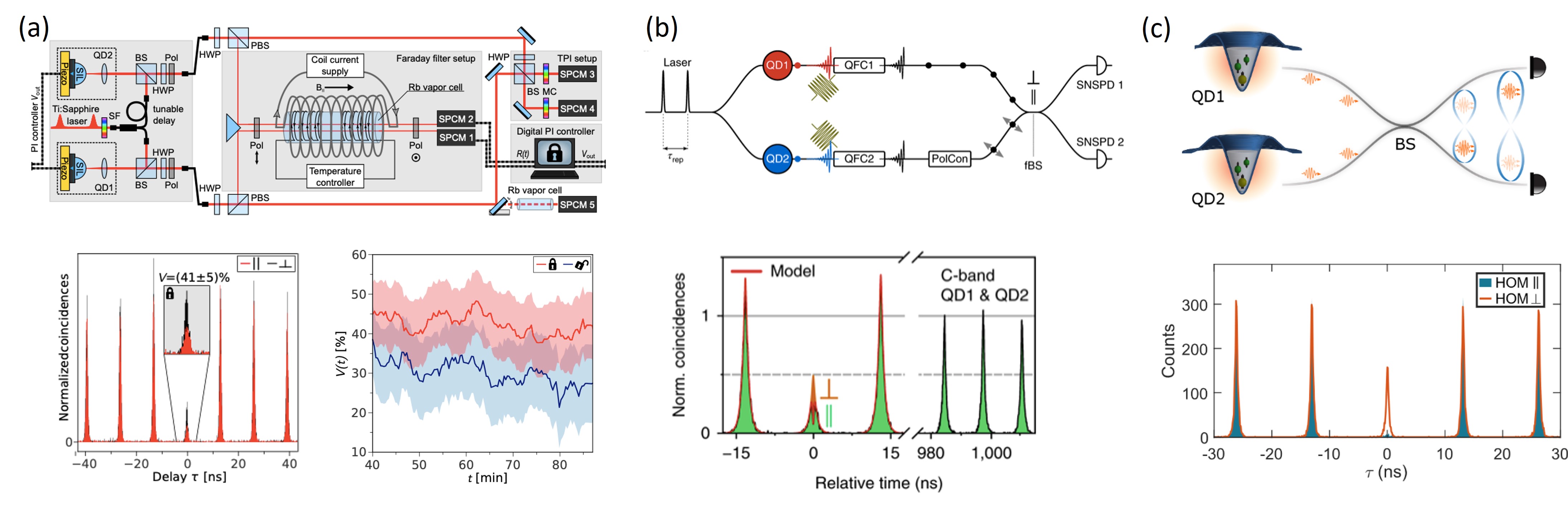}
  \caption{Progress towards the implementation of advanced QKD protocols: (a) Zopf et al. used active stabilization of QDs in remote TPI to improve the visibility \cite{Zopf2018}. (b) Weber et al. employed quantum frequency conversion to shift and spectrally match the emission of remote QDs to telecom C-band wavelengths \cite{Weber2019}. (c) Zhai et al. used low-noise GaAs QDs to demonstrate near-unity photon indistinguishability from remote solid-state quantum emitters\cite{Zhai2022}. (a) reprinted with permission from \href{https://doi.org/10.1103/PhysRevB.98.161302}{\textit{Zopf et al. 2018}} \cite{Zopf2018} Copyright 2018 by the American Physical Society, (b) reprinted with permission from Springer Nature: Nature Nanotechnology \href{http://dx.doi.org/10.1038/s41565-018-0279-8}{\textit{Weber et al. 2019}} \cite{Weber2019} Copyright 2019, (c) reprinted with permission from Springer Nature: Nature Nanotechnology \href{https://doi.org/10.1038/s41565-022-01131-2}{\textit{Zhai et al. 2022}} \cite{Zhai2022} Copyright 2022.}
  \label{fig:Fig_Advanced_QKD}
\end{figure}

To reduce spectral drifts of two QD emitters relative to each other in a dynamic fashion, active feedback routines can be employed. In the remote TPI experiments by Zopf et al. \cite{Zopf2018}, dynamic strain-tuning of both QD sources was implemented via piezo-electric actuators. Using a rubidium-vapor-cell-based Faraday filter (see \ref{fig:Fig_Advanced_QKD}(a)), spectral shifts in the emission wavelength were detected and used as input parameter for the feedback loop. Applying this technique, the remote TPI visibility was enhanced to 41\% on average compared to 31\% without stabilization.

A conceptually different technique was employed by Weber et al. using quantum frequency conversion of the single-photon emission of two spectrally unmatched QDs (see Fig. \ref{fig:Fig_Advanced_QKD}(b)) \cite{Weber2019}. To perform their remote TPI experiment at telecom C-band wavelengths, the emission of both QDs was converted from the near-infrared to 1550\,nm. During the upconversion process, the spectral mismatch between both QDs ($\approx$ 6\,GHz) was compensated for using two independently tunable lasers, resulting in a remote TPI visibility of 29\%.

The highest indistinguishability between remote QD sources was reported by Zhai et al. \cite{Zhai2022} in 2022 (see \ref{fig:Fig_Advanced_QKD}(c)). Using GaAs QDs fabricated by droplet-etching in combination with electrical gates to reduce spectral diffusion, unprecedented low levels of dephasing were obtained \cite{Zhai2020a}, resulting in remote TPI visibility of up to 93\%. The fact that this high visibility was realized without employing Purcell enhancement, tight spectral filtering, post-selection, or active stabilization, thereby rises promises for further improvements in the future. Another interesting recent experiment was reported by You et al. \cite{You2021}: by demonstrating TPI of remote QD-based QLSs separated by 300\,km of optical fiber, a remarkable distance record for the interference of QLS was set. Here, again quantum frequency conversion to 1583\,nm was used, as introduced earlier by Weber et al. \cite{Weber2019}. 

The work summarized above showed, that remote TPI visibility exceeding the classical limit of 50\% by far can nowadays be achieved using QD QLSs (see \cite{Vajner2022}, Table\,2). This lays the foundation for implementations of MDI-QKD fully exploiting the single-photon advantages possible with state-of-the-art engineered QD sources. To quantify and predict the quantum advantage possible with sub-Poissonian light sources in practical settings, however, also work from theory side is required. The secure key rates achievable in MDI-QKD with attenuated laser pulses have been studied analytically both, in the asymptotic \cite{Lo2012} and the finite-size regime \cite{Curty2014}. Theoretical studies of MDI-QKD using deterministic SPSs enabling photon indistinguishability up to unity have yet to be conducted.

Before recent advances in quantum teleportation and entanglement swapping experiments are reviewed in the following %two sections
section, it should be noted that QD-generated indistinguishable photons also enable the generation of entanglement between remote solid-state spin qubits \cite{Cabrillo1999}, as demonstrated experimentally using remote hole- \cite{Delteil2016} and electron- \cite{Stockill2017} spin qubits confined in distant QDs. This functionality opens the route towards the transfer and storage of quantum information in complex quantum network architectures.

\begin{figure}[htbp]
  \includegraphics[width=\linewidth]{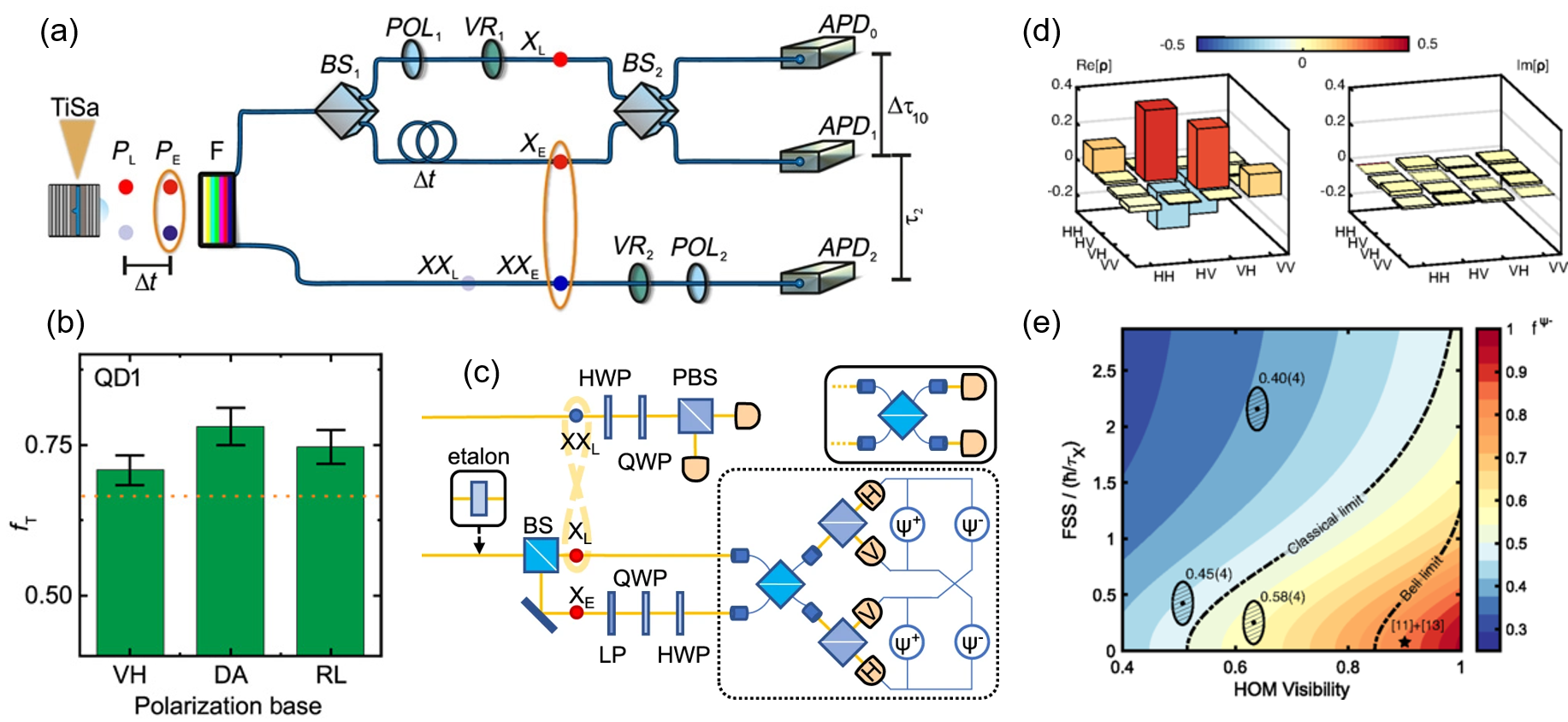}
  \caption{Quantum teleportation and entanglement swapping using photons emitted by one GaAs QD. (a) Experimental configuration used in Ref.~\cite{Reindl2018} to teleport the state of the control photon X$_L$ onto the control photon XX$_E$ using the ancilla photon X$_E$ belonging to the ''early'' (E) entangled photon pair XX$_E$-X$_E$. The photon X$_L$ is emitted by the same QD in a ``later'' (here 2~ns) excitation cycle. (b) Teleportation fidelity for three mutually unbiased basis states obtained with the experimental configuration in (a). (c) Experimental configuration used in Ref.~\cite{BassoBasset2021} to increase the efficiency of the partial Bell state measurement used for teleportation by a factor of 2 and improving the rejection of accidental coincidences. (d) Example of density matrix for a pair of photons XX$_E$-XX$_L$ belonging to two subsequence XX-X photon cascades in a GaAs QD after entanglement swapping and (e) Contour plot showing the expected entanglement swapping fidelity as a function of TPI visibility and the ratio between fine-structure-splitting and radiative lifetime. For more details see Ref.~\cite{BassoBasset2019}. (a-b) reprinted from \href{https://www.science.org/doi/full/10.1126/sciadv.aau1255}{\textit{Reindl et al. 2018}} \cite{Reindl2018} under Creative Commons CC BY license. (c) reprinted from \href{https://www.nature.com/articles/s41534-020-00356-0}{\textit{Basso Basset et al. 2021}} \cite{BassoBasset2021} under Creative Commons CC BY license. (d--e) reprinted from \href{https://journals.aps.org/prl/abstract/10.1103/PhysRevLett.123.160501}{\textit{Basso Basset et al. 2019}} \cite{BassoBasset2019} under Creative Commons CC BY license.}
  \label{fig:teleportation}
\end{figure}
\subsection{Quantum teleportation and entanglement swapping with QD photons} \label{sec:teleportation_swapping}
One of the basic elements of a BDCZ quantum repeater is a ``quantum relay''~\cite{Gisin2007}, which entangles two photons (which we denote as X$_A$ and X$_B$) of two EPR pairs (EPR$_A$ and EPR$_B$) by performing a BSM on the other two photons (which we denote as XX$_A$ and XX$_B$) and a classically-controlled rotation of the state of photon X$_B$. This ``entanglement swapping'' operation~\cite{Pan1998} is equivalent to the teleportation of the state of one of the photons (e.g. X$_A$) of the EPR$_A$ pair to one of the photon of the other pair (``target photon'', XX$_B$), resulting in a new pair of entangled photons (XX$_A$ and XX$_B$).

We have already seen that QDs can be used as EPR sources using the polarization degree of freedom of the XX and X photons emitted by the biexciton-exciton cascade. Fidelities to the ideal $\ket{\Phi^+}_{\rm XX,X}$ Bell state of up to about 98\% have been reported for GaAs QDs~\cite{Huber2018}. For a system consisting of two qubits $\ket{X}_A$ and $\ket{X}_B$, performing a BSM means projecting the state into one of the four Bell states $\ket{\Psi^\pm_X}_{\rm A,B}$ and $\ket{\Phi^\pm_X}_{\rm A,B}$, which represent a complete set of basis states for the corresponding 4-dimensional Hilbert space. For {\em indistinguishable} photons, a partial BSM can be performed by HOM interference at a 50/50 beam splitter. For a non-polarizing beam splitter, the simultaneous detection of one photon at each of the two outputs of the beam-splitter projects  the X$_A$ and X$_B$ photons onto the $\ket{\Psi^-_X}_{\rm A,B}$ Bell state. The application of a $\sigma_y$ gate on the target photon XX$_B$ completes the teleportation of the state of the control photon X$_A$ onto XX$_B$.

The first successful implementation of quantum teleportation of the polarization state of a photon emitted by a QD onto the polarization of another photon emitted by the same QD was achieved by using an InGaAs QD in a light emitting diode under DC excitation as a source of entangled photon pairs with fidelity of 0.77~\cite{Nilsson2013}. In that experiment, crossed-polarizers were inserted at the output ports of the BSM beam splitter, thus reducing by 1/2 the detection probability of the $\ket{\Psi^-_X}_{\rm A,B}$ state (and thus to 1/8 the total efficiency of the Bell state measurement). Later on, a similar experiment was performed using GaAs QD producing a stream of entangled photons with fidelity of 0.92 under pulsed resonant excitation using the arrangement shown in Fig.~\ref{fig:teleportation}(a)~\cite{Reindl2018}. In both works, photons subsequently emitted by the same QD and were delayed to meet at the Bell state measurement beam-splitter were used for teleportation. Teleportation fidelities well above the classical limit were observed in Ref.~\cite{Reindl2018} for all chosen polarization states of the control X$_L$ photon, as shown in  Fig.~\ref{fig:teleportation}(b). More recently, improved BSM efficiency and noise rejection was reported in Ref.~\cite{Basset2021} using the experimental arrangement shown in Fig.~\ref{fig:teleportation}(c). By inserting two polarizing beam-splitters at the output of the 50/50 beam-splitter, also the $\ket{\Psi^+_X}_{\rm A,B}$ becomes detectable.

By performing a thorough analysis of the sources of imperfections in the above-mentioned experiments, it is concluded that the limited photon indistinguishability mostly affects the overall fidelity of the teleportation process. Although the TPI visibility for photons emitted by resonantly driven excitons or trions can exceed 98\% (see Section~\ref{sec:SPS_Performance_NIR}), the XX-X photon pair from a cascade is not only entangled in polarization but also in time, as reported first by Moreau et al.~\cite{Moreau2001a}, so that the purity and indistinguishability of the reduced one-photon system is not unity~\cite{Huber2013, Troiani2014, Schoell2020}. This correlation, together with noise sources inherent to the solid-state environment (interaction of excitons with charges, spins, and phonons) limits the fidelity of the BSM. 

In spite of the limited indistinguishability of X photons (TPI visibility of about 0.7), two experiments have been recently performed demonstrating entanglement swapping between two XX-X pairs emitted by a QD during two independent excitation cycles~\cite{BassoBasset2019,Zopf2019}. Figure~\ref{fig:teleportation}(d) shows the two-photon polarization density matrix for the system composed of the XX$_E$-XX$_L$ photons, which are expected to be in the $\Psi^-_{\rm XX,E,L}$, after a BSM projecting the X$_E$-X$_L$ pair in the $\Psi^-_{\rm X,E,L}$ state. Although the density matrix is close to the ideal one, mixing elements are observed, as neither spectral nor temporal filtering was applied to the data, making the swapping vulnerable to the limited fidelity of the BSM. The calculated effect of the non-perfect photon indistinguishability and entanglement fidelity (produced by a finite value of the excitonic fine structure splitting) is shown in Fig.~\ref{fig:teleportation}(e). 

A route to improve the performance of QDs for entanglement swapping experiments and relying on Purcell enhanced emission and spectral filtering to alleviate the negative effect of time-correlations in the cascaded decay is sketched in Ref.~\cite{Schimpf2021a}. However, excitation-induced effects inherent to the two-photon excitation method should be considered~\cite{Seidelmann2022,Basset2023}, as discussed in Sec.~\ref{sec:entangledpair}. 
Performing entanglement swapping with photons emitted by remote QDs brings in additional challenges due to uncorrelated charge noise and blinking~\cite{Joens2017}, further reducing the TPI visibility, and the necessity of tuning the QD emission energy while maintaining $E_{\rm FSS}=0$. The first issue can be solved by embedding the QDs in charge-tunable devices~\cite{Zhai2020a, Schimpf2021c,Zhai2022} and the second by integrating the QD devices on top of multiaxial strain actuators~\cite{Trotta2015,Trotta2016,Huber2018,Lettner2021}. 

%\subsection{Quantum teleportation}

\subsection{Boson sampling}

\begin{figure}[b]
\centering\includegraphics[width= 13.35cm]{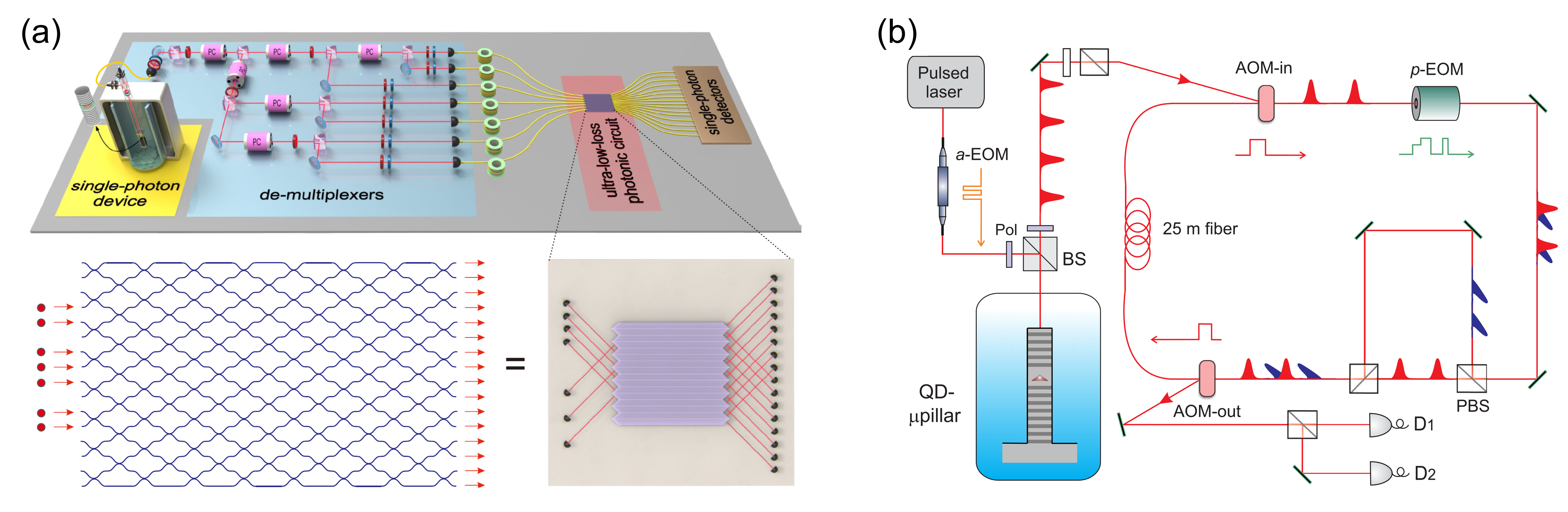}
\caption{Schematics of experimental setup for boson sampling. (a) A single InAs/GaAs QD SPS combined with a time-to-space demultiplexer consisting of Pockels cells and polarizing beam splitters. The seven single photons are then coupled with an optical fiber and fed into a 16 × 16 modes interferometer on an ultra-low-loss photonic circuit. A five-photon boson sampling rate of 4 Hz was demonstrated. (b) Time-bin-encoded boson sampling consists of a single QD-micropillar device,  two detectors, and a fiber loop-based interferometer. Three- and four-photon boson sampling rates of 18.8 and 0.2 Hz were reported. (a) Reprinted from \href{https://link.aps.org/doi/10.1103/PhysRevLett.120.230502}{\textit{Wang et al. 2018}} \cite{Wang2018}. Copyright (2018) by the American Physical Society. (b) Reprinted from \href{https://link.aps.org/doi/10.1103/PhysRevLett.118.190501}{\textit{He et al. 2017}} \cite{He2017}. Copyright (2018) by the American Physical Society.}
\label{Fig8_boson sampling}
\end{figure}

Recent advances in photonic quantum information technologies in coherent controls, scales, and integration promise practical applications of quantum states in not only communication but also in computation and simulation. While fault-tolerant universal quantum computations remain a long-term challenge, quantum machines can solve specific problems much faster than classical computers with a limited number of quantum resources and efficiency at present. Boson sampling is a well-known such computational task formulated by Aaronson and Arkhipov \cite{Aaronson2011}, and it models the probability distribution of scattered indistinguishable photons in a linear interferometer. Its usefulness has been tested in simulating molecular vibronic spectra \cite{Huh2015}. Photons and linear optics are well suited to accomplishing scalable boson devices. The implementation of boson sampling with photons highly relies on a high generation rate and indistinguishability of single photons, as well as the low multi-photon probability. Heralded single photons and squeezed states in nonlinear crystals are widely used for implementing different types of boson samplings, such as scattershot and Gaussian boson sampling \cite{Brod2019}. However, they still suffer from several difficulties in efficiency and scaling up. 

Nowadays, high-performance QD SPSs can generate trains of identical single photons at a high rate over a few tens of MHz and with high single-photon purity ($g^{(2)}(0)<0.01$) and indistinguishability (>0.99) \cite{Somaschi2016}. As the \textit{N}-photon input probability decreases exponentially with the single-photon generation rate and multi-photons are the major source of error, QDs are considered a good alternative to heralded SPSs~\cite{ Wang2018,Loredo2017}. From on-chip waveguide integrated QDs, a single-photon generation rate of 122 MHz has been demonstrated while maintaining the indistinguishability of more than 100 single photons. By combining an active time-to-space demultiplexing technique with such indistinguishable single-photon trains, QD-SPSs simulated boson sampling with fivefold coincidence detection on a 16 X 16 modes ultralow-loss photonic circuit (See Fig.~\ref{Fig8_boson sampling}(a)) \cite{Wang2018}. A different strategy that used temporal modes instead of spatial modes via a loop-based interferometer has implemented time-bin-encoded boson sampling \cite{He2017}. This approach significantly reduces experimental overhead, as it only requires a bright SPSs and two detectors with fiber delays (see Fig.~\ref{Fig8_boson sampling}(b)).

\subsection{Photonic quantum computing}
\begin{figure}[b]
\centering\includegraphics[width= 13.35cm]{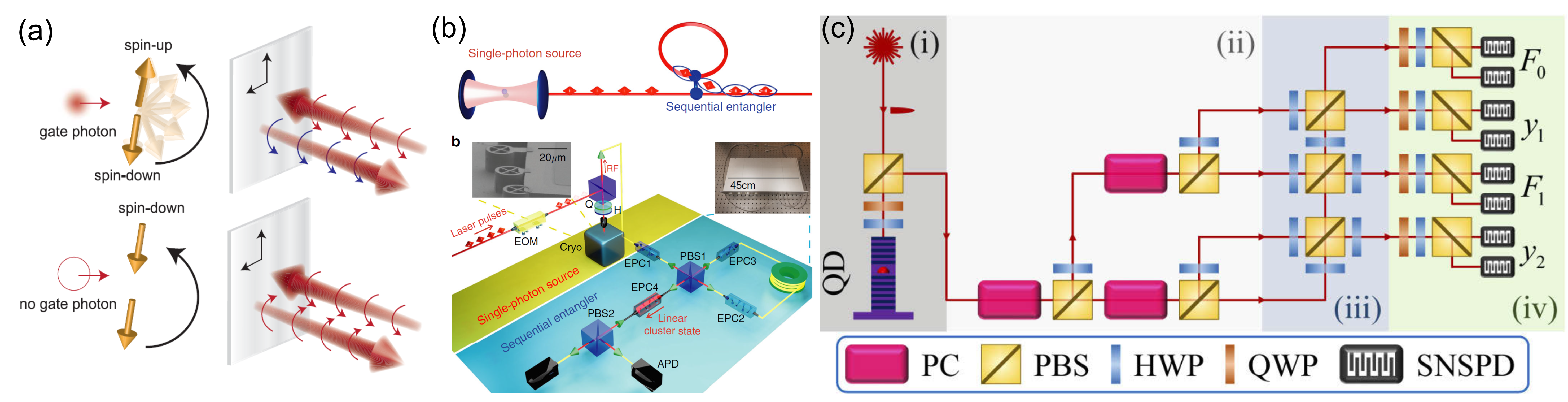}
\caption{Schematics of deterministic photon-photon interactions, compiled Shor's algorithm, and linear cluster states based on cavity-coupled single QDs. (a) A single charged QD in a photonic crystal cavity creates spin-photon interfaces. The first gate photon controls the spin state of a QD and then determines the polarization of incoming photons.  (b) A single QD in a micropillar produces indistinguishable single-photon trains. An entanglement gate consisting of a fiber delay loop, a polarizing beam splitter, and an electrically driven polarization controller creates linear cluster states encoded in the polarization degree of freedom. A delay loop stores a photon until the next photon comes in. (c) Experimental setup for Shor's algorithm, consisting of an SPS with active demultiplexer, a quantum circuit, and four-fold correlation measurement. (a) Reprinted from \cite{Sun2018} with permission from AAAS. (b) Reproduced from Ref.~\cite{Istrati2020} under Creative Commons CC BY license. (c) Adapted with permission from Ref.~\cite{Duan2020}. \textcopyright~The Optical Society.}
\label{Fig8_photonic computing}
\end{figure}

Schemes for universal quantum computations using photons were proposed in 2001 based on linear optics and measurements, known as linear optical quantum computing \cite{Knill2001} and measurement-based quantum computing \cite{Raussendorf2001}. The schemes are based on quantum interference and measurement-induced nonlinearities without direct photon-photon interactions. Most demonstrations of quantum gates and quantum algorithms typically employ photons from parametric down-conversion processes \cite{Peruzzo2014, Madsen2022}, but as QD-SPSs are beginning to meet all the important criteria (see section~\ref{sec:intro}) for ideal SPSs and outperform the existing SPSs, QDs are considered an excellent candidate for implementing photonic quantum computation. 

Furthermore, a variety of exciton complexes in QDs, including single excitons, biexcitons, and charged excitons, produce coherent single photons, entangled photon pairs, and spin-photon entanglements. As stationary spins provide local storage of quantum states \cite{Boyer2011,Huthmacher2018},  creating highly efficient QD-photon interfaces play a key role in implementing scalable photonic quantum computing architectures \cite{Uppu2021} and distributed quantum networks \cite{Cirac1999}.  

Quantum logic gates and Bell-state analyzers are basic units of quantum computations and an important prerequisite for implementing several quantum protocols such as teleportation or entanglement swapping. In measurement-based quantum optics experiments, such operations are inherently probabilistic, while introducing cavity (waveguide) QEDs with two-level atoms efficiently mediates nonlinear light-matter interactions and enables deterministic quantum operations and quantum non-demolition measurements. The cavity (waveguide)-coupled quantum emitters have been intensely studied to realize scalable quantum architecture based on deterministic quantum gates and Bell-state analysis \cite{Duan2004, Bonato2010,Cao2019}. Such schemes were experimentally demonstrated in a number of groups\cite{Sun2016, Fushman2008, Ralph2015, Javadi2015, Sun2018}, 
Fig.~\ref{Fig8_photonic computing}(a) shows deterministic photon-photon interaction mediated by QD's spin in Ref.~\cite{Sun2018}. The  result can implement single-photon transistors, which are inapplicable in conventional quantum optics.  

The capability of producing bright on-demand single-photon streams, whose polarizations are entangled with the spin state of QDs, are beneficial to create photonic cluster states, which are essential for one-way quantum computing \cite{Raussendorf2001,Hu2008,Tiurev2022}. A four-photon linear cluster state has also been demonstrated using a single quantum emitter and a single sequential entangler based on a fiber delay loop \cite{Istrati2020} (See Fig.~\ref{Fig8_photonic computing}(b)). As another approach, a proof-of-principle demonstration of Shor’s algorithm has been accomplished with a QD SPS and active multiplexers, factoring 15. Here, deterministically generated single photons with high extraction efficiency, single-photon purity, and indistinguishability are transferred into a four-photon cluster states and implement inverse quantum Fourier transform \cite{Duan2020} (See Fig.~\ref{Fig8_photonic computing}(c)). 

For a full-scale demonstration of quantum processing, a larger number of qubits, gates, and entangled states will be necessary. Furthermore, auxiliary qubits are also required to store the intermediate results, while electron or hole spins in QDs have a rather short coherence time of a few microseconds \cite{Huthmacher2018} compared to long coherence times of more than 1 second in other qubit platforms, such as atoms, ions, and color centers in crystals. This constraint could be addressed by constructing hybrid architectures with such dissimilar qubit sources. In particular, there exist a number of possible candidates, including rubidium (cesium) vapor cells \cite{Vural2018, Huang2017} and Nd$^{3+}$ doped crystals \cite{Tang2015}, which can optically interface with InGaAs or GaAs QDs. By matching their spectral frequencies, a single photon in a QD can be delayed or stored through strongly interacting hybrid quantum memories. Such hybrid architectures will provide a route for distributing entanglement and ultimately realizing distributed quantum computing. This enables us access to more advanced quantum protocols and utilizes more quantum resources from separate quantum modules.

\section{Open challenges and outlook}\label{sec:outlook} 

Based on the results presented, this section discusses open challenges and provides an outlook to future directions and required development. 

\subsection{Theory and numerical device modelling}

With most QLS design approaches, numerical simulations typically predict higher figures of merit than what is measured experimentally. While such deviations are often attributed to uncontrolled fabrication imperfection such as surface roughness or spatial QD-cavity misalignment, a careful validation of the theoretical predictions using accurate measurements and well-controlled deterministic fabrication techniques is a prerequisite to improving the device performance. A further practical challenge is the numerical difficulty in performing accurate simulations \cite{DeLasson2018} of large device geometries as well as structures including geometrical features on several length scales. 

Finally, even assuming loss-less materials and perfect nanofabrication capabilities, it is still not fully understood how to increase the product $\eta_{\rm ext} V_{\rm TPI}$ of the efficiency and the indistinguishability arbitrarily close to unity. The governing physics describing the light extraction for several QLS designs including the CBG resonator concept \cite{Sapienza2015, Liu2019,Wang2019,Yao2018} is not fully understood, posing a challenge for the device optimization. Furthermore, for the designs relying on Purcell enhancement, a fundamental trade-off persists between the achievable efficiency and indistinguishability in the presence of phonon-induced decoherence \cite{Iles-Smith2017a, Wang2020b}.

\subsection{Epitaxial growth}
The basis of any QD device is represented by epitaxially-grown heterostructures. Different material combinations and growth methods have been explored in the last three decades, resulting in QDs with steadily improving  performance. While strained InGaAs QDs obtained via the S-K growth mode on GaAs and with emission wavelength around 900~nm have been instrumental to many pioneering works~\cite{Michler2000,Akopian2006,Schwartz2016,Stockill2016,Senellart2017,Lodahl2015} and are now even commercially available, improved performance in terms of indistinguishability among photons emitted by separate QDs~\cite{Zhai2022}, polarization-entanglement fidelity~\cite{Huber2018}, and electron spin coherence~\cite{Zaporski2023} have all been achieved on almost unstrained GaAs QDs in nanoholes obtained by the local droplet etching in AlGaAs~\cite{Gurioli2019,DaSilva2021}. 

Significant deviations from the ideal figures of merit are however still observed, especial in the case of telecom-wavelength QDs. In part, these can be attributed to extrinsic effects, such as impurities, point- and extended defects, as well as surface and interface traps. Understanding the impact of these factors on the properties of QDs is a formidable challenge. A pragmatic solution to achieve the highest possible material quality should focus on the careful selection of the source materials, of the methods for conditioning epitaxial-growth-systems, and of the growth parameters, similar to what is done for the fabrication of ultrahigh mobility electron- and hole-gases. Native and processed surfaces are also a source of noise, and many different passivation methods have been developed over the years for the commonly used compound semiconductors~\cite{Hasegawa2010,Laukkanen2021,Guha2017} and in part used also for QD structures (see also next section).

There are also intrinsic factors, which should be considered and further understood, such as the role of QD structural properties (QD size, shape, strain etc.), alloy and interface disorder, heterostructure design~\cite{Houel2012}, interaction with phonons in the used materials, and the nuclear spins of the contained atoms. In this respect, a close interaction between experiment and theory will be further required to engineer the QD properties to meet the increasingly stringent requirements posed by advanced experiments and applications. As hyperfine interaction limits the performance of commonly used compound semiconductors, it would be useful to ``re-discover'' nuclear spin-free materials such as II-VI semiconductors for which, however, point defects may be the limiting factor.  

\subsection{Device nanofabrication}

The fabrication methods developed in recent years for the deterministic integration of QDs into photonic nanostructures complement established techniques such as reactive ion etching in a targeted manner. Today, they allow one to produce high-performance quantum devices with high process yield and high control of their electro-optical properties. With regard to the practical application of these devices, great progress was made concerning user-friendly on-chip fiber coupling, which made it possible to develop plug'n'play QD-SPSs and use them for QKD experiments. Despite these enormous technological advances, there are still open challenges and a need for optimization for the nanoprocessing of QD quantum devices. In addition to further optimizations in the area of QD growth mentioned above, and above all in relation to QDs with emission in the telecom O-band and C-band, further efforts are needed in the area of nanostructuring and device fabrication, for example to maximize photon indistinguishability, to enable scalability and to enhance the properties and capabilities of fiber-coupled QLSs.

\textbf{Photon indistinguishability} In this context, charge fluctuations in the vicinity of the QD are problematic, leading to spectral diffusion and thereby limiting photon indistinguishability~\cite{Vural2020}. Externally applied electric fields can counteract this problem, but this complicates device design and fabrication and is sometimes impractical. Since spectral diffusion is often induced by the charging and discharging of defect states and etched surfaces, it will be important to effectively passivate component surfaces in the future. Promising results were published in Ref.\cite{Liu2018}, where it could be shown that surfaces coated with 15 nm of Al$_2$O$_3$ using atomic layer deposition (ALD) lead to a significantly reduced blinking and spectral linewidth of QDs. ALD surface passivation using 8 nm layer of Al$_2$O$_3$ was also carried out in a QD device with open cavity design to achieve a close-to-ideal photon indistinguishability of 96.7\%~\cite{Tomm2021}.

\textbf{Scalability} Another open point concerns the scalability of single-QD devices to large-scale quantum networks and complex IQPCs. These advanced applications require a large number of QD-QLSs or QDs as single-photon emitters with identical emission wavelengths on the scale of homogeneous linewidth in the $\mu$eV range. In this context, self-organized QD growth is fundamentally problematic, since the position and spectral properties of individual QDs cannot be predetermined. In fact, when using conventional nanotechnology methods, the process yield of resonant QD devices would be drastically reduced and would take on negligibly small values as soon as one considers scaling across systems with more than one QD~\cite{Schnauber2018}(supplementary information). However, the problem of the random position can be efficiently solved by the presented deterministic nanofabrication processes in order to create single-QD devices in a very controlled manner. With simultaneous spectral selection of QDs, also devices of the same emission energy on a scale of 0.1 – 1 meV can be produced. For the necessary resonance in the area of the homogeneous linewidth, however, spectral fine-tuning is inevitable. For this purpose, spectral control via the quantum confined Stark effect and via local strain tuning are very attractive, which has already been demonstrated in experiments on individual QDs~\cite{Bennett2010,Seidl2006,Ding2010}, and more recently also for two-QD systems fabricated randomly~\cite{Petruzzella2018,Ellis2018} and deterministically~\cite{Schnauber2021}. In the future, it will be interesting and important to establish this kind of spectral fine-tuning also in a scalable way in IQPCs based on many resonant QDs and in quantum repeater networks based on BSMs on indistinguishable photons emitted by remote SPSs. For a discussion, see section~\ref{sec:teleportation_swapping}. In the case of the quantum networks, it will also be important to have an absolute wavelength reference for the spectral synchronization of the individual QLSs, for which atomic transitions could come into question, for example.

\textbf{Advanced fiber-coupling} The development of efficient solutions for on-chip fiber coupling is an important basis for real applications of single-QD devices in photonic quantum technology. In the future, it will be essential to increase the coupling efficiency. Furthermore, it will be interesting to develop advanced fiber-coupling schemes that go beyond comparatively simple single-emitter single-mode connections described above. For instance, sources of entangled photon pairs play a central role in quantum repeater networks. In this context, it is important to develop solutions through which the polarization-entangled XX and X photons of a QD can be coupled into an optical fiber while maintaining the entanglement, and which ensures that the entangled photons are transmitted to two different fiber outputs. For this purpose, the specialty fiber would have to contain wavelength-selective elements that direct XX and X photons into different fiber outputs. Another interesting approach could be the coupling of emitter arrays and multicore fibers. In this way, quantum keys could be transmitted in parallel in several fiber cores, so that the achievable QKD transmission rate would be multiplied accordingly. One could also imagine using this concept to transmit quantum channels and classical channels simultaneously with little crosstalk via different cores of the multicore fiber.

\subsection{Practical applications in quantum information}
As discussed in this review article, in recent years, QD-based QLSs clearly proved their high potential for applications in quantum information technology. Allowing for the efficient generation of single, indistinguishable, and entangled photons with excellent quantum optical properties and at high rates, QD-sources are today able to outperform probabilistic sources. While numerous proof-of-principle experiments have been already demonstrated employing QD-sources, e.g. in implementations of QKD, quantum teleportation, or entanglement swapping, major challenges remain for their practical applications in quantum networks. In this case, it is no longer sufficient to evaluate the QD source as an isolated device, rather than integrated in functional systems.

One challenge in this context concerns the efficient coupling of flying qubits not only to the "first lens" and in the controlled environment of a quantum-optical laboratory, but also to the quantum channel such as a deployed optical fiber, typically including additional losses for quantum state preparation, and in realistic environments. The direct and permanent fiber-pigtailing of carefully optimized QD-devices in combination with compact cryocoolers (cf. Sections~\ref{sec:Nanofab_FC} and \ref{sec:application_SPS_QKD}) offer promising routes to master this challenge. Experimental realizations employing this approach for stand-alone QD-sources, however, did not reach the performance level of lab-scale experiments. Thus, a crucial next step will be, to show that practical QD-devices can indeed also exploit the full potential QDs offer in terms of photon extraction efficiency, single-photon purity and photon-indistinguishability.

Another major challenge concerns the achievement of high photon indistinguishability from remote solid-state quantum emitters - a crucial prerequisite for advanced schemes of quantum communication. As discussed in Section~\ref{sec:applications_advanced_QKD}, the high fabrication quality and the degree of control possible with QD QLSs today, led to substantial advances in TPI experiments with remote sources, which resulted in numerous experiments exceeding the 50\%-limit set for the TPI visibility of sources exhibiting Poissonian statistics. To increase the reproducibility of high photon indistinguishabilities from multiple remote QLSs, also in practical scenarios outside shielded laboratories, will lead to breakthroughs in advanced implementations of quantum communication, ranging from MDI-/DI-QKD implementations with unprecedented performance, entanglement swapping and quantum teleportation of flying qubits from spatially distant sources (if combined with high entanglement fidelity), to long-haul quantum repeater links.

Furthermore, on the very applied side, to benchmark different quantum communication protocols and different technology platforms, standards must be developed or agreed upon \cite{Langer2009,alleaume2014}. Solutions can be to certify the same amount of overall $\epsilon$-security (see the discussion of security definitions in \cite{Renner2008}) or newly defined figure-of-merits such as “security-per-dollar-spent”, which considers the fact that different quantum communication architectures, that in principle promise different levels of security, also have different levels of implementation difficulty and hence costs. And, last but not least, while the achievement of unconditional security, ruling out even the most unlikely attacks (that are practically impossible but allowed by the laws of quantum mechanics), remains the ultimate aim, one might be content with a more relaxed, applied form of security in practice - whether being the result of a deliberate trade-off between security-gain and implementation-costs, or an intermediate step towards ultimate security. Noteworthy, the approach of assuming realistic restrictions on an adversary are well known and even required in the field of cryptographic primitives beyond QKD in untrusted settings, e.g. quantum oblivious transfer in the so-called noisy storage model \cite{Wehner2008}, representing other crucial building blocks for modern communication \cite{Broadbent2016} networks.

Overall, the understanding of QDs as a high-quality quantum emitters and the development of related quantum devices has reached a very high level of knowledge in science and technology, which is reflected for instance in the near-ideal emission characteristics of QD-QLSs. In addition to further optimization of the QDs and corresponding nanophotonic structures towards quantum devices with properties that come even closer to the ideal values, practical aspects in particular will play an important role in further optimizations. These certainly include the development of compact fiber-coupled QD-QLSs and a scalable QD technology for the implementation of complex quantum networks and highly functional IQPCs. In this context, the focus of the work will certainly shift away from basic research to applied research in the field of quantum engineering. This offers very attractive development opportunities in a multidisciplinary research environment that synergistically combines topics from basic physical research, nanophotonics, quantum optics, integrated photonics, network technology and quantum information science to realize highly innovative components for applications in quantum information technology.
%- Mention importance to explore different excitation methods depending on application

%\emph{This section summarizes open challenges and questions and gives an outlook on future directions.}

\section{Conclusion}~\label{sec:conclusion}
  
In summary, this article has discussed the great potential of semiconductor QDs for applications in quantum information technology. Due to their discrete energy levels, these high-quality quantum emitters form almost ideal two-, three- and four-level systems via which individual photons, entangled photon pairs and also photonic cluster states can be generated on demand. In addition, charged QDs can act as coherent spin-photon interfaces. 

From a technological point of view, these quantum emitters are very attractive because they are fundamentally compatible with established manufacturing techniques such as semiconductor epitaxy. However, as we have discussed, related techniques need to be optimized for the particular requirements relevant to the fabrication of QD-based quantum devices. For example, highly symmetrical QDs can be produced epitaxially using the droplet-etching technique, which are predestined for the generation of polarization-entangled photon pairs. Furthermore, the telecom O- and C-band can be reached via sophisticated material engineering by QDs in order to enable fiber-based quantum communication. 

Corresponding developments require special theoretical concepts and numerical methods that represent a link between device design, growth, nanofabrication and optical characterization and form an important basis for device optimization for the targeted applications in quantum information technology. Building on this, we presented modern manufacturing concepts that can be used to fabricate single-QD devices with the highest process control in a deterministic and scalable manner. These concepts include, above all, in situ lithography techniques, in which suitable QDs are first selected in order to then integrate them spatially and spectrally aligned into nanophotonic structures.

Nanophotonic structures and resonators serve to increase the photon extraction efficiency of the QDs to values beyond 70\%. On the other hand, they can exploit the Purcell effect to optimize emission dynamics and quantum optical properties, such as indistinguishability. In this way, almost ideal QD-QLSs with high multi-photon suppression, indistinguishability and entanglement fidelity in combination with high single-photon emission rates in a wide wavelength range from about 780 nm to 1550 nm have been developed in recent years. In addition, important progress towards user-friendly QLSs has been made by coupling the QD devices on-chip with optical fibers and integrating them into stand-alone cryostats for direct integration into QKD testbeds and perspectively into fiber-based quantum networks. In fact, the first QKD experiments with QD-QLSs are very promising, for example in terms of the achievable data transmission rates, while it is becoming apparent that the full potential of QD-QLS will only become apparent in advanced quantum communication networks, which are based on entanglement distribution, spin-photon entanglement and quantum state transfer.

In addition to quantum communication applications, the field of photonic quantum processors and quantum computers has generated great interest. As we discussed, QDs can also form important building blocks here, for example by being scalably integrated into integrated quantum photonic circuits. These can have a hybrid architecture in order to contain highly functional single photon detectors and elements such as ring resonators for qubit manipulation in addition to the quantum emitters themselves. In this context, 2D photonic cluster states, which promise efficient computing operations via one-way quantum computing, play a special role. While it has already been possible to generate 1D photonic cluster states, it is a great challenge to generate higher-dimensional photonic cluster states, which could be done efficiently with quantum dot molecules.

In conclusion, this article gave an insight into the fascinating advances of QD research activities in the field of photonic quantum technologies. These include a wide range of theoretical questions, technical solutions, and experimental methods, which in a very interdisciplinary environment in cooperation with experts from quantum optics and quantum information science form the basis for the application of the corresponding QD devices in quantum communication and the photonic quantum computing. The enormous dynamics in the development of QD-based quantum devices and their excellent performance parameters lead to the conclusion that semiconductor QDs will play an important role in the implementation of complex quantum networks and the future quantum internet.

%\section{Backmatter}

%Backmatter sections should be listed in the order Funding/Acknowledgment/Disclosures/Data Availability Statement/Supplemental Document section. An example of backmatter with each of these sections included is shown below.

\begin{backmatter}
\bmsection{Funding}
Content in the funding section will be generated entirely from details submitted to Prism. 

\bmsection{Acknowledgments}
T.H. acknowledges fruitful discussions with Daniel A. Vajner. N.G. acknowledges fruitful discussions with Luca Vannucci and Dara P.\ S.\ McCutcheon.

\bmsection{Disclosures}
The authors declare no conflicts of interest.

\bmsection{Data availability} No data were generated or analyzed in the presented research.

\bmsection{Abbreviations}

\begin{itemize}
    \item Bell-state measurements (BSMs)
    \item biexciton (XX)
    \item cathodoluminescnce (CL)
    \item Clauser, Horne, Shimony, and Holt (CHSH)
    \item circular Bragg grating (CBG)
    \item device-independent (DI) QKD
    \item Einstein-Podolski-Rosen (EPR)
    \item electron beam lithography (EBL)
    \item exciton (X)
    \item Frank-van-der Merwe (F-M)
    \item free-space optical (FSO)
    \item Greenberger–Horne–Zeilinger (GHZ)
    \item Hong-Ou-Mandel (HOM)
    \item integrated quantum photonic circuit (IQPC)
    \item light emitting diode (LED)
    \item measurement-device-independent QKD (MDI-QKD)
    \item metal-organic-vapor-phase-epitaxy (MOVPE)
    \item molecular beam epitaxy (MBE)
    \item numerical aperture (NA)
    \item photonic crystal (PC)
    \item quantum bit error ratio (QBER)
    \item quantum dot (QD)
    \item quantum key distribution (QKD)
    \item receiver (Bob)
    \item sender (Alice) 
    \item two-photon excitation (TPE)
    \item two-photon interference (TPI)
    \item cavity quantum electrodynamics (cQED)
    \item distributed Bragg reflector (DBR)
    \item scanning tunneling microscopy (STM)
    \item Volmer-Weber (V-W)
    \item scanning electron microscope (SEM)
    \item single-photon source (SPS)
    \item Stranski-Krastanow (S-K)
    \item single-mode fiber (SMF)
\end{itemize}

\end{backmatter}

%\section{Conclusion}
%Conclusion.

%%%%%%%%%%%%%%%%%%%%%%% References %%%%%%%%%%%%%%%%%%%%%%%%%

%Add references with BibTeX or manually.
%\cite{Zhang:14,OPTICA,FORSTER2007,Dean2006,testthesis,Yelin:03,Masajada:13,codeexample}

%%%%%%%%%% If using BibTeX:
%\bibliographystyle{unsrt}
%\bibliography{AOP_bibliography}

%%%%%%%%%% If preparing manually:
% \newcommand{\enquote}[1]{``#1''}

% \bibitem{Zhang:14}
% Y.~Zhang, S.~Qiao, L.~Sun, Q.~W. Shi, W.~Huang, L.~Li, and Z.~Yang,
%   \enquote{Photoinduced active terahertz metamaterials with nanostructured
%   vanadium dioxide film deposited by sol-gel method,}
%   {\protect\JournalTitle{Optics Express}} \textbf{22}, 11070--11078 (2014).

% \bibitem{Optica}
% {Optica}, \enquote{{Optica Publishing Group},}
%   \url{http://www.opg.optica.org}.

% \bibitem{FORSTER2007}
% P.~Forster, V.~Ramaswamy, P.~Artaxo, T.~Bernsten, R.~Betts, D.~Fahey,
%   J.~Haywood, J.~Lean, D.~Lowe, G.~Myhre, J.~Nganga, R.~Prinn, G.~Raga,
%   M.~Schulz, and R.~V. Dorland, \enquote{Changes in atmospheric consituents and
%   in radiative forcing,} in \enquote{Climate Change 2007: The Physical Science
%   Basis. Contribution of Working Group 1 to the Fourth assesment report of
%   Intergovernmental Panel on Climate Change,}  S.~Solomon, D.~Qin, M.~Manning,
%   Z.~Chen, M.~Marquis, K.~B. Averyt, M.~Tignor, and H.~L. Miler, eds.
%   (Cambridge University Press, 2007).

% \end{thebibliography}

%Tobias: please add your bio here\\
\begin{wrapfigure}{l}{25mm} 
	\includegraphics[width=1in,height=1.25in,clip,keepaspectratio]{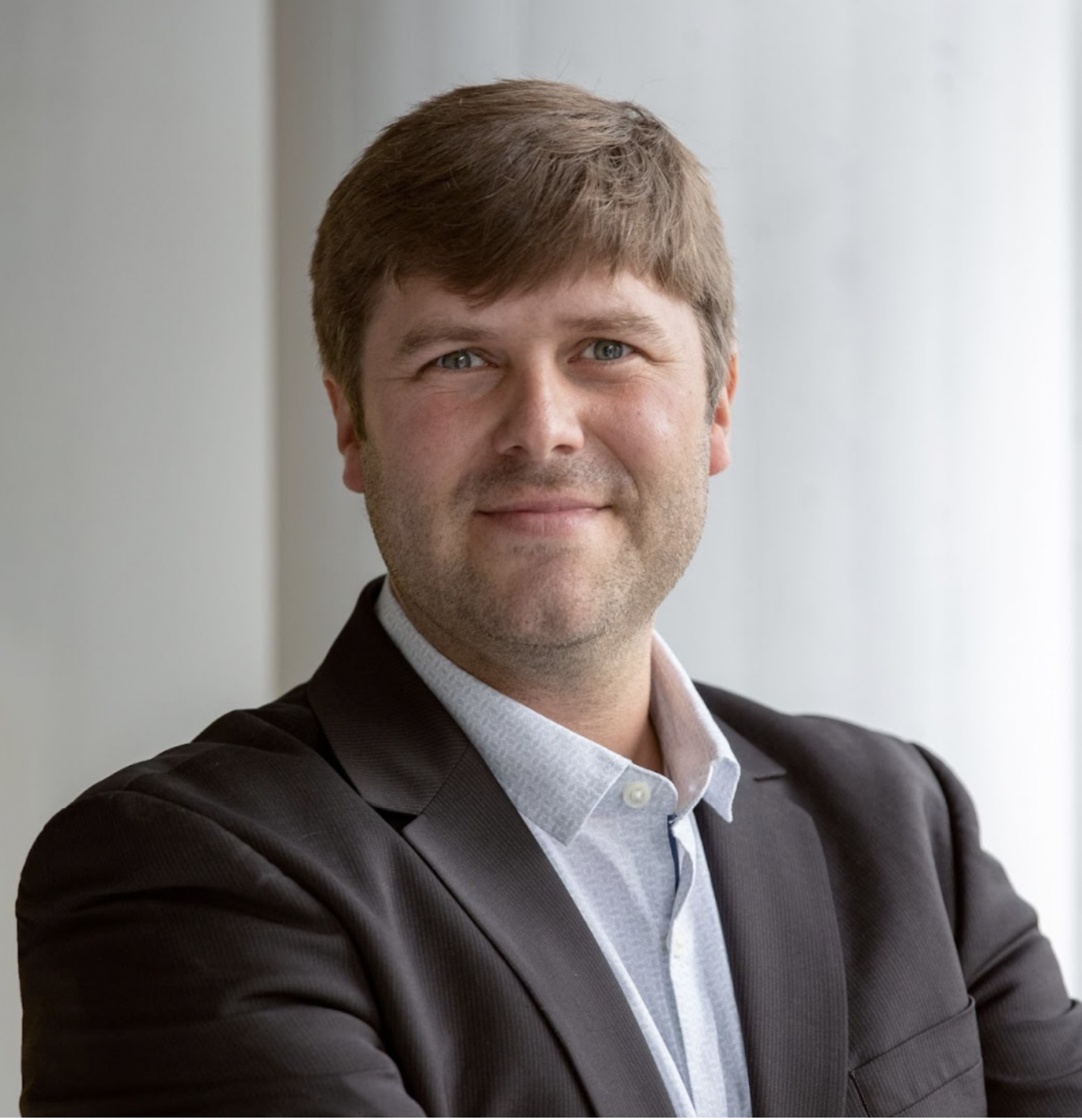}
\end{wrapfigure}\par
\textbf{T. Heindel} received his PhD degree in Physics at the Julius-Maximilians-Universität Würzburg, Germany in 2014. Afterwards he joined Technische Universität (TU) Berlin for postdoctoral studies with multiple research stays at the Technion Haifa, Israel, and Ludwigs-Maximilians-Universität Munich, Germany. In 2018, he founded the independent junior research group Quantum Communication Systems at the Institute of Solid State Physics of TU Berlin. His current research focuses on implementations of quantum communication and quantum information science using engineered solid-state quantum light sources and building blocks thereof. T. Heindel is member of the German Physical Society.\\ \par

%Jehyung: please add your bio here\\%
\begin{wrapfigure}{l}{25mm} 
	\includegraphics[width=1in,height=1.25in,clip,keepaspectratio]{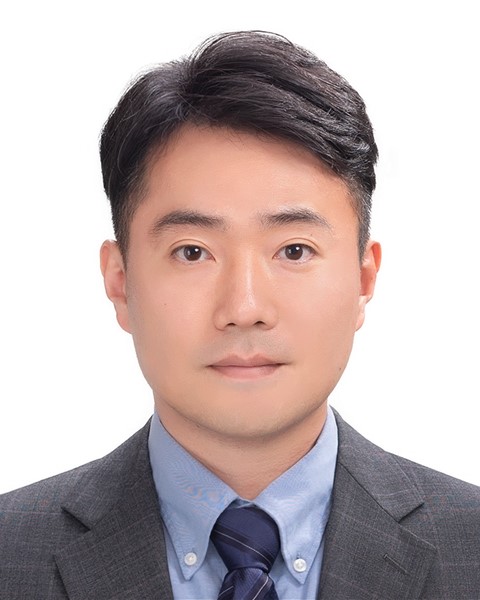}
\end{wrapfigure}\par
\textbf{J.-H. Kim} received his PhD degree in Physics at the Korea Advanced Institute of Science and Technology (KAIST), South Korea in 2014. He was a postdoc researcher at the University of Maryland from 2014 to 2017. Since 2017, he has joined the Department of Physics at the Ulsan National Institute of Science and Technology (UNIST), South Korea and  now is an associate professor at the UNIST. Major research topics of his group range are fundamental studies of quantum light-matter interactions, implantation for solid-state quantum devices, and their applications to quantum information science.\\ \par

%Niels: please add your bio here\\
\begin{wrapfigure}{l}{25mm} 	\includegraphics[width=1in,height=1.25in,clip,keepaspectratio]{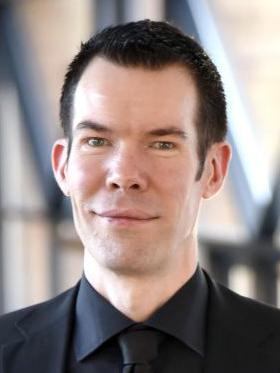}
\end{wrapfigure}\par
\textbf{N. Gregersen} received his PhD degree from the Technical University of Denmark in 2007 and has since then been with DTU Electro, Department of Electrical and Photonics Engineering, at the Technical University of Denmark. In 2021 he became group leader of the "Quantum Light Sources" group, and in 2022 he became a full Professor. His research interests are within the theory and modeling of light-matter interaction in semiconductor micro- and nano-structures, in particular QD-based quantum light sources. N. Gregersen is a Senior Member of Optica.\\ \par

%Armando: please add your bio here\\
\begin{wrapfigure}{l}{25mm} 	\includegraphics[width=1in,height=1.25in,clip,keepaspectratio]{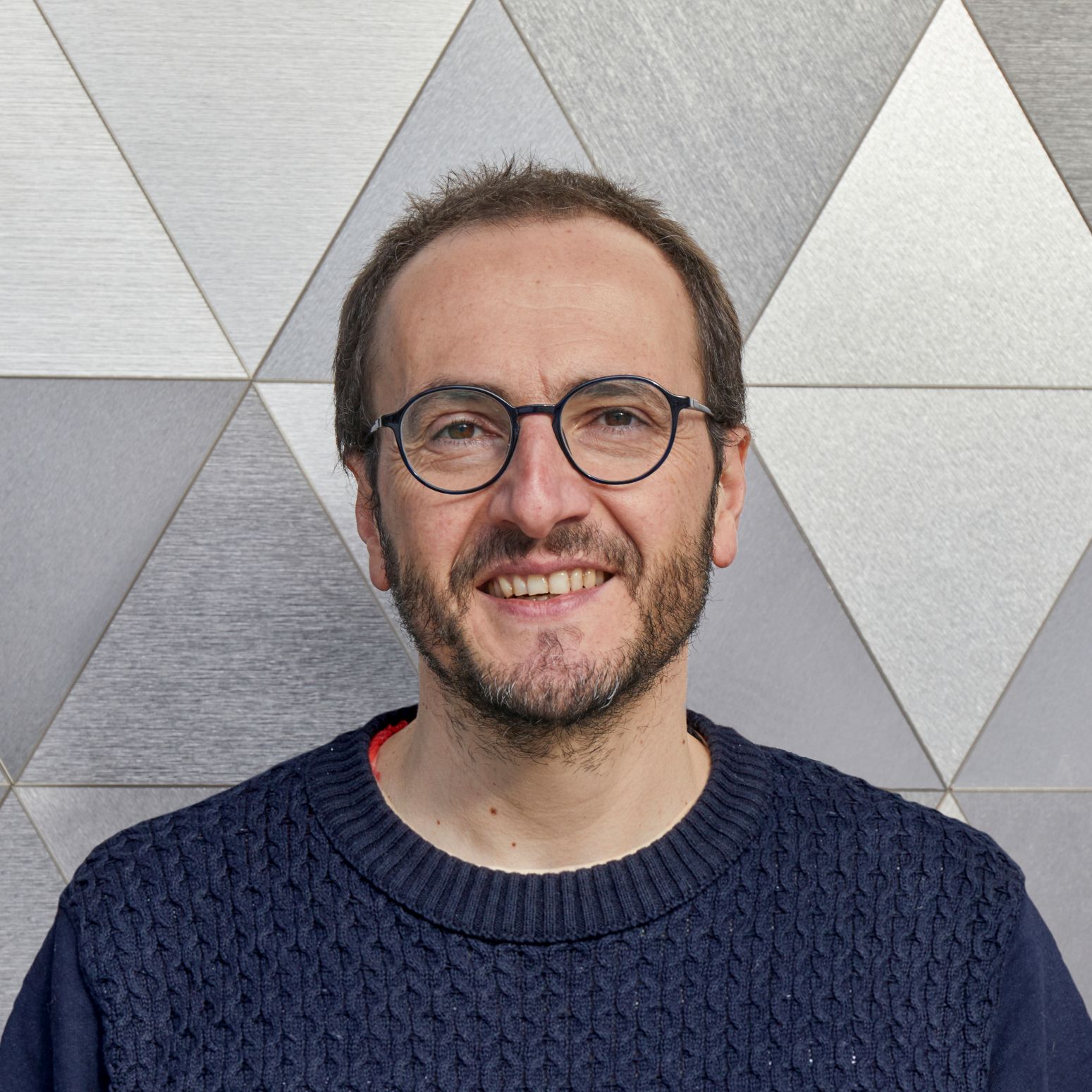}
\end{wrapfigure}\par
\textbf{A. Rastelli} received his PhD degree from the University of Pavia, Italy, in 2003. During his PhD he was research assistant at the ETH Z\"urich, Switzerland  and Marie-Curie-Fellow at the Technical University of Tampere, Finland. From 2003 to 2007 he was first PostDoc and then group leader at the Max-Planck-Institute of Stuttgart, and, till 2012, at the Leibniz Institute of Dresden, Germany. Since 2012 A. Rastelli is Professor of Semiconductor Physics and heads the Semiconductor Physics at the Johannes Kepler University of Linz, Austria. His research focuses on the development and optimization of methods to obtain, study, and control epitaxial quantum dots for quantum science and technologies. A. Rastelli is corresponding member of the Austrian Academy of Sciences and member of the German and Austrian Physical Societies.\\ \par

\begin{wrapfigure}{l}{25mm} 
    \includegraphics[width=1in,height=1.25in,clip,keepaspectratio]{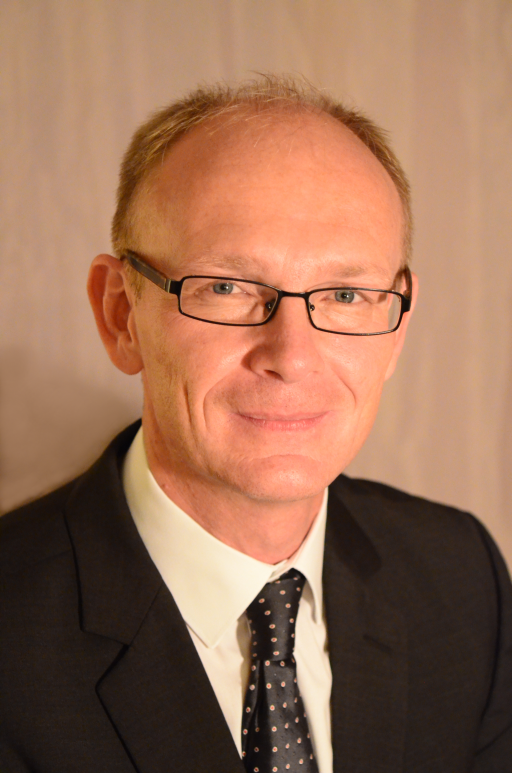}
  \end{wrapfigure}\par
  \textbf{S. Reitzenstein} received his PhD degree in physics (summa cum laude) from the University of Würzburg, Germany, in 2005. In 2010 he habilitated with studies regarding optical properties of low dimensional semiconductor systems. Since September 2011 he has been a full Professor at the Technical University of Berlin, Germany, holds the Chair of Optoelectronics and Quantum Devices and is director of the Center of Nanophotonics at the TU Berlin. His current research interests are in the area of quantum nanophotonics. They include the development of quantum light sources for applications in photonic quantum technologies. S. Reitzenstein is member of the German Physical Society and of Optica.\par

\end{document}